\documentclass[10pt,journal,twoside]{IEEEtran}
\normalsize
\usepackage[svgnames]{xcolor}
\newif\ifonecolumn
\onecolumnfalse
\usepackage{amsmath,amssymb}%
\usepackage{bbold}
\usepackage{float}
\usepackage{color}
\usepackage{csquotes}
\usepackage{hyperref}
\usepackage{graphicx}
\usepackage{amsmath}
\usepackage{amssymb}
\usepackage{algorithm}
\usepackage[noend]{algpseudocode}
\usepackage{mathdots}

\usepackage{bm}
\usepackage{cite}
\usepackage{stmaryrd}
\usepackage{array}
\usepackage{psfrag}
\usepackage{comment}
\usepackage{balance}
\usepackage{booktabs}
\usepackage{pgfplots}
\pgfplotsset{compat=newest}
\pgfplotsset{plot coordinates/math parser=false}
\usetikzlibrary{matrix,calc,backgrounds}
\usepackage{mathtools}
\newlength\figureheight
\newlength\figurewidth

\usetikzlibrary{plotmarks,arrows}
\tikzset{>=stealth'}
\tikzstyle{axis} = [gray,->]
\tikzstyle{dot} = [fill,inner sep=0mm,minimum size=1.5mm,circle]

\DeclareMathOperator{\set}{set}
\DeclareMathOperator*{\smax}{s-max}

\newcommand{\ie}{i.e.,\ }

\DeclareMathOperator{\rank}{rank}

\newtheorem{theorem}{Theorem}
\newtheorem{corollary}{Corollary}
\newtheorem{lemma}{Lemma}
\newtheorem{definition}{Definition}
\newtheorem{proposition}{Proposition}
\newtheorem{example}{Example}
\newtheorem{remark}{Remark}
\newtheorem{construction}{Construction}
\newtheorem{conjecture}{Conjecture}
\newtheorem{observation}{Observation}

\DeclareMathOperator*{\argmax}{arg\,max}
\DeclareMathOperator*{\argmin}{arg\,min}
\DeclareMathOperator{\conv}{conv}
\usepackage{siunitx}
\newcommand{\dH}{d_{\mathrm H}}

\DeclareBoldMathCommand{\boldc}{c}
\DeclareBoldMathCommand{\boldd}{d}
\DeclareBoldMathCommand{\boldb}{b}
\DeclareBoldMathCommand{\boldp}{p}
\DeclareBoldMathCommand{\bolde}{e}
\DeclareBoldMathCommand{\boldx}{x}
\DeclareBoldMathCommand{\boldy}{y}
\DeclareBoldMathCommand{\boldf}{f}
\DeclareBoldMathCommand{\bolda}{a}
\DeclareBoldMathCommand{\boldt}{t}
\DeclareBoldMathCommand{\boldH}{H}
\DeclareBoldMathCommand{\boldP}{P}
\DeclareBoldMathCommand{\boldtheta}{\theta}
\DeclareBoldMathCommand{\bolddelta}{\delta}
\DeclareBoldMathCommand{\boldzeta}{\zeta}
\DeclareBoldMathCommand{\boldgamma}{\gamma}
\DeclareBoldMathCommand{\boldlambda}{\lambda}
\DeclareBoldMathCommand{\boldxi}{\xi}

\newcommand{\f}{\mathsf f}
\newcommand{\F}{\mathbb F}
\newcommand{\Z}{\mathbb Z}
\newcommand{\R}{\mathbb R}
\newcommand{\N}{\mathbb N}
\newcommand{\Fv}{\mathsf F_{\mathrm v}}
\newcommand{\Lv}{\mathsf L}
\newcommand{\Pv}{\mathsf P}
\newcommand{\Lamv}{\mathsf\Lambda}
\newcommand{\Fm}{\mathsf F_{\mathrm m}}
\newcommand{\C}{\mathcal C}
\newcommand{\Cj}{\C_j}
\renewcommand{\P}{\mathcal P}

\newcommand{\Pd}{\P}
\newcommand\Q{\mathcal Q}

\renewcommand\S{\mathbb S}
\DeclareMathOperator{\rot}{rot}
\DeclareMathOperator{\Aut}{Aut}
\DeclareMathOperator{\GL}{GL}
\newcommand{\autFp}{\Aut{(\F_p, +)}}
\newcommand{\autFq}{\Aut{(\F_q, +)}}

\newcommand\Nc{\mathcal N_{\mathrm c}}
\newcommand\Nv{\mathcal N_{\mathrm v}}

\newcommand\LP{\mathrm{LP}}
\newcommand\ML{\mathrm{ML}}

\newcommand{\abs}[1]{\left\lvert#1\right\rvert}
\newcommand{\range}[1]{\llbracket#1\rrbracket}

\DeclareMathOperator{\aff}{aff}

\newcommand{\hi}{{\mathrm{hi}}}
\newcommand{\lo}{{\mathrm{lo}}}

\newcommand{\revone}[1]{#1}
\newcommand{\revtwo}[1]{#1}
\newcommand{\eirik}[1]{{\color{black} #1}}
\newcommand{\eiriknew}[1]{{\color{black} #1}}
\newcommand{\eiriknewnew}[1]{{\color{black} #1}}
\newcommand{\eirikNew}[1]{{\color{black} #1}}
\newcommand{\eirikfinal}[1]{{\color{black} #1}}

\begin{document}
\title{Adaptive Linear Programming Decoding of Nonbinary Linear Codes Over Prime Fields}%

\author{Eirik~Rosnes,~\IEEEmembership{Senior~Member,~IEEE} and Michael~Helmling%
\thanks{The work of E.\ Rosnes was partially funded by the Norwegian-Estonian Research Cooperation Programme (grant EMP133) and the Research Council of Norway (grant 240985/F20). The work of M.\ Helmling was partially funded by the DFG (grant RU-1524/2-1). The material in this paper was presented in part at the 2015 IEEE Information Theory Workshop, Jerusalem, Israel, Apr./May\ 2015, at the 2016 IEEE International Symposium on Information Theory, Barcelona, Spain, Jul.\ 2016, and at the 9th International Symposium on Turbo Codes \& Iterative Information Processing, Brest, France, Sep.\ 2016.
}%
\thanks{E.\ Rosnes is with Simula UiB, N-5020 Bergen, Norway. E-mail: eirikrosnes@simula.no.}%
\thanks{M.\ Helmling is with the Fraunhofer Institute for Industrial Mathematics ITWM, 67663 Kaiserslautern, Germany. He was formerly affiliated with the Mathematical Institute, University of Koblenz-Landau, 50722 Koblenz, Germany. Email: michael.helmling@itwm.fraunhofer.de.}%
}%

\newenvironment{notms}{\thickmuskip=1\thinmuskip }{}

\maketitle

\begin{notms}
\begin{abstract}
In this work, we consider adaptive linear programming (ALP) decoding of linear codes over prime fields, %
\ie the finite fields $\F_p$ of size $p$ where $p$ is a prime, \revone{when used over a $p$-ary input memoryless channel.}  In particular, we provide a general construction of valid inequalities (using no auxiliary variables) for the codeword polytope (or the convex hull) of the so-called constant-weight embedding of a single parity-check (SPC) code over any prime field. The construction is based on \revtwo{sets of vectors, called \emph{building block classes},} that are assembled to form the left-hand side of an inequality according to several rules. In the case of \emph{almost doubly-symmetric} valid classes we prove that the resulting inequalities are all \emph{facet-defining}, while we conjecture this to be true if and only if the class is valid and \emph{symmetric}. \revtwo{Valid symmetric classes impose certain symmetry conditions on the elements of the vectors from the class, while valid doubly-symmetric classes impose further technical symmetry conditions.} For $p=3$, there is only a single valid symmetric class and we prove that the resulting inequalities together with the so-called \emph{simplex} constraints give a complete and irredundant description of the codeword polytope of the embedded SPC code. For $p > 5$, we show that there are additional facets beyond those from the proposed construction. As an example, for $p=7$, we provide additional inequalities that all define facets of the embedded codeword polytope.
The resulting overall set of linear (in)equalities is conjectured to be irredundant and complete.
 Such sets of linear (in)equalities have not appeared in the literature before, have a strong theoretical interest, and we use them to develop an efficient (relaxed) ALP decoder for \emph{general} (non-SPC) linear codes over prime fields. The key ingredient is an efficient separation algorithm based on the principle of dynamic programming. Furthermore, we construct a decoder for linear codes over arbitrary fields $\F_q$ with $q =p^m$ and $m>1$ by a factor graph representation that reduces to several instances of the case $m=1$, which results, in general, in a relaxation of the original decoding polytope. Finally, we present an efficient  cut-generating algorithm to search for redundant parity-checks to further improve the performance towards maximum-likelihood decoding for short-to-medium block lengths. Numerical experiments confirm that our new decoder is very efficient compared to a static LP decoder for various field sizes, check-node degrees, and block lengths. %
\end{abstract}

\section{Introduction}

Linear programming (LP) decoding was introduced by Feldman \emph{et al.} in 2005 \cite{fel05} as an efficient, but (compared to maximum-likelihood (ML) decoding) suboptimal decoding approach for binary linear codes. Since then, LP decoding of low-density parity-check (LDPC) codes has been extensively studied by various authors, and, in particular, several low-complexity approaches have been proposed. See, for instance, \cite{von06,tag07,bur09,tag11,bar13}.
The approach was later extended to nonbinary linear codes by Flanagan \emph{et al.} \cite{fla09}, and several low-complexity approaches were proposed in 
 \cite{gol13,pun13,liu14_1,liu14}.
Nonbinary LDPC codes are especially appealing because they in general exhibit a better performance than binary codes in the important finite-length regime. %

The underlying structure of LP decoding are the codeword polytopes (or convex hulls) whose vertices correspond to the codewords of a binary image of a (nonbinary) single parity-check (SPC) code. 
By intersecting all those polytopes defined by the rows of a specific parity-check matrix of a linear code, one obtains the so-called fundamental polytope, the domain of optimization of an LP decoder. %
While an explicit description for binary codes is well known (second formulation in \cite{fel05}), all LP formulations for nonbinary codes known so far generalize the first formulation in \cite{fel05} and thus depend on auxiliary variables (one for each feasible configuration) \cite{fla09}.

As recent results have shown \cite{bar13,was19}, LP decoding based on the alternating direction method of multipliers (ADMM) for convex optimization problems \cite{boy10} is able to outperform (in terms of decoding complexity) other LP decoding approaches. The efficiency of the algorithm relies on an efficient algorithm to do Euclidean projection onto the codeword polytope of a binary SPC code. In the binary case, the so-called ``two-slice'' lemma is the main result that enables efficient Euclidean projections in time $O(d \log d)$ for a binary SPC code of length $d$. More recently,  more efficient projection algorithms have been proposed in \cite{zha13_1} and \cite{zha13}. While initial work has been done to apply ADMM to the nonbinary case \cite{liu14}, it is currently not known how this framework can be applied to codes over nonbinary fields with characteristic greater than two, one difficulty being that little is known about codeword polytopes of nonbinary codes over such fields which one would need to project on. %

In this work, we present several results on the codeword polytope and the fundamental polytope of a general nonbinary code over any finite field $\F_q$. We provide an explicit construction for valid (facet-defining) inequalities for the so-called constant-weight embedding of a nonbinary SPC code over \emph{any} prime field \revtwo{$\F_p$} without relying on auxiliary variables.  %
The construction is based on what we call \emph{classes of building blocks}. \revtwo{A building block is simply a vector of length $p$, while a class of building blocks is a set of such vectors. The vectors within a class are used to build an inequality in a block-wise manner (thus the name building block) where each block corresponds to a code coordinate. We develop rules for combining these blocks in such a way that they yield the coefficient vector of a valid inequality, whose right-hand side is also computed by a specific rule. In particular, all but one of the blocks can be freely chosen (the last block is a function of the other blocks), and  the right-hand side of the inequality  is found by imposing that  the inequality is tight for a specially designed embedded codeword (the so-called canonical codeword). The detailed construction can be found in Construction~\ref{constr:hilo} in Section~\ref{sec:hilo}.} In the
case of \emph{almost doubly-symmetric} valid classes we prove that the resulting inequalities are all \emph{facet-defining}, while we conjecture this to be true if and only if the class is valid and \emph{symmetric}.  \revtwo{Valid symmetric classes impose certain symmetry conditions on the elements of the vectors from the class (see Definition~\ref{def:symmetric_class}), while valid doubly-symmetric classes impose further technical symmetry conditions (see Definition~\ref{def:almost_double_symmetric}).} 

For the ternary case, we prove that the constructed facet-defining inequalities together with the so-called \emph{simplex} constraints give a complete and irredundant description of the embedded SPC codeword polytope. It also extends the explicit formulation of the fundamental cone by Skachek \cite{Skachek10F3} in the sense that the latter describes the convex hull at one single (namely, the all-zero) vertex. For the quinary case, we conjecture that the constructed inequalities together with the simplex constraints indeed give a complete and irredundant description of the embedded codeword polytope, while for larger $q$ we show that this is not the case. 

Such facet-defining inequalities for the nonbinary case have,  to the best of our knowledge, not appeared in the literature before, have a strong theoretical interest, and their explicit construction without the need for auxiliary variables has immediate practical consequences to LP decoding: as we show in this work, these inequalities can efficiently be separated, which allows for efficient adaptive LP (ALP) decoding of \emph{general} (non-SPC) linear codes over any prime field, thus extending the well-known ALP decoder by Taghavi and Siegel \cite{tag07} to general nonbinary linear codes over prime fields \revone{when used over a discrete-input (with alphabet size equal to the field size) memoryless channel.}  %
 Besides its computational gain over \enquote{static} (non-adaptive) LP decoding, ALP decoding is also a key component of methods for improving error-correction performance \cite{zha12,Helmling+14MLDecoding}. While not a topic of this paper, our polyhedral results might also facilitate the development of a projection algorithm for ADMM decoding of general nonbinary codes.

 Linear codes (or, in particular, linear LDPC codes) over prime fields have several application areas. For instance, such codes are a key ingredient in the construction of low-density integer lattices using the so-called Construction~A \cite{pie12}. Such lattices are referred to as LDA lattices and they perform close to capacity on the Gaussian channel, in addition to being conceptually simpler than previously proposed lattices based on multiple nested binary codes. In particular, in \cite{pie12}, several integer LDA lattices, all based on a particular $(2,5)$-regular LDPC code over the prime field of size $11$, were proposed. For dimension $5000$, the lattice attains a symbol error rate (under low-complexity iterative decoding) of less than  $10^{-6}$ at $1$ dB from capacity. Also, ternary linear codes have recently attracted some attention in the context of polar codes \cite{god15_1} and array-based spatially-coupled LDPC codes \cite{ami15}.

The remainder of this paper is organized as follows. In Section~\ref{sec:basics}, we establish notation and give a short overview of some background material. We fully characterize the relationship between the two Euclidean embeddings of finite field elements used throughout this work in Section~\ref{sec:embeddings}. Section~\ref{sec:general_results} establishes general polyhedral properties (dimension, affine hull, and box inequalities) of the codeword polytope of a nonbinary linear code and studies its symmetries by introducing \emph{rotation}. Then, in Section~\ref{sec:bb}, we present a construction method for valid inequalities for the convex hull of the constant-weight embedding of an SPC code defined over the prime field $\F_p$ for general $p$. In Section~\ref{sec:explicit_3_5_7}, we tailor the general framework developed before for $p \in \{3,5,7\}$. In particular, we prove that for $p=3$, the framework provides a complete and irredundant description of the embedded codeword polytope of a ternary SPC code under the constant-weight embedding.
A separation algorithm based on the principle of dynamic programming (DP) for efficient (relaxed) ALP decoding of general (non-SPC) nonbinary codes over \emph{any} prime field is presented in Section~\ref{sec:ALP_q}; Section~\ref{sec:ALP_q3} describes an efficient implementation of this algorithm for the special case of $p=3$. In Section~\ref{sec:ACG}, we outline an efficient method to search for cut-inducing redundant parity-check (RPC) equations using Gaussian elimination, generalizing the \emph{Adaptive Cut Generation (ACG)} algorithm from \cite{zha12}. In Section~\ref{sec:ptom}, we briefly consider the case of nonbinary codes over  the general field $\mathbb{F}_q$ of size $q = p^m$, where %
$m>1$ is integer. In particular, we adapt the relaxation method proposed in \cite{liu14,hon12} for fields of characteristic $p=2$ to any characteristic $p > 2$  \revtwo{by representing a parity-check constraint over the finite field $\mathbb{F}_{p^m}$ by a set of $p^{m}-1$ $p$-ary parity-check constraints. In particular, the convex hull of a parity-check constraint over $\mathbb{F}_{p^m}$ is relaxed by considering the intersection of the convex hulls of the $p^{m}-1$  corresponding $p$-ary parity-check constraints.}  
Numerical results for both LDPC and high-density parity-check (HDPC) codes for various field sizes and block lengths are presented in Section~\ref{sec:numerical_results}. The results show that our proposed ALP decoder outperforms (in terms of decoding complexity) the decoder from \cite{fla09} (using both the plain and the cascaded LP formulation). Also, using an appropriately generalized ACG-ALP decoding algorithm, as described in Section~\ref{sec:ACG}, near-ML  decoding performance can be achieved for short block lengths. %
\revone{For comparison purposes we also show the performance of sum-product (SP) decoding \cite{mackay98}.}
Finally, we draw some conclusions and give an outline of some future work in Section~\ref{sec:conclu}.
\end{notms}

\section{Notation and Background} \label{sec:basics}
This section establishes some basic definitions and results needed for the rest of the paper.

\subsection{General Notation}
If $x \in S$ and $A \subset S$, where $S$ is a set, we denote 
\begin{equation} \notag %
x+A \coloneqq \{x + a\colon a \in A\}
\end{equation}
 (and analogously $x\cdot A$, $x-A$, etc.). For a map $f\colon A \rightarrow B$ and a set $S \subseteq A$, $f(S) = \{f(s)\colon s\in S\}$ is the set of images of $S$ under $f$. The set of integers is denoted by $\Z$, \eiriknew{while the set of positive integers is denoted by $\N$.} For a positive integer $L \in \eiriknew{\N}$, $\range L = \{1,2,\dotsc,L\}$. \eiriknew{The binomial coefficient of $a$ over $b$, $a, b \in \{0 \} \cup \N$, is denoted by $a \choose b$ where ${a \choose b} = 0$ if $a < b$.}
 
 A multiset $S$ is a set in which an item can occur repeatedly. The size of a multiset, denoted by $|S|$, is the number of items counted with multiplicity. For example, $S= \{1,2,2,3,6,6\}\subset \Z$ is a multiset with $|S| = 6$.

\subsection{Finite Fields and Integers}
The contrast between codes, as objects living in spaces over finite fields, and polytopes in Euclidean space is a key feature of LP decoding. For the sake of mathematical rigor, we strictly separate between these two spaces. Because of the need to frequently map from one into the other, especially in indices, this section introduces succinct notation for these maps.

For any prime $p$ and integer $m \geq 1$, let $\F_q$ with $q=p^m$ denote the finite field with $q$ elements. If $m=1$ (which is assumed for most of this work), the set $\F_p=\F_q$ consists of $p$ \emph{congruence classes} of $\Z/p\Z = \{ [a]_p\colon a \in \Z\}$, where
\begin{equation}
\begin{aligned}
[\cdot]_p\colon \Z &\rightarrow \F_p\\
a &\mapsto a + p \Z = \{a+kp\colon k \in \Z\}
\label{eq:congruence}
\end{aligned}
\end{equation}
maps an integer to its congruence class modulo $p$. The \enquote{inverse} of \eqref{eq:congruence} that maps a congruence class $\zeta \in \F_p$ to its unique integer representative from $\{0,1,\dotsc,p-1\}$ is denoted by $[\cdot]_\Z$:
\begin{equation} \notag
  \begin{aligned}
    [\cdot]_\Z\colon \F_p &\rightarrow \Z\\
    \zeta = [a]_p &\mapsto a \bmod p.
  \end{aligned}
\end{equation}
In general, \emph{literal} numbers are used to denote elements of both $\F_p$ and $\Z$, depending on the context, but nonetheless designate different items; e.g., $\Z \ni 3 \neq 3 \in \F_7$ (we may explicitly use the explicit form $[3]_7$ if there is risk of ambiguity or confusion). Note also that operators like \enquote{$+$} are defined for both $\Z$ and $\F_p$, such that $3+5=8$ in $\Z$, but $3+5=1$ in $\F_7$. If $a, b\in \Z$, the expression $a \equiv b \pmod p$ is an alternative notation for $[a]_p = [b]_p$, which we use especially if $a$ and $b$ are compound expressions. For $a \in \Z$, $[[a]_p]_\Z = a \bmod p$.

In the general case $m \geq 1$, each element $\zeta \in \F_q = \F_{p^m}$ can be represented by a polynomial $\zeta(x) = \sum_{i=1}^m p_i x^{i-1}$, where $p_i \in \F_p$, and we will use the integer representation $\zeta(p) = \sum_{i=1}^m [p_i]_\Z p^{i-1}$ for representing $\zeta$. Furthermore, let $\mathsf p(\zeta) = (p_1,\dotsc,p_m)$ be the $p$-ary vector representation of $\zeta$.

For any finite set $\mathcal{A} = \{\zeta_1,\dotsc,\zeta_{|\mathcal{A}|}\}$, $\zeta_i \in \F_q$, $i \in \range{|\mathcal{A}|}$,
we use the short-hand notation $\sum \mathcal{A} = \sum_{a \in \mathcal{A}} a$ for the sum (in $\F_q$) of the elements in $\mathcal A$.

\subsection{Linear Codes Over Finite Fields}
Let $\mathcal{C}$ denote a linear code of length $n$ and dimension $k$  over the finite field $\mathbb{F}_q$ with $q$ elements. The code $\mathcal{C}$ can be defined by an $r \times n$ parity-check matrix $\boldH = (h_{j,i})$, where $r \geq n-k$ and  each matrix entry $h_{j,i} \in \F_q$, $i \in \mathcal{I}$ and $j \in \mathcal{J}$, where $\mathcal I = \range n$ and $\mathcal J = \range r$ are the column and row index sets, respectively, of $\bm H$. Then, $\mathcal{C}  = \C(\bm H) = \{\bm{c}=(c_1,\dotsc,c_n)^T \in \mathbb{F}_q^n\colon  \boldH \bm{c} = \bm{0} \}$, where $(\cdot)^T$ denotes the transpose of its vector argument. When represented by a factor  (or Tanner) graph \cite{fla09}, $\mathcal{I}$ is also the variable node index set and $\mathcal{J}$ is the check node index set. In the following, let $\Nv(i)$ (resp.\ $\Nc(j)$) denote the set of neighboring nodes of variable node $i$ (resp.\ check node $j$). Finally, call $\mathcal C$ an $(n, k, d)$ code if $d$ denotes the minimum Hamming weight of its codewords. \eirik{The Hamming distance between two codewords $\bm c, \bm c' \in \mathcal{C}$ is denoted by $d_{\rm H}(\bm c,\bm c')$.}

\revtwo{In the original work by Feldman \emph{et al.} \cite{fel05}, the ML decoding problem was  stated as an integer linear program in the real space by using (yet not explicitly discussing) the above-defined $[\cdot]_2$ as the embedding of $\F_2$ into $\R$, where $\R$ denotes the real numbers, and then relaxed into a linear program using vectors that live in $[0,1]^n$. In the nonbinary case, one might be tempted to embed $\zeta \in \F_q$ into the reals by using its integer representation $\zeta(p) \in \Z$, which however does not work out for several reasons. Instead, the following mapping $\f(\cdot)$ (see \cite{kautz1964nonrandom,hon12,liu14}) embeds elements of $\F_q$ into the Euclidean space of dimension $q$ by using unit vectors of length $q$.}
\begin{definition}[Constant-Weight Embedding]\label{def:Constant}
 Let
  \begin{align*}
  \f\colon \mathbb{F}_q &\rightarrow \{0,1\}^q \subseteq \R^q\\
  \zeta &\mapsto \bm{x} = (x_0,\dotsc,x_{q-1})
  \end{align*}
  where $x_\delta = 1$ if $\delta = \zeta(p)$ is the integer representation of $\zeta$ and $x_\delta=0$ otherwise, and further the constant-weight embedding of \emph{vectors} from $\F_q^n$ as
  \begin{align*}
    \Fv\colon \F_q^n &\rightarrow \{0,1\}^{nq} \\
    \bm\zeta = (\zeta_1,\dotsc,\zeta_n)\eiriknewnew{^T} &\mapsto \left( \f(\zeta_1) \mid \dotsc \mid \mathsf{f}(\zeta_n) \right)^T
  \end{align*}
where $(\bm v_1\mid\dotsc\mid\bm v_n)$ denotes the concatenation of row vectors $\bm v_1,\dotsc,\bm v_n$.
\end{definition}
\begin{remark}\label{rem:zeroIndexing}
Motivated by the definition of $\f$, we identify, for any ground set $A$ (above, $A=\R$), $A^q$ with $A^{\F_q}$, \ie use elements from $\F_q$ and their integer representation interchangeably for indexing such vectors. As a consequence, the index starts at $0$ when its integer representation is used, as opposed to normal vectors which we index starting from $1$.

More generally, a space $A^{nq}$ which is related to $n$ embedded elements of $\F_q$ (such as $\R^{nq}$ in the above definition of $\Fv$) is identified with $\left(A^{\F_q}\right)^n$, which is why we usually employ double-indexing to emphasize on the $q$-blocks $\bm v_i$ of a vector $\bm v \in A^{nq}$, as in
\begin{equation} \notag
\bm v = (\bm v_1,\dotsc, \bm v_n)\eiriknew{^T} = (v_{1,0},\dotsc,v_{1,q-1},\dotsc,v_{n,0},\dotsc,v_{n,q-1})\eiriknew{^T}
\end{equation}
where $\bm v_i \in A^{q} = A^{\F_q}$.
\end{remark}

Observe that $\f$ defined in Definition~\ref{def:Constant} maps the elements of $\F_q$ to the vertices of the full-dimensional standard $(q-1)$-simplex embedded in $\R^q$, $S_{q-1} \coloneqq \conv( \{\bolde^i\}_{i=1}^q)$, \eirik{where $\conv(\cdot)$ denotes the convex hull in the real space of its argument  and} $\bolde^i$ is the $i$-th unit vector in $\R^q$.  Hence, $\Fv$ maps $\F_q^n$ onto the vertices of $S_{q-1}^n = S_{q-1} \times \dotsm \times S_{q-1}$ ($n$ times). \eiriknew{With some abuse of notation, in the rest of the paper, we will use the same notation $\bolde^i$ for the $i$-th unit vector in different vector spaces. It will be clear from the context in which vector space $\bolde^i$ lies. Moreover, the entries of $S_{q-1}^n$ are represented as column vectors, while the entries of $S_{q-1}$ are row vectors.}

Flanagan \emph{et al.} \cite{fla09} have proposed the following, slightly more compact embedding.

\revtwo{\begin{definition}[Flanagan's Embedding]\label{rem:Flanagan}
 Let
\begin{align*}
\f'\colon \F_q &\rightarrow \{0,1\}^{q-1} \subseteq \mathbb{R}^{q-1}\\
\zeta &\mapsto \bm{x} = (x_1,\dotsc,x_{q-1} )
\end{align*}
where $x_\delta = 1$ if $\delta=\zeta(p)$ and $x_\delta=0$ otherwise, with the analog vector-embedding $\Fv'$.
\end{definition}}

\revtwo{Flanagan's embedding has the advantage of using one less dimension per entry of $\F_q$, but exhibits less symmetry since the zero element is distinguished: $\|\f'(0)\|_1=0$ while $\|\f'(\zeta)\|_1 =1$ for $\zeta \neq 0$, where $\|\cdot\|_1$ denotes the standard $\ell_1$-norm of a vector. Because of the latter, $\f$ turned out to be better suitable for presenting the results of this paper, while $\f'$ is advantageous in some proofs. In Section~\ref{sec:embeddings}, we characterize the close relationship between $\f$ and $\f'$ and, in particular, show how any result stated using $\f$ can be transformed into its respective form under $\f'$.}

\subsection{LP Decoding of Nonbinary Codes}\label{sec:lpdecoding}

In this subsection, we review the LP decoding formulation proposed by Flanagan \emph{et al.} in \cite{fla09}, where in contrast to \cite{fla09} we use constant-weight embedding. Let $\F_q$ and $\Sigma$, respectively, denote the input and output alphabets of a memoryless channel with input $X$ and output $Y$, and define  for each $y \in \Sigma$ and $\delta \in \F_q$ the value $\gamma_{\delta} = \log \left( \frac{ {\rm Pr} \left(Y = y \mid X= 0 \right)}{{\rm Pr} \left(Y = y \mid X= \delta \right)} \right)$. Then, the function $\lambda\colon \Sigma \mapsto \left( \mathbb{R} \cup \left\{ \pm \infty \right\} \right)^q$ is defined as
\begin{displaymath}
\lambda(y) = \boldgamma = \left(\gamma_0,\dotsc,\gamma_{q-1} \right).
\end{displaymath}
Furthermore, we define $\Lamv(\bm{y}) = \left( \lambda(y_1) \mid \dotsc \mid  \lambda(y_n) \right)^T$ for $\bm{y}=(y_1,\dotsc,y_n)^T$. Now, the ML decoding problem can be written as \cite{fla09}
\begin{equation} \label{eq:MLformulation}
\begin{split}
\hat\boldc_\ML &= \argmin_{\bm{c} \in \mathcal{C}} \sum_{i=1}^n \log \left( \frac{ {\rm Pr} \left(Y=y_i \mid X=0 \right)\hfill}{{\rm Pr} \left( Y=y_i \mid X=c_i \right)} \right) \\
&= \argmin_{\bm{c} \in \mathcal{C}} \sum_{i=1}^n  \mathsf\lambda(y_i)  \mathsf{f}(c_i)^T \\
&= \argmin_{\bm{c} \in \mathcal{C}} \Lamv(\bm{y})^T  \mathsf{F}_{\rm v}(\bm{c})
\end{split}
\end{equation}
%
%
where $y_1,\dotsc,y_n$ are the channel outputs. The problem in \eqref{eq:MLformulation} can be relaxed into a linear program using the embedding from Definition~\ref{def:Constant}  as follows \cite{fla09}:
\begin{equation} \label{eq:LPformulation}
\begin{split}
\hat\boldx_{\LP} = \argmin\quad &  \Lamv(\boldy)^T \boldx \\
\text{s.\,t.}\quad & \boldx^{(j)} = \boldP_j \boldx \in   \conv(\Fv(\Cj)),\, \forall j \in \mathcal{J}
\end{split}
\end{equation}
where $ \boldx^{(j)}=(\boldx^{(j)}_{1},\dotsc,\boldx^{(j)}_{|\Nc(j)|})^T$, $\boldx^{(j)}_i = (x^{(j)}_{i,0},\dotsc,x^{(j)}_{i,q-1})$ for all $i \in \eirik{\range{|\Nc(j)|}}$, and  $\bm P_j$ is a binary %
\emph{indicator} matrix that selects the variables from $\bm{x}$ that participate in the $j$-th check node. %
In (\ref{eq:LPformulation}), $\mathcal{C}_j$ represents the SPC  code defined by the $j$-th check node, and $\conv(\mathsf{F}_{\rm v}(\mathcal{C}_j))$ is the convex hull of $\mathsf{F}_{\rm v}(\mathcal{C}_j)$ in $\mathbb{R}^{\abs{\mathcal{N}_{\rm c}(j)}\cdot q}$.

LP decoding, \ie using the LP relaxation \eqref{eq:LPformulation} as a decoder (which is defined to output a decoding failure if the optimal solution  $\hat\boldx_{\LP}$ does not happen to be integral) has several desirable properties. Most importantly, the so-called \emph{ML certificate} property \cite{fel05,fla09} assures that, if $\hat\boldx_\LP$ is a codeword, then $\hat\boldx_\LP = \hat\boldc_{\ML}$, where $\hat\boldc_{\ML}$ is the ML decoded codeword.

Note that the ML certificate property remains to hold if $\conv(\Fv(\Cj))$ is replaced by a relaxation $\mathcal Q_j \supseteq \conv(\Fv(\Cj))$. We will use the term \emph{LP decoding} also when such a further relaxation is used, which is the case  because (except for $q \in \{2,3\}$) the inequalities constructed in this paper may describe only a \revtwo{superset}  of $\conv(\Fv(\Cj))$.

\revone{As noted in \cite{fla09}, the performance of LP decoding is independent of the transmitted codeword only under a certain channel symmetry condition. For details, we refer the interested reader to \cite[Thm.~5.1]{fla09}. See also \cite[Def.~1]{hof09} where a very similar symmetry condition which guarantees codeword-independent performance under ML decoding was introduced.}

\subsection{Background on Polyhedra} \label{sec:polyhedra}
The convex hull of a finite number of points in $\R^n$ is called a \emph{polytope}. It can be alternatively characterized as the (bounded) intersection of a finite number of halfspaces, \ie the solution space of a finite number of linear inequalities. 

Let $\P\subseteq \R^n$ be a polytope. An inequality $\boldtheta^T \boldx \leq \kappa$ with $\boldtheta \in \R^n$ and $\kappa \in \R$ is \emph{valid} if it holds for any $\boldx \in \P$. Every valid inequality defines a \emph{face} $F = \{\boldx \in \P\colon \boldtheta^T \boldx = \kappa\}$ of $\P$, which is itself a polytope. For notational convenience, we will identify a face $F$ with its defining inequality $\boldtheta^T \boldx \leq \kappa$ as long as there is no risk of ambiguity.

The \emph{dimension} of a face (or polytope) $F$ is defined as the dimension of its affine hull $\aff(F)$, which is calculated as one less than the maximum number of affinely independent vectors in $F$. Recall that a set of $k$ vectors $\{\bm v^1,\dotsc, \bm v^k\} \subseteq \R^n$ is affinely independent if and only if the vectors $\{\bm v^2 - \bm v^1, \dotsc, \bm v^k - \bm v^1\}$ are linearly independent.

A face $F$ with $\dim(F) = \dim(\P) - 1$ is called a \emph{facet}, while a zero-dimensional face is a \emph{vertex} of $\P$. It is a basic result of polyhedral theory that a face $F$ of dimension $\dim(F)$ actually contains $\dim(F)$ affinely independent \emph{vertices} of $\P$. Conversely, a face $F$ is uniquely determined by $\dim(F)$ affinely independent vertices of $\P$ that are contained in $F$.

Facets are important because every \enquote{minimal} representation of a polytope $\P$ is of the form
\[ \P = \left\{ \boldx \in \R^n\colon \bm A \bm x = \bm b, \bm C\bm x \leq \bm d\right\}\]
where $\bm A$ is an $r \times n$ matrix of rank $r = n -\dim(\P)$ such that $\aff(\P) = \{\boldx\colon \bm A \boldx = \boldb\}$, and $\bm C$ is an $s\times n$ matrix such that the rows of $\bm C \bm x \leq \bm d$ are in one-to-one correspondence with the $s$ facets of $\P$. For a more rigorous treatment of this topic, see, e.g., \cite{NemhauserWolsey88}.

\revtwo{\subsection{Some Final Remarks}}

\revtwo{We list two intermediate results (Lemmas~\ref{lem:affInd} and \ref{lem:bb-hilo-plusi}) in Appendix~\ref{app:intermediate_results}.   To have a more modular presentation and since these two lemmas are used only in the proofs in the appendices, they are only presented there.}

\revtwo{Finally, in Table~\ref{table:notation} we present a  summary of the most important terms and notation introduced in Section~\ref{sec:basics} and used throughout the paper.}

\begin{table}
 \revtwo{\caption{A summary of basic notation and terms introduced in Section~\ref{sec:basics} and used throughout the paper.}
  \label{table:notation}
  \centering
 \vskip -2.0ex
  \ifonecolumn
  \begin{tabular}{lp{13.8cm}}
  \else
  \begin{tabular}{lp{6.85cm}}
  \fi
    \toprule
    $p$ & Prime number \\
    $\Z$ & The set of integers\\
    $\R$ & The set of real numbers\\
    $\range L$  & The set of integers $\{1,2,\dotsc,L\}$\\
    $\mathbb{F}_q$ & Finite field with $q$ elements \\
    $[\cdot]_p$ & The mapping of an integer to its congruence class modulo $p$\\
    $[\cdot]_\Z$ & The  \enquote{reverse} mapping of $[\cdot]_p$ \\
    $\zeta(p)$ & The integer representation of $\zeta \in \F_{p^m}$ in the reals \\
    $\mathsf p(\zeta)$ & The $p$-ary vector representation of $\zeta$\\
    $\mathcal{C}$ & A linear code over $\mathbb{F}_q$\\
    $\f$, $\Fv$ & Constant-weight embedding from Definition~\ref{def:Constant}\\
    $\f'$, $\Fv'$ & Flanagan's embedding from Definition~\ref{rem:Flanagan}\\
    $\Sigma$ & The output alphabet of the channel \\
    $S_{q-1}$ & The full-dimensional standard $(q-1)$-simplex  in $\R^q$ \\
    $\bolde^i$ & The $i$-th unit vector in \eiriknew{some vector space} \\  %
    $\conv(\cdot)$ & The convex hull of its argument    \\
    $\P$ & The convex hull of a finite number of points in $\R^n$ (polytope) \\
    $\boldtheta^T \boldx \leq \kappa$ & An inequality defining a (potential) face of a polytope\\
    $F$ & A face of a polytope $\P$  \\
    $\aff(F)$ & The  affine hull  of a face (or polytope) $F$\\
    $\dim(\cdot)$ & The dimension of its argument (face or polytope)\\
    Facet & A face $F$ of $\P$ of maximum dimension ($\dim(F) = \dim(\P) - 1$)\\
    Vertex & A zero-dimensional face \\
    \bottomrule
  \end{tabular}}
\end{table}

\section{Comparison of Embeddings of $\F_q$} \label{sec:embeddings}
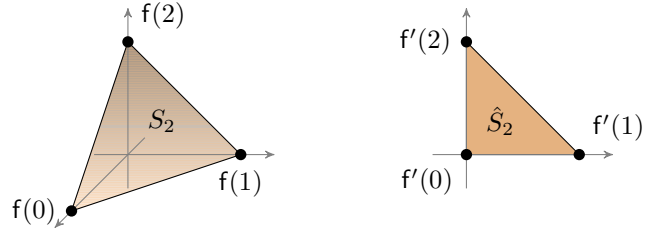
\begin{figure}
  \centering
  \begin{tikzpicture}[scale=1.5,z=-5mm]

    \filldraw[shade,fill=orange,fill opacity=.5] (1,0,0) -- (0,1,0) -- (0,0,1) -- cycle;
    \draw[axis] (-.3, 0, 0) -- (1.3, 0, 0);
    \draw[axis] (0, -.3, 0) -- (0, 1.3, 0);
    \draw[axis] (0, 0, -.3) -- (0, 0, 1.3);
    
    \node[dot,label=below:$\f(1)$] at (1,0,0) {};
    \node[dot,label=above right:$\f(2)$] at (0,1,0) {};
    \node[dot,label=left:$\f(0)$] at (0,0,1) {};
    \node at (.3, .3) {$S_2$};

    \begin{scope}[xshift=3cm]
        \filldraw[fill=orange!80!black,fill opacity=.5] (0,0) -- (1,0) -- (0,1) -- cycle;
        
        \draw[axis] (-.3, 0) -- (1.3, 0);
        \draw[axis] (0, -.3) -- (0, 1.3);
        
        \node at (.3, .3) {$\hat S_2$};
        
        \node[dot,label=below left:$\f'(0)$] at (0,0) {};
        \node[dot,label=above right:$\f'(1)$] at (1, 0) {};
        \node[dot,label=left:$\f'(2)$] at (0, 1) {};
    \end{scope}
  \end{tikzpicture}
  \vspace{-2ex}
  \caption{Constant-weight embedding $\f$ (left) and Flanagan's embedding $\f'$ (right) of $\F_3$ into $\R^3$ and $\R^2$, respectively.  
  Note that $S_2$ is an equilateral triangle, while $\hat S_2$ is not.}
  \label{fig:fvsfprime}
  \vskip -3ex
\end{figure}

In this section, we establish a close relationship between the convex hull of a nonbinary code under the embeddings $\Fv$ (Definition~\ref{def:Constant}) and $\Fv'$ (Definition~\ref{rem:Flanagan}), respectively, of $\F_q^n$ into the Euclidean space. Note that, while $\Fv$ maps $\F_q^n$ to the vertices of $S_{q-1}^n \subseteq \R^{nq}$, the embedding $\Fv'$ maps to the vertices of $\hat S_{q-1}^n \subseteq \R^{(q-1)n}$, where
\[ \hat S_{q-1} = \conv( \{\bm 0\} \cup \{\bm e^i\}_{i=1}^{q-1}) \subset \R^{q-1} \]
is the full-dimensional embedding of the $(q-1)$-simplex. \eiriknew{Again, as for $S_{q-1}^n$, the entries of $\hat S_{q-1}^n$ are represented as column vectors, while the entries of $\hat S_{q-1}$ are row vectors.} Geometrically, $S_{q-1}^n$ exhibits a higher symmetry (cf.\ Fig.~\ref{fig:fvsfprime}). For this reason, we found the constant-weight embedding more helpful for grasping the geometry of nonbinary linear codes. In addition, several formulas turned out more compact under this embedding. On the other hand, the following results show that the choice of the embedding does not affect the key polyhedral properties. Note that these results hold for arbitrary finite fields $\F_q = \F_{p^m}$.

\begin{lemma}\label{lem:embeddings}
  Let 
  \begin{align*}
   &\Pv\colon S_{q-1}^n \rightarrow \hat S_{q-1}^n\\
   &(\Pv(\boldx))_{i,j} = x_{i,j}&\text{for }i\in \range n, j\in \F_q \setminus \{0\}
  \end{align*}
  be the map that \enquote{projects out} the entries $x_{i,0}$, and let
  \begin{align*}
    &\Lv\colon \hat S_{q-1}^n \rightarrow S_{q-1}^n \\
    &(\Lv(\boldx'))_{i,j} = \begin{cases}
      1 - \sum_{k=1}^{q-1} x'_{i,k}&\text{if }j=0,\\
      x'_{i,j} &\text{otherwise}
      \end{cases} \quad \text{(for }i\in\range n\text{)}
  \end{align*}
  \enquote{lift} $\hat S_{q-1}^n$ onto $S_{q-1}^n$. Then, $\Pv = \Lv^{-1}$ and $\Lv = \Pv^{-1}$. In particular, both maps are bijective. Furthermore, $\Pv(\Fv(\bm\xi)) = \Fv'(\bm\xi)$ and $\Lv(\Fv'(\bm\xi)) = \Fv(\bm\xi)$ for any $\bm\xi \in \F_q^n$.
\end{lemma}
\begin{IEEEproof}
The statements can be easily verified by running through the cases. For example, $\Lv(\Pv(\bm x))_{i,0} = 1 - \sum_{k \neq 0} \Pv(x_{i,k}) = 1 - \sum_{k \neq 0} x_{i,k} = x_{i,0}$, where the last step holds because $\bm x_i \in S_{q-1}$ and hence $\sum_{k \in \F_q} x_{i,k} = 1$.
\end{IEEEproof}

\revtwo{
\begin{example}
Let $q=3$, $n=3$, and 
\begin{displaymath}
\begin{split}
\boldx &= (x_{1,0}, x_{1,1},x_{1,2}, x_{2,0}, x_{2,1},x_{2,2}, x_{3,0}, x_{3,1},x_{3,2})\eiriknew{^T} \\
&= (1/2,1/3,1/6, 1/10,8/10,1/10,1,0,0)\eiriknew{^T} \in S_2^3.
\end{split}
\end{displaymath}
 Then, $\boldx' = \Pv(\boldx) = (1/3,1/6,8/10,1/10,0,0)\eiriknew{^T} \in \hat{S}_2^3$ (the map  $\Pv(\cdot)$ \enquote{projects out} the entries $x_{i,0}$, $i \in \range 3$), while 
 \ifonecolumn
 \begin{displaymath}
 \begin{split}
  \Lv(\boldx') &= \Pv^{-1}(\boldx') \\
  &= (1-(1/3+1/6),1/3,1/6,1-(8/10+1/10),8/10,1/10,1-0-0,0,0)\\
  &=  (1/2,1/3,1/6, 1/10,8/10,1/10,1,0,0) = \boldx.
  \end{split}
\end{displaymath}  
\else
\begin{displaymath}
 \begin{split}
  \Lv(\boldx') &= \Pv^{-1}(\boldx') \\
  &= (1-(1/3+1/6),1/3,1/6,\\
  &\;\;\;\;\;\;1-(8/10+1/10),8/10,1/10,1-0-0,0,0)\eiriknew{^T}\\
  &=  (1/2,1/3,1/6, 1/10,8/10,1/10,1,0,0)\eiriknew{^T} = \boldx.
  \end{split}
\end{displaymath} 
\fi
\end{example}
}

Let $\C$ be a linear code of length $n$ defined over the finite field $\F_q$, $q=p^m$, with $p$ prime and $m \geq 1$. Let $\P= \conv(\Fv(\C))$ and $\P' = \conv(\Fv'(\C))$. From the above lemma, it follows immediately that $\P' = \Pv(\P)$ and $\P= \Lv(\P')$. Because $\Pv$ and $\Lv$ are affine linear and bijective, we also get the following.
\begin{lemma} \label{lem:affinely}
  The vectors $\boldx^1,\dotsc, \boldx^k \in S_{q-1}^n$ are affinely independent if and only if $\Pv(\boldx^1),\dotsc, \Pv(\boldx^k) \in \hat S_{q-1}^n$ are affinely independent.
\end{lemma}
\begin{corollary}\label{cor:facetEquiv}
  $F$ is a face of $\P$ with $\dim(F)=\delta$ if and only if $\Pv(F)$ is a face of $\P'$ also with $\eirik{\dim(\Pv(F))}=\delta$. In particular, $\dim(\P) = \dim(\P')$, and the facets of both polytopes are in one-to-one correspondence.
\end{corollary}

\revtwo{The particular form of $\Pv$ allows to immediately convert a  description of $\P$ by means of linear (in)equalities into such a description of $\P'$. In Appendix~\ref{app:embeddings}, we present the construction and give additional remarks. }

\section{Dimension of $\P$ and Rotational Symmetry}
\label{sec:general_results}
\subsection{Simplex Constraints and Dimension of $\P$}
In this subsection, we determine the dimension of $\P = \conv(\Fv(\C))$, where $\C$ is an \enquote{all-ones} SPC code (\ie its parity-check matrix contains only ones) of length $d$ defined over the field $\F_p$, $p$ prime, \revtwo{which is crucial for proving that inequalities are facets of $\P$ (cf.\ Section~\ref{sec:polyhedra})}. We then show that the linear equations and inequalities describing $S_{p-1}^d$ define the affine hull and are facets, respectively, of $\P$.

\begin{definition}
  \label{def:spx}
  For the finite field $\F_p$ and $d \geq 1$, let $\Delta_p^d$ denote the set of $p\times d$ inequalities and $d$ equations in $\R^{dp}$ that define $S_{p-1}^d$, \ie the inequalities
  \begin{subequations}
    \label{eq:spx}
    \begin{align}
      &x_{i,j} \geq 0 &&\text{for }i \in \range d\text{ and } j \in \F_p \label{eq:spx-geq}\\
      \text{and}\quad&\sum_{j \in \F_p} x_{i,j} = 1 &&\text{for }i \in \range d \label{eq:spx-eq};
    \end{align}
  \end{subequations}
  we call the (in)equalities in $\Delta_p^d$ \emph{simplex constraints}.
\end{definition}

\begin{proposition}\label{prop:Pjdim}
Let $\C$ be a length-$d$ \enquote{all-ones} SPC code over the finite field $\F_p$ and $\Pd = \conv(\Fv(\C))$.  For $d \geq 3$ (if $p=2$, for $d \geq 4$),
  \begin{enumerate}
    \item $\dim(\Pd) = d(p-1)$, \label{prop:Pjdim-1}
    \item the affine hull of $\Pd$ is $\aff(\Pd) = \{\boldx \colon \eqref{eq:spx-eq} \text{ holds for }i \in \range d\}$, and
    \item \eqref{eq:spx-geq} defines a facet of $\Pd$ for $i \in \range d$ and $j \in \F_p$.
  \end{enumerate}
\end{proposition}

\begin{IEEEproof}
  The results for $p=2$ are already known; see, e.g., \cite[Thm.~III.2]{hel12}. The proof for $p \neq 2$ is given in Appendix~\ref{app:proofPjdim}.
\end{IEEEproof}

The simplex constraints $\Delta_p^d$ can be interpreted as generalized box constraints that restrict, for $i \in \range d$, the $p$ variables representing $\mathsf f(c_i)$ to the simplex $S_{p-1}$, where $(c_1,\dotsc,c_d)^T$ denotes a codeword of the SPC code. As they are independent of $\boldH$, an arbitrary code $\C$ of length $n$ thus has only $n(p+1)$ simplex constraints ($n$ equations and $pn$ inequalities) in total. These will be denoted by $\Delta_p^\C$.

\subsection{Symmetries of $\P$ and General SPC Codes}
\label{sec:rotation}
In this subsection, we develop a notion of \emph{rotating} the simplex $S_{q-1} = \conv(\f(\F_q))$  according to a permutation of the elements of $\F_q$. This allows both to reduce the study of general SPC codes to \enquote{all-ones} SPC codes and to derive many valid inequalities from a single one by using automorphisms of the code.

Rotation is based on a work by Liu and Draper (see \cite[Sec.~IV.C]{liu14}). The results in this section hold for an arbitrary finite field $\F_q = \F_{p^m}$ \revtwo{and reduce to the ones from \cite[Sec.~IV.C]{liu14} for $p=2$ (only fields of characteristic two are considered in \cite{liu14}).}

\begin{definition}\label{def:rotation1}
Let $\S_q$ denote the group of permutations of the numbers $\{0,\dotsc,q-1\}$, which we will identify, via integer representation, with the permutations of $\F_q$.
For each $\pi \in \mathbb S_q$, the \emph{rotation operation} $\rot_\pi$ on  $\R^q$ is defined as
\[ \rot_\pi(\bm a = (a_0,\dotsc, a_{q-1})) = (a_{\pi(0)}, \dotsc, a_{\pi(q-1)}).\]
\end{definition}

Note that $\rot_\pi$ is a simple coordinate permutation and hence a vector-space automorphism of $\R^q$ that, in particular, maps the simplex $S_{q-1} \subset \R^q$ onto itself.

We now extend the above definitions to vectors in $\F_q^l$ and $\R^{lq}$, respectively.
\begin{definition}\label{def:rotation2}
Let $\bm\pi = (\pi_1,\dotsc, \pi_l) \in \S_q^l$ be a vector of $l$ permutations and $\bm\zeta=(\zeta_1,\dotsc,\zeta_l)\eiriknewnew{^T} \in \F_q^l$. Define
\[ \bm\pi(\bm\zeta) = (\pi_1(\zeta_1),\dotsc,\pi_l(\zeta_l))\eiriknewnew{^T}\]
and the corresponding rotation of $\R^{lq}$ as
\begin{align*}
  \rot_{\bm\pi}\colon  \R^{lq} &\rightarrow \R^{lq}\\
  (\bm a_1,\dotsc, \bm a_l)\eiriknew{^T} &\mapsto (\rot_{\pi_1}(\bm a_1),\dotsc,\rot_{\pi_l}(\bm a_l))\eiriknew{^T}
\end{align*}
by applying the $l$ individual rotations of $S_{q-1}$ component-wise to the $\bm a_i$.
\end{definition}

The following lemma shows how $\bm\pi$ and $\rot_{\bm\pi}$ relate.
\begin{lemma}\label{lem:rotation-embedding-swap}
Let $\bm\pi =(\pi_1,\dotsc,\pi_l) \in \mathbb S_q^l$ and $\bm\zeta = (\zeta_1,\dotsc, \zeta_l)\eiriknewnew{^T} \in \F_q^l$. Then,
$\rot_{\bm\pi}(\Fv(\bm \zeta)) = \Fv(\bm\pi^{-1}(\bm\zeta))$,
where $\bm\pi^{-1} = (\pi_1^{-1},\dotsc,\pi_l^{-1})$.
\end{lemma}
\begin{IEEEproof}
  We show for \eirikNew{$i \in \range l$} that $\rot_{\pi_i}(\f(\zeta_i)) = \f(\pi_i^{-1}(\zeta_i))$. Choose \eirikNew{$i \in \range l$} and let $\pi_i \in \S_q$ and $\zeta_i \in \F_q$, and denote $\bm x_i = \f(\zeta_i) \in \R^q$. By Definition~\ref{def:Constant}, $x_{i,\pi_i(j)} = 1 \Leftrightarrow \zeta_i(p) = \pi_i(j) \Leftrightarrow \pi_i^{-1}(\zeta_i(p)) = j \Leftrightarrow (\pi_i^{-1}(\zeta_i))(p) = j$. Hence, the $j$-th component of $\f(\pi_i^{-1}(\zeta_i))$ is equal to $1$ if and only if $x_{i,\pi_i(j)} = 1$, \ie if and only if the $\pi_i(j)$-th component of $\rot_{\pi_i}(\f(\zeta_i))$ is equal to $1$, which shows the claim and hence (by component-wise application) immediately proves the vector case of the lemma.
\end{IEEEproof}

\eirikNew{
In the rest of the paper, we will write permutations using \emph{one-line notation} as $\pi = (\pi(0),\dotsc,\pi(q-1))$, i.e.,  $0$ is permuted to $\pi(0)$, $1$ to $\pi(1)$, etc.
}

\revtwo{
\begin{example}
Let $q=3$, $l=3$, $\pi_1=(0,2,1)$,  $\pi_2=(0,1,2)$, $\pi_3=(2,1,0)$, and $\boldsymbol{\zeta} = (2,2,1)\eiriknewnew{^T}$. Then,
\begin{align*}
\boldsymbol{\pi}(\boldsymbol{\zeta}) &= (\pi_1(2),\pi_2(2),\pi_3(1))\eiriknewnew{^T} = (1,2,1)\eiriknewnew{^T}, \\
\Fv(\bm\zeta) &=  (0,0,1,0,0,1,0,1,0)\eiriknew{^T}.
\end{align*}
Furthermore,
\[ \boldsymbol{\pi}^{-1} = (\pi_1^{-1},\pi_2^{-1},\pi_3^{-1}) = ((0,2,1),(0,1,2),(2,1,0)), \]
$\bm\pi^{-1}(\bm\zeta)  = (1,2,1)\eiriknewnew{^T}$, and 
\[ \Fv(\bm\pi^{-1}(\bm\zeta))  = \Fv(\eiriknew{(1,2,1)}\eiriknewnew{^T}) = \eiriknew{(0,1,0,0,0,1,0,1,0)^T}. \]
Finally,
\begin{align*}
\rot_{\bm\pi}(\Fv(\bm \zeta)) &= (\rot_{\pi_1}(0,0,1),\rot_{\pi_2}(0,0,1),\rot_{\pi_3}(0,1,0))\eiriknew{^T} \\
&= (0,1,0, 0,0,1,0,1,0)\eiriknew{^T} \\
&= \Fv(\bm\pi^{-1}(\bm\zeta)).
\end{align*}
\end{example}}

\begin{theorem}\label{thm:rotation-general}
  For any set $\mathcal S \subseteq \F_q^l$ and any $\bm\pi \in \S_q^l$, let $\P = \conv(\Fv(\mathcal S))$ and $\tilde\P = \conv(\Fv(\bm\pi(\mathcal S)))$. Then,
  \begin{enumerate}
    \item for $\bm x\in \R^{lq}$, $\bm x \in \P$ if and only if $\rot_{\bm\pi}^{-1}(\bm x) \in \tilde\P$.
    \label{part:rotation-polytope}
    \item The inequality $\bm a^T \bm x \leq b$ with $\bm a \in \R^{lq}$, $b \in \R$ is valid for $\P$ and defines the face $F$ of $\P$, if and only if $\eiriknew{\rot^{-1}_{\bm\pi}}(\bm a)^T\bm x \leq b$ is valid for $\tilde\P$ and defines the face $\rot_{\bm\pi}^{-1}(F)$ of $\tilde\P$. In particular, both $F$ and $\rot_{\bm\pi}^{-1}(F)$ have the same dimension.
    \label{part:rotation-inequality}
  \end{enumerate}
  \revtwo{In other words, from a description of $\P$, as the embedding of $\mathcal S$, descriptions of the embeddings of arbitrary rotations of $\mathcal S$ can be derived.}
\end{theorem}

To prove Theorem~\ref{thm:rotation-general}, we need an auxiliary result.
\begin{lemma}\label{lem:permScalProd}
  If $\bm a,\bm b \in \R^{lq}$ and $\bm\pi \in \S_q^l$, then \begin{enumerate}
    \item $\rot_{\bm\pi}(\bm a)^T \rot_{\bm\pi}(\bm b) = \bm a^T\bm b$, and
    \item $\rot_{\bm\pi}(\bm a)^T \bm b = \bm a^T \rot_{\bm\pi}^{-1}(\bm b)$.
  \end{enumerate}
\end{lemma}
\begin{IEEEproof}
  The first claim is obvious because both sums contain the same elements, only in a different order. To show 2), apply 1) with $\rot_{\bm\pi}^{-1}$: 
  $\rot_{\bm\pi}(\bm a)^T \bm b = \rot_{\bm\pi}^{-1}(\rot_{\bm\pi}(\bm a))^T \rot_{\bm\pi}^{-1}(\bm b) = \bm a^T\rot_{\bm\pi}^{-1}(\bm b)$.
\end{IEEEproof}
\begin{IEEEproof}[Proof of Theorem~\ref{thm:rotation-general}]\\
  Part~\ref{part:rotation-polytope}): For $\bm\zeta = (\zeta_1,\dotsc,\zeta_l)\eiriknewnew{^T} \in \mathcal S$,
  \[
  \Fv(\bm \zeta) \in \P
  \Leftrightarrow \Fv(\bm\pi(\bm \zeta)) = \rot_{\bm\pi}^{-1}(\Fv(\bm\zeta)) \in \tilde\P \]
  (where the equality is by Lemma~\ref{lem:rotation-embedding-swap}), which shows the claim for all vertices of $\P$ and $\tilde\P$. But since $\rot_{\bm\pi}^{-1}$ is linear, the result extends to convex combinations, which proves the first part.\\
  Part~\ref{part:rotation-inequality}): It holds that
  \begin{align*}
    &\bm a^T \bm x \leq b&&\text{for all }\bm x \in \P\\
    \Leftrightarrow\;&\bm a^T \rot_{\bm\pi}(\tilde{\bm x}) \leq b&&\text{for all } \tilde{\bm x} \in \tilde\P
    \intertext{because, by Part~\ref{part:rotation-polytope}, $\bm x$ equals $\rot_{\bm \pi}(\tilde{\bm x})$ for some $\tilde{\bm x}\in \tilde\P$,}
    \Leftrightarrow\;&\rot_{\bm \pi}^{-1}(\bm a)^T \bm x \leq b &&\text{for all }\bm x \in \tilde\P\text{ (by Lemma~\ref{lem:permScalProd})}
  \end{align*}
  which shows the first claim of Part~\ref{part:rotation-inequality}). If now $F = \{\bm x \colon \bm a^T \bm x = b\text{ and }\bm x \in \P\}$, then 
  \ifonecolumn
  \begin{align*}
    \rot_{\bm\pi}^{-1}(F) &= \{\rot_{\bm\pi}^{-1}(\bm x)\colon \bm a^T \bm x = b\text{ and }\bm x \in \P\} \\
    &= \{\rot_{\bm \pi}^{-1}(\bm x)\colon \bm a^T \bm x = b \text{ and }\rot_{\bm \pi}^{-1}(\bm x) \in \tilde\P\}\\
    &= \{\rot_{\bm \pi}^{-1}(\bm x)\colon \rot_{\bm\pi}^{-1}(\bm a)^T \rot_{\bm\pi}^{-1}(\bm x) = b \text{ and }
    \rot_{\bm\pi}^{-1}(\bm x) \in \tilde\P\}\\
    &=\{\tilde{\bm x}\colon \rot_{\bm\pi}^{-1}(\bm a)^T \tilde{\bm x} = b\text{ and }\tilde{\bm x} \in \tilde\P\}
  \end{align*}
  \else
  \begin{align*}
    \rot_{\bm\pi}^{-1}(F) &= \{\rot_{\bm\pi}^{-1}(\bm x)\colon \bm a^T \bm x = b\text{ and }\bm x \in \P\} \\
    &= \{\rot_{\bm \pi}^{-1}(\bm x)\colon \bm a^T \bm x = b \text{ and }\rot_{\bm \pi}^{-1}(\bm x) \in \tilde\P\}\\
    &= \{\rot_{\bm \pi}^{-1}(\bm x)\colon \rot_{\bm\pi}^{-1}(\bm a)^T \rot_{\bm\pi}^{-1}(\bm x) = b \text{ and }\\
    &\qquad\qquad\rot_{\bm\pi}^{-1}(\bm x) \in \tilde\P\}\\
    &=\{\tilde{\bm x}\colon \rot_{\bm\pi}^{-1}(\bm a)^T \tilde{\bm x} = b\text{ and }\tilde{\bm x} \in \tilde\P\}
  \end{align*}
  \fi
  where we have again used Part~\ref{part:rotation-polytope} and Lemma~\ref{lem:permScalProd}. The last line is the definition of the face of $\tilde\P$ induced by $\rot_{\bm\pi}^{-1}(\bm a)^T \bm x \leq b$. Because $\rot_{\bm\pi}^{-1}$ does not influence affine independence, both $F$ and $\rot_{\bm\pi}^{-1}(F)$ have the same dimension.
\end{IEEEproof}

In the following, Theorem~\ref{thm:rotation-general} is applied to special cases to derive important results for the remainder of this work. %
\begin{definition}\label{def:rotMultByConstant}
For $0 \neq h \in \F_q$, define $\varphi_h \in \S_q$ by $\varphi_h(\zeta) = h\cdot\zeta$ for $\zeta \in \F_q$ (note again the identification of $\S_q$ and the bijections on $\F_q$), and denote by
$\GL(\F_q) = \{ \varphi_h\colon 0 \neq h \in \F_q\}$
the general linear group of $\F_q$ (as a 1-dimensional vector space over $\F_q$).

For $\bm h = (h_1,\dotsc, h_d) \in (\F_q \setminus \{0\})^d$, the corresponding map from $\GL(\F_q)^d$ is named
$\bm\varphi_{\bm h} = (\varphi_{h_1},\dotsc,\varphi_{h_d})$. For convenience, we will abbreviate $\rot_{\bm \varphi_{\bm h}}$ as $\rot_{\bm h}$. In the event that $h_1 = \dotsm = h_d = h$, we abbreviate $\bm\varphi_{\bm h} = \bm\varphi_h$ and $\rot_{\bm h} = \rot_h$.
\end{definition}
\begin{corollary}\label{cor:rotation-generalSPC}
  Let $\C$ be an  \enquote{all-ones} SPC code of length $d$, and $\C(\bm h)$ an arbitrary SPC code defined by the parity-check vector $\bm h = (h_1,\dotsc, h_d)$ with $h_i \neq0$ for $i \in \range d$.
  Further, let $\P = \conv(\Fv(\C))$ and $\P(\bm h) = \conv(\Fv(\C(\bm h))$. Then,
  \begin{enumerate}
  \item $\P(\bm h) = \rot_{\bm h}(\P)$ and $\P = \rot_{\bm h}^{-1}(\P(\bm h))$, and
  \item $\bm a^T \bm x \leq b$ is valid for $\P$ if and only if $\rot_{\bm h}(\bm a)^T \bm x \leq b$ is valid for $\P(\bm h)$.
  \end{enumerate}
\end{corollary}
\begin{IEEEproof}
  Follows from Theorem~\ref{thm:rotation-general}  with $\mathcal S = \C$ and $\bm\pi = \bm\varphi_{\bm h}^{-1}$ because $\C(\bm h) = \bm\varphi_{\bm h}^{-1}(\C)$ by definition.
\end{IEEEproof}

\revtwo{
\begin{example}
Let $\C$ denote a ternary \enquote{all-ones} SPC code of length $d=3$, and let $\bm h = (1,2,2)$. It follows that 
\ifonecolumn
\begin{align*}
\C &= \{ (0,0,0)\eiriknewnew{^T}, (0,1,2)\eiriknewnew{^T},(1,0,2)\eiriknewnew{^T},(1,1,1)\eiriknewnew{^T},(0,2,1)\eiriknewnew{^T},(2,0,1)\eiriknewnew{^T},(1,2,0)\eiriknewnew{^T},(2,1,0)\eiriknewnew{^T},(2,2,2)\eiriknewnew{^T} \}, \\
\C(\bm h) &= \{ (0,0,0)\eiriknewnew{^T}, (0,1,2)\eiriknewnew{^T},(1,0,1)\eiriknewnew{^T},(1,1,0)\eiriknewnew{^T},(0,2,1)\eiriknewnew{^T},(2,0,2)\eiriknewnew{^T},(1,2,2)\eiriknewnew{^T},(2,1,1)\eiriknewnew{^T},(2,2,0)\eiriknewnew{^T} \}.
\end{align*}
\else
\begin{align*}
\C &= \{ (0,0,0)\eiriknewnew{^T}, (0,1,2)\eiriknewnew{^T},(1,0,2)\eiriknewnew{^T},(1,1,1)\eiriknewnew{^T},(0,2,1)\eiriknewnew{^T},\\
&\;\;\;\;\;\;\;(2,0,1)\eiriknewnew{^T},(1,2,0)\eiriknewnew{^T},(2,1,0)\eiriknewnew{^T},(2,2,2)\eiriknewnew{^T} \}, \\
\C(\bm h) &= \{ (0,0,0)\eiriknewnew{^T}, (0,1,2)\eiriknewnew{^T},(1,0,1)\eiriknewnew{^T},(1,1,0)\eiriknewnew{^T},(0,2,1)\eiriknewnew{^T},\\
&\;\;\;\;\;\;\;(2,0,2)\eiriknewnew{^T},(1,2,2)\eiriknewnew{^T},(2,1,1)\eiriknewnew{^T},(2,2,0)\eiriknewnew{^T} \}.
\end{align*}
\fi
Furthermore,
\ifonecolumn
\begin{align*}
\Fv(\C) &=  \{ (1,0,0,1,0,0,1,0,0)\eiriknew{^T}, (1,0,0,0,1,0,0,0,1)\eiriknew{^T},
                (0,1,0,1,0,0,0,0,1)\eiriknew{^T}, (0,1,0,0,1,0,0,1,0)\eiriknew{^T},\\
                &\;\;\;\;\;\;\;(1,0,0,0,0,1,0,1,0)\eiriknew{^T}, (0,0,1,1,0,0,0,1,0)\eiriknew{^T},(0,1,0,0,0,1,1,0,0)\eiriknew{^T}, (0,0,1,0,1,0,1,0,0)\eiriknew{^T},\\
                &\;\;\;\;\;\;\;(0,0,1,0,0,1,0,0,1)\eiriknew{^T} \}
\end{align*}
\else
\begin{align*}
\Fv(\C) &=  \{ (1,0,0,1,0,0,1,0,0)\eiriknew{^T}, (1,0,0,0,1,0,0,0,1)\eiriknew{^T},\\
                &\;\;\;\;\;\;\;(0,1,0,1,0,0,0,0,1)\eiriknew{^T}, (0,1,0,0,1,0,0,1,0)\eiriknew{^T},\\
                &\;\;\;\;\;\;\;(1,0,0,0,0,1,0,1,0)\eiriknew{^T}, (0,0,1,1,0,0,0,1,0)\eiriknew{^T},\\
                &\;\;\;\;\;\;\;(0,1,0,0,0,1,1,0,0)\eiriknew{^T}, (0,0,1,0,1,0,1,0,0)\eiriknew{^T},\\
                &\;\;\;\;\;\;\;(0,0,1,0,0,1,0,0,1)\eiriknew{^T} \}
\end{align*}
\fi
and
\ifonecolumn
\begin{align*}
\Fv(\C(\bm h)) &=  \{ (1,0,0,1,0,0,1,0,0), (1,0,0,0,1,0,0,0,1),
                            (0,1,0,1,0,0,0,1,0), (0,1,0,0,1,0,1,0,0), \\
                            &\;\;\;\;\;\;\;(1,0,0,0,0,1,0,1,0), (0,0,1,1,0,0,0,0,1),(0,1,0,0,0,1,0,0,1),(0,0,1,0,1,0,0,1,0),\\
                            &\;\;\;\;\;\;\;(0,0,1,0,0,1,1,0,0) \}.
\end{align*}
\else
\begin{align*}
\Fv(\C(\bm h)) &=  \{ (1,0,0,1,0,0,1,0,0)\eiriknew{^T}, (1,0,0,0,1,0,0,0,1)\eiriknew{^T},\\
                            &\;\;\;\;\;\;\;(0,1,0,1,0,0,0,1,0)\eiriknew{^T}, (0,1,0,0,1,0,1,0,0)\eiriknew{^T}, \\
                            &\;\;\;\;\;\;\;(1,0,0,0,0,1,0,1,0)\eiriknew{^T}, (0,0,1,1,0,0,0,0,1)\eiriknew{^T},\\
                            &\;\;\;\;\;\;\;(0,1,0,0,0,1,0,0,1)\eiriknew{^T},(0,0,1,0,1,0,0,1,0)\eiriknew{^T},\\
                            &\;\;\;\;\;\;\;(0,0,1,0,0,1,1,0,0)\eiriknew{^T} \}.
\end{align*}
\fi
Also,
\begin{align*}
\rot_{\bm h} &=  (\rot_{\varphi_1},\rot_{\varphi_2},\rot_{\varphi_2}) = ((0,1,2),(0,2,1),(0,2,1)),\\
\rot_{\bm h}^{-1} &=  (\rot_{\varphi_1},\rot_{\varphi_2},\rot_{\varphi_2}) = ((0,1,2),(0,2,1),(0,2,1)),
 \end{align*}
 from which it follows that $\rot_{\bm h}(\Fv(\C))=$
 \ifonecolumn
 \begin{align*}
\rot_{\bm h}(\Fv(\C)) &= \{ (1,0,0,1,0,0,1,0,0), (1,0,0,0,0,1,0,1,0),
                	                   (0,1,0,1,0,0,0,1,0), (0,1,0,0,0,1,0,0,1),\\
                                    &\;\;\;\;\;\;\;(1,0,0,0,1,0,0,0,1), (0,0,1,1,0,0,0,0,1),
                                    (0,1,0,0,1,0,1,0,0), (0,0,1,0,0,1,1,0,0),\\
                                    &\;\;\;\;\;\;\;(0,0,1,0,1,0,0,1,0) \} =   \Fv(\C(\bm h))
 \end{align*}
 \else
  \begin{align*}
&\{ (1,0,0,1,0,0,1,0,0)\eiriknew{^T}, (1,0,0,0,0,1,0,1,0)\eiriknew{^T},\\
                	                   &\;\;(0,1,0,1,0,0,0,1,0)\eiriknew{^T}, (0,1,0,0,0,1,0,0,1)\eiriknew{^T},\\
                                    &\;\;(1,0,0,0,1,0,0,0,1)\eiriknew{^T}, (0,0,1,1,0,0,0,0,1)\eiriknew{^T},\\
                                    &\;\;(0,1,0,0,1,0,1,0,0)\eiriknew{^T}, (0,0,1,0,0,1,1,0,0)\eiriknew{^T},\\
                                    &\;\;(0,0,1,0,1,0,0,1,0)\eiriknew{^T} \} =   \Fv(\C(\bm h))
 \end{align*}
 \fi
 and $\rot_{\bm h}^{-1}(\eirik{\Fv}(\C(\bm h)))=$
  \ifonecolumn
  \begin{align*}
&\{ (1,0,0,1,0,0,1,0,0), (1,0,0,0,0,1,0,1,0),
                            (0,1,0,1,0,0,0,0,1), (0,1,0,0,0,1,1,0,0), \\
                            &\;\;\;\;\;\;\;(1,0,0,0,1,0,0,0,1), (0,0,1,1,0,0,0,1,0),
                            (0,1,0,0,1,0,0,1,0),(0,0,1,0,0,1,0,0,1),\\
                            &\;\;\;\;\;\;\;(0,0,1,0,1,0,1,0,0) \} = \Fv(\C).
 \end{align*}
 \else
   \begin{align*}
&\{ (1,0,0,1,0,0,1,0,0)\eiriknew{^T}, (1,0,0,0,0,1,0,1,0)\eiriknew{^T},\\
                            &\;\;(0,1,0,1,0,0,0,0,1)\eiriknew{^T}, (0,1,0,0,0,1,1,0,0)\eiriknew{^T}, \\
                            &\;\;(1,0,0,0,1,0,0,0,1)\eiriknew{^T}, (0,0,1,1,0,0,0,1,0)\eiriknew{^T},\\
                            &\;\;(0,1,0,0,1,0,0,1,0)\eiriknew{^T},(0,0,1,0,0,1,0,0,1)\eiriknew{^T},\\
                            &\;\;(0,0,1,0,1,0,1,0,0)\eiriknew{^T} \} = \Fv(\C).
 \end{align*}
 \fi
\end{example}
}

The corollary shows that $\P$ and $\P(\bm h)$ are equivalent up to an index permutation; in particular, they coincide in most interesting structural properties such as dimension, number of facets, volume, etc. By the second part, a description (or a relaxation) of $\P$ by means of linear (in)equalities immediately leads to a description (or an equally tight relaxation) of $\P(\bm h)$.

Another special case of Theorem~\ref{thm:rotation-general} reveals symmetries \emph{within} the SPC codeword polytope $\P(\bm h)$. First, we need another subclass of the permutations of $\S_q$.
\begin{definition}\label{def:glq}
  By $\autFq$ we denote the set of automorphisms of the additive group $(\F_q,+)$, \ie bijections $\varphi$ on $\F_q$ that satisfy $\varphi(\zeta+\eta) = \varphi(\zeta) + \varphi(\eta)$ for all $\zeta, \eta \in \F_q$ (note that this implies that $\varphi(0) = 0$).
\end{definition}

\begin{corollary}\label{cor:rotation-autfq}
  Let $\C$, $\P$, $\C(\bm h)$, and $\P(\bm h)$ as above, $\varphi \in \autFq$, and let $\bm\varphi = (\varphi,\dotsc,\varphi)$ ($d$ times). Then,
  \begin{enumerate}
    \item $\bm a^T \bm x \leq b$ valid for $\P$ $\Leftrightarrow$ $\rot_{\bm\varphi}(\bm a)^T \bm x \leq b$ valid for $\P$, and
    \item $\bm a^T \bm x \leq b$ valid for $\P(\bm h)$ $\Leftrightarrow$ $\rot_{\bm\varphi_{\bm h}\circ \bm\varphi \circ \bm\varphi_{\bm h}^{-1}}(\bm a)^T \bm x \leq b$ valid for $\P(\bm h)$.
  \end{enumerate}
\end{corollary}
\begin{IEEEproof}
For the first statement, we can apply Theorem~\ref{thm:rotation-general} with $\mathcal S=\C$ and $\bm\pi = \eiriknew{\bm\varphi^{-1}}$ because $\C = \eiriknew{\bm\varphi^{-1}}(\C)$: $\bm c \in \C \Leftrightarrow \sum_{i=1}^d c_i = 0 \Leftrightarrow \eiriknew{\varphi^{-1}}(\sum_i c_i) = \eiriknew{\varphi^{-1}}(0) = 0 \Leftrightarrow \sum_i \eiriknew{\varphi^{-1}}(c_i) = 0 \Leftrightarrow \eiriknew{\bm\varphi^{-1}}(\bm c) \in \C$. The second statement then follows by applying Corollary~\ref{cor:rotation-generalSPC} twice and the first statement in between:
\begin{align*}
  &\bm a ^T \bm x \leq b&&\text{valid for $\P(\bm h)$}\\
  \Leftrightarrow\;
  &\rot_{\bm h}^{-1}(\bm a)^T \bm x \leq b&&\text{valid for $\P$}\\
  \Leftrightarrow\;
  &\rot_{\bm\varphi}(\rot_{\bm h}^{-1}(\bm a))^T\bm x \leq b&&\text{valid for $\P$}\\
  \Leftrightarrow\;
  &\rot_{\bm h}(\rot_{\bm\varphi}(\rot_{\bm h}^{-1}(\bm a)))^T\bm x \leq b&&\text{valid for $\P(\bm h)$}.\quad\IEEEQEDhere
\end{align*}
\end{IEEEproof}

\begin{remark}\label{rem:rotation-autfq-allones}
  For $q=p^1=p$, $\autFp = \GL(\F_p)$ equals the set of multiplications with a nonzero constant as defined in Definition~\ref{def:rotMultByConstant}. By the distributive law, this in particular implies that $\autFp$ is commutative, such that $\bm\varphi_{\bm h}\circ \bm\varphi \circ \bm\varphi_{\bm h}^{-1}$ in the above corollary reduces to $\bm\varphi$.
  
  In general, it can be shown that $\autFq =  \GL(\F_p^m)$ (where the $m$-dimensional $\F_p$-space 
  $\F_p^m$ is, as a vector space, isomorphic to $\F_q$), which for $m>1$ is a strict superset of $\GL(\F_q)$ and not commutative.
\end{remark}

\revtwo{In Table~\ref{table:notation2} we present a summary of the most important notation introduced in Sections~\ref{sec:embeddings} and \ref{sec:general_results}.}

\begin{table}
\revtwo{\caption{A summary of notation introduced in Sections~\ref{sec:embeddings} and \ref{sec:general_results}.}
  \label{table:notation2}
  \centering
 \vskip -2.0ex
  \ifonecolumn
  \begin{tabular}{lp{12.4cm}}
  \else
  \begin{tabular}{lp{6.4cm}}
  \fi
    \toprule
    $\hat S_{q-1}$ & The convex hull of  $\{\bm 0\} \cup \{\bm e^i\}_{i=1}^{q-1}$ in $\R^{q-1}$ \\    
   $\Pv$ & Mapping from $S_{q-1}^n$ to $\hat S_{q-1}^n$ \enquote{projecting out}  $x_{i,0}$\\
   $\Lv$ & The \enquote{inverse} of $\Pv$ that \enquote{lifts} $\hat S_{q-1}^n$ onto $S_{q-1}^n$ \\
   $\Delta_p^d$, $\Delta_p^\C$ & Simplex constraints (see Definition~\ref{def:spx})\\
   $\S_q$ & The permutation group of $\{0,\dotsc,q-1\}$  (see Definition~\ref{def:rotation1})\\
$\rot_\pi$ & Rotation operator  on $\R^q$ for $\pi \in \mathbb S_q$  (see Definition~\ref{def:rotation1})\\
$\rot_{\bm\pi}$ & Vector version of $\rot_\pi$ (see Definition~\ref{def:rotation2}) \\
$\varphi_h$ & Permutation in $\S_q$ defined by $\varphi_h(\zeta) = h\cdot\zeta$ for $\zeta \in \F_q$ \\
$\GL(\F_q)$ & The general linear group of $\F_q$ (see Definition~\ref{def:rotMultByConstant}) \\
$\rot_{\bm h}$, $\rot_{h}$ & Short-hand notation for rotation operator (see  Definition~\ref{def:rotMultByConstant}) \\
  $\autFq$ & The set of automorphisms of the additive group $(\F_q,+)$ (see Definition~\ref{def:glq})\\
    \bottomrule
  \end{tabular}}
\end{table}

\tikzset{dot/.style={fill,inner sep=0mm,minimum size=1.5mm,circle}}
\tikzset{rbrace/.style={decorate,decoration={brace,mirror}}}
\tikzset{>=stealth}
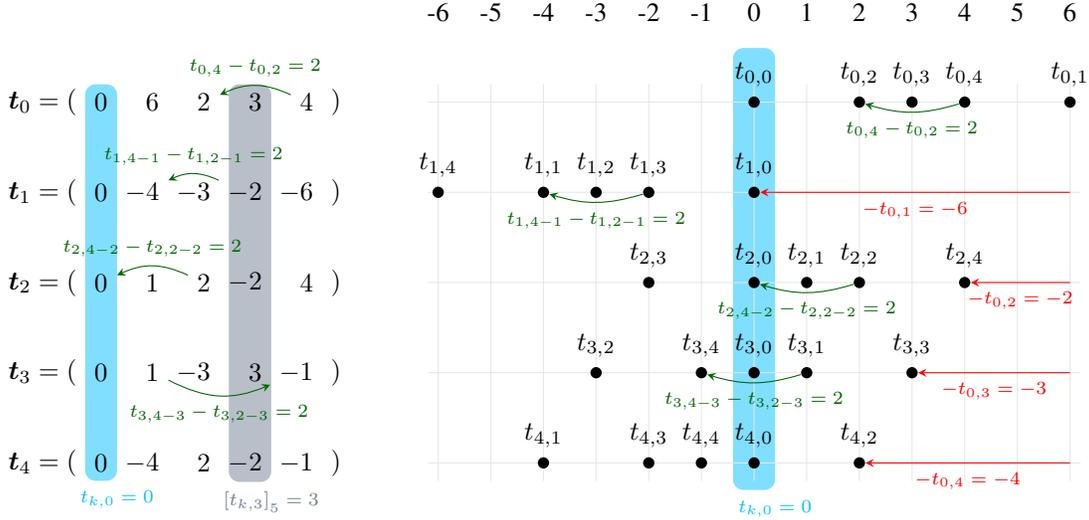
\begin{figure*}
  \centering
  \begin{tikzpicture}[xscale=.7,yscale=1.2]
    \draw[black!10,very thin] (-6.2, .2) grid (6.2, -4.2);
    \begin{scope}[xshift=-11cm]
      \matrix (T) [matrix of math nodes,row sep={1.2cm,between origins},anchor=center,every node/.append style={anchor=east}] at (0, -2) 
      { \boldt_0=( &0 & 6 & 2 & 3 & 4 &)\\
        \boldt_1=( &0 & -4 & -3 & -2 & -6&) \\
        \boldt_2=( &0 & 1 & 2 & -2 & 4 &)\\
        \boldt_3=( &0 & 1 & -3 & 3 & -1 &)\\
        \boldt_4=( &0& -4 & 2 & -2 & -1 &) \\
      };
      
    \end{scope}
    \foreach \i in {-6,...,6}
      \node at (\i,1) {\i};
    \foreach \i [count=\j from 0] in {0,6,2,3,4}
      \node[dot,label=$t_{0,\j}$] (t0\j) at (\i, 0) {};
    \foreach \i [count=\j from 0] in {0,-4,-3,-2,-6}
          \node[dot,label=$t_{1,\j}$] (t1\j) at (\i, -1) {};
    \foreach \i [count=\j from 0] in {0,1,2,-2,4}
          \node[dot,label=$t_{2,\j}$] (t2\j) at (\i, -2) {};
    \foreach \i [count=\j from 0] in {0,1,-3,3,-1}
          \node[dot,label=$t_{3,\j}$] (t3\j) at (\i, -3) {};
    \foreach \i [count=\j from 0] in {0,-4,2,-2,-1}
          \node[dot,label=$t_{4,\j}$] (t4\j) at (\i, -4) {};
    
    \begin{scope}[font=\scriptsize]
    \draw[red,->] (6,-1) -- node[below] {$-t_{0,1}=-6$} (t10);
    \draw[red,->] (6,-2) -- node[below] {$-t_{0,2}=-2$} (t24);
    \draw[red,->] (6,-3) -- node[below] {$-t_{0,3}=-3$} (t33);
    \draw[red,->] (6,-4) -- node[below] {$-t_{0,4}=-4$} (t42);

    \begin{scope}[DarkGreen,->]
      \draw (t04) to[bend left=10] node[below] {$t_{0,4} - t_{0,2}=2$} (t02);
      \draw (T-1-6) to[bend right=15] node[above] {$t_{0,4} - t_{0,2}=2$} (T-1-4);
      \draw (t13) to[bend left=10] node[below] {$t_{1,4-1} - t_{1,2-1}=2$} (t11);
      \draw (T-2-5) to[bend right=15] node[above] {$t_{1,4-1} - t_{1,2-1}=2$} (T-2-3);
      \draw (t22) to[bend left=10] node[below] {$t_{2,4-2} - t_{2,2-2}=2$} (t20);
      \draw (T-3-4) to[bend right=15] node[above] {$t_{2,4-2} - t_{2,2-2} = 2$} (T-3-2);
      \draw (t31) to[bend left=10] node[below] {$t_{3,4-3} - t_{3,2-3}=2$} (t34);
      \draw (T-4-3) to[bend right=15] node[below] {$t_{3,4-3} - t_{3,2-3}=2$} (T-4-6);
    \end{scope}
    
    \begin{pgfonlayer}{background}
    \fill[thick,DeepSkyBlue!50,rounded corners] (-.4,.6) rectangle (.4,-4.3) node[below,DeepSkyBlue] {$t_{k,0} = 0$};
    \fill[thick,DeepSkyBlue!50,rounded corners] (T-1-2.north west) rectangle (T-5-2.south east) node[below,DeepSkyBlue] {$t_{k,0} =0$};
    \fill[thick,SlateGray!50,rounded corners] ($ (T-1-5.north west) + (-.2, 0) $) rectangle (T-5-5.south east) node[below,SlateGray] {$\left[t_{k,3}\right]_5 = 3$};
    \end{pgfonlayer}
    \end{scope}
    
  \end{tikzpicture}
  \vspace{-2ex}
  \caption{Example of the statements of Lemma~\ref{lem:bb-properties} for $p=5$ and $\bm m = (0,1,0,0,0)$. The five building blocks in $\mathcal T^{\bm m}$ are shown on the left. On the right, the entries of each block are placed according to their respective value. \textcolor{SlateGray}{Property~\ref{part:bb-property1}} is shown on the left for $j=3$. \textcolor{DeepSkyBlue}{Property~\ref{part:bb-property2}} is apparent in both figures. The statement of \textcolor{red}{Property~\ref{part:bb-property3}} becomes obvious on the right, while \textcolor{DarkGreen}{Property~\ref{part:bb-property4}} is again shown in both figures (for $k=0, i=4$, and $j=2$).}
  \label{fig:bb-properties}
\vskip -3ex
\end{figure*}

\section{Construction of Valid Inequalities From Building Blocks}\label{sec:bb}
In this section, we establish a construction of valid inequalities for the polytope $\P = \conv(\Fv(\C))$, where $\C$ is an \enquote{all-ones} SPC code of length $d$ over the finite field $\F_p$ with $p$ prime; the symbols $\P$, $\C$, $d$, and $\F_p$ will be used, with the above meaning, throughout the entire section. The construction is based on classes of \emph{building blocks} that are assembled to form the left-hand side of an inequality according to several rules developed in the following.

First, a set of building block classes is defined in Section~\ref{sec:bb-const}, \revtwo{where a building block class is just a set of vectors of length $p$.} The method for constructing inequalities from a building block class is described in Section~\ref{sec:hilo}. Based on these inequalities, we derive necessary and sufficient conditions for a building block class to induce only valid inequalities in Section~\ref{sec:validInvalid}, and some necessary conditions for inequalities to define \emph{facets} in Section~\ref{sec:facet-defining}. Section~\ref{sec:automorph} discusses the application of Corollary~\ref{cor:rotation-autfq} in order to obtain a set of additional inequalities from each of the previously constructed ones. Finally, the issue of redundancy within the set of valid inequalities constructed by the method of this section is addressed in Section~\ref{sec:redundant}.

\revtwo{In Table~\ref{table:notation3} we present a summary of the most important terms and notation introduced in this section.}

\subsection{Building Block Construction}\label{sec:bb-const}

\revtwo{A basic building block class is defined by an integer vector $\bm m = (m_0, m_1, \dotsc,m_{p-1}) \in \{0,1\}^{p}$ with $m_0 = 0$. From the vector $\bm m$ the individual blocks (or vectors) from the class can be constructed as detailed in the following definition. Each class contains $p$ vectors.}

\begin{definition}[Basic Building Block Class]\label{def:bb}
  For any $\bm m = (m_0, m_1, \dotsc,m_{p-1}) \in \{0,1\}^{p}$ with $m_0 = 0$, define $p$ vectors $\{\boldt_k^{\bm m}\}_{k \in \F_p} \subset \R^p$ %
  by
  \begin{enumerate}
    \item $t^{\bm m}_{0,j} = [j]_\Z + m_j p$ for $j\in \F_p$, and
    \item $t^{\bm m}_{k,j} = t^{\bm m}_{0, j + k} - t^{\bm m}_{0,k}$ for $k \in \F_p \setminus \{0\}$ and $j\in \F_p$.
  \end{enumerate}
  Each $\bm t_k^{\bm m}$ is called a \emph{basic building block}, and the set $\mathcal T^{\bm m} = \{ \bm t_k^{\bm m} \}_{k \in \mathbb{F}_p}$ %
of building blocks constructed in this way is called the \emph{basic building block class} induced by $\bm m$.
\end{definition}

In the sequel, we will sometimes omit the prefix \enquote{basic} when it is clear from the context that we talk about a basic building block class.

Note that Remark~\ref{rem:zeroIndexing} applies to the above definition, such that, e.g., $t^{\bm m}_{0,j+k} = t^{\bm m}_{0,[j+k]_\Z}$, \ie there is no need to take the index modulo $p$ because it is an element of $\F_p$.

\begin{example}\label{ex:bb-p3}
Let $p=3$ and $\bm m = (0,1,1)$. Then, the class $\mathcal T^{\bm m}$ consists of $\boldt^{\bm m}_0 = (0, 4, 5)$, $\boldt^{\bm m}_1 = (0, 1, -4)$, and $\boldt^{\bm m}_2 = (0, -5, -1)$. 
\end{example}
\begin{example}\label{ex:bb-p7}
Let $p=7$ and $\bm m = (0,1,1,0,0,1,0)$. Then, 
\ifonecolumn
\begin{align}
\boldt^{\bm m}_0 &= (0, 8, 9,3,4,12,6),\; 
\boldt^{\bm m}_1 = (0, 1, -5,-4,4,-2,-8),\; 
\boldt^{\bm m}_2 = (0, -6, -5,3,-3,-9,-1), \notag \\
\boldt^{\bm m}_3 &= (0, 1, 9,3,-3,5,6),\; 
\boldt^{\bm m}_4 = (0, 8, 2,-4,4,5,-1),\; 
\boldt^{\bm m}_5 = (0, -6, -12,-4,-3,-9,-8), \notag \\
\boldt^{\bm m}_6 &= (0, -6, 2,3,-3,-2,6). \notag
\end{align}
\else
\begin{align}
\boldt^{\bm m}_0 &= (0, 8, 9,3,4,12,6), \notag \\
\boldt^{\bm m}_1 &= (0, 1, -5,-4,4,-2,-8), \notag \\
\boldt^{\bm m}_2 &= (0, -6, -5,3,-3,-9,-1), \notag \\
\boldt^{\bm m}_3 &= (0, 1, 9,3,-3,5,6), \notag \\
\boldt^{\bm m}_4 &= (0, 8, 2,-4,4,5,-1), \notag \\
\boldt^{\bm m}_5 &= (0, -6, -12,-4,-3,-9,-8), \notag \\
\boldt^{\bm m}_6 &= (0, -6, 2,3,-3,-2,6). \notag
\end{align}
\fi
\end{example}

When $\bm m$ is fixed in the respective context, we will  frequently omit the superscript, i.e., write $\boldt_k$ instead of $\boldt^{\bm m}_k$.

\revtwo{The remainder of this subsection establishes several technical properties of $\mathcal T^{\bm m}$ which are used later to construct inequalities and show that these are valid for $\P$.}
\begin{lemma}\label{lem:bb-properties}
\begin{enumerate}
  \item For any $k,j \in \F_p$, $[t_{k,j}]_p = j$ (\ie $t_{k,j} \bmod p = [j]_\Z $).
    \label{part:bb-property1}
  \item $t_{k,0} = 0$ for all $k \in \mathbb{F}_p$.
    \label{part:bb-property2}
  \item For $k \in \F_p$, let
    $\set(\bm t_k) = \{t_{k,j}\colon j \in \F_p\}$ be the unordered set of entries of $\boldt_k$. Then,
    \begin{equation} 
      \set(\bm t_k) = \set(\bm t_0) - t_{0,k} = \{ i - t_{0,k}\colon i \in \set(\bm t_0)\}. \label{eq:property-3}
    \end{equation}
    Hence, the entries (regardless of order) of different building blocks within each basic class differ only by a constant.
    \label{part:bb-property3}
  \item For all $k, i, j, l \in \F_p$ holds
  \begin{equation} \notag
    t_{k,i} - t_{k,j} = t_{k+l, i-l} - t_{k+l, j-l}
  \end{equation}
  \ie the relative placement of entries $i$ and $j$ of the building block $\boldt_k$ is the same as that of entries $i-l$ and $j-l$ of $\boldt_{k+l}$. \label{part:bb-property4}
\end{enumerate}
\end{lemma}
\begin{IEEEproof}
  For $k=0$, the first three statements are immediate for all $j \in \F_p$ by definition. For general $k \in \F_p$:
    \begin{enumerate}
      \item We have
      $[t_{k,j}]_p = [t_{0,j+k} - t_{0, k}]_p = j + k - k = j$.
      \item $t_{k,0} = t_{0, k} - t_{0,k} = 0$.
    \item  By definition of $t_{k,j}$,
    \ifonecolumn
      \begin{displaymath} \set(\bm t_k) = \{t_{k,j}\}_{j \in \F_p} = \{t_{0,j+k}\}_{j \in \F_p} - t_{0,k} = \{t_{0,j}\}_{j \in \F_p} - t_{0,k} = \set(\bm t_0) - t_{0,k}\end{displaymath}
      \else
      \begin{multline*} \set(\bm t_k) = \{t_{k,j}\}_{j \in \F_p} = \{t_{0,j+k}\}_{j \in \F_p} - t_{0,k} \\= \{t_{0,j}\}_{j \in \F_p} - t_{0,k} = \set(\bm t_0) - t_{0,k}\end{multline*}
      \fi
    where we have used that $\{j+k\}_{j \in \F_p} = \{j\}_{j \in \F_p}$.
  \end{enumerate}
  Finally, Property~\ref{part:bb-property4} holds because
  \begin{align*}
    t_{k+l, i-l} - t_{k+l,j-l} &= t_{0,k+i} - t_{0,k+l} - (t_{0,k+j} - t_{0,k+l}) \\
    &=t_{0,k+i} - t_{0,k+j} \\
    &=t_{0,k+i} - t_{0,k} - (t_{0,k+j} - t_{0,k}) \\
    &=t_{k,i} - t_{k,j}
  \end{align*}
  where both the first and last step are by definition of $t_{k,j}$.
\end{IEEEproof}

\begin{example}\label{ex:bb-properties}\begin{notms}
  For $p=5$ and $\bm m = (0,1,0,0,0)$, the statements of Lemma~\ref{lem:bb-properties} are shown in Fig.~\ref{fig:bb-properties} for some example values.\end{notms}
\end{example}
\begin{definition}\label{def:hilo-notation}
For any basic building block class $\mathcal T^{\bm m}$ and $k \in \F_p$, define
\ifonecolumn
\begin{displaymath}
  t_{k,\uparrow} = \argmax_{\zeta \in \F_p} t_{k,\zeta} \in \F_p 
\text{ and } t_{k,\downarrow} = \argmin_{\zeta \in \F_p} t_{k,\zeta} \in \F_p
\end{displaymath}
\else
\begin{align*}
  t_{k,\uparrow} &= \argmax_{\zeta \in \F_p} t_{k,\zeta} \in \F_p \\
\text{and}\quad  t_{k,\downarrow} &= \argmin_{\zeta \in \F_p} t_{k,\zeta} \in \F_p
\end{align*}
\fi
which is the (congruence class of the) \emph{index} of the largest and smallest entry, respectively, of $\bm t_k$. Further define, for given $\zeta \in \F_p$, the inverses of the above expressions as
\ifonecolumn
\begin{displaymath}
  t_{\uparrow, \zeta} \in \F_p \text{ (with } t_{\uparrow, \zeta} = k \Leftrightarrow\zeta = t_{k,\uparrow}\text{)} 
  \text{ and } t_{\downarrow, \zeta} \in \F_p \text{ (with }t_{\downarrow,\zeta} = k \Leftrightarrow\zeta = t_{k,\downarrow}\text{)}
\end{displaymath}
\else
\begin{align*}
  &t_{\uparrow, \zeta} \in \F_p \quad\text{(with } t_{\uparrow, \zeta} = k \Leftrightarrow\zeta = t_{k,\uparrow}\text{)}\\
  \text{and}\quad &t_{\downarrow, \zeta} \in \F_p \quad\text{(with }t_{\downarrow,\zeta} = k \Leftrightarrow\zeta = t_{k,\downarrow}\text{)}
\end{align*}
\fi
that tell, for given $\zeta$, in which block the $\zeta$-th entry is the maximizer (resp. minimizer) of $\set(\bm t_k)$. Finally,
\begin{equation} \notag
  \sigma = \sigma^{\bm m} =  t_{0,\uparrow} = \argmax_{j \in \F_p} t_{0,j} = \argmax_{j \in \F_p}([j]_\Z+m_j p) \neq 0
  \label{eq:sigma}
\end{equation}
which is a constant in $\F_p$ for $\bm m$ fixed.
\end{definition}
\begin{lemma}\label{lem:hilo-formulas}
  The definitions from Definition~\ref{def:hilo-notation} are indeed well-defined, \ie every building block has a unique largest and smallest entry and these are at different positions for the different building blocks within a class, and further admit the following explicit formulas for $k,\zeta \in \F_p$ (hence subtraction is in $\F_p$):
  \begin{align}
    t_{k,\uparrow} &= \sigma - k,
      \label{eq:zeta-hi}\\
    t_{k,\downarrow} &= -k,
      \label{eq:zeta-lo}\\
   t_{\uparrow,\zeta} &=\sigma - \zeta,
      \label{eq:k-hi}\\
   t_{\downarrow,\zeta} &=-\zeta.
      \label{eq:k-lo}
  \end{align}
  In addition, the largest and smallest value, respectively, within the building block $\bm t_k$ are explicitly given by
  \begin{align}
    &\max(\bm t_k) = \max (\set(\bm t_k)) = t_{0,\sigma} - t_{0, k}\label{eq:val-k-hi}\\
    \text{and}\quad&\min(\bm t_k) = \min(\set(\bm t_k)) = - t_{0,k}\label{eq:val-k-lo}
  \end{align}
  where $t_{0,\sigma} = \max(\bm t_0)$ by definition of $\sigma$.
\end{lemma}

\begin{IEEEproof}
  We consider the \enquote{$\max$}-cases (\eqref{eq:zeta-hi}, \eqref{eq:k-hi}, and \eqref{eq:val-k-hi}) only.
  By \eqref{eq:property-3},
  \[\max(\bm t_k) = \max (\bm t_0) - t_{0,k} = t_{0,\sigma} - t_{0,k}\]
  which shows \eqref{eq:val-k-hi}.
  Applying Property~\ref{part:bb-property1} of Lemma~\ref{lem:bb-properties} to the above equation shows that  $[\max(\bm t_k)]_p  = \sigma - k$, hence (by another application of that property) the maximizer is unique and must equal $\sigma - k$, i.e., \eqref{eq:zeta-hi} is correct. As this value is different for distinct values of $k\in\F_p$, the map $\F_p \ni k \mapsto t_{k,\uparrow} \in \F_p$ is bijective and hence admits an inverse. By resolving the expression for $k$, that inverse is seen to be $t_{\uparrow, \zeta} = \sigma - \zeta$, which proves \eqref{eq:k-hi}.
  
  The proofs for \eqref{eq:zeta-lo}, \eqref{eq:k-lo}, and \eqref{eq:val-k-lo} are completely analogous (note that the constant $t_{0,\downarrow}$ (which is the analog of $\sigma$) is zero and hence does not appear in the formulas).
\end{IEEEproof}
\begin{example}
  Let $p=5$ and $\bm m = (0,1,0,0,0)$, then $\sigma=1$ and $t_{0,\sigma} = 6$ is the largest entry of $\bm t_0$. For e.g.\ $k=3$, we have $t_{k,\uparrow} = [1]_5 - [3]_5 = [3]_5$ by \eqref{eq:zeta-hi} and $\max(\bm t_3) = t_{3,(t_{3,\uparrow})} = 6 - 3 = 3$ by \eqref{eq:val-k-hi}, while $t_{k,\downarrow} = -[3]_5 = [2]_5$ by \eqref{eq:zeta-lo} and $\min(\bm t_3) = t_{3,(t_{3,\downarrow})} = -3$ by \eqref{eq:val-k-lo}. These statements can be easily verified using the left-hand side of Fig.~\ref{fig:bb-properties}.
\end{example}

\begin{definition} \label{def:symmetric_class}
  A basic building block class $\mathcal T^{\bm m}$ is called \emph{symmetric} if  $\set(\bm t_0)= t_{0,\sigma} - \set(\bm t_0)$, i.e., if
  \begin{equation} i \in \set(\bm t_0) \Leftrightarrow t_{0,\sigma} - i \in \set(\bm t_0).\label{eq:symmetry} \end{equation}
\end{definition}

The building block classes from Examples~\ref{ex:bb-p7} and \ref{ex:bb-properties} are symmetric (the latter becomes obvious on the right of Fig.~\ref{fig:bb-properties}), while the one from Example~\ref{ex:bb-p3} is not: here, $\set(\bm t_0) = \{0,4,5\}$, $4 \in \set(\bm t_0)$, but $t_{0,\sigma} - 4 = 5-4 = 1$ is not.

\begin{lemma} \label{lem:bb-symmetric_properties}
Let $\mathcal{T}^{\bm m}$ be a basic building block class for a prime $p$. The following are equivalent:
\begin{enumerate}
  \item $\mathcal T^{\bm m}$ is symmetric. \label{lem:bb-symmetric-propertybase}
  \item $t_{\sigma, j} = -t_{0,-j}$ for all $j \in \F_p$.
    \label{lem:bb-symmetric-property1}
  \item If $p>2$, either $\bm m = \bm 0 = (0,\dotsc, 0)$, or $[\sigma]_\Z$ is odd and $m_i + m_{\sigma - i} = 1$ for $[i]_\Z \leq [\sigma]_\Z$.
    \label{lem:bb-symmetric-property3}
\end{enumerate}
\end{lemma}
\begin{IEEEproof}
By Property~\ref{part:bb-property1} of Lemma~\ref{lem:bb-properties}, \eqref{eq:symmetry} is equivalent to the condition that, for $i \in \F_p$,
\begin{equation}
  t_{0,\sigma} - t_{0,i} = t_{0,\sigma - i}.
  \label{eq:symmetry-alt}
\end{equation}
By Definition~\ref{def:bb}, the left-hand side equals $-t_{\sigma,i-\sigma}$, which shows the equivalence of \ref{lem:bb-symmetric-propertybase}) and \ref{lem:bb-symmetric-property1}) (using $j=i-\sigma$).

To show that \ref{lem:bb-symmetric-propertybase}) implies \ref{lem:bb-symmetric-property3}), let $p\geq 3$ and $\bm m \neq \bm 0$, hence $m_\sigma = 1$. Assume first that $[\sigma]_\Z = 2s$ is even, hence $t_{0,\sigma} = [\sigma]_\Z + pm_\sigma = [\sigma]_\Z + p = 2s + p$. Because this number is odd, $a \neq t_{0,\sigma} - a$ for all $a \in \set(\bm t_0)$, so that \eqref{eq:symmetry} partitions $\set(\bm t_0)$ into disjoint pairs, which contradicts  $\abs{\set(\bm t_0)} = p$ being odd.
Now, expand \eqref{eq:symmetry-alt} by Definition~\ref{def:bb} to obtain
\begin{align}
  &[\sigma]_\Z + p m_\sigma - [i]_\Z - p m_i = [\sigma - i]_\Z + p m_{\sigma - i} \notag \\
  \Leftrightarrow\quad
  &[\sigma]_\Z - [i]_\Z - [\sigma - i]_\Z + p(m_\sigma - m_i - m_{\sigma -i}) = 0 \notag \\
  \Leftrightarrow\quad
  &m_\sigma - m_i - m_{\sigma - i} - \left\{\begin{matrix}1 &\text{if }[\sigma]_\Z < [i]_\Z\\
  0 &\text{otherwise}
  \end{matrix}\right\} = 0,\label{eq:symmetry-alt2}
\end{align}
hence $m_i + m_{\sigma-i} = m_\sigma$ for $[i]_\Z \leq [\sigma]_\Z$, which shows \ref{lem:bb-symmetric-property3}).

\ref{lem:bb-symmetric-property3})$\Rightarrow$\ref{lem:bb-symmetric-propertybase}): For $\bm m =\bm 0$, $[\sigma]_\Z = p-1$, such that the braced expression in \eqref{eq:symmetry-alt2} is $0$ for all $i \in \F_p$, hence \eqref{eq:symmetry-alt2} holds for all $i$ and the class is symmetric. For $p=2$, there is only one additional basic class defined by $\bm m=(0,1)$, which implies $\set(\bm t_0) = \{0, 2\}$ and hence fulfills Definition~\ref{def:symmetric_class}. Thus assume that $[\sigma]_\Z$ is odd and $m_i + m_{\sigma - i} = 1$ for $[i]_\Z \leq [\sigma]_\Z$.

For $[i]_\Z \leq [\sigma]_\Z$, then, \eqref{eq:symmetry-alt2} holds by assumption. On the other hand, $[i]_\Z > [\sigma]_\Z$ implies $[\sigma - i]_\Z > [\sigma]_\Z$, such that $m_i = m_{\sigma - i} = 0$ by definition of $\sigma$, and \eqref{eq:symmetry-alt2} holds as well, which concludes the proof.
\end{IEEEproof}

\begin{corollary}
  For $p \geq 3$, there are $2^{(p-1)/2}$ symmetric building block classes.
\end{corollary}
\begin{IEEEproof}
  Besides $\mathcal T^{\bm 0}$ which is always symmetric, the first statement of Property~\ref{lem:bb-symmetric-property3} of the above lemma implies that symmetric classes exist only for $[\sigma]_\Z = 2s+1$ with $0 \leq s \leq (p-3)/2$. By the second part, only half of the $m_i$ for $[0]_\Z < [i]_\Z < [\sigma]_\Z$ are free to choose, which gives $2^s = 2^{([\sigma]_\Z - 1)/2}$ symmetric classes for a single $\sigma$. In total (and including $\mathcal T^{\bm 0}$) this results in
  $1 + \sum_{s=0}^{(p-3)/2} 2^s = 2^{(p-1)/2}$
  classes.
\end{IEEEproof}

We now define \emph{almost doubly-symmetric classes} which will become important in Section~\ref{sec:facet-defining}. \revtwo{These classes possess certain technical symmetry properties that makes it possible to prove that the resulting inequalities are facet-defining.}
\begin{definition} \label{def:almost_double_symmetric}
A symmetric basic building block class $\mathcal{T}^{\bm m}$ for $p \geq 3$ is called \emph{almost doubly-symmetric} if its ``projection'' onto the ``interior'' entries, \ie the building block class obtained by removing both the smallest and largest entries from each building block of the class, has the property that there exists a subset $\tilde{T}_0^{\text{proj}} \subset T_0^{\text{proj}}$ such that 
\begin{subequations}
\begin{align}
&\abs{\tilde{T}_0^{\text{proj}}} \geq (p-3)/2, \label{eq:condition_dsymmetrix_1} \\
&\max \left( \tilde{T}_0^{\text{proj}}\right) \leq \left\lfloor \smax(\bm t_0)/2 \right\rfloor, \label{eq:condition_dsymmetrix_2} \\
\text{and}\quad&i \in \tilde{T}_0^{\text{proj}} \Rightarrow \smax(\bm t_0) - i \in T_0^{\text{proj}}   \label{eq:condition_dsymmetrix_3}
\end{align} \label{eq:condition_dsymmetrix}%
\end{subequations}%
where $T_0^{\text{proj}} =  \set(\bm t_0) \setminus \{ 0, t_{0,\sigma}\}$ (\ie the projection of $\set(\bm t_0)$ onto the interior entries) and $\smax(\bm t_k)$ is the second-largest entry of $\boldt_k$, \ie
\begin{displaymath}
 \smax(\bm t_k)= \max ( \set(\bm t_k) \setminus \{ \max(\boldt_k) \}).
\end{displaymath}
\end{definition}

\begin{example} \label{ex:doubly-symmetric}
The building block class from Example~\ref{ex:bb-properties} is almost doubly-symmetric: here, $\set(\bm t_0) = \{0,2,3,4,6\}$, $T_0^{\text{proj}} = \{2,3,4\}$, and $\smax(\bm t_0) = 4$. It is easy to check that $\tilde T_0^{\text{proj}} = \{2\}$ fulfills \eqref{eq:condition_dsymmetrix}. In contrast, the class from Example~\ref{ex:bb-p7} is not almost doubly-symmetric: here, $\set(\bm t_0)  = \{0,3,4,6,8,9,12\}$, $T_0^{\text{proj}}  = \{3,4,6,8,9\}$, and the largest subset $\tilde{T}_0^{\text{proj}} \subset  T_0^{\text{proj}}$ that satisfies \eqref{eq:condition_dsymmetrix_2} and \eqref{eq:condition_dsymmetrix_3} is $\tilde{T}_0^{\text{proj}} = \{3\}$ which does not satisfy \eqref{eq:condition_dsymmetrix_1}.
\end{example}

\subsection{Deriving Inequalities From Building Blocks}\label{sec:hilo}
\revtwo{In this section, we present a construction of inequalities derived from a given basic building block class $\mathcal T^{\bm m}$; these inequalities will later be shown to be valid or even facet-defining for $\P$, given that certain conditions on $\mathcal T^{\bm m}$ are satisfied.

For each such inequality $\boldtheta^T \boldx \leq \kappa$, $\boldtheta \in \R^{dp}$ is of the form $\boldtheta = (\boldt_{k_1} \mid \dotsc \mid \boldt_{k_d} )^T$, where each $\boldt_{k_i}  \in \mathcal T^{\bm m}$  for some fixed $\bm m$. In particular, $\boldt_{k_1},\dotsc, \boldt_{k_{d-1}}$ can be arbitrarily chosen, while  $\boldt_{k_d}$ is a function of $\boldt_{k_1},\dotsc, \boldt_{k_{d-1}}$. %
The right-hand side $\kappa$ of the inequality is found by imposing that  the inequality is tight for a specially designed embedded codeword $\Fv(\boldc)$ (the so-called canonical codeword). } For any codeword $\boldc =(c_1,\dotsc, c_d)\eiriknewnew{^T} \in \C$, the left-hand side of $\boldtheta^T \Fv(\boldc) \leq \kappa$ is
\begin{equation}
  \sum_{i=1}^d \boldt_{k_i} \f(c_i)^T = \sum_{i=1}^d t_{k_i,c_i}
  \label{eq:ineqVals}
\end{equation}
because $\f(c_i)$ is the $c_i$-th unit vector \eiriknew{in $\R^p$} by Definition~\ref{def:Constant}. %

\begin{construction}\label{constr:hilo}
Choose and fix $k_1,\dotsc,k_{d-1}$ arbitrarily from $\F_p$, \revtwo{which define the first $d-1$ parts of $\boldtheta$ in the above form. Then, the \emph{canonical codeword} $\boldc$ for $k_1,\dotsc,k_{d-1}$ is defined as follows. The first $d-1$ entries of $\boldc$ are chosen to maximize $t_{k_i,c_i}$}, \ie
\begin{subequations}
\begin{align}
  &c_i = t_{k_i,\uparrow} = \sigma - k_i \label{eq:hilocons-ci}\\
  \text{(and hence}\;&k_i = t_{\uparrow, c_i} = \sigma - c_i  \text{)}\label{eq:hilocons-ki}
\end{align}
\end{subequations}
for $i \in \range{d-1}$. The condition $\boldc \in \C$ then uniquely specifies the last entry $c_d$ of the codeword. Now, $k_d$ is chosen such that $t_{k_d, c_d}$ is minimized, \ie
\begin{subequations}
\begin{align}
  &k_d = t_{\downarrow, c_d} = - c_d \label{eq:hilocons-kd}\\
  \text{(and hence}\;&c_d = t_{k_d,\downarrow} = -k_d\text{)} \label{eq:hilocons-cd}
\end{align}
\end{subequations}
\revtwo{by which $\boldtheta = (\boldt_{k_1} \mid \dotsc \mid \boldt_{k_d} )^T$ is completely specified}. Finally, the right-hand side $\kappa$ is defined as $\boldtheta^T \Fv(\boldc)= \sum_{i=1}^d t_{k_i,c_i}$ which ensures that $\Fv(\boldc)$ is tight for the resulting inequality $\boldtheta^T \boldx \leq \kappa$.
\end{construction}
\begin{remark}\label{rem:hilo-arbitrary}
  The choice of $d$ in the above construction is arbitrary; using any other position $i \in \range d$ to define the last entry $c_i$ of $\bm c$ and the last building block $\bm t_i$ leads to a different mapping of inequalities to canonical codewords, but the set of inequalities constructed in total remains the same.
\end{remark}
\begin{corollary}\label{cor:kappaModp}
  For any inequality $\boldtheta^T \boldx \leq \kappa$ obtained from Construction~\ref{constr:hilo}, $[\kappa]_p = 0$.
\end{corollary}
\begin{IEEEproof}
  By construction, $[\kappa]_p = \sum_{i=1}^d [t_{k_i, c_i}]_p = \sum_{i=1}^d c_i$ by Property~\ref{part:bb-property1} of Lemma~\ref{lem:bb-properties}, and $\sum c_i = 0$ because $\boldc \in \C$.
\end{IEEEproof}

\begin{example}
  Let  $p=5$ and $\bm m = (0,1,0,0,0)$ as before, $d=6$, and choose  $(k_1,\dotsc,k_{5}) = (0,1,2,3,4)$. Then, the first $5$ entries of the canonical codeword $\boldc$ are $(t_{0,\uparrow}, \dotsc, t_{4,\uparrow}) = (1, 0, 4, 3, 2)$ by \eqref{eq:zeta-hi}. As $\sum_{i=1}^{d-1} t_{k_i,\uparrow} = 0$, this implies $c_6 = 0$, such that $\boldc=(1,0,4,3,2,0)\eiriknewnew{^T} \in \C$, and we set $k_6 = t_{\downarrow, c_6} = 0$  by \eqref{eq:zeta-lo}, hence $\boldtheta = (\bm t_0^{\bm m} \mid \bm t_1^{\bm m} \mid \bm t_2^{\bm m} \mid \bm t_3^{\bm m} \mid \bm t_4^{\bm m} \mid  \bm t_0^{\bm m})^T$. Finally, we can compute $\kappa = \boldtheta^T \Fv(\boldc) = t_{0,1} + t_{1,0} + t_{2,4} + t_{3,3} + t_{4,2} + t_{0,0} = 6 + 0 + 4 + 3 + 2 + 0 = 15$ and obtain the inequality $(\boldt^{\bm m}_0 \mid \boldt^{\bm m}_1 \mid \boldt^{\bm m}_2 \mid \boldt^{\bm m}_3 \mid\boldt^{\bm m}_4 \mid \boldt^{\bm m}_0) \boldx \leq 15$.
\end{example}

Note that by Construction~\ref{constr:hilo}, a total of $p^{d-1}$ inequalities can be constructed per class $\mathcal T^{\bm m}$. We denote the set of these inequalities by $\Theta^{\bm m}$. The following lemma states two alternative characterizations of the elements of $\Theta^{\bm m}$.
\begin{lemma}\label{lem:conditions}
  An inequality
  $\boldtheta^T \boldx \leq \kappa$ with $\boldtheta=(\bm t_{k_1} \mid \dotsc \mid \bm t_{k_d})^T$ is in $\Theta^{\bm m}$ if and only if
  \begin{subequations} \label{eq:hilo-conditions}
  \begin{align}
    &\sum_{i=1}^d k_i = [d-1]_p \cdot \sigma\label{eq:Thetacondition}\\
    \text{and}\quad&\kappa = (d-1) t_{0,\sigma} - \sum_{i=1}^d t_{0,k_i} \label{eq:kappacondition}
  \end{align}
  \end{subequations}
  which is in turn equivalent to the condition that
    \begin{subequations} \label{eq:hilo-conditions'}
    \begin{align}
      &\sum_{k\in \F_p} \left[\abs{V^\boldtheta_k}\right]_p (\sigma - k) = \sigma \label{eq:Thetacondition'}\\
      \text{and}\quad&\kappa = \sum_{k\in \F_p} \abs{V^\boldtheta_k} \max(\bm t_k) - \max(\bm t_0)\label{eq:kappacondition'}
    \end{align}
    \end{subequations}
    where, for $k \in \F_p$, $V^\boldtheta_k = \{i \in \range d\colon k_i = k\}$ denotes the index set of entries in $\boldtheta$ that equal $\bm t_k$.
\end{lemma}
\begin{IEEEproof}
  See Appendix~\ref{app:proofConditions}.
\end{IEEEproof}

\begin{remark} \label{rem:Thetamunique}
  Note that no two inequalities constructed by Construction~\ref{constr:hilo} from distinct $(k_1^1,\dotsc,k^1_{d-1}) \neq (k_1^2,\dotsc,k^2_{d-1})$ for the same class $\mathcal T^{\bm m}$ are equivalent, in the sense that none is a positive scalar multiple of another. Assume for the contrary that $(\boldtheta^1 = (\bm t_{k^1_1} \mid \dotsc \mid \bm t_{k^1_d})^T, \kappa^1)$ and $(\boldtheta^2=(\bm t_{k^2_1} \mid \dotsc \mid \bm t_{k^2_d})^T,\kappa^2)$ are constructed from Construction~\ref{constr:hilo} with $\boldtheta^1 = a \boldtheta^2$, $\kappa^1 = a \kappa^2$, and $a \geq 0$. If $a=1$, then $k^1_i = k^2_i$ for $i \in \range{d-1}$. Otherwise, $\boldt_{k^1_1} = a \boldt_{k^2_1}$ with $a \neq 1$, which is a contradiction: since by Property~\ref{part:bb-property3} of Lemma~\ref{lem:bb-properties}, the difference between the largest and smallest element of a building block is constant among a fixed class, no building block can be a proper scalar multiple of another.
\end{remark}

\subsection{Valid and Invalid Building Block Classes}\label{sec:validInvalid}
In this subsection, we show that in a class $\Theta^{\bm m}$ of inequalities, either \emph{all} inequalities are valid or \emph{all} are invalid for $\P$, and that this distinction depends only on $\bm m$, not on the code's length $d$. The key argument is given by the following lemma. The proofs of all results of this subsection \eirik{(except Proposition~\ref{prop:specialMvalid})} are given in Appendix~\ref{app:validInvalid}.
\begin{lemma}\label{lem:validIndependent}
  Let $\boldtheta^T \boldx \leq \kappa$ with $\boldtheta = (\bm t_{k_1} \mid \dotsc \mid \bm t_{k_d})^T$ be an inequality in $\Theta^{\bm m}$ and let $\boldc \in \C$ be the corresponding canonical codeword (cf.\ Construction~\ref{constr:hilo}). Then, for any $\bm\xi \in \F_p^d$,
  \begin{equation} \label{eq:tightness}
  \boldtheta^T \Fv(\bm c + \bm\xi) - \kappa = \sum_{i=1}^{d-1} t_{\sigma, \xi_i} + t_{0, \xi_d}.
  \end{equation}
  In particular, the change to the left-hand side of the inequality $\boldtheta^T\Fv(\boldc) \leq \kappa$ induced by adding $\bm\xi$ to the canonical codeword only depends on $\bm\xi$; it is independent of $\boldtheta$, $\kappa$, and $\boldc$, \ie independent of which inequality was chosen.
\end{lemma}

\begin{corollary}\label{cor:allValidOrNot}
  Let $\mathcal T^{\bm m} = \{ \boldt_k^{\bm m}\}_{k \in \F_p}$ be a basic building block class.
  \begin{enumerate}
    \item If, for all $\bm c \in \mathcal{C}$,
    \begin{equation}
      \sum_{i=1}^{d-1} t_{\sigma, c_i} + t_{0,c_d} \leq 0,
      \label{eq:allValidCondition}
    \end{equation}
    then all inequalities in $\Theta^{\bm m}$ are valid for $\P$. \label{cor:allValidOrNot-1}
    \item Conversely, if there is a codeword $\boldc \in \C$  such that $\sum_{i=1}^{d-1} t_{\sigma, c_i} + t_{0, c_d} > 0$, then no inequality in $\Theta^{\bm m}$ is valid for $\P$.
  \end{enumerate}
\end{corollary}

\revtwo{In the following, we establish a simpler condition for all inequalities in $\bm\Theta^{\bm m}$ being valid.}
\begin{definition}\label{def:valid-class}
  The basic building block class $\mathcal T^{\bm m} = \{ \boldt_k^{\bm m}\}_{k \in \F_p}$ is called \emph{valid} if the equation
  \begin{equation}
    \sum_{i\in I} n_i t_{\sigma,i} + [r]_\Z = 0
    \label{eq:valid-class-condition}
  \end{equation}
  with $I = \{i \in \F_p\colon 0 > t_{\sigma,i} \geq -[\sigma]_\Z\}$, nonnegative integer variables $n_i$, and
   $ r = - \sum_{i\in I} [n_i]_p\cdot i$
  has no solution for which $m_r = 1$.
\end{definition}
\begin{theorem}\label{thm:newValidProgram}
  If the class $\mathcal T^{\bm m}$ is valid, then all inequalities in $\Theta^{\bm m}$ are valid for $\P$ (independently of $d$). If $\mathcal T^{\bm m}$ is not valid, there is a $d_0 \leq [\sigma]_\Z+1$ such that all inequalities in $\Theta^{\bm m}$ are invalid for $\P$ if $d \geq d_0$.
\end{theorem}
\begin{example}
  Let $p=3$ and $\bm m = (0,1,1)$. Then, $\boldt^{\bm m}_0 = (0, 4, 5)$, $\boldt^{\bm m}_1 = (0, 1, -4)$, and $\boldt^{\bm m}_2 = (0, -5, -1)$. Thus, $\sigma = t_{0,\uparrow} = 2$ and $\boldt^{\bm m}_{\sigma} = (0, -5, -1)$.  Now, $I = \{2\}$, and with $n_2=2$, we obtain $r=-[2]_p\cdot [2]_p = [2]_p$, which satisfies both $m_r = m_2 = 1$ and \eqref{eq:valid-class-condition} because $n_2\cdot t_{2,2} + [r]_\Z = 2 \cdot (-1) + 2 = 0$.    Hence, Definition~\ref{def:valid-class} and Theorem~\ref{thm:newValidProgram} tell us that this class is invalid for $d \geq 3$. Indeed, for $d=3$ and the canonical codeword $\bm 0 \in \C$, we obtain $\boldtheta = (0, -5, -1, 0, -5, -1, 0, 4, 5)^T$ and $\kappa = 0$. However, the resulting inequality is violated by the codeword $\boldc = (2,2,2)\eiriknewnew{^T} \in \C$, as $\boldtheta^T \Fv(\boldc) = -1 + (-1) + 5 = 3 > 0 = \kappa$.
\end{example}
\begin{remark} \label{rem:search_m1}
  The condition that $m_r=1$ in Definition~\ref{def:valid-class} implies that $[r]_\Z \leq [\sigma]_\Z$ in \eqref{eq:valid-class-condition}. Since also $t_{\sigma,i} < 0$ for $i \in I$, the number of potential solutions to \eqref{eq:valid-class-condition} is relatively small. We have verified that a simple enumeration runs in negligible time for all  $p$ of reasonable size. The first row of Table~\ref{table:classNumbers} lists the number of classes that pass the test for all primes $p \leq 19$.
\end{remark}

\ifonecolumn
\begin{table}
  \caption{Number of valid basic building block classes for different values of $p$. Also, the numbers of unique (cf. Section~\ref{sec:redundant}), unique symmetric, almost doubly-symmetric, and facet-defining  valid classes are given.}
  \label{table:classNumbers}
  \centering
 \vskip -2.0ex
  \begin{tabular}{lcccccccc}
    \toprule
    $p$: &2 & 3 & 5 & 7 & 11 & 13 & 17 & 19 \\ \midrule
    valid classes: & 2 & 3 & 7 & 17 & 109 & 261 & 1621 & 4085 \\
    unique valid: & 1 & 2 & 6 & 16 & 108 & 260 & 1620 & 4084\\
    of which are \dots \\
    - symmetric: & 1 & 1 & 2 & 4 & 10 & 16 & 31 & 46\\
    - almost doubly-symmetric: & 0 & 1 & 2 & 3 & 5 & 6 & 8 & 9
    \\ 
    - facet-defining: & 1 & 1 & 2 & 4 & 10 & 16 & 31 & 46\\
    \bottomrule
  \end{tabular}
\end{table}
\else
\begin{table}
  \caption{Number of valid basic building block classes for different values of $p$. Also, the numbers of unique (cf. Section~\ref{sec:redundant}), unique symmetric, almost doubly-symmetric, and facet-defining  valid classes are given.}
  \label{table:classNumbers}
  \centering
 \vskip -2.0ex
  \begin{tabular}{lcccccccc}
    \toprule
    $p$: &2 & 3 & 5 & 7 & 11 & 13 & 17 & 19 \\ \midrule
    valid classes: & 2 & 3 & 7 & 17 & 109 & 261 & 1621 & 4085 \\
    unique valid: & 1 & 2 & 6 & 16 & 108 & 260 & 1620 & 4084\\
    of which are \dots \\
    - symmetric: & 1 & 1 & 2 & 4 & 10 & 16 & 31 & 46\\
    - almost doubly-\\
    \phantom{- }symmetric: & 0 & 1 & 2 & 3 & 5 & 6 & 8 & 9
    \\ 
    - facet-defining: & 1 & 1 & 2 & 4 & 10 & 16 & 31 & 46\\
    \bottomrule
  \end{tabular}
\end{table}
\fi

\revtwo{For symmetric classes, the conditions can be further simplified.}
\begin{lemma}\label{lem:valid-symmetric-condition}
  If $\mathcal T^{\bm m}$ is symmetric, then $\mathcal T^{\bm m}$ is valid if and only if the equation
  \begin{equation} \notag
    \sum_{j \in J} \nu_j\cdot [j]_\Z  = [\rho]_\Z
  \end{equation}
  with $J = \{j \in \F_p\colon m_j = 0\text{ and }0 < [j]_\Z < [\sigma]_\Z\}$, \eirik{nonnegative integer variables $\nu_j$, and $\rho = - \sum_{j\in J} [\nu_j]_p\cdot j$
  has no solution for which $m_\rho = 1$.}  
\end{lemma}
\begin{remark} \label{rem:search_m2}
  The conditions of Lemma~\ref{lem:valid-symmetric-condition} depend on $\sigma$ and $m_0,\dotsc,m_\sigma$ only; in particular, they are independent of $p$. Hence, once such an $\bm m$-vector prefix has been determined to be valid, a valid symmetric class is obtained for \emph{any} prime $p > \sigma$ by appending an appropriate number of zeros. Table~\ref{table:facetClasses} lists the valid prefixes for $\sigma \leq 17$.
\end{remark}

\begin{proposition}\label{prop:specialMvalid}
  For any $p$, let $\bm m$ be of the form $(0,1,\dotsc,0,1,0,\dotsc,0)$, \ie consisting of $s$ copies of $(0,1)$, for an arbitrary $0 \leq s \leq \lceil(p-1)/2\rceil$, followed by zeros (note that this includes the all-zero $\bm m$-vector). Then, $\mathcal T^{\bm m}$ is a valid symmetric class, which additionally is almost doubly-symmetric for $p\geq 3$.\footnote{Note that a similar result with an upper bound of $(p-1)/2$ on $s$ (without the ceiling operator) was stated in \cite[Prop.~2]{ros16_2}. That result is slightly weaker since it does not include $s=1$ for $p=2$. However, the resulting class turns out to be redundant (see Section~\ref{sec:redundant} below).}
\end{proposition}
\begin{IEEEproof}
  \eirik{Assume $p \geq 3$.} Every class of the above form is symmetric by Item~\ref{lem:bb-symmetric-property3} of Lemma~\ref{lem:bb-symmetric_properties}, such that Lemma~\ref{lem:valid-symmetric-condition} can be applied. Assume $\mathcal T^{\bm m}$ is invalid. The condition $m_\rho = 1$ implies that $s \neq 0$, hence $[\sigma]_\Z$ is odd, while $J$ contains entries with even integer representations only; hence $\sum_{j \in J} \nu_j \cdot [j]_\Z$ is even and the condition in Lemma~\ref{lem:valid-symmetric-condition} is not satisfiable. \eirik{For $p=2$, $\bm m = (0,0)$ or $\bm m = (0,1)$. In both cases, $\mathcal{T}^{\bm m}$ is symmetric since \eqref{eq:symmetry} of Definition~\ref{def:symmetric_class} is satisfied, and we can apply Lemma~\ref{lem:valid-symmetric-condition}.} \eirikNew{For $\bm m=(0,1)$, the same argument above shows that $\mathcal{T}^{(0,1)}$ is valid. For $\bm m = (0,0)$, the condition $m_\rho=1$ of Lemma~\ref{lem:valid-symmetric-condition} cannot be fulfilled, and thus it follows that $\mathcal{T}^{(0,0)}$ is valid as well.}

  For $p \geq 3$ and $\bm 0 \neq \bm m$ of the above form, let $\tilde T_0^{\text{proj}}$ consist of the smallest $(p-3)/2$ elements of $\set(\bm t_0) \setminus \{0\}$. By the form of $\bm m$, this implies $\tilde T_0^{\text{proj}} \subseteq \{t_{0,i}\colon m_i=0\}$, and one can show (details omitted) that this set fulfills  \eqref{eq:condition_dsymmetrix_2} and \eqref{eq:condition_dsymmetrix_3}, hence $\mathcal T^{\bm m}$ is almost doubly-symmetric. 
  
  \begin{notms}
  \eirik{For $p \geq 3$ and $\bm m = \bm 0$, $T_0^{\text{proj}} = \{1,\dotsc,p-2\}$, and we let $\tilde T_0^{\text{proj}}=\{1,\dotsc,(p-3)/2\}$. Since $\left\lfloor \smax(\bm t_0)/2 \right\rfloor = \left\lfloor (p-2)/2 \right\rfloor = (p-3)/2 = \max\left(\tilde T_0^{\text{proj}} \right)$ and $\left\{\smax(\bm t_0)-i: i \in \tilde T_0^{\text{proj}} \right\}= \{(p-1)/2,\dotsc,p-3\} \subset T_0^{\text{proj}}$, it follows that both \eqref{eq:condition_dsymmetrix_2} and \eqref{eq:condition_dsymmetrix_3} are fulfilled, hence $\mathcal T^{\bm m}$ is almost doubly-symmetric, which concludes the proof.  
}  \end{notms}
\end{IEEEproof}

\ifonecolumn
\begin{table}
  \caption{Prefixes of $\bm m$-vectors for which the corresponding building block class $\mathcal T^{\bm m}$ is symmetric and valid for any prime $p$. For $p \leq 19$, these are exactly the prefixes that lead to valid facet-defining classes. The entries corresponding to Proposition~\ref{prop:specialMvalid} are printed in bold.}
  \label{table:facetClasses}
 \vskip -2.0ex
  \begin{tabular}{rlrl}
    \toprule
    $\sigma$ &valid $\bm m$ prefixes &$\sigma$&valid $\bm m$ prefixes\\ \midrule
    1 &$\mathbf{01}$ & 3 & $\mathbf{0101}$\\
    5 & $\mathbf{010101}$, $011001$  & 7 & $\mathbf{01010101}$, $01101001$, $01110001$\\
    9 & \multicolumn{3}{l}{
    $\mathbf{0101010101}$,
    $0111010001$,
    $0111100001$}\\
    11 & \multicolumn{3}{l}{
    $\mathbf{010101010101}$,
    $011011001001$,
    $011100110001$,
     $011101010001$,
     $011110100001$,
     $011111000001$}\\
   13 & \multicolumn{3}{l}{
   $\mathbf{01010101010101}$,
   $01101101001001$,
   $01110101010001$,
   $01110110010001$,
   $01111001100001$,
   $01111010100001$,}\\
   &\multicolumn{3}{l}{
   $01111101000001$,
   $01111110000001$}\\
   15&\multicolumn{3}{l}{
   $\mathbf{0101010101010101}$,
   $0111010101010001$,
   $0111011100010001$,
   $0111110011000001$,
   $0111110101000001$,}\\
   &\multicolumn{3}{l}{
   $0111111010000001$,
   $0111111100000001$}\\
   17&\multicolumn{3}{l}{
   $\mathbf{010101010101010101}$,
   $011011011001001001$,
   $011101010101010001$,
   $011101100110010001$,$011101110100010001$}\\
   &\multicolumn{3}{l}{
   $011110110100100001$,
   $011110111000100001$,
   $011111000111000001$,$011111001011000001$,
   $011111010101000001$,}\\
   &\multicolumn{3}{l}{
   $011111011001000001$,
   $011111100110000001$,
   $011111101010000001$,
   $011111110100000001$,$011111111000000001$}\\
   $p-1$&$\bm 0$ (all-zero $\bm m$-vector) \\
    \bottomrule  
  \end{tabular}
\end{table}
\else
\begin{table}
  \caption{Prefixes of $\bm m$-vectors for which the corresponding building block class $\mathcal T^{\bm m}$ is symmetric and valid for any prime $p$. For $p \leq 19$, these are exactly the prefixes that lead to valid facet-defining classes. The entries corresponding to Proposition~\ref{prop:specialMvalid} are printed in bold.}
  \label{table:facetClasses}
 \vskip -2.0ex
  \begin{tabular}{rlrl}
    \toprule
    $\sigma$ &valid $\bm m$ prefixes &$\sigma$&valid $\bm m$ prefixes\\ \midrule
    1 &$\mathbf{01}$ & 3 & $\mathbf{0101}$\\
    5 & $\mathbf{010101}$, $011001$  & 7 & $\mathbf{01010101}$, $01101001$, \\
      & & & $01110001$\\
    9 & \multicolumn{3}{l}{
    $\mathbf{0101010101}$,
    $0111010001$,
    $0111100001$}\\
    11 & \multicolumn{3}{l}{
    $\mathbf{010101010101}$,
    $011011001001$,
    $011100110001$,}\\
     & \multicolumn{3}{l}{
     $011101010001$,
     $011110100001$,
     $011111000001$}\\
   13 & \multicolumn{3}{l}{
   $\mathbf{01010101010101}$,
   $01101101001001$,
   $01110101010001$,}\\
   &\multicolumn{3}{l}{
   $01110110010001$,
   $01111001100001$,
   $01111010100001$,}\\
   &\multicolumn{3}{l}{
   $01111101000001$,
   $01111110000001$}\\
   15&\multicolumn{3}{l}{
   $\mathbf{0101010101010101}$,
   $0111010101010001$,}\\
   &\multicolumn{3}{l}{
   $0111011100010001$,
   $0111110011000001$,}\\
   &\multicolumn{3}{l}{
   $0111110101000001$,
   $0111111010000001$,}\\
   &\multicolumn{3}{l}{$0111111100000001$}\\
   17&\multicolumn{3}{l}{
   $\mathbf{010101010101010101}$,
   $011011011001001001$,}\\ 
   &\multicolumn{3}{l}{
   $011101010101010001$,
   $011101100110010001$,}\\
   &\multicolumn{3}{l}{
   $011101110100010001$,
   $011110110100100001$,}\\
   &\multicolumn{3}{l}{
   $011110111000100001$,
   $011111000111000001$,}\\
   &\multicolumn{3}{l}{
   $011111001011000001$,
   $011111010101000001$,}\\
   &\multicolumn{3}{l}{
   $011111011001000001$,
   $011111100110000001$,}\\
   &\multicolumn{3}{l}{
   $011111101010000001$,
   $011111110100000001$,}\\
   &\multicolumn{3}{l}{$011111111000000001$}\\
   $p-1$&$\bm 0$ (all-zero $\bm m$-vector) \\
    \bottomrule  
  \end{tabular}
\end{table}
\fi

\revtwo{The following lemma provides counting formulas for the number of embeddings of vectors that are not codewords of an \enquote{all-ones} SPC code $\C$ cut by any inequality from a valid basic building block class and also for the number of embedded codewords of $\C$  for which the inequality is tight.}

\begin{lemma} \label{lem:counting_formulas}
Let $\C$ be an \enquote{all-ones} SPC code over $\mathbb{F}_p$ of length $d \geq 2$, and let $\mathcal{T}^{\bm m}$ be a basic valid building block class. Define
\begin{displaymath}
I^>_{c,J} = \begin{cases}
1 & \text{if $t_{0,c} + \sum_{j \in J} t_{\sigma,j} > 0$},\\
0 & \text{otherwise}
\end{cases}
\end{displaymath}
where $c \in \F_p$ and $J$ is a multiset of $\F_p \setminus \{ 0 \}$ (including the empty set). Furthermore, let 
\begin{displaymath}
I^=_{c,J} = \begin{cases}
1 & \text{if $c + \| J \|_1 = 0$}\\
& \text{and $t_{0,c} + \sum_{j \in J} t_{\sigma,j} = 0$},\\
0 & \text{otherwise}
\end{cases}
\end{displaymath}
where $\| J \|_1$ denotes the sum (in $\mathbb{F}_p$) of the entries of $J$. 
Now, any inequality $\boldtheta^T \boldx \leq \kappa$ from $\Theta^{\bm m}$
\begin{enumerate}
\item cuts the embeddings of
\begin{equation} \label{eq:noncodewords_q}
\sum_{\substack{c \in \F_p,\,J \text{ multiset of }\F_p  \setminus \{ 0 \}, \\ \eirik{|J| \leq d-1}}} I^>_{c,J}  {{d-1} \choose {|J|}} \frac{|J|!}{n_1^J! \cdots n_{k^J}^J!}
\end{equation}
elements $\boldzeta \in \mathbb{F}_p^d \setminus \C$, where $k^J$ is the number of \emph{types} (number of different elements) of the multiset $J$ with
repetition numbers  $n_1^J,\dotsc,n_{k^J}^J$, and %
\label{part:valid2_q}
    \item is tight for the embeddings of exactly
\begin{equation} \label{eq:codewords_q}
\sum_{\substack{c \in \F_p,\,J \text{ multiset of }\F_p  \setminus \{ 0 \}, \\ \eirik{|J| \leq d-1}}} I^=_{c,J}  {{d-1} \choose {|J|}}  \frac{|J|!}{n_1^J! \cdots n_{k^J}^J!}
\end{equation}
codewords of $\C$. %
\label{part:valid3_q}
\end{enumerate}
\end{lemma}

\revtwo{
\begin{example} \label{ex:11}
Let $\C$ be an \enquote{all-ones} SPC code over $\mathbb{F}_3$ of length $d = 3$. In this case, $\mathcal{T}^{\bm 0} = \{(0, 1, 2),(0, 1,-1),(0, -2, -1)\}$, $\sigma=2$, and the possible mulitsets $J$ are
\begin{align*}
\{ \}, 
\{ 1\}, \{ 2\}, 
\{1,1\},\{1,2\},\{2,2\}.
\end{align*}
It follows that
\begin{displaymath}
\begin{array}{lll}
I^>_{0,\{\;\,\}} = 0, &I^>_{1,\{\;\,\}} = 1, &I^>_{2,\{\;\,\}} = 1,\\
I^>_{0,\{1\}} = 0, &I^>_{1,\{1\}} = 0, &I^>_{2,\{1\}} = 0,\\
I^>_{0,\{2\}} = 0, &I^>_{1,\{2\}} = 0, &I^>_{2,\{2\}} = 1,\\
\end{array}
\end{displaymath}
and
\begin{displaymath}
I^>_{c,\{1,1\}} = I^>_{c,\{1,2\}} = I^>_{c,\{2,2\}}=0,\; \forall c \in \mathbb{F}_3.
\end{displaymath}
Applying the counting formula in \eqref{eq:noncodewords_q}, we get that any inequality $\boldtheta^T \boldx \leq \kappa$ from $\Theta^{\bm m}$
 cuts the embeddings of
\begin{equation*}
2 + (d-1) = d+1 = 4
\end{equation*}
elements $\boldzeta \in \mathbb{F}_3^3 \setminus \C$. Similarly, we get
\begin{displaymath}
I^=_{0,\{\}}=I^=_{1,\{2\}} =I^=_{2,\{1\}}=I^=_{2,\{2,2\}}=1
\end{displaymath}
while all other $I^=_{c,J}$ are zero. Hence, from \eqref{eq:codewords_q}, any inequality $\boldtheta^T \boldx \leq \kappa$ from $\Theta^{\bm m}$ is tight for the embeddings of exactly
\begin{equation*}
1 + 2(d-1) + {d-1 \choose {2} } = \frac{d(d+1)}{2} = 6
\end{equation*}
codewords of $\C$. %
\end{example}
}

\subsection{Facet-Defining Valid Building Block Classes}\label{sec:facet-defining}

In this subsection, we state several results regarding the dimension of the face defined by an inequality from $\Theta^{\bm m}$ when $\mathcal{T}^{\bm m}$ is a valid \emph{symmetric} building block class. \revtwo{Recall from Section~\ref{sec:polyhedra} that facets, i.e., faces of dimension $\dim(\P)-1$ (where $\dim(\P) = d(p-1)$ by Proposition~\ref{prop:Pjdim}), are of most interest.}

\begin{lemma}\label{lem:facets}
Let $\mathcal{T}^{\bm m}$ denote a valid \emph{symmetric} basic building block class. Then, any inequality from $\Theta^{\bm m}$ defines a face of $\mathcal{P}$ of dimension at least $d(p-1)-1 - \frac12(p-3)$, when $d \geq 3$ and $p \geq 3$.
\end{lemma}

\begin{IEEEproof}
  See Appendix~\ref{app:prooffacets}.
\end{IEEEproof}
\begin{conjecture}\label{conj:facetsymmetric}
  A valid building block class $\mathcal T^{\bm m}$ is facet-defining if and only if it is symmetric.
\end{conjecture}

The \enquote{if}-part of the conjecture was verified numerically (cf. Remark~\ref{rem:numerical-check-facet} in Appendix~\ref{app:prooffacets}) for all primes $p \leq 19$; see Table~\ref{table:classNumbers} and also Table~\ref{table:facetClasses} from which the corresponding $\bm m$-vectors can be derived. The \enquote{only if}-part is supported by complementary numerical experiments for all primes $p \leq 19$;  see Appendix~\ref{app:only-if-part}. %

While we are not able to prove Conjecture~\ref{conj:facetsymmetric} for general $p$, the following stronger result (compared to Lemma~\ref{lem:facets}) holds for valid almost doubly-symmetric building block classes.

\begin{proposition}\label{prop:facets_modified}
  Let $\mathcal{T}^{\bm m}$ denote a valid \emph{almost doubly-symmetric} basic building block class. Then, any inequality from $\Theta^{\bm m}$ defines a facet of $\mathcal{P}$ for $d \geq 3$ and for any prime $p \geq 3$. In particular, all inequalities derived from the building block classes in Proposition~\ref{prop:specialMvalid} define facets of $\P$ for any prime $p \geq 2$.
\end{proposition}
\begin{IEEEproof}
  For $p\geq 3$, see Appendix~\ref{app:prooffacets_modified}. The statement for $p=2$ (which is included by the \enquote{in particular}-part of the proposition) has been shown previously in LP decoding literature, as pointed out in Section~\ref{sec:q2}.
\end{IEEEproof}

Proposition~\ref{prop:facets_modified} is an important result since it shows that almost doubly-symmetric classes give rise to linear inequalities that provably define facets of $\P$ for any prime \eirik{$p \geq 3$}.

\subsection{More Inequalities by Rotation}\label{sec:automorph}
For a set $\Theta^{\bm m}$ of inequalities and $0 \neq h \in \F_p$,  denote by $\varphi_h(\Theta^{\bm m})$ the set of all inequalities derived by replacing each $\boldtheta^T \boldx \leq \kappa$ in $\Theta^{\bm m}$ by $\rot_{\bm h}(\boldtheta)^T \boldx \leq \kappa$, where $\bm h = (h,\dotsc,h)$. By Corollary~\ref{cor:rotation-autfq}, the inequalities in $\varphi_h(\Theta^{\bm m})$ are valid for $\P$ if and only if those in $\Theta^{\bm m}$ are.

Concludingly, the set of all inequalities obtained from a class $\mathcal T^{\bm m}$ is denoted by
\[ \Phi(\Theta^{\bm m}) = \bigcup_{0 \neq h \in \F_p} \varphi_h(\Theta^{\bm m}).\]
\begin{remark}\label{rem:phiThetaunique}
  The uniqueness statement of Remark~\ref{rem:Thetamunique} remains valid among $\Phi(\Theta^{\bm m})$: if $\boldtheta^{1T}\boldx \leq \kappa^1$ and $\boldtheta^{2T} \boldx \leq \kappa^2$ are from $\varphi_h(\Theta^{\bm m})$ and $\varphi_{h'}(\Theta^{\bm m})$, respectively, where $\boldtheta^1 = a \boldtheta^2$, $\kappa^1 = a \kappa^2$, and $h,h'\in \F_p \setminus\{0\}$, then the same argument as in Remark~\ref{rem:Thetamunique} shows that $a \neq 1$ is impossible. For $a=1$, Property~\ref{part:bb-property1} of Lemma~\ref{lem:bb-properties} implies that $h=h'$, hence this case reduces to  Remark~\ref{rem:Thetamunique}.
\end{remark}
\begin{remark}\label{rem:inequalityCount}
Since $\abs{\F_p \setminus \{0\}} = p-1$, $\Phi(\Theta^{\bm m})$ contains $(p-1) p^{d-1}$ unique inequalities.
\end{remark}

\subsection{Redundant Inequalities}\label{sec:redundant}
The procedure described so far leads to the set
\begin{equation} %
  \Theta(d)  = \Delta_p^d \cup \bigcup_{\bm m \text{ valid}} \Phi(\Theta^{\bm m})
  \label{eq:allConstraints}
\end{equation}
with $\Delta_p^d$ as defined in Definition~\ref{def:spx}, of (in)equalities that are valid for $\Pd$; \ie they describe a relaxation of $\Pd$ and can thus be used for (relaxed) LP decoding as described in Section~\ref{sec:lpdecoding}.

In general, however, some entries of $\Theta(d)$ might be \emph{redundant}: an inequality $\boldtheta^T \boldx \leq \kappa$ in $\Theta(d)$ is redundant if it is a linear combination of other (in)equalities in $\Theta(d)$, where the coefficients of all \emph{in}equalities are nonnegative:
\begin{equation}
  \label{eq:redundancy}
  \boldtheta = \sum_{j=1}^N \lambda_j \boldtheta^j + \sum_{i=1}^d \mu_i \bm s^i \quad\text{and}\quad \kappa = \sum_{j=1}^N \lambda_j \kappa^j + \sum_{i=1}^d \mu_i
\end{equation}
where $\{\boldtheta^{jT} \boldx \leq \kappa^j\}_{j=1}^N$ are the remaining \emph{in}equalities in $\Theta(d)$, $\lambda_j \geq 0$ for $j \in \range N$, and $\{\bm s^{iT} \boldx = 1\}_{i=1}^d$ are the $d$ equations from \eqref{eq:spx-eq} that describe $\aff(\P)$ (which are the only equations in $\Theta(d)$  by construction). %
\begin{observation}
  If an inequality $\boldtheta^T \boldx \leq \kappa$ of $\Theta(d)$ is redundant for $d \geq 3$, then all $\mu_i = 0$ in \eqref{eq:redundancy}, \ie the equations \eqref{eq:spx-eq} are not necessary in the above representation of the inequality.
\end{observation}
\begin{IEEEproof}
  Assume $\boldtheta^T \boldx \leq \kappa$ in $\Theta(d)$ is redundant and satisfies \eqref{eq:redundancy} for $\{\lambda_j\}_{j=1}^N$ and $\{\mu_i\}_{i=1}^d$, where $\mu_{i^*} \neq 0$ for some $i^* \in \range d$. By Property~\ref{part:bb-property2} of Lemma~\ref{lem:bb-properties}, $\theta^j_{i,0}=0$ for $i \in \range d$. On the other hand, $s^{i}_{i,0}=1$ by \eqref{eq:spx-eq}, while $s^{i}_{k,0}=0$ for $i \neq k$, from which follows that $\theta_{i^*,0}=\mu_{i^*} \neq 0$. Again by Property~\ref{part:bb-property2} of Lemma~\ref{lem:bb-properties}, this means that the redundant inequality cannot be contained in $\Phi(\Theta^{\bm m})$.
  
 But then it must be in $\Delta_p^d$, \ie one of \eqref{eq:spx-geq}, say $-x_{k,l} \leq 0$ for some $k \in \range d, l \in \F_p$. By Proposition~\ref{prop:Pjdim}, it is a facet, which by basic polyhedral theory implies $\rank(\{\boldtheta^j\colon \lambda_j \neq 0\}) = 1$. Hence, we can assume wlog.\ $\lambda_j \neq 0$ for only one $j=i^*$, \ie $\boldtheta = \lambda_{i^*} \boldtheta^{i^*} + \sum_i \mu_i \bm s^i$. This is obviously impossible for both the case that the corresponding inequality $\boldtheta^{i^*T}\boldx \leq \kappa^{i^*}$ is also of type \eqref{eq:spx-geq} or obtained from Construction~\ref{constr:hilo}.
\end{IEEEproof}

Hence, for the study of redundancy within $\Theta(d)$ we can reduce \eqref{eq:redundancy} to the condition that 
\begin{equation}\boldtheta = \sum_{j=1}^N \lambda_j \boldtheta^j\text{ and }\kappa = \sum_{j=1}^N \lambda_j \kappa^j, \quad\text{all } \lambda_j \geq 0.
\label{eq:redundancy-2}
\end{equation}

\revtwo{Geometrically, \eqref{eq:redundancy-2} implies that the face $\bm\theta^T \bm x \leq \kappa$ is the intersection of all faces $\bm\theta^{jT}\bm x\leq \kappa^j$ for which $\lambda_j > 0$. In particular, if $\boldtheta^T\bm x \leq \kappa$ induces a facet, \ie a \enquote{maximal} nontrivial face, then all $(\bm\theta^j, \kappa^j)$ for which $\lambda_j>0$ must be positive scalar multiples of $(\boldtheta, \kappa)$. By Remark~\ref{rem:phiThetaunique}, this requires all involved inequalities to originate from different building block classes.

The following proposition shows that this is possible only for one specific pair of building block classes.}

\begin{proposition}[Equivalent Inequalities]\label{prop:equivalent}
  Let the inequalities $\boldtheta^T \boldx \leq \kappa$ and $\boldtheta'^T \boldx \leq \kappa'$ be contained in $\varphi(\Theta^{\bm m})$ and $\varphi'(\Theta^{\bm m'})$, respectively, where $\mathcal T^{\bm m} \neq \mathcal T^{\bm m'}$ and $\varphi,\varphi' \in \GL(\F_p)$. Assume that $\boldtheta = a \boldtheta'$ and $\kappa = a \kappa'$ for some $a \geq 0$, where wlog.\ $a \leq 1$ (swap $\bm m$ and $\bm m'$ otherwise). Then, $\bm m = (0,\dotsc,0)$, $\bm m' = (0,1,0,1,\dotsc)$, and $a = 1/3$ (for $p=2$) or $a=1/2$ (for $p > 2$).
  
  Conversely, $\Phi(\Theta^{(0,\dotsc,0)})$ and $\Phi(\Theta^{(0,1,0,1,\dotsc)})$ are equivalent:
  \[  \boldtheta^T \boldx \leq \kappa \in \Phi(\Theta^{(0,1,0,1,\dotsc)}) \Leftrightarrow a\boldtheta^T \boldx \leq a\kappa \in \Phi(\Theta^{(0,\dotsc,0)})\]
  with $a$ as above.
\end{proposition}
\begin{IEEEproof}
  See Appendix~\ref{app:equivalent}.
\end{IEEEproof}
\revtwo{\begin{corollary}
Except for $\mathcal T^{\bm m}$ with $\bm m = (0,1,\dotsc)$, each building block class that was shown to induce facets in Section~\ref{sec:facet-defining} gives rise to, by Remark~\ref{rem:inequalityCount}, $(p-1)p^{d-1}$ unique irredandant facets of $\P$, \ie inequalities that are necessary in the polyhedral description of $\Pd$.

In particular, using $\left\lceil(p-1)/2\right\rceil$ building block classes explicitly constructed in Proposition~\ref{prop:specialMvalid} for $s < \left\lceil(p-1)/2\right\rceil$ and together with Proposition~\ref{prop:Pjdim}, we are able to explicitly construct a total of $\left\lceil(p-1)^2/2\right\rceil p^{d-1} + dp$ unique facets of $\P$, for any \eiriknew{$p \geq 3$ and $d \geq 3$, and $2^{d-1}+2d$ unique facets for $p=2$ and $d \geq 4$}.
\end{corollary}}

\subsection{Valid Inequalities for General SPC Codes } \label{sec:generalSPC}

This section has so far focused on \enquote{all-ones} SPC codes. As noted in Corollary~\ref{cor:rotation-generalSPC}, however, each of the facets constructed in this section can be translated into a facet of the codeword polytope of a general SPC code by applying the appropriate rotation operation to the respective $\boldtheta$-vector. For an SPC code $\C(\bm h)$ with parity-check matrix $\bm h$ and $\varphi_r \in \GL(\F_p)$, we denote the respective equivalent of $\varphi_r\left(\Theta^{\bm m}\right)$ by \[\varphi_{\bm h,r}(\Theta^{\bm m}) = \{ (\rot_{\bm h}\circ \rot_r)(\boldtheta)^T \boldx \leq \kappa)\colon (\boldtheta^T \boldx \leq \kappa) \in \Theta^{\bm m}\}.\]

\begin{figure*}[!t]
\begin{displaymath}
\begin{array}{llclcl}
& -x_{1,1} -2x_{1,2} +x_{2,1} +2x_{2,2} -2x_{3,1} -x_{3,2} & \leq 0, &  -2x_{1,1} -x_{1,2} +2x_{2,1} +x_{2,2} -x_{3,1} -2x_{3,2} &\leq 0, \notag \\
& -x_{1,1} -2x_{1,2} +x_{2,1} - x_{2,2} +x_{3,1} - x_{3,2} &\leq 0, & -2x_{1,1} -x_{1,2} -x_{2,1} +x_{2,2} -x_{3,1} +x_{3,2} &\leq  0,  \notag \\
& -x_{1,1} -2x_{1,2} -2x_{2,1} -x_{2,2} +x_{3,1} +2x_{3,2} &\leq 0, & -2x_{1,1} -x_{1,2} -x_{2,1} -2x_{2,2} +2x_{3,1} +x_{3,2} &\leq 0,  \notag \\
& -x_{1,1} +x_{1,2} +x_{2,1} -x_{2,2} -2x_{3,1} -x_{3,2} &\leq 0,  & x_{1,1} -x_{1,2} -x_{2,1} +x_{2,2} -x_{3,1} -2x_{3,2} &\leq 0, \notag \\
& -x_{1,1} +x_{1,2} -2x_{2,1} -x_{2,2} +x_{3,1} -x_{3,2} &\leq 0,  & x_{1,1} -x_{1,2} -x_{2,1} -2x_{2,2} -x_{3,1} +x_{3,2} &\leq 0, \notag \\
& 2x_{1,1} +x_{1,2} -2x_{2,1} -x_{2,2} -2x_{3,1} -x_{3,2} &\leq 0, & x_{1,1} + 2x_{1,2} -x_{2,1} -2x_{2,2} -x_{3,1} -2x_{3,2} &\leq 0, \notag \\
& -x_{1,1} +x_{1,2} +x_{2,1} +2x_{2,2} +x_{3,1} +2x_{3,2} &\leq 3, &x_{1,1} -x_{1,2} +2x_{2,1} +x_{2,2} +2x_{3,1} +x_{3,2} &\leq 3, & \notag \\
& 2x_{1,1} +x_{1,2} +x_{2,1} -x_{2,2} +x_{3,1} +2x_{3,2} &\leq 3, & x_{1,1} +2x_{1,2} -x_{2,1} +x_{2,2} +2x_{3,1} +x_{3,2} &\leq 3, & \notag \\
 & 2x_{1,1} +x_{1,2} +x_{2,1} +2x_{2,2} +x_{3,1} -x_{3,2} &\leq 3,  & x_{1,1} +2x_{1,2} +2x_{2,1} +x_{2,2} -x_{3,1} +x_{3,2} &\leq 3. & \notag
\end{array}
\end{displaymath}
 \hrulefill
\vspace*{-2mm}
\end{figure*}

\begin{table}
\revtwo{
  \caption{A summary of notation and terms introduced in Section~\ref{sec:bb}.}
  \label{table:notation3}
  \centering
 \vskip -2.0ex
  \ifonecolumn
  \begin{tabular}{p{4.5cm}p{10.5cm}}
  \else
  \begin{tabular}{p{1.5cm}p{6.5cm}}
  \fi
   \toprule
  $\bm m$ & Integer vector in $\{0,1\}^{p}$ inducing a building block class \\
  $d$ & Length of an \enquote{all-ones} SPC code\\
     $\bm t_k^{\bm m}$ & Basic building block  induced by $\bm m$ (see Definition~\ref{def:bb}) \\
   $\mathcal T^{\bm m}$ & Basic building block class induced by $\bm m$ (see Definition~\ref{def:bb}) \\  
   $\set(\bm t_k)$ & The set of unordered entries of $\boldt_k$ \\
   $t_{k,\uparrow}$, $t_{k,\downarrow}$ & The \emph{index} of the largest and smallest entry, respectively, of $\bm t_k$ (see Definition~\ref{def:hilo-notation}) \\
  $t_{\uparrow, \zeta}$, $t_{\downarrow, \zeta}$ &  The block for which the $\zeta$-th entry is the maximizer (resp. minimizer) of $\set(\bm t_k)$ (see Definition~\ref{def:hilo-notation})\\ 
  $\sigma^{\bm m}$ &   The index of the largest entry of $\bm t_0^{m}$ (see Definition~\ref{def:hilo-notation})\\
  $T_0^{\text{proj}}$ &  The projection of $\set(\bm t_0)$ onto the interior entries, \ie $\set(\bm t_0) \setminus \{ 0, t_{0,\sigma}\}$\\
    Sym.\ $\mathcal T^{\bm m}$ & See Definition~\ref{def:symmetric_class} \\
  Almost doubly-sym.\ $\mathcal T^{\bm m}$ & See Definition~\ref{def:almost_double_symmetric}\\
 $\smax(\bm t_k)$ & The second-largest entry of $\boldt_k$ \\
 $V^\boldtheta_k$ & The index set of entries in $\boldtheta$ that equal $\bm t_k$\\
 Valid $\mathcal T^{\bm m}$ & See Definition~\ref{def:valid-class}\\
 $I^>_{c,J}$, $I^=_{c,J}$ & See Lemma~\ref{lem:counting_formulas} (counting formulas)\\
$\Phi(\Theta^{\bm m})$ &  The set of all inequalities obtained from a class $\mathcal T^{\bm m}$\\
$\Theta(d)$ & Set of valid (in)equalities for $\Pd$ (see \eqref{eq:allConstraints})\\
   \bottomrule
  \end{tabular}}
\end{table}

\section{Explicit Building Block Classes for Small Values of $p$} \label{sec:explicit_3_5_7}

In this section, we present explicit building block classes for $p = 3,5$, and $7$ obtained from the general construction of the previous section. For $p=3$ (resp.\ $5$) we prove (resp.\ conjecture) that these classes together with $\Delta_p^d$ give a complete and irredundant set of linear (in)equalities describing the convex hull of an embedded SPC code of length $d$. However, for $p=7$, this is not the case, and we present two  (up to additive automorphisms) additional nonbasic classes. Based on numerical experiments, we conjecture that these classes (basic and nonbasic) together with $\Delta_7^d$ give a complete and irredundant set of linear (in)equalities describing the convex hull of an embedded length-$d$ SPC code over $\F_7$. For completeness, we also consider the binary case $p=2$.

\subsection{The Case $p=2$} \label{sec:q2}
For $p=2$, there is only a single vector $\bm m=(0,0)$ that gives a valid basic building block class (the  alternative, $\bm m'=(0,1)$, is redundant by Proposition~\ref{prop:equivalent}). In particular, $\bm t^{\bm m}_0=(0,1)$ and $\bm t^{\bm m}_1=(0,-1)$, with $[\sigma]_\Z = t_{0,\sigma} = 1$. %
Thus, the conditions from \eqref{eq:hilo-conditions'} state that the inequalities obtained from Construction~\ref{constr:hilo} are exactly those for which $\left[\abs{V_0^\boldtheta}\right]_2 = [1]_2$, \ie $\abs{V_0^\boldtheta}$ is odd, and $\kappa = \abs{V_0^\boldtheta} - 1$.

As $\GL(\F_2) = \{ \varphi_1\}$ consists of only the identity, there is a single class of facets plus $\Delta_2^d$ which describes $S_1^d$, where $S_1 = \conv\{ (1,0), (0,1) \}$.

Note that the existing literature on binary LP decoding uses Flanagan's embedding instead of constant-weight embedding. Using the procedure from Appendix~\ref{app:embeddings}, we obtain corresponding building blocks $\boldt'_0 = (1)$ and $\boldt'_1= (-1)$, which exactly leads to the so-called \enquote{forbidden-set} or \enquote{Feldman} inequalities \cite{fel05} which date back to an earlier work of Jeroslow \cite{jer75}. Furthermore, the procedure replaces the simplex constraints $\Delta_2^d$ by the usual box constraints $x_{i,1} \geq 0$ (by \eqref{eq:hsimplex-0}) and $x_{i,1} \leq 1$ (by \eqref{eq:hsimplex-1}), which shows that the inequalities constructed in this paper are indeed a generalization of the well-known binary case.

\subsection{The Case $p=3$} \label{sec:q3}

For $p=3$, there is also only a single vector $\bm m=(0,0,0)$ that gives a valid, irredundant facet-defining basic building block class (see Table~\ref{table:facetClasses}). In particular, $\bm t^{\bm m}_0=(0,1,2)$, $\boldt^{\bm m}_1 = (0,1, -1)$, and $\boldt^{\bm m}_2 = (0,-2,-1)$. %
In the following, the explicit dependence of $\bm m =(0,0,0)$ will be omitted to simplify notation.

The building blocks $\boldt_{k}$ and values from Definition~\ref{def:hilo-notation} and Lemma~\ref{lem:hilo-formulas} are presented in Table~\ref{table:values_tk}.
\begin{table}
\centering
  \caption{Structural properties of the building blocks $\boldt^{\bm m}_0$, $\boldt^{\bm m}_1$, and $\boldt^{\bm m}_2$ for $p=3$ and $\bm m=(0,0,0)$. For this $\bm m$, $\sigma = 2$ (cf.\ Definition~\ref{def:hilo-notation}).} %
 \vskip -2.0ex
  \label{table:values_tk}
  \begin{tabular}{*5{>{$}c<{$}}}
    \toprule
    k   & 0 & 1 & 2 & \text{expression}\\
    \midrule
    \boldt_k & (0, 1, 2) & (0, 1,-1) & (0, -2, -1) \\
    t_{k,\uparrow} & 2 & 1 & 0 & 2-k\\
    t_{k,\downarrow} & 0 & 2 & 1 & -k\\
    \max\left(\bm t_k\right) & 2 & 1 & 0 & 2-[k]_\Z\\
    \min\left(\bm t_k\right) & 0 & -1 & -2 &-[k]_\Z\\
    \bottomrule
  \end{tabular}
\end{table}

\begin{proposition}\label{prop:Theta1Facets}
  Every inequality $\boldtheta^T \boldx \leq \kappa$ in  $\Theta^{\bm m}$ defines a facet of $\Pd$, $d \geq 3$. Moreover,
  \begin{enumerate}
    \item there are $d+1$ elements $\boldzeta \in \mathbb{F}_3^d \setminus \C$ whose embeddings are cut by the inequality, i.e., $\boldtheta^T\Fv(\boldzeta) > \kappa$.\label{part:valid1_q3}
    \item The inequality is tight for the embeddings of exactly $\frac12 d(d+1)$ codewords of $\C$.\label{part:valid2_q3}
  \end{enumerate}
\end{proposition}
\begin{IEEEproof}
The class $\mathcal{T}^{\bm m}$ is almost doubly-symmetric (see Proposition~\ref{prop:specialMvalid}) and then it follows from Proposition~\ref{prop:facets_modified} that any inequality from $\Theta^{\bm m}$ defines a facet of $\mathcal{P}$ for $d \geq 3$.
The remaining counting formulas are special cases of the general counting formulas of Lemma~\ref{lem:counting_formulas} (details omitted for brevity). \revtwo{See also  Example~\ref{ex:11}.}
\end{IEEEproof}

\begin{theorem} \label{thm:facetComplete}
  Let $\C$ be the ternary \enquote{all-ones} SPC code of length $d \geq 3$ and $\Pd = \conv(\Fv(\C))$. Then, $\Theta(d) = \Theta^{\bm m} \cup \varphi_2(\Theta^{\bm m}) \cup \Delta_3^d = \Delta_3^d \cup \Phi(\Theta^{\bm m})$ with $\abs{\Theta(d)} = 2\cdot 3^{d-1} + 4d$ gives a complete and irredundant description of $\Pd$ (note that $\varphi_2$ is the only entry of $\GL(\F_3)$ besides the identity).
\end{theorem}

\begin{IEEEproof}
See Appendix~\ref{app:proofFacetComplete3}.
\end{IEEEproof}

\begin{example}
Consider a ternary SPC code $\mathcal{C}$ of length $d=3$, defined by the parity-check matrix $\boldH = (1,2,2)$. In the ternary case, the constant-weight embedding (from Definition~\ref{def:Constant}) is as follows: $0 \mapsto (1,0,0)$, $1 \mapsto (0,1,0)$, and $2 \mapsto (0,0,1)$. %
The image $\mathsf{F}_{\rm v}(\mathcal{C})$ (which is a nonlinear binary code) has length $3 d = 9$ and contains nine codewords as follows:
\ifonecolumn
\begin{align}
\{ &(1 0 0 1 0 0 1 0 0), (1 0 0 0 1 0 0 0 1), (1 0 0 0 0 1 0 1 0), 
(0 1 0 1 0 0 0 1 0), (0 1 0 0 1 0 1 0 0), (0 1 0 0 0 1 0 0 1), 
(0 0 1 1 0 0 0 0 1), (0 0 1 0 1 0 0 1 0), \notag \\
&(0 0 1 0 0 1 1 0 0) \}. \notag
\end{align}
\else
\begin{align}
\{ &(1 0 0 1 0 0 1 0 0)\eiriknew{^T}, (1 0 0 0 1 0 0 0 1)\eiriknew{^T}, (1 0 0 0 0 1 0 1 0)\eiriknew{^T}, \notag \\
&(0 1 0 1 0 0 0 1 0)\eiriknew{^T}, (0 1 0 0 1 0 1 0 0)\eiriknew{^T}, (0 1 0 0 0 1 0 0 1)\eiriknew{^T}, \notag \\
&(0 0 1 1 0 0 0 0 1)\eiriknew{^T}, (0 0 1 0 1 0 0 1 0)\eiriknew{^T}, (0 0 1 0 0 1 1 0 0)\eiriknew{^T} \}. \notag
\end{align}
\fi
By Theorem~\ref{thm:facetComplete}, the linear (in)equalities obtained from Section~\ref{sec:generalSPC} and $\Delta_3^d$ give a complete and irredundant description %
of the convex hull $\Pd = \conv(\mathsf{F}_{\rm v}(\mathcal{C}))$. These inequalities (except for the ones from  $\Delta_3^d$) %
are shown at the top of 
\ifonecolumn
the previous page.
\else
the page.
\fi
\end{example}
\subsection{The Case $p=5$} \label{sec:q5}

\begin{table*}
\centering
  \caption{Structural properties for $p=5$ of the basic building block classes  $\mathcal T^{\bm m = (0,0,0,0,0)}$ and $\mathcal T^{\bm m' = (0,1,0,0,0)}$ with $\sigma^{\bm m} = 4$ and $\sigma^{\bm m'} = 1$.}
 \vskip -2.0ex
  \label{table:values_tk_q5_1}
  \begin{tabular}{*7{>{$}c<{$}}}
    \toprule
    k   & 0 & 1 & 2 & 3 & 4\\
    \midrule
    \boldt^{\bm m}_k &  (0,1,2,3,4) &  (0,1,2,3,-1) &  (0,1,2,-2,-1) &  (0,1,-3,-2,-1) &  (0,-4,-3,-2,-1) \\

    \midrule
    t^{\bm m}_{k,\uparrow} & 4 & 3 & 2 & 1 & 0  \\ %
    t^{\bm m}_{k,\downarrow} & 0 & 4 & 3 & 2 & 1  \\ %
    \max\left(\bm t^{\bm m}_k\right) & 4 & 3 & 2 & 1 & 0 \\
    \min\left(\bm t^{\bm m}_k\right) & 0 & -1 & -2 & -3 & -4 \\
    \midrule
    \boldt^{\bm m'}_k &  (0,6,2,3,4) & (0,-4,-3,-2,-6) & (0,1,2,-2,4)  & (0,1,-3,3,-1) &  (0,-4,2,-2,-1) \\
  \midrule
    t^{\bm m'}_{k,\uparrow} & 1 & 0 & 4 & 3 & 2 \\ %
    t^{\bm m'}_{k,\downarrow} & 0 & 4 & 3 & 2 & 1 \\ %
    \max\left(\bm t^{\bm m'}_k\right) & 6 & 0 & 4 & 3 & 2 \\
    \min\left(\bm t^{\bm m'}_k\right) & 0 & -6 & -2  & -3 & -4 \\
    \bottomrule
  \end{tabular}
\end{table*}

For $p = 5$, there are two vectors $\bm m = (0, 0, 0,0,0)$ and $\bm m' = (0,1,0,0,0)$ that give valid, irredundant facet-defining basic building block classes (see Table~\ref{table:facetClasses}). The five building blocks and some properties of each class are tabulated in Table~\ref{table:values_tk_q5_1}.

\begin{example} \label{lem:conditions_q5}
  Using Lemma~\ref{lem:conditions}, we find that an inequality $\boldtheta^T \boldx \leq \kappa$ with $\boldtheta = (\bm t^{\bm m}_{k_1} \mid \dotsc \mid \bm t^{\bm m}_{k_d})^T$ and $\kappa \in \R$ is in $\Theta^{\bm m}$ if and only if 
  \ifonecolumn
    \begin{displaymath}
 \left[4\abs{V^\boldtheta_0}+ 3 \abs{V^\boldtheta_1} +2 \abs{V^\boldtheta_2} + \abs{V^\boldtheta_3}\right]_5  = [4]_5 %
      \text{ and } \kappa =4\abs{V^\boldtheta_0}+3\abs{V^\boldtheta_1}+2\abs{V^\boldtheta_2}+\abs{V^\boldtheta_3}-4  \notag %
    \end{displaymath}
    \else
    \begin{align}
& \left[4\abs{V^\boldtheta_0}+ 3 \abs{V^\boldtheta_1} +2 \abs{V^\boldtheta_2} + \abs{V^\boldtheta_3}\right]_5  = [4]_5 \notag \\
      \text{and\quad}&\kappa =4\abs{V^\boldtheta_0}+3\abs{V^\boldtheta_1}+2\abs{V^\boldtheta_2}+\abs{V^\boldtheta_3}-4  \notag %
    \end{align}
    \fi
while an inequality $\boldtheta'^T \boldx \leq \kappa'$ with $\boldtheta' = (\bm t^{\bm m'}_{k_1} \mid  \dotsc \mid \bm t^{\bm m'}_{k_d})^T$ and $\kappa' \in \R$ is in $\Theta^{\bm m'}$ if and only if 
  \ifonecolumn
    \begin{displaymath}
 \left[\abs{V^{\boldtheta'}_0} + 4 \abs{V^{\boldtheta'}_2} +3 \abs{V^{\boldtheta'}_3}+ 2\abs{V^{\boldtheta'}_1}\right]_5  =  [1]_p %
      \text{ and } \kappa' =6\abs{V^{\boldtheta'}_0}+4\abs{V^{\boldtheta'}_2}+3\abs{V^{\boldtheta'}_3}+2\abs{V^{\boldtheta'}_4}-6.  \notag %
    \end{displaymath}
    \else
    \begin{align}
& \left[\abs{V^{\boldtheta'}_0} + 4 \abs{V^{\boldtheta'}_2} +3 \abs{V^{\boldtheta'}_3}+ 2\abs{V^{\boldtheta'}_1}\right]_5  =  [1]_p \notag \\
      \text{and}\quad&\kappa' =6\abs{V^{\boldtheta'}_0}+4\abs{V^{\boldtheta'}_2}+3\abs{V^{\boldtheta'}_3}+2\abs{V^{\boldtheta'}_4}-6.  \notag %
    \end{align}
    \fi
\end{example}

The following proposition is an adapted version of Proposition~\ref{prop:Theta1Facets} for the case $p=5$.

\begin{proposition}\label{prop:validIneq_q5}
  Every inequality $\boldtheta^T \boldx \leq \kappa$ in $\Theta^{\bm m} \cup \Theta^{\bm m'}$ defines a facet of $\Pd$, $d \geq 3$. Moreover, %
  \begin{enumerate}
    \item there are 
\begin{displaymath}
\begin{cases}
4 + 6(d-1) + 4 \binom{d-1}{2} + \binom{d-1}{3} & \text{for $\Theta^{\bm m}$},\\
4 + 6(d-1) + 3 \binom{d-1}{2} & \text{for $\Theta^{\bm m'}$} \end{cases}
\end{displaymath}
elements $\boldzeta \in \mathbb{F}_5^d \setminus \C$ whose embeddings are cut by the inequality, i.e., $\boldtheta^T\Fv(\boldzeta) > \kappa$.\label{part:valid2_q5_theta}
    \item The inequality is tight for the embeddings of exactly
\begin{displaymath}
\begin{cases}
1+4(d-1) + 6 \binom{d-1}{2} + 4\binom{d-1}{3} + \binom{d-1}{4} & \text{for $\Theta^{\bm m}$},\\
1+4(d-1) + 4 \binom{d-1}{2} + \binom{d-1}{3} & \text{for $\Theta^{\bm m'}$} \end{cases}
\end{displaymath}
codewords of $\C$.\label{part:valid3_q5_theta}
  \end{enumerate}
\end{proposition}

\begin{IEEEproof}
Both classes $\mathcal{T}^{\bm m}$ and $\mathcal{T}^{\bm m'}$ are almost doubly-symmetric (see Proposition~\ref{prop:specialMvalid}) and then it follows from Proposition~\ref{prop:facets_modified} that any inequality from $\Theta^{\bm m}$ or $\Theta^{\bm m'}$ defines a facet of $\mathcal{P}$ for $d \geq 3$. The remaining counting formulas are special cases of the general counting formulas of Lemma~\ref{lem:counting_formulas} (details omitted for brevity).
\end{IEEEproof}

Now, for each of the three nontrivial automorphisms $\varphi_2, \varphi_3, \varphi_4 \in \GL(\F_5)$ we obtain additional inequalities (as described in Section~\ref{sec:automorph}) if we apply the corresponding permutation to each building block $\boldt_k$ in an inequality. %

\begin{example}
By applying the automorphism $\varphi_4$ to the building blocks from  of $\mathcal T^{\bm m'}$ (see Table~\ref{table:values_tk_q5_1}), we get the building blocks
\ifonecolumn
\begin{align*}
\varphi_4(\bm t^{\bm m'}_0) &= (0,4,3,2,6),\;
\varphi_4(\bm t^{\bm m'}_1) = (0, -6, -2, -3, -4),\;
\varphi_4(\bm t^{\bm m'}_2) = (0,4,-2,2,1),\\
\varphi_4(\bm t^{\bm m'}_3) &= (0,-1,3,-3,1),\;
\varphi_4(\bm t^{\bm m'}_4) = (0,-1,-2,2,-4).
\end{align*}
\else
\begin{align*}
\varphi_4(\bm t^{\bm m'}_0) &= (0,4,3,2,6),\\
\varphi_4(\bm t^{\bm m'}_1) &= (0, -6, -2, -3, -4),\\
\varphi_4(\bm t^{\bm m'}_2) &= (0,4,-2,2,1),\\
\varphi_4(\bm t^{\bm m'}_3) &= (0,-1,3,-3,1),\\
\varphi_4(\bm t^{\bm m'}_4) &= (0,-1,-2,2,-4).
\end{align*}
\fi
\end{example}

\begin{conjecture} \label{conj:facetComplete_q5}
  Let $\C$ be the quinary \enquote{all-ones} SPC code of length $d \geq 3$ and $\Pd = \conv(\Fv(\C))$. Then, 
\[
\Theta(d) = \Phi(\Theta^{\bm m}) \cup \Phi(\Theta^{\bm m'}) \cup \Delta_5^d
\]
 with $\abs{\Theta(d)} = 8\cdot 5^{d-1} + 6d$  gives a complete and irredundant description of $\Pd$. %
\end{conjecture}

We have verified numerically that the conjecture is true for $d=3$, $4$, and $5$.

\subsection{The Case $p=7$} \label{sec:q7}

For $p=7$, there are four vectors $\bm m = (0,0,0,0,0,0,0)$, $\bm m = (0,1,0,1,0,0,0)$, $\bm m = (0,1,0,0,0,0,0)$, and $\bm m = (0,1,1,0,0,1,0)$ that give valid, irredundant facet-defining basic building block classes (see Table~\ref{table:facetClasses}).  The elements of the resulting basic building block classes,  denoted by $\mathcal{T}_1$ through $\mathcal{T}_4$,  are summarized in the first four rows of Table~\ref{table:q7classes}. The corresponding sets of inequalities are denoted by $\Theta_1$ through $\Theta_4$. %

Furthermore, there is a set $\Theta_5$ of \enquote{hybrid} facet-defining inequalities built from \emph{two} basic building block classes, namely the (valid) class $\mathcal T^{(0010000)}$ and the (invalid) basic class $\mathcal T^{(0110000)}$, both of which are not symmetric; we will describe below how to construct inequalities in $\Theta_5$ using a modification of Construction~\ref{constr:hilo}.

Finally, there is a sixth structurally distinct set $\Theta_6$ of facet-defining inequalities that are built from \emph{three} different classes of building blocks, which however are not conforming to Definition~\ref{def:bb} but instead are of a completely different form. Still, the construction of $\Theta_6$, which is outlined below, shares several aspects with the one developed in Section~\ref{sec:bb}.

\begin{table*}
\centering
  \caption{Collection of classes (up to rotational symmetry) of building blocks for the case $p=7$. $\mathcal T_1$ to $\mathcal T_4$ are basic classes in the sense of Definition~\ref{def:bb}. $\mathcal T_5^{\rm b}$ and $\mathcal T_5^{\rm nb}$ are used to construct $\Theta_5$, while $\mathcal T_6^{\rm b}$, $\mathcal T_6^{\rm lo}$, and $\mathcal T_6^{\rm hi}$ are the building blocks of inequalities in $\Theta_6$.}
  \label{table:q7classes}
 \vskip -2.0ex
 \ifonecolumn
{\scriptsize{
\else
\fi
  \begin{tabular}{*6{>{$}c<{$}}}
    \toprule
    \mathcal{T}_1 &  (0, 1, 2, 3, 4, 5, 6) & (0,1, 2, 3, 4, 5, -1) & (0,1, 2, 3, 4, -2, -1) & (0,1, 2, 3, -3, -2, -1) \\ & (0,1, 2, -4, -3, -2, -1) 
             & (0,1, -5, -4, -3, -2, -1) &  (0,-6, -5, -4, -3, -2, -1) \\
    \midrule
    \mathcal{T}_2 & (0,8, 2, 10, 4, 5, 6) & (0,-6, 2, -4, -3, -2, -8) & (0,8, 2, 3, 4, -2, 6) & (0,-6, -5, -4, -10, -2, -8) \\ & (0,1, 2, -4, 4, -2, 6) 
             & (0,1,-5, 3, -3, 5, -1) & (0,-6, 2, -4, 4, -2, -1) \\
    \midrule
    \mathcal{T}_3 & (0,8, 2, 3, 4, 5, 6) & (0,-6, -5, -4, -3, -2, -8) & (0,1, 2, 3, 4, -2, 6) & (0,1, 2, 3, -3, 5, -1) \\ & (0,1, 2, -4, 4, -2, -1) 
             & (0,1, -5, 3, -3, -2, -1) & (0,-6, 2, -4, -3, -2, -1) \\
    \midrule
    \mathcal{T}_4 & (0, 8, 9, 3, 4, 12,6) & (0,1, -5, -4, 4, -2, -8) & (0,-6, -5, 3, -3, -9, -1) & (0, 1, 9, 3, -3, 5,6) \\ & (0, 8, 2, -4, 4, 5,-1) 
             & (0, -6, -12, -4, -3, -9,-8) & (0,-6, 2, 3, -3, -2, 6) \\
    \midrule
\mathcal{T}_5^{\rm b} & \bm{(0,1, 9, 3, 4, 5, 6)} & \bm{(0,8, 2, 3, 4, 5, -1)} & \bm{(0,-6, -5, -4, -3, -9, -8)}  & \bm{(0,1, 2, 3, -3, -2, 6)} \\
&  \bm{(0,1, 2, -4, -3, 5, -1)}& \bm{(0,1,-5, -4, 4, -2, -1)}&  \bm{(0,-6, -5, 3, -3, -2, -1)}& \\
\mathcal T_5^{\rm nb} & (0,8, 9, 3, 4, 5, 6) & (0,1, -5, -4, -3, -2, -8) &  (0,-6, -5, -4, -3, -9, -1) & (0,1, 2, 3, -3, 5, 6)  \\
& (0,1, 2, -4, 4, 5, -1)  &  (0,1, -5, 3, 4, -2, -1) &(0,-6, 2, 3, -3, -2, -1) \\
    \midrule
    \mathcal{T}_6^{\rm b} & \bm{(0,-1, -1, -1, -1, -1, -1)} & \bm{(0, 0, 0, 0, 0, 0, 1)} & \bm{(0, 0, 0, 0, 0, 1, 0)} & \bm{(0,0, 0, 0, 1, 0, 0)} \\
& \bm{(0, 0, 0, 1, 0, 0, 0)} & \bm{(0, 0, 1, 0, 0, 0, 0)} & \bm{(0, 1, 0, 0, 0, 0, 0)} \\
\mathcal T_6^{\rm hi} & (0, 0, 0, -1, 0, -1, -1) & (0, 0, -1, 0, -1, -1, 0) & (0, -1, 0, -1, -1, 0, 0) &  (0, 1, 0, 0, 1, 1, 1)  \\
& (0, -1, -1, 0, 0, 0, -1) & (0, 0, 1, 1, 1, 0, 1)  & (0, 1, 1, 1, 0, 1, 0)  \\
\mathcal T_6^{\rm lo}& \it{(0, 1, 1, 0, 1, 0, 0)} & \it{(0, 0, -1, 0, -1, -1, -1)} & \it{(0, -1, 0, -1, -1, -1, 0)} & \it{(0, 1, 0, 0, 0, 1, 1)}  \\
 & \it{(0, -1, -1, -1, 0, 0, -1)} & \it{(0, 0, 0, 1, 1, 0, 1)} &\it{(0, 0, 1, 1, 0, 1, 0)}  \\
    \bottomrule
  \end{tabular}
   \ifonecolumn
}}
\else
\fi
\end{table*}

The following proposition is an adapted version of Proposition~\ref{prop:Theta1Facets} for the case $p=7$.

\begin{proposition}\label{prop:validIneq_q7_1to4}
  Every inequality $\boldtheta^T \boldx \leq \kappa$ in $\Theta_1 \cup \cdots \cup \Theta_4$ defines a facet of $\Pd$, $d \geq 3$. Moreover, %

  \begin{enumerate}
    \item there are 
    \ifonecolumn
\begin{displaymath}
\begin{cases}
6 + 15(d-1) + 20 \binom{d-1}{2} + 15\binom{d-1}{3} + 
6\binom{d-1}{4}+ 
\binom{d-1}{5} & \text{for $\Theta_1$},\\
6 + 15(d-1) + 17 \binom{d-1}{2} + 8\binom{d-1}{3} +\binom{d-1}{4} & \text{for $\Theta_2$},\\
6 + 15(d-1) + 14 \binom{d-1}{2} + 4\binom{d-1}{3} & \text{for $\Theta_3$},\\
6 + 15(d-1) + 17 \binom{d-1}{2} + 7\binom{d-1}{3} & \text{for $\Theta_4$} \end{cases}
\end{displaymath}
\else
\begin{displaymath}
\begin{cases}
6 + 15(d-1) + 20 \binom{d-1}{2} + 15\binom{d-1}{3} + \\
6\binom{d-1}{4}+ 
\binom{d-1}{5} & \text{for $\Theta_1$},\\
6 + 15(d-1) + 17 \binom{d-1}{2} + 8\binom{d-1}{3} +\binom{d-1}{4} & \text{for $\Theta_2$},\\
6 + 15(d-1) + 14 \binom{d-1}{2} + 4\binom{d-1}{3} & \text{for $\Theta_3$},\\
6 + 15(d-1) + 17 \binom{d-1}{2} + 7\binom{d-1}{3} & \text{for $\Theta_4$} \end{cases}
\end{displaymath}
\fi
elements $\boldzeta \in \mathbb{F}_7^d \setminus \C$ whose embeddings are cut by the inequality, i.e., $\boldtheta^T\Fv(\boldzeta) > \kappa$.\label{part:valid2_q5_1to4}
    \item The inequality is tight for the embeddings of exactly
    \ifonecolumn
\begin{displaymath}
\begin{cases}
1+6(d-1) + 15 \binom{d-1}{2} + 20\binom{d-1}{3} + 
15\binom{d-1}{4} + 
6\binom{d-1}{5} + \binom{d-1}{6}& \text{for $\Theta_1$},\\
1+6(d-1) + 11 \binom{d-1}{2} + 10\binom{d-1}{3} + 
5\binom{d-1}{4} + \binom{d-1}{5} & \text{for $\Theta_2$},\\
1+6(d-1) + 11 \binom{d-1}{2} + 7\binom{d-1}{3} + \binom{d-1}{4}  & \text{for $\Theta_3$},\\
1+6(d-1) + 9 \binom{d-1}{2} + 4\binom{d-1}{3} + \binom{d-1}{4}  & \text{for $\Theta_4$} \end{cases}
\end{displaymath}
\else
\begin{displaymath}
\begin{cases}
1+6(d-1) + 15 \binom{d-1}{2} + 20\binom{d-1}{3} + \\
15\binom{d-1}{4} + 
6\binom{d-1}{5} + \binom{d-1}{6}& \text{for $\Theta_1$},\\
1+6(d-1) + 11 \binom{d-1}{2} + 10\binom{d-1}{3} + \\
5\binom{d-1}{4} + \binom{d-1}{5} & \text{for $\Theta_2$},\\
1+6(d-1) + 11 \binom{d-1}{2} + 7\binom{d-1}{3} + \binom{d-1}{4}  & \text{for $\Theta_3$},\\
1+6(d-1) + 9 \binom{d-1}{2} + 4\binom{d-1}{3} + \binom{d-1}{4}  & \text{for $\Theta_4$} \end{cases}
\end{displaymath}
\fi
codewords of $\C$.\label{part:valid3_q7_1to4}
  \end{enumerate}
\end{proposition}

\begin{IEEEproof}
By Proposition~\ref{prop:specialMvalid}, $\mathcal{T}_1$ through $\mathcal{T}_3$ are almost doubly-symmetric and hence facet-defining by Proposition~\ref{prop:facets_modified}. The class $\mathcal{T}_4$ is not almost doubly-symmetric (see Example~\ref{ex:doubly-symmetric}), but it is symmetric, and it can be easily verified that the $\frac12(p-3)=2$ length-$3$ vectors $(1,4,2)\eiriknewnew{^T}$ and $(3,3,1)\eiriknewnew{^T} \in \F_7^3$ satisfy the conditions in Remark~\ref{rem:numerical-check-facet}  in Appendix~\ref{app:prooffacets}, which proves that $\mathcal T_4$ is facet-defining.

The remaining counting formulas are special cases of the general counting formulas of Lemma~\ref{lem:counting_formulas} (details omitted for brevity).
\end{IEEEproof}

We now construct the inequalities in $\Theta_5$. To that end, let $\mathcal{T}_5^{\rm b} = \mathcal T^{(0010000)}$ and $\mathcal T_5^{\rm nb} = \mathcal T^{(0110000)}$ which correspond to the bold (resp. nonbold) entries in Table~\ref{table:q7classes}.
\begin{construction}[Construction of $\Theta_5$]\label{constr:hilo7}
  Choose an index $i^{\rm nb} \in \range d$. For $i \in \range d \setminus \{i^{\rm nb}\}$, choose $k_i \in \F_7$ arbitrarily.
  This choice results in a \emph{canonical codeword} $\bm c \in \C$ by defining, for $i \neq i^{\rm nb}$, $c_i = t^{\rm b}_{k_i, \uparrow}$, and the condition $\bm c \in \C$ specifies the remaining entry $c_{i^{\rm nb}}$.
  Now, set $k_{i^{\rm nb}} = t^{\rm nb}_{\downarrow, c_{i^{\rm nb}}}$.
  
  The resulting inequality is $\boldtheta^T \bm x \leq \kappa$ with $\boldtheta_i = \bm t^{\rm b}_{k_i}$ for $i \neq i^{\rm nb}$, $\boldtheta_{i^{\rm nb}} = \bm t^{\rm nb}_{k_{i^{\rm nb}}}$, and $\kappa = \boldtheta^T \Fv(\bm c)$, and the set of all inequalities obtained by this construction is denoted by $\Theta_5$.
\end{construction}
\begin{remark}
  The above construction differs from Construction~\ref{constr:hilo} in two points. First, here one building block is chosen from $\mathcal T_5^{\rm nb}$, while all others are from $\mathcal T_5^{\rm b}$. Secondly, in Construction~\ref{constr:hilo7}, $i^{\rm nb}$ is the index that is not free to choose (instead of $d$ in Construction~\ref{constr:hilo}), \ie Remark~\ref{rem:hilo-arbitrary} has been incorporated. The latter is necessary because each choice of $i^{\rm nb}$ leads to a \emph{different} set of inequalities; consequently, $\abs{\Theta_5} = d \cdot 7^{d-1} = d \cdot \abs{\Theta_r}$ for $r\in\{1,2,3,4\}$.
\end{remark}

Lemma~\ref{lem:conditions} (and its proof) can straightforwardly be adapted to Construction~\ref{constr:hilo7}, which leads to:
\begin{lemma}\label{lem:conditions_theta5}
  An inequality $\boldtheta^T \boldx \leq \kappa$ with $\boldtheta_{i^{\rm nb}} = \bm t^{\rm nb}_{k_{i^{\rm nb}}}$ for some $i^{\rm nb} \in \range d$ and $\boldtheta_i = \bm t^{\rm b}_{k_i}$ for $i \neq i^{\rm nb}$ is in $\Theta_5$ if and only if
\begin{subequations}
  \begin{align}
    &\sum_{i=1}^d k_i = [d-1]_7 \cdot \sigma^{\rm b} \label{eq:thetaconditions-Theta5}\\
    \text{and}\quad&\kappa = (d-1) t^{\rm b}_{0,\sigma^{\rm b}} - \sum_{i\neq i^{\rm nb}} t^{\rm b}_{0,k_i} - t^{\rm nb}_{0,k_{i^{\rm nb}}}
  \end{align}
\end{subequations}
  where $\sigma^{\rm b} = 2$ and $t_{0,\sigma}^{\rm b} = 9$ by definition.
\end{lemma}
\begin{proposition} \label{prop:Theta5_p7}
All inequalities from $\Theta_5$ are valid and facet-defining for $\Pd = \conv(\Fv(\C))$, where $\C$ is an  \enquote{all-ones}  septenary SPC code of length $d \geq 3$.
\end{proposition}
\begin{IEEEproof}
  The result follows because the relevant results from Sections~\ref{sec:validInvalid} and \ref{sec:facet-defining} can be easily generalized to the situation of Construction~\ref{constr:hilo7}, \ie that one building block within $\boldtheta$ is chosen from a different class. See Appendix~\ref{app:proofTheta5} for an outline.
\end{IEEEproof}

We remark that, because the class $\mathcal T_5^{\rm nb}$ is invalid, the presence of more than one building block from  $\mathcal{T}_5^{\rm nb}$ in an invalid inequality does not result in a valid inequality. In other words, it is crucial for $\Theta_5$ that \emph{exactly} one entry of $\bm\theta$ is picked from $\mathcal T_5^{\rm nb}$.

For $\Theta_6$, there are three classes of building blocks named $\mathcal T_6^{\rm b}$, $\mathcal T_6^{\rm lo}$, and $\mathcal T_6^{\rm hi}$, which are listed in Table~\ref{table:q7classes}. To construct inequalities, Construction~\ref{constr:hilo7} is further extended, as shown shortly. Since the structure of the building blocks (which do not satisfy Definition~\ref{def:bb}) results in the \enquote{$\argmax$} expressions in Definition~\ref{def:hilo-notation} not being well-defined, we need the following definitions to choose specific maximizers.
\begin{definition}\label{def:hilo-theta6}
  For $k \in \F_7$, let $t^{\rm b}_{k,\uparrow} = -k$, $t^{\rm hi}_{k,\uparrow} = -k$, and $t^{\rm lo}_{\downarrow, k} = -k$. Further, define $\sigma^{\rm b} = \sigma^\hi = 0$.
\end{definition}
\begin{construction}[Construction of $\Theta_6$]\label{constr:hilo7_Theta6}
  Choose $i^{\rm hi},i^{\rm lo} \in \range d$ with $i^{\rm lo} \neq i^{\rm hi}$. For notational purposes, we introduce the label
  \[ l_i = \begin{cases}\lo&\text{if }i = i^\lo,\\
   \hi&\text{if }i=i^\hi,\\
   \mathrm b&\text{otherwise}.\end{cases}\]
  For $i \in \range d \setminus \{i^{\rm lo}\}$, choose $k_i \in \F_7$ arbitrarily and define the canonical codeword $\bm c \in \C$ by $c_i = t^{l_i}_{k_i,\uparrow} = -k_i$; this specifies $c_{i^{\rm lo}}$. Now, set $k_{i^{\rm lo}} = t^{\rm lo}_{\downarrow, c_{i^{\rm lo}}} = -c_{i^{\rm lo}}$.
  
  The resulting inequality is $\boldtheta^T \boldx \leq \kappa$ with $\boldtheta_i = \boldt^{l_i}_{k_i}$ for $i \in \range d$, and $\kappa = \boldtheta^T \Fv(\bm c)$, and the set of all inequalities obtained by this construction is denoted by $\Theta_6$.
\end{construction}

\begin{lemma} \label{lem:conditions_theta6}
 For fixed $i^\hi,i^\lo \in \range d$ with $i^\hi \neq i^\lo$, the inequality $\boldtheta^T \boldx \leq \kappa$ with $\boldtheta_i = \bm t^{l_i}_{k_i}$ is in $\Theta_6$ if and only if
  \begin{subequations} \label{eq:hilo-conditions_3}
  \begin{align}
    &\sum_{i=1}^d k_i = 0 \label{eq:kcondition-theta6}\\
\text{and}\quad&\kappa = -\sum_{i=1}^{d} t^{l_i}_{0,k_i}.\label{eq:kappacondition-theta6}
  \end{align}
  \end{subequations}
\end{lemma}
\begin{IEEEproof}
  We again argue similar to the proof of Lemma~\ref{lem:conditions}. Let $\bm\theta^T \bm x \leq \kappa$ be constructed from Construction~\ref{constr:hilo7_Theta6} with canonical codeword $\bm c$. For \eqref{eq:kcondition-theta6},
  \[ \bm c \in \C \Leftrightarrow 0 = \sum_{i=1}^d c_i = \sum_{l_i \neq \lo} t^{l_i}_{k_i,\uparrow} + -k_{i^\lo} = -\sum_{i=1}^d k_i \]
  where the last step is due to Definition~\ref{def:hilo-theta6}. Now, \eqref{eq:kappacondition-theta6} holds because
  \[
    \kappa = \sum_{i=1}^d t^{l_i}_{k_i,c_i} = \sum_{l_i \neq \lo} \max(t^{l_i}_{k_i}) + \min(t^\lo_{k_{i^\lo}})
    =-\sum_{i=1}^d t^{l_i}_{0,k_i}
  \]
  where the last step can be verified by inspection of the building blocks of $\mathcal T_6^{l_i}$, $l_i = \mathrm b, \hi, \lo$. Finally, it is easily seen (as in the proof of Lemma~\ref{lem:conditions}) that the above arguments work in both directions, which concludes the proof.
\end{IEEEproof}
\begin{example}
  Let $d=3$ and choose $i^\hi=1$ and $i^\lo=3$. Then, choose $k_2=1$ (corresponding to the building block $\bm\theta_2 = \bm t^{\rm b}_1 = (\mathbf{0,0,0,0,0,0,1}) \in \mathcal T_6^{\rm b}$) and $k_1=5$ (corresponding to $\bm\theta_1 = \bm t^\hi_5 = (0,0,1,1,1,0,1) \in \mathcal T_6^\hi$). This results in $c_1=[-5]_7=[2]_7$, $c_2=[-1]_7=[6]_7$,  hence $c_3=[6]_7$ and thus $k_3 = [-6]_7 = [1]_7$, such that $\bm\theta_3 = \bm t^\lo_1 = (0,0,-1,0,-1,-1,-1) \in \mathcal T_6^\lo$.
\end{example}
\begin{proposition} \label{prop:Theta6_p7}
All inequalities from $\Theta_6$ are valid and facet-defining for $\Pd = \conv(\Fv(\C))$, where $\C$ is an \enquote{all-ones}  septenary SPC code of length $d \geq 3$.
\end{proposition}
\begin{IEEEproof}
  See Appendix~\ref{app:proofTheta6}.
\end{IEEEproof}

\begin{conjecture}\label{conj:facetComplete_q7}
  Let $\C$ be the septenary  \enquote{all-ones} SPC code of length $d \geq 3$ and $\Pd = \conv(\Fv(\C))$. %
Then, 
\begin{displaymath}
\Theta(d) = \left( \cup_{\varphi \in \GL(\F_7)} \varphi \left( \cup_{i=1}^6 \Theta_i  \right) \right) \cup \Delta_7^d
\end{displaymath}
with $\abs{\Theta(d)} = \left(4  {d \choose 2} + 24 + 6d\right) \cdot 7^{d-1} + 8d$ gives a complete and irredundant description of $\P$.
\end{conjecture}

We remark that when applying the $6$ distinct elements of $\GL(\F_7)$ to $\Theta_6$, only two distinct classes of inequalities occur, i.e., $\abs{\cup_{\varphi \in \GL(\F_7)} \varphi\left(\Theta_6\right)}=2 \cdot \abs{\Theta_6}$, which explains the multiplication by $4$ in the expression for $\abs{\Theta(d)}$.

We have verified numerically that the conjecture is true for $d=3$ and $4$.

\begin{remark}
Note that for $p > 7$, several \enquote{hybrid} classes of inequalities of the same form as $\Theta_5$ will exist. To identify these classes, one can loop through all possible choices for a bold
building block class (there are $2^{p-1}$ possible choices, not
considering the all-zero $m$-vector). The corresponding nonbold class can be identified be generalizing the results and arguments of Appendix~\ref{app:proofTheta5} for $p=7$ to the general case.
This procedure can identify $21$, $60$, $405$, and $967$ additional (some of which may be redundant) valid facet-defining classes for $p=11$, $13$, $17$, and $19$, respectively. Note that the procedure will also identifiy the valid facet-defining basic classes, which are considered degenerated cases (the bold
class and the nonbold class are the same) here. The numbers above refer to nondegenerated cases.
\end{remark}

\section{Adaptive Linear Programming Decoding} \label{sec:ALP_q}

\revtwo{While the (in)equalities in $\Theta(d)$ constructed in Section~\ref{sec:bb} could in theory be subsumed in one large LP for (relaxed) LP decoding as described in Section~\ref{sec:lpdecoding}, this approach is pratically infeasible because $\abs{\Theta(d)}$ is exponential in $d$.}

In this section, we show how to overcome that issue by means of an efficient separation algorithm, which allows for efficient relaxed ALP decoding of general codes over $\F_p$. It thus generalizes the well-known \emph{Adaptive LP Decoder} for binary codes \cite{tag07}. Throughout the section, let $\C$ denote a general $p$-ary code defined by the parity-check matrix $\bm H$, the $j$-th row of which is denoted by $\bm h_j$.

The main loop of our ALP decoder (Algorithm~\ref{alg:ALP}) is similar to \cite[Alg.~2]{tag07}, except that the $n(p + 1)$ \emph{simplex constraints} $\Delta_p^\C$ are present from start. Denote by $\mathcal{M}$ the set of vectors $\bm m$ corresponding to valid irredundant facet-defining classes $\Theta^{\bm m}$.  \revtwo{Note that valid $\bm m$-vectors can be found by a simple enumeration based on Corollary~\ref{cor:allValidOrNot} and Definition~\ref{def:valid-class} that runs in negligible time for all $p$ of reasonable size (see also Remarks~\ref{rem:search_m1} and \ref{rem:search_m2}). In Table~\ref{table:facetClasses}, we have listed valid $\bm m$-vector prefixes for $p \leq 19$. Furthermore, Proposition~\ref{prop:specialMvalid} gives a construction for valid $\bm m$-vectors for any value of $p$. }

\begin{algorithm}
  \caption{ALP Decoder for $p$-Ary Codes}\label{alg:ALP}
  \textbf{Input:} $p$-ary code $\C$ of length $n$, channel output $\Lamv(\boldy)$\\
  \textbf{Output:} Solution $\boldx$ of \eqref{eq:LPformulation}
  \begin{algorithmic}[1]
    \State Initialize a linear program with variables $\boldx \in \R^{np}$, constraints from $\Delta_p^\C$, and objective function $\min\,\Lamv(\boldy)^T\boldx$\label{algALP:init}%
    \While{\texttt{True}}\label{algALP:outerWhile}
      \State $\boldx^{\mathrm{LP}} \leftarrow$ optimal LP solution
      \ForAll{$j\in \mathcal J$, $\bm m \in \mathcal{M}$, and $r \in \F_p \setminus \{0\}$} \label{algALP:jandi}
        \State $(\bm\theta,\kappa) \leftarrow$ \Call{Separate}{$\bm m, r, \bm h_j, P_j\boldx^\LP$} \label{algALP:separate}
        \If{$(\boldtheta,\kappa) \neq \texttt{Null}$}
          \State insert $\boldtheta^T \boldx \leq \kappa$ into the LP model
        \EndIf
      \EndFor
      \If{no cut was added in the above loop}
        \State \textbf{return} $\boldx^{\rm LP}$
      \EndIf
    \EndWhile
  \end{algorithmic}
\end{algorithm}

\begin{notms}
The main issue that needs to be addressed is that of efficient separation (line~\ref{algALP:separate} of Algorithm~\ref{alg:ALP}) of the sets of inequalities $\varphi_{\bm h, r}(\Theta^{\bm m})$, for some $\varphi_r \in \GL(\F_p)$, valid $\bm m \in \mathcal{M}$, and an SPC code with parity-check vector $\bm h = (h_1,\dotsc,h_d)$, \ie to develop an efficient cut-search algorithm that finds, for given $\bm x \in \R^{dp}$, an inequality in $\varphi_{\bm h, r}(\Theta^{\bm m})$ that is violated by $\bm y$ or concludes that no such inequality exists.
\end{notms}

At first, we reduce separation to the case of $\bm h = (1,\dotsc,1)$ (\ie $\Cj$ is an \enquote{all-ones} SPC code) and $r=1$ (\ie $\varphi_r$ is the identity). This reduction, which is implemented by Algorithm~\ref{alg:separateGeneralSPC}, is based on the following corollary of Theorem~\ref{thm:rotation-general}.

\begin{corollary}
  \label{cor:rotation-separation}
  Let $\bm h$, $\P$, and $\P(\bm h)$ be defined as in Corollary~\ref{cor:rotation-generalSPC} and $\bm y \in \R^{dq}$. Then, the inequality
  $(\rot_{\bm h}\circ \rot_r)(\bm \theta)^T \bm x \leq \kappa$
  from $\varphi_{\bm h, r}(\Theta^{\bm m})$
  separates $\bm y$ from $\P(\bm h)$ if and only if the inequality 
  $\bm \theta^T \bm x \leq \kappa$ from $\Theta^{\bm m}$ separates $(\rot_r^{-1} \circ \rot^{-1}_{\bm h})(\bm y)$ from $\P$.
\end{corollary}
\begin{IEEEproof}
  Starting from the second statement, it holds that
  \ifonecolumn
  \begin{align*}
    &\bm \theta^T \bm x \leq \kappa \text{ separates } (\rot_r^{-1}\circ \rot_{\bm h}^{-1})(\bm y)\text{ from }\P\\
    \Leftrightarrow\;&
    \bm \theta^T \bm x \leq \kappa\text{ for }\bm x \in \P\text{ and }\bm \theta^T(\rot_r^{-1}\circ \rot_{\bm h}^{-1})(\bm y) > \kappa\\
    \Leftrightarrow\;&
    \rot_r(\bm\theta)^T\bm x \leq \kappa\text{ for }\bm x\in \P\text{ and }\rot_r(\bm \theta)^T \rot_{\bm h}^{-1}(\bm y) > \kappa\\
    \Leftrightarrow\;&
    (\rot_{\bm h}\circ \rot_r)(\bm \theta)^T\bm x \leq \kappa \text{ for }\bm x \in\P(\bm h)
    \text{ and }(\rot_{\bm h}\circ\rot_r)(\bm \theta)^T \bm y > \kappa
  \end{align*}
  \else
  \begin{align*}
    &\bm \theta^T \bm x \leq \kappa \text{ separates } (\rot_r^{-1}\circ \rot_{\bm h}^{-1})(\bm y)\text{ from }\P\\
    \Leftrightarrow\;&
    \bm \theta^T \bm x \leq \kappa\text{ for }\bm x \in \P\text{ and }\bm \theta^T(\rot_r^{-1}\circ \rot_{\bm h}^{-1})(\bm y) > \kappa\\
    \Leftrightarrow\;&
    \rot_r(\bm\theta)^T\bm x \leq \kappa\text{ for }\bm x\in \P\text{ and }\rot_r(\bm \theta)^T \rot_{\bm h}^{-1}(\bm y) > \kappa\\
    \Leftrightarrow\;&
    (\rot_{\bm h}\circ \rot_r)(\bm \theta)^T\bm x \leq \kappa \text{ for }\bm x \in\P(\bm h)\\
    &\quad\quad\text{ and }(\rot_{\bm h}\circ\rot_r)(\bm \theta)^T \bm y > \kappa
  \end{align*}
  \fi
  where we have used, on the left side, Corollaries~\ref{cor:rotation-generalSPC} and \ref{cor:rotation-autfq} and Lemma~\ref{lem:permScalProd} on the right. Now, the last line exactly states that $(\rot_{\bm h}\circ \rot_r)(\bm \theta)^T\bm x \leq \kappa$ separates $\bm y$ from $\P(\bm h)$, which concludes the proof.
\end{IEEEproof}

{\revtwo{From now, hence, $\bm h = (1,\dotsc,1)$ and $r=1$ are assumed.}}

\begin{algorithm}
  \caption{\textsc{Separate}($\bm m$, $r$, $\bm h$, $\bm x$)}
  \label{alg:separateGeneralSPC}
  \textbf{Input:} $\bm m \in \mathcal M$, $r \in \F_p \setminus \{0\}$, $\bm h =(h_1,\dotsc,h_d)$ with nonzero entries, and current (projected) LP solution $\bm x$ \\
  \textbf{Output:} Inequality in $\varphi_{\bm h, r}(\Theta^{\bm m})$ violated by $\bm x$, if such exists; \texttt{Null} otherwise
  \begin{algorithmic}[1]
    \State $\tilde \boldx \leftarrow (\rot_r^{-1}\circ \rot_{\bm h}^{-1})(\boldx)$
    \State $\bm\theta \leftarrow$ \Call{Separate}{$\bm m$, $\tilde{\bm x}$}
    \If{$\bm\theta \neq$ \texttt{Null}}
      \State Compute $\kappa$ from \eqref{eq:kappacondition}
      \State \textbf{return} $((\rot_{\bm h}\circ \rot_r)(\bm \theta), \kappa)$
    \Else
      \State \textbf{return} \texttt{Null}
    \EndIf
    \end{algorithmic}
\end{algorithm}
We now describe the separation of inequalities in $\Theta^{\bm m}$, for some $\bm m \in \mathcal{M}$. Any such inequality $\bm\theta^T \bm x \leq \kappa$ with $\boldtheta = (\bm t^{\bm m}_{k_1}\mid \dotsc \mid \bm t^{\bm m}_{k_d})^T$ can be rewritten (using \eqref{eq:kappacondition}) as
\begin{equation}
  \Psi(\boldtheta, \boldx)= \sum_{i=1}^d v^{k_i}(\boldx_i)
  \geq t^{\bm m}_{0,\sigma}  \label{eq:facetAltForm_p}
\end{equation}
where $\boldx = (\boldx_1,\dotsc,\boldx_d)^T$, $\boldx_i = (x_{i,0},\dotsc,x_{i,p-1})$ for all $i \in \range d$, and $v^k(\boldx_i) = t^{\bm m}_{0,\sigma} - t^{\bm m}_{0,k} -\boldt^{\bm m}_k \boldx_i^T$. %
Thus, $\Theta^{\bm m}$ contains a cut for $\boldx$ (for some $j \in \mathcal{J}$) if and only if $\Psi(\bm\theta,\bm x) < t^{\bm m}_{0,\sigma}$ for some $\boldtheta$ from $\Theta^{\bm m}$, \ie if and only if the optimization problem (in $d$ variables $k_1,\dotsc,k_d \in \F_p$)
\begin{subequations}
\begin{align}
  \psi^*=\min\;& \Psi(\boldtheta,\boldx)\\
\text{s.t.}\;&\boldtheta = (\bm t^{\bm m}_{k_1} \mid \dotsc \mid  \bm t^{\bm m}_{k_d})^T\\
&\sum_{i=1}^d k_i = [d-1]_p\sigma \label{eq:sepOptFacet_generic}
\end{align}\label{eq:sepOpt_generic}%
\end{subequations}
where the condition \eqref{eq:sepOptFacet_generic} is due to \eqref{eq:Thetacondition},
has an optimal solution that satisfies $\psi^* < t^{\bm m}_{0,\sigma}$.

We now describe how to solve \eqref{eq:sepOpt_generic} using a DP approach with linear running time in $d$. For $s \in \range d$ and $\zeta \in \F_p$, define
\ifonecolumn
\begin{displaymath}\psi(\boldx,s,\zeta) = \min\left\{ \sum_{i=1}^s v^{k_i}(\boldx_i)\colon k_i \in \F_p\text{ for }i \in \range s  
\text{ and } \sum_{i=1}^s k_i  = \zeta \right\}.
\end{displaymath}
\else
\begin{multline*}\psi(\boldx,s,\zeta) =\\ \min\left\{ \sum_{i=1}^s v^{k_i}(\boldx_i)\colon k_i \in \F_p\text{ for }i \in \range s  
\text{ and } \sum_{i=1}^s k_i  = \zeta \right\}.
\end{multline*}
\fi
It holds that $\psi^* = \psi(\boldx,d,[d-1]_p\sigma)$ and the obvious recursion
\begin{equation}
  \psi(\boldx, s, \zeta) = \min_{\beta \in \F_p} \left\{v^{\beta}(\boldx_s) + \psi(\boldx, s-1, \zeta - \beta)\right\}\label{eq:sepRecursion}
\end{equation}
for $s \geq 2$ and $\zeta \in \F_p$ 
allows us to compute a $d\times \F_p$ table $\texttt{T}$ with entries $\texttt{T}[s,\zeta] = \psi(\boldx, s, \zeta)$: initialize the first row of $\texttt T$ with
$\texttt T[1,\zeta] = \psi(\boldx, 1, \zeta) = v^{\zeta}(\boldx_1)$ for $\zeta \in \F_p$.
Then, use \eqref{eq:sepRecursion} to proceed from top to bottom until reaching row $d$, where only the entry $\texttt T[d, [d-1]_p\sigma]$ is needed. Because the expression in \eqref{eq:sepRecursion} can be calculated in time $\mathcal O(p)$, the overall time needed to obtain $\psi^*$ is $\mathcal O(dp^2)$, which is linear for fixed $p$. Observe that the actual solution $\boldtheta^*$ of \eqref{eq:sepOpt_generic} can be obtained within the same asymptotic running time by storing the minimizing $\beta$'s from \eqref{eq:sepRecursion} in a second $d \times \F_p$ table $\texttt S$. The complete algorithm is outlined in Algorithm~\ref{alg:SAgeneral} (note that, in an actual implementation, one has to use the integer representations of field elements everywhere, and insert \enquote{mod $p$} statements in the appropriate places).

\begin{algorithm}
  \caption{\textsc{Separate}($\bm m, \bm x$)}\label{alg:SAgeneral}
  \textbf{Input:} $\bm m \in \mathcal M$ and $\bm x = (\bm x_1,\dotsc,\bm x_d)^T \in \R^{dp}$ \\
  \textbf{Output:} Solution $\boldtheta^*$ of \eqref{eq:sepOpt_generic}, if $\psi^* < t^{\bm m}_{0,\sigma}$; \texttt{Null} otherwise
  \begin{algorithmic}[1]
    \State Let \texttt{T}, \texttt{v},  and \texttt S be $d \times \F_p$ arrays, and let \texttt k be a length-$d$ array 
    \For{$\zeta \in \F_p$}
      \For{$i \in \range d$}
        \State{$\texttt{v}[i,\zeta] \leftarrow t^{\bm m}_{0,\sigma} - t^{\bm m}_{0,\zeta} - \bm t_\zeta^{\bm m} \bm x_i^T$}\Comment{initialize \texttt{v}}
      \EndFor
      \State $\texttt{T}[1,\zeta] \leftarrow \texttt v[1,\zeta]$\Comment{initialize $\texttt{T}[1,:]$}
      \State $\texttt{S}[1,\zeta] \leftarrow \zeta$\Comment{initialize $\texttt{S}[1,:]$}
    \EndFor
    \For{$i =2,\dotsc, d$}
      \For {$\zeta \in \F_p$}
        \State $\texttt S[i,\zeta] \leftarrow -1$
        \State $\texttt T[i,\zeta] \leftarrow \infty$
        \For {$\beta \in \F_p$}\Comment{find min.\ from \eqref{eq:sepRecursion}}
          \State $\texttt{val} \leftarrow \texttt v[i,\beta] + \texttt T[i-1,\zeta - \beta]$
          \If{$\texttt{val} < \texttt T[i,\zeta]$}
            \State $\texttt T[i,\zeta] \leftarrow \texttt{val}$
            \State $\texttt S[i,\zeta] \leftarrow \beta$
          \EndIf
        \EndFor
      \EndFor
    \EndFor
    \If{$\texttt T[d,[d-1]_p\sigma] < t_{0,\sigma}^{\bm m}$}
      \State $\texttt k[d] \leftarrow \texttt S[d,[d-1]_p \sigma]$
      \State $\texttt{next} \leftarrow [d-1]_p\sigma - \texttt k[d]$
      \For{$i = d-1,\dotsc, 1$}
        \State $\texttt k[i] \leftarrow \texttt S[i,\texttt{next}]$
        \State $\texttt{next} \leftarrow \texttt{next} - \texttt k[i]$
      \EndFor
      \State \textbf{return} \eirik{$(\bm t_{{\texttt k}[1]} \mid \dotsc \mid \bm t_{{\texttt k}[d]})^T$}
    \EndIf
    \State \textbf{return} \texttt{Null}
  \end{algorithmic}
\end{algorithm}

For the binary case, the well-known separation algorithm  \cite[Alg.~1]{zha12} is more efficient than the above general approach: there, problem \eqref{eq:sepOpt_generic} is first solved without the constraint \eqref{eq:sepOptFacet_generic} by setting $k_i^* = 0 \Leftrightarrow x_{i,1} > 1/2$. If this solution does not happen to fulfill \eqref{eq:sepOptFacet_generic}, the constraint is restored by altering a single $k_i^*$ with minimal corresponding $\abs{x_{i,1} - 1/2}$ (see \cite[Alg.~1]{zha12} for details).

For $p=3$ a more efficient algorithm can be derived as we will show below in Section~\ref{sec:ALP_q3}. However, as the number of possible combinations for restoring \eqref{eq:sepOptFacet_generic} grows rapidly with increasing $p$, the DP approach is preferable in the general case.

\begin{remark}\label{rem:SAtweaks}
 Algorithm~\ref{alg:SAgeneral} can be tweaked in several ways:
  \begin{itemize}
    \item In the $d$-th row of $\texttt T$, only the single value $\texttt T[d,[d-1]_p\sigma]$ has to be computed.
    \item If $d$ and/or $p$ are large, one could first minimize $\Psi(\boldtheta,\boldx)$ without the constraint \eqref{eq:sepOptFacet_generic} (which is possible in time $\mathcal O(dp)$). If the result satisfies \eqref{eq:sepOptFacet_generic} (optimum found) or fulfills $\psi^* \geq t^{\bm m}_{0,\sigma}$ (no cut can be included), we are done.
    \item Because $v^k(\boldx_i) \geq 0$ for all $k \in \F_p$ and $i \in \range{d}$, $\psi^* \geq \min_{\zeta \in \F_p} \psi(\boldx, i, \zeta)$ holds for \emph{any} $i \in \range d$. Hence, the search can be stopped as soon as all entries in a single row of $\texttt T$ are larger than or equal to $t^{\bm m}_{0,\sigma}$.    
  \end{itemize}
\end{remark}

\subsection{Efficient Implementation for $p=3$} \label{sec:ALP_q3}

In this subsection, we explicitly develop an optimized version of Algorithm~\ref{alg:SAgeneral} for the case of $p=3$. Note that there is only one relevant class $\mathcal T^{\bm m = (0,0,0)}$ (cf. Section~\ref{sec:q3}), such that the \enquote{$\bm m$}-parameter is omitted in the sequel.

The optimization problem in \eqref{eq:sepOpt_generic} simplifies to the following form for $p=3$:
\begin{subequations}
\begin{align}
  \psi^*=\min\;& \Psi(\boldtheta,\boldx) \label{eq:sepOptP3-obj}\\
  \text{s.\,t.}\quad&\boldtheta = (\bm t_{k_1} \mid \dotsc \mid \bm t_{k_d})^T\\ &\sum_{i=1}^d k_i = [2(d-1)]_3.\label{eq:sepOptFacet}
\end{align}\label{eq:sepOpt}%
\end{subequations}

To solve \eqref{eq:sepOpt}, we first ignore \eqref{eq:sepOptFacet}, \ie find $\hat\boldtheta$ that unconditionally minimizes $\Psi(\boldtheta,\boldx)$ by computing, for $i \in \range d$,
\begin{equation} \notag
  \hat k_i = \argmin_{k\in \F_3} v^k(\boldx_i),
\end{equation}
breaking ties arbitrarily.

If $\Psi(\hat\boldtheta,\boldx)\geq2$, then $\Theta$ does not contain a cut. If otherwise $\hat\boldtheta$ happens to fulfill \eqref{eq:sepOptFacet}, it is the optimal solution of \eqref{eq:sepOpt} and hence leads to a cut; in both cases we are done.

In the remaining case, $\eta = \sum k_i - [2(d-1)]_3 \neq [0]_3$. Let $I \subseteq \range d$ be the set of positions in which \eirik{the unconstrained optimal solution} $(\hat k_1,\dotsc, \hat k_d)$ differs from the optimal solution $(k_1^*,\dotsc, k_d^*)$ of \eqref{eq:sepOpt}, which we now need to find. By definition of the $\hat k_i$, we can assume that no subset of $I' \subseteq I$ satisfies $\sum_{i \in I'} \hat k_i - k_i^* = [0]_3$; otherwise, one could replace $k_i^*$ by $\hat k_i$ for $i \in I'$ while maintaining \eqref{eq:sepOptFacet} without increasing the objective value \eqref{eq:sepOptP3-obj}. In particular, this shows that $\abs I \leq 2$.

Let
\ifonecolumn
\begin{displaymath}
  \psi_i^1 = v^{\hat k_i + [1]_3}(\bm x_i) - v^{\hat k_i}(\bm x_i)
  \text{ and } \psi_i^2= v^{\hat k_i + [2]_3}(\bm x_i) - v^{\hat k_i}(\bm x_i)
\end{displaymath}
\else
\begin{align*}
  &\psi_i^1 = v^{\hat k_i + [1]_3}(\bm x_i) - v^{\hat k_i}(\bm x_i)\\
  \text{and }&\psi_i^2= v^{\hat k_i + [2]_3}(\bm x_i) - v^{\hat k_i}(\bm x_i)
\end{align*}
\fi
denote the increase of \eqref{eq:sepOptP3-obj} incurred by replacing $\hat k_i$ with $\hat k_i + [1]_3$ and $\hat k_i + [2]_3$, respectively. Furthermore, define $i^1 \neq j^1$ and $i^{2} \neq j^{2} \in \range d$ such that
  \begin{subequations}
  \begin{align}
    &\psi^1_{i^1} \leq \psi^1_{j^1} \leq \psi^1_l&&\text{for all }l \notin \{i^1,j^1\}
    \\
    \text{and }&\psi^{2}_{i^{2}} \leq \psi^{2}_{j^{2}} \leq \psi^{2}_l&&\text{for all }l \notin \{i^{2},j^{2}\}
  \end{align}%
  \label{eq:psiPlusMinus}%
  \end{subequations}%
  \ie $\psi^\zeta_{i^\zeta}$ and $\psi^\zeta_{j^\zeta}$ correspond to the two optimal positions in which to add $\zeta$ to $\hat k_i$.

\begin{lemma}\label{lem:psiPlusMinus_q3}
  Let $\boldtheta^1 = (\bm t_{k^1_1} \mid \dotsc \mid \bm t_{k^1_d})\eiriknew{^T}$ \eirik{and $\boldtheta^2 = (\bm t_{k^2_1} \mid \dotsc \mid \bm t_{k^2_d})\eiriknew{^T}$} be defined by
  \begin{align*}
   k^1_i &=
    \begin{cases}
      \hat k_i - \eta &\text{if }i = i^{-\eta},\\
      \hat k_i&\text{otherwise}
    \end{cases}\\
    \text{and}\quad k^2_i &=
      \begin{cases}
        \hat k_i + \eta &\text{if }i \in \{i^{\eta},j^{\eta}\},\\
        \hat k_i&\text{otherwise}.
      \end{cases}
  \end{align*}
  If $\psi^{-\eta}_{i^{-\eta}} < \psi^{\eta}_{i^{\eta}} + \psi^{\eta}_{j^{\eta}}$, then $\boldtheta^1$ is the optimal solution of \eqref{eq:sepOpt}, otherwise $\boldtheta^2$.
\end{lemma}
\begin{IEEEproof}
  By the above discussion, $\boldtheta^1$ minimizes $\Psi(\boldtheta, \boldx)$ among all possibilities that differ in only one entry from $(\hat k_1,\dotsc, \hat k_d)$, while $\boldtheta^2$ is optimal for two different positions. As $\abs I \leq 2$, one of both is the optimal solution of \eqref{eq:sepOpt}, which concludes the proof.
\end{IEEEproof}

This completes the description of \textsc{Separate}($\bm x$) for ternary codes; the pseudocode is shown in Algorithm~\ref{alg:SAternary}.
\begin{algorithm}
  \caption{\textsc{Separate}($\bm x$) for $p=3$}\label{alg:SAternary}
  \textbf{Input:} Current LP solution $\boldx \in \R^{3d}$ \\
  \textbf{Output:} Solution $\boldtheta^*$ of \eqref{eq:sepOpt}, if $\psi^\ast < t_{0,\sigma}= 2$; \texttt{Null} otherwise
  \begin{algorithmic}[1]
    \State Initialize arrays $\bm k$, $\bolda$, $\boldb$, $\boldsymbol{\psi}^1$, $\boldsymbol{\psi}^{2}$ of length $d$ each
    \State $\Psi \leftarrow 0$, $\eta \leftarrow [-2(d-1)]_3$
    \For{$i \in \range d$}
      \State Compute $v^0(\bm x_i)$, $v^1(\bm x_i)$, and $v^2(\bm x_i)$
      \If{$v^0(\bm x_i) \leq v^1(\bm x_i)$ and $v^0(\bm x_i) \leq v^2(\bm x_i)$}
        \State $k_i \leftarrow 0$
      \ElsIf{$v^1(\bm x_i) \leq v^0(\bm x_i)$ and $v^1(\bm x_i) \leq v^2(\bm x_i)$}
        \State $k_i \leftarrow 1$
      \Else
        \State $k_i \leftarrow 2$
      \EndIf
      \State $\eta \leftarrow \eta + k_i$
      \State $\Psi \leftarrow \Psi + v^{k_i}(\bm x_i)$
    \EndFor
    \If{$\Psi < 2$ and $\eta = [0]_3$}
      \State \textbf{return} $(\bm t_{k_1} \mid \dotsc \mid \bm t_{k_d})^T$
    \ElsIf{$\Psi < 2$}
      \State compute $\psi^1_i$ and $\psi^2_i$ for $i \in \range d$, and
      \State compute \eirik{$i^{-\eta}, i^{\eta}, j^{\eta}$} defined in \eqref{eq:psiPlusMinus}
      \If{$\psi^{-\eta}_{i^{-\eta}} < \psi^{\eta}_{i^{\eta}} + \psi^{\eta}_{j^{\eta}}$}
        \State $k_{i^{-\eta}} \leftarrow k_{i^{-\eta}} - \eta$
        \State $\Psi \leftarrow \Psi + \psi^{-\eta}_{i^{-\eta}}$
      \Else
        \State $k_{i^{\eta}} \leftarrow k_{i^{\eta}} + \eta$
        \State $k_{j^{\eta}} \leftarrow k_{j^{\eta}} + \eta$
        \State $\Psi \leftarrow \Psi + \psi^{\eta}_{i^{\eta}} + \psi^{\eta}_{j^{\eta}}$
      \EndIf
      \If{$\Psi < 2$}
        \State \textbf{return} $(\bm t_{k_1} \mid \dotsc \mid \bm t_{k_d})^T$
      \EndIf
    \EndIf
    \State \textbf{return} \texttt{Null}
  \end{algorithmic}
\end{algorithm}

\subsection{Implementation for $p=7$} \label{sec:ALP_q7}

As shown in Section~\ref{sec:q7} (the case $p=7$), there are two nonbasic classes $\Theta_5$ and $\Theta_6$ of inequalities. The similarity of Lemmas~\ref{lem:conditions_theta5} and \ref{lem:conditions_theta6} with Lemma~\ref{lem:conditions} allows to use the same separation algorithm as above for these classes.

An inequality in $\Theta_5$ admits the form \eqref{eq:facetAltForm_p} (with $t_{0,\sigma}^{\bm m}$ replaced by $t^{\rm b}_{0,\sigma^{\rm b}} = 9$) when one defines $v^k(\bm x_i) = 9 - t^{\rm b}_{0,k} - \bm t^{\rm b}_k \bm x_i^T$ for $i \neq i^{\rm nb}$ and $v^k(\bm x_{i^{\rm nb}}) = 9 - t^{\rm nb}_{0,k} - \bm t^{\rm nb}_k \bm x_{i^{\rm nb}}^T$, and hence the same optimization problem \eqref{eq:sepOpt_generic} (with \eqref{eq:sepOptFacet_generic} replaced by \eqref{eq:thetaconditions-Theta5}) can be used for separation and solved with the DP approach described by Algorithm~\ref{alg:SAgeneral}.

Analogously, \eqref{eq:facetAltForm_p} holds for $\Theta_6$ (with the right-hand side $t_{0,\sigma}^{\bm m}$ replaced by $0$) when we define $v^k(\bm x_i) = - t^{l_i}_{0,k} - \bm t^{l_i}_k \bm x_i^T$ (with $l_i \in \{\rm b, hi, lo\}$ as defined in Section~\ref{sec:q7}).

In both cases, the \enquote{special} indices $i^{\rm nb}$ (for $\Theta_5$) and $i^\hi$ / $i^\lo$ (for $\Theta_6$) are assumed to be fixed in advance, which can be implemented by one ($\Theta_5$) or two ($\Theta_6$) extra \texttt{for}-loops around the actual separation algorithm.

\section{Redundant Parity-Check Cuts} \label{sec:ACG}

\revtwo{In this section, we outline an efficient algorithm to improve the error-correcting performance of the ALP decoder from Section~\ref{sec:ALP_q} by considering RPC constraints of the code.}

Assume that ALP decoding of a $p$-ary linear code $\C$ of length $n$ (using Algorithm~\ref{alg:ALP}) has returned a \emph{fractional} pseudocodeword $\bm p=(\bm p_1,\dotsc,\bm p_n)\eiriknewnew{^T}$, \ie the ALP decoding algorithm (in Algorithm~\ref{alg:ALP}) has returned $\bm p = \bm x^{\rm LP}$ with some fractional entries. Due to the generalized box constraints (from the individual constituent codes), $p_{i,0}+\cdots+p_{i,p-1} = 1$, for all $i \in \range n$. Now, let 
\begin{displaymath}
\mathcal{F}_{\bm p} =\{ i \in \range n: \text{ $p_{i,0},\dotsc,p_{i,p-2}$, or $p_{i,p-1}$  is fractional} \}
\end{displaymath}
denote the set of fractional \emph{positions} in $\bm p$. We can prove the following theorem which generalizes \cite[Thm.~3]{zha12} and \cite[Thm.~3.3]{tan10} to the $p$-ary case.

\begin{theorem} \label{thm:RPC}
Let $\bm h = (h_1,\dotsc,h_n)$ denote a valid (redundant) parity-check constraint for a $p$-ary linear code $\C$ of length $n$, let $I = \{i \in \range n\colon h_i \neq 0\}$ and assume that $I \cap \mathcal F_{\bm p} = 1$ for a given pseudocodeword $\bm p$. Then, the inequalities constructed in Section~\ref{sec:bb} contain a cut that separates $\bm p$ from $\P = \conv(\Fv(\C))$.
\end{theorem}

\begin{IEEEproof}
As the entries $h_j$ and $\bm p_j$ for $j \notin I$ are not relevant, we can assume that $I = \range n$. Furthermore, by Corollary~\ref{cor:rotation-separation} and because rotating $\bm p$ does not change  $\mathcal F_{\bm p}$, we can assume without loss of generality that $\bm h = (1,\dotsc, 1)$.
Now, note that  when $\bm p_i$ is not fractional, then $\bm p_i \in \{\bm e^1,\dotsc,\bm e^p\}$ (due to the generalized box constraints), where, for $i \in \range p$, $\bm e^i = (0,\dotsc,0,1,0,\dotsc,0)$ is the $i$-th unit vector \eiriknew{in $\R^p$}.

We show that $\mathcal{T}^{\bm m = (0,\dotsc,0)}$, which is a valid class for any prime $p$ (see Proposition~\ref{prop:specialMvalid}), leads to a cut for $\bm p$.  The corresponding building blocks are
\begin{equation} \label{eq:Tm000}
\begin{split} 
\bm t^{\bm m}_0 &= (0,1,\dotsc,p-3,p-2,p-1), \\
\bm t^{\bm m}_1 &= (0,1,\dotsc,p-3,p-2,-1), \\
\bm t^{\bm m}_2 &= (0,1,\dotsc,p-3,-2,-1), \\
\vdots\quad&\phantom{=}\quad \vdots \\
\bm t^{\bm m}_{p-2} &= (0,1,-p+2,\dotsc,-2,-1), \\
\bm t^{\bm m}_{p-1} &= (0,-p+1,-p+2,\dotsc,-2,-1).
\end{split}
\end{equation}
In the optimization problem in \eqref{eq:sepOpt_generic}, $v^k(\bm p_i)=t^{\bm m}_{0,\sigma} - t^{\bm m}_{0,k}-\boldt_k^{\bm m} \boldp_i^T = p-1-\left[k\right]_\Z-\boldt_k^{\bm m} \boldp_i^T$, $k \in \F_p$ and $\boldt_k^{\bm m}$ is one of the $p$ possible building blocks of (\ref{eq:Tm000}). When $\bm p_i$ is not fractional, 
\begin{equation} \notag %
\begin{split} 
v^{0}(\bm p_i) &\in \{p-1,p-2,p-3,\dotsc,2,1,0\}, \\
v^{1}(\bm p_i) &\in \{p-2,p-3,p-4,\dotsc,1,0,p-1\}, \\
v^{2}(\bm p_i) &\in \{p-3,p-4,p-5,\dotsc,0,p-1,p-2\}, \\
\vdots\quad&\phantom{\in}\quad \vdots \\
v^{p-2}(\bm p_i) &\in \{1,0,p-1,\dotsc,4,3,2\}, \\
v^{p-1}(\bm p_i) &\in \{0,p-1,p-2,\dotsc,3,2,1\} 
\end{split}
\end{equation} 
where the ordering of elements is according to $\bm e^1,\dotsc,\bm e^p$.  Furthermore, it can easily be verified that when $\bm p_i$ is fractional, \ie $\bm p_{i} = (p_{i,0},\dotsc,p_{i,p-1})$  where $0 \leq p_{i,j} \leq 1$, $j=0,\dotsc,p-1$, $\sum_{j=0}^{p-1} p_{i,j} = 1$, and $p_{i,0},\dotsc,p_{i,p-1}$ are not all integers, then $v^{k}(\bm p_i) < p-1$, for $k \in \F_p$, \ie \emph{strictly} smaller than $p-1$. Now, we build a valid inequality from $\Theta^{\bm m = (0,\dotsc,0)}$ in the following way, assuming that $n$ is the fractional position. If $\bm p_i=\bm e^j$, $j \in \range{p}$, choose $k_i=-\left[j\right]_p$, where $i \in \range{n-1}$. \eiriknew{Otherwise, i.e., $\bm p_i = \bm 0$, choose $k_i$ arbitrarily.} This will give an overall contribution of zero to the objective function $\Psi(\boldtheta,\bm x)$ in the  optimization problem in \eqref{eq:sepOpt_generic}. Finally, we need to choose $k_n$ such that the constraint in \eqref{eq:sepOptFacet_generic} is fulfilled. However, since $\bm p_n$ is fractional (by assumption), the contribution to $\Psi(\boldtheta,\bm x)$ is strictly less than $p-1$ (independent of the choice for $k_n$) and it follows that in the optimization problem in \eqref{eq:sepOpt_generic} the optimal objective value is indeed strictly less than $p-1$, while $t^{\bm m}_{0,\sigma} = p-1$, and thus $\Theta^{\bm m = (0,\dotsc,0)}$ indeed contains a cut  (the one that we constructed) for the pseudocodeword $\bm p$. More formally, all $v^k(\bm p_i)$ are linear maps of the simplex onto the interval $[0,p-1]$, and we know that the images of the extreme points ${\bm e^1,\dotsc,\bm e^p}$ of the simplex under that map are $\{0,\dotsc,p-1\}$. By linearity, the image of a fractional \eiriknew{$\bm p_i$}, which is a nontrivial convex combination (at least two nonzero coefficients) of those extreme points, equals the corresponding convex combination (with at least two nonzero coefficients) of the images $\{0,\dotsc,p-1\}$, and is thus strictly less than $p-1$.

\end{IEEEproof}

As in the binary case, cut-inducing RPC equations can be found, for instance, by reducing the parity-check matrix $\bm H$ using Gaussian elimination to \emph{reduced row echelon} form, where columns are processed in the order of ``fractionality'' (or \emph{closeness} to $(\frac{1}{p},\dotsc,\frac{1}{p})$) of the corresponding coordinate of $\bm p$. For instance, since $\bm p_i$, $i \in \range{n}$, is a probability vector, the entropy can be computed and compared to the entropy of the uniform distribution $(\frac{1}{p},\dotsc,\frac{1}{p})$. Although Theorem~\ref{thm:RPC} guarantees a cut only for rows $\tilde{\bm h}$ of $\pi^{-1}(\tilde{\bm H}$), where $\tilde{\bm H}$ denotes the \emph{reduced row echelon} form of $\pi(\bm H$) and $\pi$ is a permutation (of length $n$) which reorders the columns of $\bm H$ in order of closeness (or entropy) to $(\frac{1}{p},\dotsc,\frac{1}{p})$, such that $\tilde{\bm h}_{\mathcal{F}_{\bm p}}$ \eirik{(containing the coordinates of $\tilde{\bm h}$ indexed by  ${\mathcal{F}_{\bm p}}$)} has weight one, rows of larger weight  may also provide a cut. Thus, in a practical decoding situation all rows $\tilde{\bm h}$ of $\pi^{-1}(\tilde{\bm H})$ should be processed, using the separation algorithm described in Section~\ref{sec:ACG}, in order to locate redundant cut-inducing parity-check equations. In \cite{zha12}, this is called \emph{adaptive cut generation} or ACG and can be combined with ALP decoding as outlined in \cite[Alg.~2]{zha12}. In Section~\ref{sec:numerical_results}, we provide simulation results for the decoding algorithm obtained by appropriately generalizing, as outlined above, the ACG-ALP procedure described in \cite[Alg.~2]{zha12} (without considering removal of constraints) to nonbinary linear codes, showing that near-ML decoding performance can be obtained for a ternary Reed-Muller (RM) code. %

\section{The Case $q=p^m$} \label{sec:ptom}

In this section, we consider the problem of efficient LP decoding of linear codes over the field $\F_q = \F_{p^m}$, where $m > 1$ is a positive integer and $p$ is any prime. 

For any nonzero $h \in \mathbb{F}_{p^m}$, $\emptyset \neq \mathcal{K} \subseteq \range{m}$, and $\boldsymbol{\gamma} \in (\F_{p} \setminus \{0\})^{|\mathcal{K}|}$, let
\begin{displaymath}
\mathcal{B}^{(\beta)}(\mathcal{K},\boldsymbol{\gamma},h) = \left \{\zeta \in \mathbb{F}_{p^m}\colon \sum_{k \in \mathcal{K}} \gamma_k \cdot \mathsf{p}(h \zeta)_k = \beta \right\}
\end{displaymath}
for $\beta \in \F_p$, where $(\cdot)_k$ denotes the $k$-th entry of its vector argument (note that summation and multiplication (except for the $h\zeta$ term) above are in $\F_p$). 
Now, let $\mathcal{C}$ denote a nonbinary SPC code over $\mathbb{F}_{p^m}$ of length $d$ defined by a parity-check vector $\bm h = (h_1,\dotsc,h_d)$. Furthermore, for any vector
\[ \bm f = (\bm f_1,\dotsc, \bm f_d)^T  \in \R^{dq}\]
define
\begin{displaymath}
g_j^{(\mathcal{K},\boldsymbol{\gamma},\bm f)} = \sum_{\beta \in \F_p \setminus \{0\}} \sum_{i \in \mathcal{B}^{(\beta)}(\mathcal{K},\boldsymbol{\gamma},h_j)} [\beta]_\Z \cdot f_{j,i} 
\end{displaymath}
where $\emptyset \neq \mathcal{K} \subseteq \range{m}$, $\boldsymbol{\gamma} \in (\F_{p} \setminus \{0\})^{|\mathcal{K}|}$, and $j \in \range d$, and  where the summation is in the real space.

We have the following proposition, which generalizes \cite[Lem.~12]{liu14} to any field (only $p=2$ was considered in \cite{liu14}) under the constant-weight embedding.\footnote{Note that a \emph{weaker} version of the proposition appeared in \cite{ros15} (proof omitted) in which $\boldsymbol{\gamma}$ was constrained to $(1,\dotsc,1)$ which results in a potentially weaker relaxation. Also, Flanagan's embedding from Remark~\ref{rem:Flanagan} was considered in \cite{ros15}.}

\begin{proposition} \label{prop:pm}
Let $\C$ be the SPC code over the field $\mathbb{F}_{p^m}$ defined by the parity-check vector $\bm h=(h_1,\dotsc,h_d)$. Then, $\Fv(\C)$ is equal to the set of vectors $\bm f \in \R^{dq}$ that \eirik{satisfy} the following three conditions (and will be denoted by $\mathcal E$ in the sequel):
\begin{enumerate}
\item $f_{j,i} \in \{0,1\}$ for $j \in \range d$ and $i \in \F_q$,%
\item $\sum_{i \in \F_q} f_{j,i}= 1$ for $j \in \range d$, and
\item for any  $\emptyset \neq \mathcal{K} \subseteq \range{m}$ and $\boldsymbol{\gamma} \in (\F_{p} \setminus \{0\})^{|\mathcal{K}|}$ holds 
$\sum_{j=1}^{d} \left[ g_j^{(\mathcal{K},\boldsymbol{\gamma},\bm f)} \right]_p = \left[ 0 \right]_p$.
\end{enumerate}
\end{proposition}

\begin{IEEEproof}
See Appendix~\ref{app:proofproppm}.
\end{IEEEproof}

Now, we can write $g_j^{(\mathcal{K},\boldsymbol{\gamma},\bm f)}$ as
\ifonecolumn
\begin{displaymath}
g_j^{(\mathcal{K},\boldsymbol{\gamma},\bm f)} = 1 \cdot \sum_{i \in \mathcal{B}^{(1)}(\mathcal{K},\boldsymbol{\gamma},h_j)} f_{j,i} + \cdots + (p-1) \cdot  \sum_{i \in \mathcal{B}^{(p-1)}(\mathcal{K},\boldsymbol{\gamma},h_j)} f_{j,i}
= \sum_{s=1}^{p^{m-1}} \sum_{\beta \in \F_p \setminus \{0\}} \left[\beta \right]_\Z \cdot  f_{j,i_{s,\beta}} 
\end{displaymath}
\else
\begin{displaymath}
\begin{split}
g_j^{(\mathcal{K},\boldsymbol{\gamma},\bm f)} &= 1 \cdot \sum_{i \in \mathcal{B}^{(1)}(\mathcal{K},\boldsymbol{\gamma},h_j)} f_{j,i} + \cdots + \\
&\;\;\;\;(p-1) \cdot  \sum_{i \in \mathcal{B}^{(p-1)}(\mathcal{K},\boldsymbol{\gamma},h_j)} f_{j,i}\\
&= \sum_{s=1}^{p^{m-1}} \sum_{\beta \in \F_p \setminus \{0\}} \left[\beta \right]_\Z \cdot  f_{j,i_{s,\beta}} 
\end{split}
\end{displaymath}
\fi
where  $i_{s,\beta}$ is the $s$-th element (under some arbitrary ordering) of $\mathcal{B}^{(\beta)}(\mathcal{K},\boldsymbol{\gamma},h_j)$. The last equality follows since $|\mathcal{B}^{(0)}(\mathcal{K},\boldsymbol{\gamma},h_j)| = \cdots = |\mathcal{B}^{(p-1)}(\mathcal{K},\boldsymbol{\gamma},h_j)| = p^{m-1}$ (details omitted for brevity). It follows (from the third condition of Proposition~\ref{prop:pm}) that
\begin{equation} \label{eq:gjk}
 \sum_{j=1}^d \left[ g_j^{(\mathcal{K},\boldsymbol{\gamma},\bm f)} \right]_p = \sum_{j=1}^d  \sum_{s=1}^{p^{m-1}} \sum_{\beta \in \F_p \setminus \{0\}} \beta \cdot  \left[ f_{j,i_{s,\beta}} \right]_p = \left[0\right]_p.
\end{equation}
The constraint in (\ref{eq:gjk})  can be written as the $p$-ary parity-check constraint
\begin{equation} \label{eq:SPCp-ary}
\sum_{j=1}^d  \sum_{s=1}^{p^{m-1}} \tilde{f}_{j,s} = \sum_{j=1}^d  \tilde{f}_{j} = [0]_p 
\end{equation}
where $\tilde{f}_{j,s} = \sum_{\beta \in \F_p \setminus \{0\}} \beta  \cdot  \left[  f_{j,i_{s,\beta}}  \right]_p  \in \mathbb{F}_p$ 
and $\tilde{f}_j =  \sum_{s=1}^{p^{m-1}} \tilde{f}_{j,s} \in \F_p$. 

Thus, in summary, a length-$d$ parity-check constraint over the finite field $\mathbb{F}_{p^m}$ can be written as a set of $p^{m}-1$ length-$d$ $p$-ary parity-check constraints ($p^m-1$ is the number of combinations of possible nonempty subsets of $\range{m}$ and vectors $\boldsymbol{\gamma}$).  As pointed out in \cite{liu14} (for the case $p=2$) some of these constrains are redundant, and it is sufficient to consider $\mathcal{K} \in \{ \{1\},\dotsc,\{m\} \}$ and $\boldsymbol{\gamma}=(1,\dotsc,1)$. Now, each of these parity-check equations (including the redundant ones) can be considered separately in Algorithm~\ref{alg:ALP}, which results in an efficient relaxed ALP decoding algorithm for nonbinary codes over $\mathbb{F}_{p^m}$. %
The following lemma is a key ingredient of the proposed relaxation.

\begin{figure*}[!t]
\normalsize 
\begin{equation} \label{eq:convex_hull}
\mathcal{P}^{(\mathcal{K},\boldsymbol{\gamma})} = \conv\left( \left\{ \left( f_{1,0},\dotsc,f_{1,q-1},\dotsc,
f_{d,0},\dotsc,f_{d,q-1} \right)^T
 \in \{0,1\}^{dq}  \colon %
 \sum_{j=1}^d  \tilde{f}_{j} = [0]_p \text{ and }  \sum_{i=0}^{q-1} f_{j,i}  = 1,\; j \in \range{d} \right\} \right)
\end{equation}
 \hrulefill
\vspace*{-2mm}
\end{figure*}

\begin{proposition} \label{lem:relaxation}
Let $\boldtheta^T \boldx \leq \kappa$ be a valid facet-defining inequality for $\conv(\Fv(\C))$ where $\C$ is a nonbinary SPC code over $\F_p$ of  length $d > 0$,  where $p$ is any prime, $ \boldtheta = (\boldt_{k_1} \mid \dotsc \mid \boldt_{k_d})^T$, $k_i \in \F_p$, and $\boldt_k = (t_{k,0},\dotsc,t_{k,p-1})$ for all $k \in \F_p$. Then, the inequality $\left(\tilde\boldtheta^{(\mathcal{K},\boldsymbol{\gamma})}\right)^T \boldx \leq \kappa$ where $\tilde\boldtheta^{(\mathcal{K},\boldsymbol{\gamma})} = \left(\tilde{\boldt}_{k_1}^{(\mathcal{K},\boldsymbol{\gamma})} \mid \dotsc \mid \tilde{\boldt}_{k_d}^{(\mathcal{K},\boldsymbol{\gamma})}\right)^T$ and  %
$\tilde{t}_{k,\eta}^{(\mathcal{K},\boldsymbol{\gamma})} = t_{k,\beta}$ for all $\eta \in \mathcal{B}^{(\beta)}(\mathcal{K},\boldsymbol{\gamma},\bm h)$ and $\beta \in \F_p$ (each entry of $\boldt_k$ is repeated $p^{m-1}$ times in $\tilde{\boldt}_{k}^{(\mathcal{K},\boldsymbol{\gamma})}$) %
is valid %
for  the convex hull $\mathcal{P}^{(\mathcal{K},\boldsymbol{\gamma})}$,  described in \eqref{eq:convex_hull} at the 
\ifonecolumn
top of the page, 
\else
top of the page, 
\fi
for all $\emptyset \neq \mathcal{K} \in \range{m}$ and $\boldsymbol{\gamma} \in \left(\F_p \setminus \{0\} \right)^{|\mathcal{K}|}$.
\end{proposition}

\begin{IEEEproof}
Because of the first two conditions of Proposition~\ref{prop:pm} and since $\mathcal{B}^{(\beta_1)}(\mathcal{K},\boldsymbol{\gamma},h_j) \cap \mathcal{B}^{(\beta_2)}(\mathcal{K},\boldsymbol{\gamma},h_j) = \emptyset$, for any $\beta_1 \neq \beta_2$, $\beta_1, \beta_2 \in \F_p \setminus \{0\}$,  $(f_{j,i_{s,1}},\dotsc,f_{j,i_{s,p-1}})$ can only take values in the set
\ifonecolumn
\begin{displaymath}
\{(0,0,0,\dotsc,0,0), 
(1,0,0,\dotsc,0,0),(0,1,0,\dotsc,0,0),\dotsc,(0,0,0,\dotsc,0,1) \}.
\end{displaymath}
\else
\begin{displaymath}
\begin{split}
\{&(0,0,0,\dotsc,0,0), \\
&(1,0,0,\dotsc,0,0),(0,1,0,\dotsc,0,0),\dotsc,(0,0,0,\dotsc,0,1) \}.
\end{split}
\end{displaymath}
\fi
Hence, $\tilde{f}_{j,s}$ is either equal to $[0]_p$ or to a single $f_{j,i_{s,\beta}}$ times $\beta$  (for some $\beta \in \F_p \setminus \{0\}$). Furthermore, since at most one $\tilde{f}_{j,s}$, $s \in \range{p^{m-1}}$, is nonzero, $\tilde{f}_{j}$ is either equal to $[0]_p$ or to a single $f_{j,i_{s,\beta}}$ times $\beta$ (for some $\beta \in \F_p \setminus \{0\}$ and $s \in \range{p^{m-1}}$). 
From this observation and the fact that $\tilde{t}_{k,\eta}^{(\mathcal{K},\boldsymbol{\gamma})} = t_{k,\beta}$ for all $\eta \in \mathcal{B}^{(\beta)}(\mathcal{K},\boldsymbol{\gamma},\bm h)$ and $\beta \in \F_p$, it can readily be seen that for any binary vector $\bm{f}=(f_{1,0},\dotsc,f_{1,q-1},\dotsc,f_{d,0},\dotsc,f_{d,q-1})^T \in \mathcal{P}^{(\mathcal{K},\boldsymbol{\gamma})}$, $\left(\tilde\boldtheta^{(\mathcal{K},\boldsymbol{\gamma})}\right)^T \bm f = t_{k_1,\tilde{f}_1} + \cdots + t_{k_d,\tilde{f}_d}$. Since $\sum_{j=1}^d  \tilde{f}_{j} = [0]_p$, $t_{k_1,\tilde{f}_1} + \cdots + t_{k_d,\tilde{f}_d} \leq \kappa$, and it follows that the inequality $\left(\tilde\boldtheta^{(\mathcal{K},\boldsymbol{\gamma})}\right)^T \bm x \leq \kappa$ is valid for any binary vector from $\mathcal{P}^{(\mathcal{K},\boldsymbol{\gamma})}$ and hence valid for $\mathcal{P}^{(\mathcal{K},\boldsymbol{\gamma})}$.
\end{IEEEproof}

%
%
%
%

%

%

%

%

%

\begin{remark}
According to Proposition~\ref{lem:relaxation}, for a given valid facet-defining inequality $\boldtheta^T \boldx \leq \kappa$, we can derive a valid inequality for $\mathcal{P}^{(\mathcal{K},\boldsymbol{\gamma})}$ using the interleaving scheme of Proposition~\ref{lem:relaxation}. Note that varying  $\emptyset \neq \mathcal{K} \in \range{m}$ and $\boldsymbol{\gamma} \in \left(\F_p \setminus \{0\} \right)^{|\mathcal{K}|}$ corresponds to permuting the building block entries of the building blocks for a given $\emptyset \neq \mathcal{K} \in \range{m}$ and $\boldsymbol{\gamma} \in \left(\F_p \setminus \{0\} \right)^{|\mathcal{K}|}$. This corresponds exactly to applying all permutations from $\GL(\F_{p^m})$ %
to the entries of the building blocks.
\end{remark}

From Propositions~\ref{prop:pm} and \ref{lem:relaxation}, and \eqref{eq:gjk} and \eqref{eq:SPCp-ary}, it follows that 
\begin{equation} \label{eq:relaxation}
\conv(\Fv(\mathcal{C})) \subseteq  \bigcap_{\emptyset \neq \mathcal{K} \in \range{m}, \boldsymbol{\gamma} \in \F_p \setminus \{0\}} \mathcal{P}^{(\mathcal{K},\boldsymbol{\gamma})}  
\end{equation}
where $\mathcal{C}$ is a nonbinary SPC code of length $d > 0$ over $\F_{p^m}$ where $p$ is any prime and $m$ is a positive integer. The relaxation from \eqref{eq:relaxation} can be used for (relaxed) ALP decoding of general nonbinary codes over  $\F_{p^m}$.

\begin{example} \label{ex:gf9}
Consider a nonbinary SPC code of length $d=3$ over
$\mathbb{F}_{3^2} = \{0,1,2,\alpha,1+\alpha,2+\alpha,2\alpha,1+2\alpha,2+2\alpha \}$, 
where $\alpha$ is a primitive element in $\mathbb{F}_{3^2}$, defined by the parity-check vector  $\bm h=(1,1,1)$. In this case, the constant-weight embedding  (from Definition~\ref{def:Constant}) is as follows:
\ifonecolumn
\begin{align}
\mathsf{f}(0) &= (1,0,0,0,0,0,0,0,0),\; 
\mathsf{f}(1) = (0,1,0,0,0,0,0,0,0), \;
\mathsf{f}(2) = (0,0,1,0,0,0,0,0,0), \notag \\
\mathsf{f}(\alpha) &= (0,0,0,1,0,0,0,0,0), \;
\mathsf{f}(1+\alpha) = (0,0,0,0,1,0,0,0,0), \;
\mathsf{f}(2+\alpha) = (0,0,0,0,0,1,0,0,0), \notag \\
\mathsf{f}(2\alpha) &= (0,0,0,0,0,0,1,0,0), \;
\mathsf{f}(1+2\alpha) = (0,0,0,0,0,0,0,1,0), \;
\mathsf{f}(2+2\alpha) = (0,0,0,0,0,0,0,0,1). \notag 
\end{align}
\else
\begin{align}
\mathsf{f}(0) &= (1,0,0,0,0,0,0,0,0), \notag \\
\mathsf{f}(1) &= (0,1,0,0,0,0,0,0,0), \notag \\
\mathsf{f}(2) &= (0,0,1,0,0,0,0,0,0), \notag \\
\mathsf{f}(\alpha) &= (0,0,0,1,0,0,0,0,0), \notag \\
\mathsf{f}(1+\alpha) &= (0,0,0,0,1,0,0,0,0), \notag \\
\mathsf{f}(2+\alpha) &= (0,0,0,0,0,1,0,0,0), \notag \\
\mathsf{f}(2\alpha) &= (0,0,0,0,0,0,1,0,0), \notag \\
\mathsf{f}(1+2\alpha) &= (0,0,0,0,0,0,0,1,0), \notag \\
\mathsf{f}(2+2\alpha) &= (0,0,0,0,0,0,0,0,1). \notag 
\end{align}
\fi
Furthermore, we have
\ifonecolumn
\begin{align}
&\mathcal{B}^{(1)}(\{1\},1,1) =\{1,1+\alpha,1+2\alpha \}, \;
\mathcal{B}^{(2)}(\{1\},1,1) =\{2,2+\alpha,2+2\alpha \}, \notag \\
&\mathcal{B}^{(1)}(\{1\},2,1) =\{2,2+\alpha,2+2\alpha \} = \mathcal{B}^{(2)}(\{1\},1,1), \;
\mathcal{B}^{(2)}(\{1\},2,1) =\{1,1+\alpha,1+2\alpha \} = \mathcal{B}^{(1)}(\{1\},1,1), \notag \\
&\mathcal{B}^{(1)}(\{2\},1,1) =\{\alpha,1+\alpha,2+\alpha \}, \;
\mathcal{B}^{(2)}(\{2\},1,1) =\{2\alpha,1+2\alpha,2+2\alpha \}, \notag\\
&\mathcal{B}^{(1)}(\{2\},2,1) =\{2\alpha,1+2\alpha,2+2\alpha \}=\mathcal{B}^{(2)}(\{2\},1,1), \;
\mathcal{B}^{(2)}(\{2\},2,1) =\{\alpha,1+\alpha,2+\alpha \} = \mathcal{B}^{(1)}(\{2\},1,1), \notag\\
&\mathcal{B}^{(1)}(\{1,2\},(1,1),1) =\{1,\alpha,2+2\alpha \}, \;
\mathcal{B}^{(2)}(\{1,2\},(1,1),1) =\{2,1+\alpha,2\alpha \}, \notag\\
&\mathcal{B}^{(1)}(\{1,2\},(1,2),1) =\{1,2+\alpha,2\alpha \}, \;
\mathcal{B}^{(2)}(\{1,2\},(1,2),1) =\{2,\alpha,1+2\alpha \}, \notag\\
&\mathcal{B}^{(1)}(\{1,2\},(2,1),1) =\{2,\alpha,1+2\alpha \} = \mathcal{B}^{(2)}(\{1,2\},(1,2),1),\notag \\
&\mathcal{B}^{(2)}(\{1,2\},(2,1),1) =\{1,2+\alpha,2\alpha \}= \mathcal{B}^{(1)}(\{1,2\},(1,2),1), \notag\\
&\mathcal{B}^{(1)}(\{1,2\},(2,2),1) =\{2,1+\alpha,2\alpha\}= \mathcal{B}^{(2)}(\{1,2\},(1,1),1),\notag \\
&\mathcal{B}^{(2)}(\{1,2\},(2,2),1) =\{1,\alpha,2+2\alpha \}= \mathcal{B}^{(1)}(\{1,2\},(1,1),1).\notag
\end{align}
\else
\begin{align}
&\mathcal{B}^{(1)}(\{1\},1,1) =\{1,1+\alpha,1+2\alpha \}, \notag \\
&\mathcal{B}^{(2)}(\{1\},1,1) =\{2,2+\alpha,2+2\alpha \}, \notag \\
&\mathcal{B}^{(1)}(\{1\},2,1) =\{2,2+\alpha,2+2\alpha \} = \mathcal{B}^{(2)}(\{1\},1,1), \notag \\
&\mathcal{B}^{(2)}(\{1\},2,1) =\{1,1+\alpha,1+2\alpha \} = \mathcal{B}^{(1)}(\{1\},1,1), \notag \\
&\mathcal{B}^{(1)}(\{2\},1,1) =\{\alpha,1+\alpha,2+\alpha \}, \notag \\
&\mathcal{B}^{(2)}(\{2\},1,1) =\{2\alpha,1+2\alpha,2+2\alpha \}, \notag\\
&\mathcal{B}^{(1)}(\{2\},2,1) =\{2\alpha,1+2\alpha,2+2\alpha \}=\mathcal{B}^{(2)}(\{2\},1,1), \notag \\
&\mathcal{B}^{(2)}(\{2\},2,1) =\{\alpha,1+\alpha,2+\alpha \} = \mathcal{B}^{(1)}(\{2\},1,1), \notag\\
&\mathcal{B}^{(1)}(\{1,2\},(1,1),1) =\{1,\alpha,2+2\alpha \}, \notag \\
&\mathcal{B}^{(2)}(\{1,2\},(1,1),1) =\{2,1+\alpha,2\alpha \}, \notag\\
&\mathcal{B}^{(1)}(\{1,2\},(1,2),1) =\{1,2+\alpha,2\alpha \}, \notag \\
&\mathcal{B}^{(2)}(\{1,2\},(1,2),1) =\{2,\alpha,1+2\alpha \}, \notag\\
&\mathcal{B}^{(1)}(\{1,2\},(2,1),1) =\{2,\alpha,1+2\alpha \} \notag \\
&\;\;\;\;\;\;\;\;\;\;\;\;\;\;\;\;\;\;\;\;\;\;\;\;\;\;\;\;\;\;\; = \mathcal{B}^{(2)}(\{1,2\},(1,2),1),\notag \\
&\mathcal{B}^{(2)}(\{1,2\},(2,1),1) =\{1,2+\alpha,2\alpha \} \notag \\
&\;\;\;\;\;\;\;\;\;\;\;\;\;\;\;\;\;\;\;\;\;\;\;\;\;\;\;\;\;\;\;= \mathcal{B}^{(1)}(\{1,2\},(1,2),1), \notag\\
&\mathcal{B}^{(1)}(\{1,2\},(2,2),1) =\{2,1+\alpha,2\alpha\} \notag \\
&\;\;\;\;\;\;\;\;\;\;\;\;\;\;\;\;\;\;\;\;\;\;\;\;\;\;\;\;\;\;\;= \mathcal{B}^{(2)}(\{1,2\},(1,1),1),\notag \\
&\mathcal{B}^{(2)}(\{1,2\},(2,2),1) =\{1,\alpha,2+2\alpha \} \notag\\
&\;\;\;\;\;\;\;\;\;\;\;\;\;\;\;\;\;\;\;\;\;\;\;\;\;\;\;\;\;\;\;= \mathcal{B}^{(1)}(\{1,2\},(1,1),1).\notag
\end{align}
\fi
As an example, we can write out the constraint $\sum_{j=1}^3 \left[g_j^{(\mathcal{K},\boldsymbol{\gamma},\bm f)} \right]_3 = [0]_3$ for $\mathcal{K}=\{1,2\} = \range{2}$ and $\boldsymbol{\gamma}=(2,1)$ as follows:
\ifonecolumn
\begin{displaymath}
\sum_{j=1}^3 \left[ f_{j,2} + f_{j,\alpha} + f_{j,1+2\alpha}  \right]_3 
+ 2 \left[ f_{j,1} + f_{j,2+\alpha} + f_{j,2\alpha}  \right]_3 
= \sum_{j=1}^3 \left( \tilde{f}_{j,1} + \tilde{f}_{j,2} + \tilde{f}_{j,3} \right) 
= \tilde{f}_{1} + \tilde{f}_{2} +  \tilde{f}_{3}  = [0]_3
\end{displaymath}
\else
\begin{displaymath}
\begin{split}
&\sum_{j=1}^3 \left[ f_{j,2} + f_{j,\alpha} + f_{j,1+2\alpha}  \right]_3 
+ 2 \left[ f_{j,1} + f_{j,2+\alpha} + f_{j,2\alpha}  \right]_3 \\
&= \sum_{j=1}^3 \left( \tilde{f}_{j,1} + \tilde{f}_{j,2} + \tilde{f}_{j,3} \right) 
= \tilde{f}_{1} + \tilde{f}_{2} +  \tilde{f}_{3}  = [0]_3
\end{split}
\end{displaymath}
\fi
where 
\ifonecolumn
\begin{displaymath}
\tilde{f}_{j,1} = \left[ f_{j,2} + 2 f_{j,1} \right]_3, \;
\tilde{f}_{j,2} = \left[ f_{j,\alpha} + 2 f_{j,2+\alpha}\right]_3, \;
\tilde{f}_{j,3} = \left[ f_{j,1+2\alpha} + 2 f_{j,2\alpha} \right]_3 
\end{displaymath}
\else
\begin{align*}
\tilde{f}_{j,1} &= \left[ f_{j,2} + 2 f_{j,1} \right]_3, \notag \\
\tilde{f}_{j,2} &= \left[ f_{j,\alpha} + 2 f_{j,2+\alpha}\right]_3, \notag \\
\tilde{f}_{j,3} &= \left[ f_{j,1+2\alpha} + 2 f_{j,2\alpha} \right]_3 \notag
\end{align*}
\fi
and where $\tilde{f}_{j} = \tilde{f}_{j,1} + \tilde{f}_{j,2} + \tilde{f}_{j,3} \in \F_3$.
\end{example}

\begin{example}
For $p=3$, there is a single vector $\bm m = (0,0,0)$ that gives a valid, irredundant basic building block class (see Section~\ref{sec:q3}). In particular, $\boldt_0^{\bm m} = (0,1,2)$, $\boldt_1^{\bm m} = (0,1,-1)$, and $\boldt_2^{\bm m} = (0,-2,-1)$.  From the construction of Proposition~\ref{lem:relaxation} (and Example~\ref{ex:gf9}), $\tilde\boldt_k^{(\{1,2\},(2,1))} = \left(t_{k,0},t_{k,2},t_{k,1},t_{k,1},t_{k,0},t_{k,2},t_{k,2},t_{k,1},t_{k,0} \right)$ for all $k \in \F_3$. From Construction~\ref{constr:hilo} and Proposition~\ref{prop:specialMvalid}, the inequality $(0,1,2,\,0,1,2,\,0,1,-1) \boldx \leq 3$ is valid and facet-defining 
for an \enquote{all-ones} SPC code of length $3$ over $\F_3$. Thus, according to Proposition~\ref{lem:relaxation}, the inequality
\ifonecolumn
\begin{displaymath}
\left(0,2,1,1,0,2,2,1,0,\,  0,2,1,1,0,2,2,1,0,\, 
0,-1,1,1,0,-1,-1,1,0\right) \boldx \leq 3 
\end{displaymath}
\else
\begin{displaymath}
\begin{split}
&\left(0,2,1,1,0,2,2,1,0,\,  0,2,1,1,0,2,2,1,0,\, \right. \\
&\;\left. 0,-1,1,1,0,-1,-1,1,0\right) \boldx \leq 3 
\end{split}
\end{displaymath}
\fi
is valid %
for $\mathcal{P}^{(\{1,2\},(2,1))}$.
\end{example}

\begin{example}
For the case $p=2$ and $m=2$, the number of facets (and the corresponding sets of inequalities) can be computed numerically (using, for instance, the software package \emph{Polymake} \cite{polymake}) for $d=3$, $4$, $5$, and $6$. The number of facets is $24$, $40$, $68$, and $120$, respectively, and the sets of inequalities match perfectly with the sets derived using the relaxation method presented above. Note that in \cite[Conj.~62]{liu14}, it was conjectured that the relaxation is indeed tight. Also, in \cite{hon12}, the same observation was made, but no proof was given. If we increase $m$ to $3$, the number of facets is $2740$ and $35928$ for $d=3$ and $4$, respectively, while the number of inequalities using the relaxation method presented above is only $7 \cdot 2^{d-1} + 8 \cdot d$, which is equal to $52$ and $88$ for $d=3$ and $4$, respectively, and the relaxation is not tight.
\end{example}

\begin{example}
For the case $p=3$ and $m=2$, the number of facets (and the corresponding sets of inequalities) can be computed numerically for $d=3$. The number of facets is $73323$, while the number of inequalities using the relaxation method presented above is only $8 \cdot 2 \cdot 3^{3-1} + 9 \cdot 3 = 144 + 27 = 171$,  and the relaxation is not tight. Note that using the weaker Proposition~2 from \cite{ros15} (or, equivalently, Proposition~\ref{prop:pm} constrained with $\boldsymbol{\gamma}=(1,\dotsc,1)$), we only get $3 \cdot 2 \cdot 3^{3-1} + 9 \cdot 3 = 54 + 27 = 81$ inequalities.
\end{example}

\section{Numerical Results} \label{sec:numerical_results}
\pgfplotsset{grid style={gray!40,thin}}
\pgfplotsset{every axis/.append style={font=\scriptsize}}
\tikzset{
  curveRPC/.style = {mark=asterisk,Yellow!60!Black},
  curveALP/.style = {mark=x,Red,mark options={fill=Red}},
  curvePLP/.style = {mark=diamond*,Blue,mark options={fill=white}},
  curveT1/.style = {mark=|,Violet,mark options={fill=white}},
  curveML/.style = {mark=*,Green,mark options={scale=.6}},
  curveSP/.style = {mark=*,black,mark options={scale=.6,fill=white}},
  curveSPMod/.style = {mark=o,Green, mark options={scale=.6,fill=white}},
  curveALPMod/.style = {mark=x,Green, mark options={scale=.6}},
}

\ifonecolumn
\newlength{\myheight}\setlength{\myheight}{.6\columnwidth}
\newlength{\mywidth}\setlength{\mywidth}{.75\columnwidth}
\else
\newlength{\myheight}\setlength{\myheight}{.7\columnwidth}
\newlength{\mywidth}\setlength{\mywidth}{\columnwidth}
\fi
        
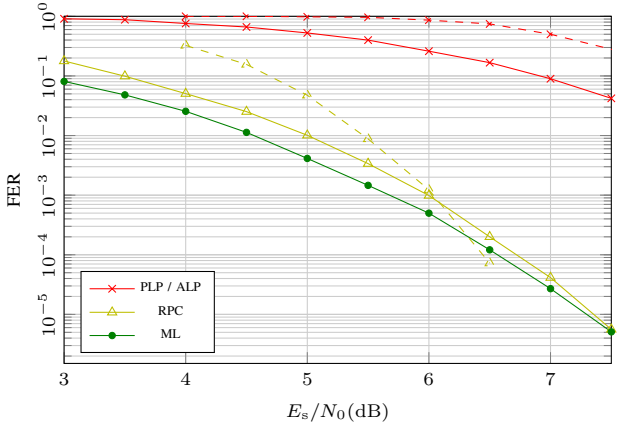
\begin{figure}
  \centering
  \begin{tikzpicture}
    \begin{semilogyaxis}[
        xlabel=$E_{\mathrm s}/N_0 (\si{\decibel})$,
        ylabel=FER,
        grid=both,
        ymax=1,
        ymin=1e-6,
        ytick={1e-6,1e-5,1e-4,1e-3,1e-2,1e-1,1},
        xmin=3, xmax=7.5,
        xtick={3,3.5,4,4.5,5,5.5,6,6.5,7,7.5},
        height=\myheight,
        width=\mywidth,
        y tick label style={rotate=90},
        grid style={gray,opacity=0.5,dotted},
        legend style={legend pos=south west,font=\tiny}]
      \addplot[curveALP] plot coordinates {
        (3.00, 9.09e-01)   %
        (3.50, 8.77e-01)   %
        (4.00, 7.58e-01)   %
        (4.50, 6.62e-01)   %
        (5.00, 5.26e-01)   %
        (5.50, 3.99e-01)   %
        (6.00, 2.60e-01)   %
        (6.50, 1.67e-01)   %
        (7.00, 8.97e-02)   %
        (7.50, 4.21e-02)   %
      };
      \addlegendentry{PLP / ALP}
      \addplot[curveRPC] plot coordinates {
        (3.00, 1.78e-01)   %
        (3.50, 9.90e-02)   %
        (4.00, 5.05e-02)   %
        (4.50, 2.52e-02)   %
        (5.00, 1.01e-02)   %
        (5.50, 3.38e-03)   %
        (6.00, 9.86e-04)   %
        (6.50, 2.00e-04)   %
        (7.00, 4.14e-05)   %
        (7.50, 5.60e-06)   %
      };
      \addlegendentry{RPC}

      \addplot[curveML] plot coordinates {
        (3.00, 8.09e-02)   %
        (3.50, 4.81e-02)   %
        (4.00, 2.55e-02)   %
        (4.50, 1.13e-02)   %
        (5.00, 4.13e-03)   %
        (5.50, 1.46e-03)   %
        (6.00, 4.99e-04)   %
        (6.50, 1.21e-04)   %
        (7.00, 2.70e-05)   %
        (7.50, 5.09e-06)   %
      };
      \addlegendentry{ML}
    \addplot[curveALP,dashed] plot coordinates {
      (4.00, 1.00e+00)   %
      (4.50, 1.00e+00)   %
    (5.00, 9.80e-01)   %
    (5.50, 9.62e-01)   %
    (6.00, 8.62e-01)   %
    (6.50, 7.46e-01)   %
    (7.00, 5.00e-01)   %
    (8.00, 1.57e-01)   %
    (10.50, 1.22e-04)   %

      };
      \addplot[curveRPC,dashed] plot coordinates {
        (4.00, 3.25e-01)   %
        (4.50, 1.56e-01)   %
        (5.00, 4.74e-02)   %
        (5.50, 8.78e-03)   %
        (6.00, 1.23e-03)   %
        (6.50, 7.25e-05)   %
      };
    \end{semilogyaxis}
  \end{tikzpicture}
  \vspace{-2ex}
  \caption{FER performance of the ternary RM codes $\C_{\mathrm{RM}(1)}^{(3)}$ (solid lines) and $\C_{\mathrm{RM}(2)}^{(3)}$ (dashed lines) as a function of $E_{\rm s}/N_0$.}
  \label{fig:ternaryRM}
  \vskip -3ex
\end{figure}

\ifonecolumn
\newlength{\myheightt}\setlength{\myheightt}{.6\textwidth}
\else
\newlength{\myheightt}\setlength{\myheightt}{.4\textwidth}
\fi

\begin{figure*}[t]
  \centering
  \begin{tikzpicture}
    \begin{semilogyaxis}[
        xlabel=$E_{\mathrm s}/N_0 (\si{\decibel})$,
        ylabel=FER,
        grid=both,
        ytick={1e-8,1e-7,1e-6,1e-5,1e-4,1e-3,1e-2,1e-1,1},
        ymax=1,
        ymin=1e-8,
        xmin=4, xmax=12,
        height=\myheightt,
        width=\textwidth,
        y tick label style={rotate=90},
        legend columns=1,
        grid style={gray,opacity=0.5,dotted},
	legend style={legend pos=south west,font=\tiny}]
      \addplot[curvePLP] plot coordinates {
        (5.00, 9.26e-01)   %
        (5.50, 6.92e-01)   %
        (6.00, 3.07e-01)   %
        (6.50, 9.02e-02)   %
        (7.00, 1.22e-02)   %
        (7.50, 6.78e-04)   %
        (8.00, 1.31e-05)   %
      };
      \addlegendentry{PLP / CLP}
      \addplot[curveALP] plot coordinates {
        (5.00, 9.26e-01)   %
        (5.50, 6.92e-01)   %
        (6.00, 3.07e-01)   %
        (6.50, 9.02e-02)   %
        (7.00, 1.22e-02)   %
        (7.50, 6.78e-04)   %
        (8.00, 1.31e-05)   %
        (8.50, 8.94679078058721e-8) %
      };
      \addlegendentry{ALP}
      \addplot[curveT1] plot coordinates {
        (5.00, 9.26e-01)   %
        (5.50, 6.99e-01)   %
        (6.00, 3.11e-01)   %
        (6.50, 9.24e-02)   %
        (7.00, 1.26e-02)   %
        (7.50, 7.38e-04)   %
        (8.00, 1.46e-05)   %
        (8.50, 9.97215597371549e-8) %
      };
      \addlegendentry{ALP$|\Phi(\Theta^{\bm 0})$}
      \addplot[curveRPC] plot coordinates {
        (5.00, 9.26e-01)   %
        (5.50, 6.92e-01)   %
        (6.00, 3.07e-01)   %
        (6.50, 9.02e-02)   %
        (7.00, 1.21e-02)   %
        (7.50, 6.76e-04)   %
        (8.00, 1.29e-05)   %
        (8.50, 7.66390452077366e-8) %
      };
      \addlegendentry{RPC}
       \addplot[curveSP] plot coordinates {
      (4.50, 0.709219872952) %
      (5.00, 0.456620991230) %
      (5.50, 0.157977879047) %
      (6.00, 0.040080159903) %
      (6.50, 0.005582225975) %
      (7.00, 0.000716291310) %
      (7.50, 0.000037421549) %
      (8.00, 0.000002146946) %
      (8.50, 0.000000065625) %
      };
      \addlegendentry{SP}
      \addplot[curvePLP] plot coordinates {
        (4.00, 1.04e-01)   %
        (4.50, 2.21e-02)   %
        (5.00, 2.61e-03)   %
        (5.50, 1.81e-04)   %
        (6.00, 5.10e-06)   %
      };
      \addplot[curveALP] plot coordinates {
        (4.00, 1.04e-01)   %
        (4.50, 2.21e-02)   %
        (5.00, 2.61e-03)   %
        (5.50, 1.81e-04)   %
        (6.00, 5.10e-06)   %
        (6.25, 8.18804621074539e-7) %
      };
      \addplot[curveRPC] plot coordinates {
        (4.00, 6.99e-02)   %
        (4.50, 1.09e-02)   %
        (5.00, 9.80e-04)   %
        (5.50, 3.82e-05)   %
        (6.00, 4.8e-07)   %
      };
      \addplot[curveSP] plot coordinates {
      (3.00, 0.374531835318) %
      (3.50, 0.135869562626) %
      (4.00, 0.042158514261) %
      (4.50, 0.008078849874) %
      (5.00, 0.001329186256) %
      (5.50, 0.000132728092) %
      (6.00, 0.000012212075) %
      (6.50, 0.000000704532) %
      }; 
      \addplot[curvePLP] plot coordinates {
        (6.50, 0.980392156862745) %
        (7.00, 0.819672131147541) %
        (7.50, 0.495049504950495) %
        (8.00, 1.97e-01)   %
        (8.50, 4.13e-02)   %
        (9.00, 2.73e-03)   %
        (9.50, 7.64e-05)   %
      };
      \addplot[curveALP] plot coordinates {
        (6.50, 0.980392156862745) %
        (7.00, 0.819672131147541) %
        (7.50, 0.495049504950495) %
        (8.00, 1.97e-01)   %
        (8.50, 4.13e-02)   %
        (9.00, 2.73e-03)   %
        (9.50, 7.64e-05)   %
        (10.0, 1.16170724705874e-6) %
      };
      \addplot[curveT1] plot coordinates {
        (6.50, 0.980392156862745) %
        (7.00, 0.847457627118644) %
        (7.50, 0.568181818181818) %
        (8.00, 2.32e-01)   %
        (8.50, 5.22e-02)   %
        (9.00, 3.91e-03)   %
        (9.50, 1.37e-04)   %
        (10.0, 2.32584148450049e-6) %
      };
      \addplot[curveRPC] plot coordinates {
        (6.50, 0.980392156862745) %
        (7.00, 0.819672131147541) %
        (7.50, 0.495049504950495) %
        (8.00, 1.97e-01)   %
        (8.50, 4.31e-02)   %
        (9.00, 2.73e-03)   %
        (9.50, 7.64e-05)   %
        (10.0, 1.28553805681153e-6) %
      };
      \addplot[curveSP] plot coordinates {
      (6.50,  0.396825402975) %
      (7.00,  0.184842884541) %
      (7.50, 0.077399380505) %
      (8.00, 0.019409937784) %
      (8.50, 0.002457847819) %
      (9.00, 0.000363765983) %
      (9.50, 0.000031430725) %
      (10.0,2.14607066907702e-6) %
      };
      \addplot[curvePLP] plot coordinates {
        (10.00, 3.58e-01)   %
        (10.50, 1.04e-01)   %
        (11.00, 1.81e-02)   %
        (11.50, 1.81e-03)   %
      };
      \addplot[curveALP] plot coordinates {
        (8.00, 1.00e+00)   %
        (8.50, 9.95e-01)   %
        (9.00, 9.48e-01)   %
        (9.50, 7.60e-01)   %
        (10.00, 3.74e-01)   %
        (10.50, 1.20e-01)   %
        (11.00, 2.00e-02)   %
        (11.50, 1.95e-03)   %
        (12.00, 9.86e-05)   %
      };
      \addplot[curveT1] plot coordinates {
        (8.00, 1.00e+00)   %
        (8.50, 1.00e+00)   %
        (9.00, 9.76e-01)   %
        (9.50, 8.47e-01)   %
        (10.00, 5.54e-01)   %
        (10.50, 2.19e-01)   %
        (11.00, 4.47e-02)   %
        (11.50, 5.45e-03)   %
        (12.00, 3.05e-04)   %
      };
      \addplot[curveRPC] plot coordinates {
        (8.00, 1.00e+00)   %
        (8.50, 9.95e-01)   %
        (9.00, 9.48e-01)   %
        (9.50, 7.60e-01)   %
        (10.00, 3.74e-01)   %
        (10.50, 1.20e-01)   %
        (11.00, 2.00e-02)   %
        (11.50, 1.93e-03)   %
        (12.00, 1.05e-04)   %
      };
       \addplot[curveSP] plot coordinates {
      (8.00, 0.990099012852) %
      (8.50, 0.884955763817) %
      (9.00, 0.724637687206) %
      (9.50, 0.549450576305) %
     (10.00, 0.331125825644) %
     (10.50, 0.177935943007) %
     (11.00, 0.077821008861) %
     (11.50, 0.020738283172) %
     (12.00, 0.004750142340) %
     };
    \end{semilogyaxis}
  \end{tikzpicture}
  \vspace{-2ex}
  \caption{FER performance of $\mathcal{C}_{\rm Tan}^{(3)}$, $\mathcal{C}_{\rm Tan}^{(5)}$, $\mathcal{C}_{\rm Tan}^{(7)}$, and $\mathcal{C}_{\rm Tan}^{(11)}$ (left to right) as a function of $E_{\rm s}/N_0$. %
For $p=3$, $\Delta^d_p \cup \Phi(\Theta^{\bm 0})$ gives a complete and irredundant description of $\mathcal{P}$ (see Theorem~\ref{thm:facetComplete}); thus only three curves are displayed \revone{in addition to the SP FER decoding curve.}}
  \label{fig:Tannercode}
  \vskip -3ex
\end{figure*}
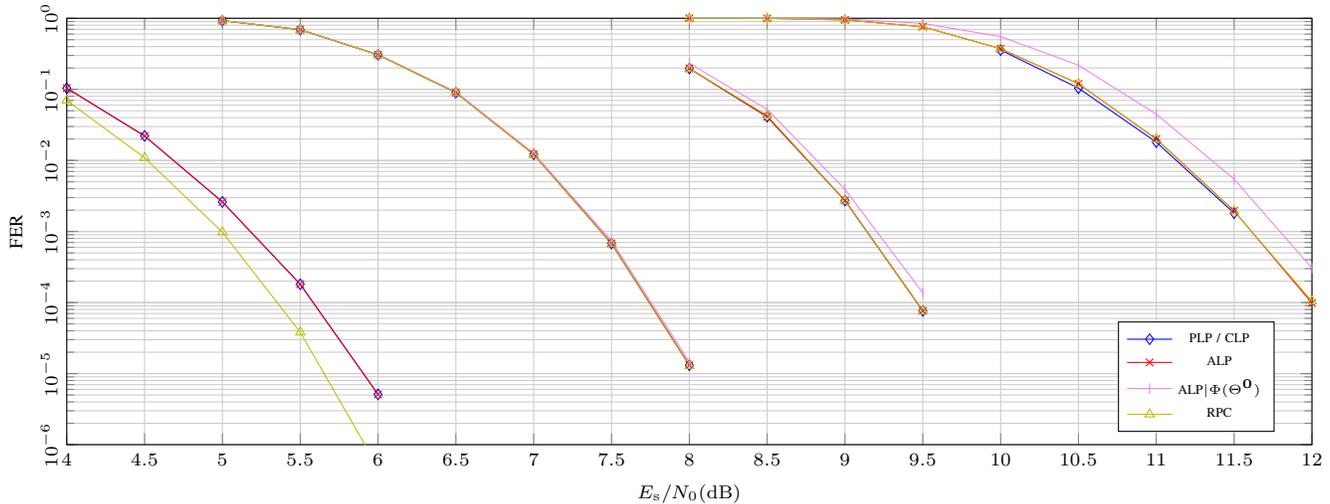

\begin{notms}
In this section, we compare the proposed ALP decoding algorithm from Section~\ref{sec:ALP_q} and the version augmented by RPC cuts (named RPC in the following) from Section~\ref{sec:ACG} with both the plain and cascaded \enquote{static} (nonadaptive) approaches from \cite{fla09} (named PLP and CLP, respectively). We present frame error-rate (FER) performance results for various codes over additive white Gaussian noise channels with different signal-to-noise ratios (SNRs), using $p$-phase-shift keying modulation (for codes over $\F_p$). The code symbols $\zeta \in \F_p$ are mapped to constellation points according to $\zeta \mapsto \exp(\sqrt{-1}(2[\zeta]_\Z+1)\pi / p)$. In addition to the decoding performance, we present several performance figures (for selected combinations of code and SNR value) in Table~\ref{table:resultsNumbers}: the average CPU time, the average number of iterations of the simplex algorithm that is used to solve the LPs, and, for ALP and RPC, the average number of cuts added (\ie number of constraints in the final linear program) and the average number of LPs solved (\ie iterations of the main loop of Algorithm~\ref{alg:ALP}).

For $p=3$, simulations were performed using
a $(27, 10, 9)$  and an $(81, 31, 18)$ ternary RM code ($\C_{\mathrm{RM}(1)}^{(3)}$ and $C_{\mathrm{RM}(2)}^{(3)}$, respectively).
Note that the respective parity-check matrices are rather dense, with a nonzeros ratio of $0.23$ and $0.14$, respectively. %

In view of Theorem~\ref{thm:facetComplete}, it is clear that ALP, CLP, and PLP show the same error-correction performance  for $p=3$. As shown in Fig.~\ref{fig:ternaryRM}, the RPC cut-search algorithm drastically improves decoding performance for the dense codes and, for $\C_{\mathrm{RM}(1)}^{(3)}$, nearly achieves ML performance. Note that this is in line with the well-known observation that, in the binary case, LP decoding without RPC search performs \eirik{poorly} for dense codes (see, e.g., \cite{tan10}). The ML curve in Fig.~\ref{fig:ternaryRM} was computed using an integer programming formulation (and the commercial Gurobi solver \cite{Gurobi600}) of the nonbinary ML decoding problem that is based on the compact binary IPD formulation first presented in \cite{tan10}. Except for the small RM code, the complexity of this approach however is intractable for all codes considered in this section.
\end{notms}

\begin{notms}
To study the effect of increasing $p$ on the ALP algorithm, we employ the $(3,5)$-regular $(155,64)$ Tanner code \cite{tan01}  over the fields $\F_p$, $p \in \{3,5,7,11\}$ (denoted by $\C_{\mathrm{Tan}}^{(p)}$, respectively). To construct the codes, we have replaced the $5$ ones in each row of the binary $\bm H$ by the patterns $(1,2,2,1,1)$, $(1,2,4,3,1)$, $(1,2,4,6,1)$, and $(1,2,6,10,1)$, for  $\C_{\mathrm{Tan}}^{(3)}$ to $\C_{\mathrm{Tan}}^{(11)}$, respectively. The results are shown in Fig.~\ref{fig:Tannercode}. Because of Theorem~\ref{thm:facetComplete} and the numerical verification of Conjecture~\ref{conj:facetComplete_q5} for $d \leq 5$, the ALP and PLP/CLP curves are identical for $p\in \{3,5\}$, and the fact that they also are for $p=7$ supports Conjecture~\ref{conj:facetComplete_q7}. Interestingly, the class $\Phi(\Theta^{\bm 0})$ (the pink $+$-marked curve) is sufficient for achieving close-to-exact LP decoding performance \eirikfinal{for small $p$} (especially for $p=5$; for $p=3$, this is the only class, as detailed in Section~\ref{sec:q3}). This can only mean that the facets induced by the other classes somehow cut off only \enquote{smaller} parts of the polytope. This can be explained by the counting formulas in Lemma~\ref{lem:counting_formulas}. For instance, for $p=7$, Proposition~\ref{prop:validIneq_q7_1to4} shows that the number of codewords that each inequality is tight for decreases strictly from $\Theta_1$ to $\Theta_4$. This proves that the $\Theta_1$-facets are \enquote{larger} than the others. For $p=11$, we observe an increasing gap between ALP and PLP/CLP which shows that the inequalities proposed in this paper specify only a strict relaxation of the LP decoder as in \eqref{eq:LPformulation} for $p\geq 11$.

In contrast to the binary case (see \cite[Fig.~3]{zha12}) and the dense codes above, RPCs lead to a noteworthy improvement only for $p=3$ for the Tanner codes. In Table~\ref{table:resultsNumbers}, one can observe that again ALP decoding is much more efficient than both PLP and CLP.

As a remark, no single cut from the special class $\Phi(\Theta_6)$ for $p=7$ (see Section~\ref{sec:q7}) was found during all of our simulations, hence they appear not to influence decoding performance in practice. On the other  \revone{hand} the complexity of the cut-search algorithm is $d^2$ times higher for $\Theta_6$ than for the basic classes due to the additional loop that sets $i^\lo$ and $i^\hi$. Especially in conjunction with RPC search (where RPCs are generally dense even with LDPC codes), it may not be worthwhile to search for $\Theta_6$ inequalities at all.

\revone{For comparison, the FER performance of the SP decoder \cite{mackay98} with a maximum of $100$ iterations for all four Tanner codes is also depicted in Fig.~\ref{fig:Tannercode}. From the figure we observe that SP decoding exhibits an \enquote{error floor} (in the sense that the FER decays more slowly with the SNR) for all codes, except possibly for $\C_{\mathrm{Tan}}^{(5)}$ (there might be a crossing also for this code but at a lower error rate). For $\C_{\mathrm{Tan}}^{(11)}$ (and also for $\C_{\mathrm{Tan}}^{(3)}$ at high SNRs), ALP decoding shows a clear advantage over SP decoding. For $\C_{\mathrm{Tan}}^{(5)}$ and $\C_{\mathrm{Tan}}^{(7)}$ there is a performance loss with ALP decoding compared to SP decoding for low SNRs. However,  the performance gap diminishes as the SNR increases, \eirik{and for $\C_{\mathrm{Tan}}^{(7)}$ we observe a crossing at low error rates.}} %

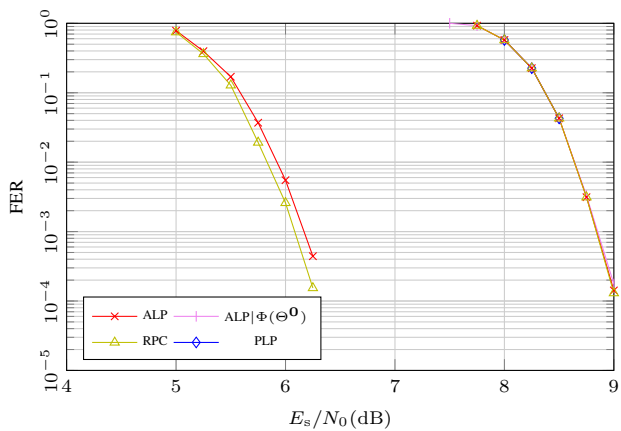
\begin{figure}
  \centering
  \begin{tikzpicture}
    \begin{semilogyaxis}[
        xlabel=$E_{\mathrm s}/N_0 (\si{\decibel})$,
        ylabel=FER,
        grid=both,
        ymax=1,
        ymin=1e-7,
        ytick={1e-7,1e-6,1e-5,1e-4,1e-3,1e-2,1e-1,1},
        xtick={5,5.5,6,6.5,7,7.5,8,8.5,9,9.5},
        xmin=5, xmax=9.5,
        height=\myheight,
        width=\mywidth,
        y tick label style={rotate=90},
        legend columns=5,
        grid style={gray,opacity=0.5,dotted},
        transpose legend,
        legend style={at={(0.005,0.2)},anchor=west,font=\tiny}]
        \addplot[curvePLP] plot coordinates {
      (8.00, 5.75e-01)   %
      (8.25, 2.26e-01)   %
      (8.50, 4.28e-02)   %
          };
      \addlegendentry{PLP / CLP}

      \addplot[curveALP] plot coordinates {
      (5.00, 7.84e-01)   %
        (5.25, 3.97e-01)   %
        (5.50, 1.70e-01)   %
        (5.75, 3.70e-02)   %
        (6.00, 5.50e-03)   %
        (6.25, 4.42e-04)   %
        (6.50,0.0000278188395440631) %
        (6.75, 2.67185354018527e-6) %
        (7.00, 5.36703606567060e-7) %
      };
      \addlegendentry{ALP}
      \addplot[curveT1] plot coordinates {
          (7.75, 9.26e-01)   %
          (8.00, 5.75e-01)   %
          (8.25, 2.26e-01)   %
          (8.50, 4.28e-02)   %
          (8.75, 3.31e-03)   %
          (9.00, 1.76e-04)   %
          (9.25, 4.13886436266849e-6) %
          (9.50, 1.04724394393700e-7) %
      };
      \addlegendentry{ALP$|\Phi(\Theta^{\bm 0})$}
      \addplot[curveRPC] plot coordinates {
      (5.00, 7.46e-01)   %
        (5.25, 3.62e-01)   %
        (5.50, 1.28e-01)   %
        (5.75, 1.92e-02)   %
        (6.00, 2.59e-03)   %
        (6.25, 1.54e-04)   %
        (6.50, 6.54186960091325e-6) %
        (6.75, 1.01246134689375e-6) %
        (7.00, 4.43633852630348e-7) %
      };
      \addlegendentry{RPC}

       \addplot[curveSP] plot coordinates {
      (5.00, 0.613496959209) %
      (5.25, 0.251256287098) %
      (5.50, 0.059136606753) %
      (5.75, 0.009812579490) %
      (6.00, 0.001120209694) %
      (6.25, 0.000157455521) %
      (6.50, 0.000025188765) %
      (6.75, 6.76164387192554e-6) %
      (7.00, 0.000001571608) %
      };
      \addlegendentry{SP}
      \addplot[curveALP] plot coordinates {
        (7.75, 9.26e-01)   %
        (8.00, 5.75e-01)   %
        (8.25, 2.26e-01)   %
        (8.50, 4.28e-02)   %
        (8.75, 3.13e-03)   %
        (9.0, 1.42e-4) %
        (9.25, 3.69036528273484e-6) %
        (9.50, 9.81586362797758e-8) %
      };
      \addplot[curveRPC] plot coordinates {
        (7.75, 9.26e-01)   %
        (8.00, 5.75e-01)   %
        (8.25, 2.26e-01)   %
        (8.50, 4.28e-02)   %
        (8.75, 3.13e-03)   %
        (9.0, 1.30e-4) %
        (9.25, 3.42748192020424e-6) %
        (9.50, 8.39528493932896e-8) %
      };
      \addplot[curveSP] plot coordinates {
      (7.75, 0.584795296192) %
      (8.00, 0.165837481618) %
      (8.25, 0.023391813040) %
      (8.50, 0.002017104998) %
      (8.75, 0.000091720314) %
      (9.00, 0.000007532696) %
      (9.25, 1.38737172633538e-6) %
      (9.50, 3.09633768005391e-7) %
      };
    \end{semilogyaxis}
  \end{tikzpicture}
  \vspace{-2ex}
  \caption{FER performance of the length-$999$ MacKay 999.111.3.5543 code $\mathcal{C}_{\rm MacKay}^{(3)}$ (left) and $\mathcal{C}_{\rm MacKay}^{(5)}$ (right).}
  \label{fig:MacKay}
  \vskip -3ex
\end{figure}

In order to examine the scalability of the proposed algorithm, we present numerical results for two sets of larger LDPC codes. The first one is based on MacKay's random $(3,27)$-regular $(999, 888)$ code 999.111.3.5543 from \cite{mac-web}, from which we derive a ternary ($\C^{(3)}_{\text{MacKay}}$) and a quinary ($\C^{(5)}_{\text{MacKay}}$) %
code, respectively, by iteratively replacing the nonzeros in each row of the parity-check matrix by the pattern $1,2,\dotsc, p-1,1,2,\dotsc$. The error-rate results are shown in Fig.~\ref{fig:MacKay}. Because of the large row-weight of this code, both PLP and CLP are intractable to run \eirikfinal{in practice}. As before, it can be observed that the RPC approach as stated in Section~\ref{sec:ACG} is helpful only for $p=3$. For $p=5$, the decoding results for ALP and PLP coincide exactly, providing strong evidence that Conjecture~\ref{conj:facetComplete_q5} holds also for larger $d$, because here each row code has length $d=27$. \revone{For comparison, we also plot the FER performance of the SP decoder with a maximum of $100$ iterations for both codes. From the figure we observe that ALP decoding performs almost as good as SP decoding, especially for $\mathcal{C}_{\rm MacKay}^{(3)}$, \eirik{at low-to-medium SNRs. For both $\mathcal{C}_{\rm MacKay}^{(3)}$ and $\mathcal{C}_{\rm MacKay}^{(5)}$ we observe a crossing of the ALP and RPC curves with the SP curve at low error rates.}} %

\begin{figure}
  \centering
  \begin{tikzpicture}
    \begin{semilogyaxis}[
        xlabel=$E_{\mathrm s}/N_0 (\si{\decibel})$,
        ylabel=FER,
        grid=both,
        ymax=1,
        ymin=1e-6,
        ytick={1e-6,1e-5,1e-4,1e-3,1e-2,1e-1,1},
        xtick={2,2.5,3,3.5,4,4.5,5,5.5,6,6.5,7,7.5,8,8.5,9},
        xmin=2, xmax=9,
        height=\myheight,
        width=\mywidth,
        y tick label style={rotate=90},
        legend columns=1,
        grid style={gray,opacity=0.5,dotted},
        legend style={legend pos=south west,font=\tiny}]
      \addplot[curveALP] plot coordinates {
        (6.00, 5.80e-01)   %
        (6.25, 1.89e-01)   %
        (6.50, 2.40e-02)   %
        (6.75, 7.47e-04)   %
        (7.00, 0.000010943491812) %
      };
      \addlegendentry{ALP}
      \addplot[curveT1] plot coordinates {
      (6.00, 5.83e-01)   %
      (6.25, 1.97e-01)   %
      (6.50, 2.44e-02)   %
      (6.75, 9.08e-04)   %
      (7.00, 0.0000109237422293960) %
          };
      \addlegendentry{ALP$|\Phi(\Theta^{\bm 0})$}

      \addplot[curveRPC] plot coordinates {
       (6.00, 5.76e-01)   %
        (6.25, 1.86e-01)   %
        (6.50, 2.40e-02)   %
        (6.75, 7.47e-04)   %
        (7.00, 9.97053604857482e-6) %
      };
      \addlegendentry{RPC}
             \addplot[curveSP] plot coordinates {
       (5.00, 0.740740716457) %
       (5.25, 0.268817216158) %
       (5.50, 0.047415837646) %
       (5.75, 0.003977250308) %
       (6.00, 0.000152728651) %
       (6.25, 0.000022039470) %
       (6.50, 0.000008284494) %
       (6.75, 0.000002729681) %
       (7.00, 0.000001072945) %
       };
      \addlegendentry{SP}
        \addplot[curveALP] plot coordinates {
          (3.50, 5.29e-01)   %
          (3.75, 1.74e-01)   %
          (4.00, 3.92e-02)   %
          (4.25, 3.75e-03)   %
          (4.50, 1.40e-04)   %
          (4.75, 6.94432822339571e-6) %
        };
        \addplot[curveRPC] plot coordinates {
          (3.50, 5.13e-01)   %
          (3.75, 1.74e-01)   %
          (4.00, 3.43e-02)   %
          (4.25, 2.87e-03)   %
          (4.50, 1.00e-04)   %
          (4.75, 2.86923697044441e-6) %
        };
        \addplot[curveSP] plot coordinates {
        (2.75, 0.666666686535) %
        (3.00, 0.413223147392) %
        (3.25, 0.106157109141) %
        (3.50, 0.030759766698) %
        (3.75, 0.007919537835) %
        (4.00, 0.001600640244) %
        (4.25, 0.000772773637) %
        (4.50, 0.000331417745) %
        (4.75, 0.000142472258) %
        };
      \addplot[curveALP] plot coordinates {
     (7.75, 8.77e-01)   %
      (8.00, 4.08e-01)   %
      (8.25, 7.99e-02)   %
      (8.50, 5.84e-03)   %
      (8.75, 0.0000537192384662805) %
      };
    \addplot[curveT1] plot coordinates {
        (7.75, 8.77e-01)   %
        (8.00, 4.31e-01)   %
        (8.25, 7.99e-02)   %
        (8.50, 7.1e-03)   %
        (8.75, 0.0000599966761841394) %
        
    };
    \addplot[curveSP] plot coordinates {
    (6.75, 0.568181812763) %
    (7.00, 0.172117039561) %
    (7.25, 0.025960540399) %
    (7.50, 0.001757376594) %
    (7.75, 0.000035395635) %
    (8.00, 0.0000003881956) %
    };
    \end{semilogyaxis}
  \end{tikzpicture}
  \vspace{-2ex}
  \caption{FER performance of $\C_{(3,6)}^{(3)}$ (left), $\C_{(3,6)}^{(5)}$ (center), and $\C_{(3,6)}^{(7)}$ (right).}
  \label{fig:RandomLDPC}
  \vskip -3ex
\end{figure}
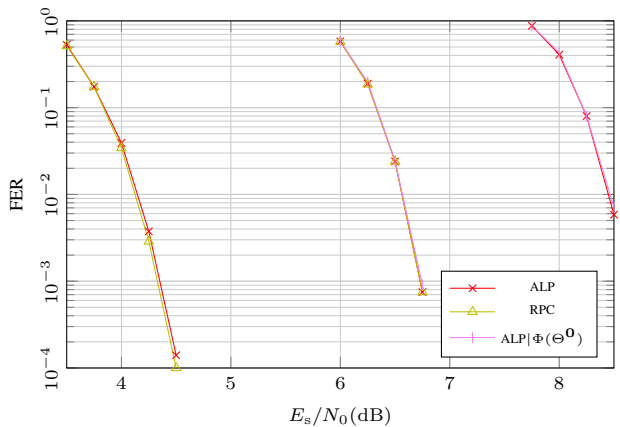

\begin{table*}
\caption{Numerical comparison of the proposed algorithms ALP and RPC (ALP with RPC search) with static LP decoding.
CPU times are specified in seconds$ \times 10^{-2}$.
Missing entries have been skipped because of intractable complexity.
}
 \vskip -2.0ex
\label{table:resultsNumbers}\centering
$\begin{array}{c|cc|c|*4c|*4c|*2c|*2c}
\toprule
p & \text{code} & \text{dimensions}&  \scriptsize \text{SNR} & \multicolumn{4}{c|}{\text{CPU time}\, (s\times 10^{-2})}  & \multicolumn{4}{c|}{\text{simplex iterations}} & \multicolumn{2}{c|}{\text{\# cuts}} & \multicolumn{2}{c}{\text{\# LPs}} \\
&&\text{$(N,K)$}& \text{(dB)} & \text{ALP} & \text{RPC}& \text{CLP} &  \text{PLP} &  \text{ALP}&\text{RPC} &  \text{CLP} &  \text{PLP} & \text{ALP}&\text{RPC} & \text{ALP}&\text{RPC} \\ \midrule
3&\C_{\mathrm{RM}(1)}^{(3)}   &(27,10)   &3
&.19&1.3&2.3  &7.9  %
&34.7&142&447  &1043 %
&64&165 %
&5.3&12.3 %
\\
&  &&7
&.11& .13&1.2  & 5.6 %
&8.6 &9.9&236  &694 %
&25&27.3 %
&2.9&2.98 %
\\
\midrule
3&\C_{\mathrm{RM}(2)}^{(3)}   &(81,31)   &8
&.37&.43& 39 &- %
&27.8&29.8&2009  &- %
&99.4&103 %
&3.8&3.8 %
\\
&&&10.5
&.1&.14&  11 &- %
&14.1&14.1& 1171 &- %
&12.9&12.9 %
&1.66&1.65 %
\\
\midrule
3&\C_{\mathrm{Tan}}^{(3)}   &(155,64)   &5
&.5&.8&16&14%
&66&70&1035&1962 %
&138&140 %
&3.5&3.6 %
\\
\midrule
5&\C_{\mathrm{Tan}}^{(5)}   &(155,64)   &7.5
&2.4&3.5&114&144%
&155&155&4178&7105%
&792&792 %
&3.3&3.3 %
\\
\midrule
7&\C_{\mathrm{Tan}}^{(7)}   &(155,64)   &9 
&20&25&561&1726%
&377&377&11399&16037%
&3188&3188 %
&3.3&3.3 %
\\
\midrule
11&\C_{\mathrm{Tan}}^{(11)}   &(155,64)   &11.5
&115&169&3508&-%
&864&864&35299&-%
&17358&17363 %
&3.6&3.6 %
\\
\midrule
3&\C_{\mathrm{MacKay}}^{(3)}   &(999,888)   &6
&.14&.42&341&-%
&73&83&1207&-%
&133&135 %
&3.5&3.7 %
\\
\midrule
5&\C_{\text{MacKay}}^{(5)}   &(999,888)   &8.5
&19&31&12509&-%
&391&392&53457&-%
&1226&1226 %
&4.3&4.4 %
\\
\midrule
3&\C_{(3,6)}^{(3)}   &(1000,500)   &4
&39&122& 2685  &- %
&1439&1581&14206  &- %
&1337&1372 %
&5.7&7.7 %
\\
\midrule
5&\C_{(3,6)}^{(5)}   &(1000,500)   &6.5
&655&744& 39155  &- %
&5279&5284&88432  &- %
&7668&7669 %
&4.7&4.7 %
\\
\midrule
7&\C_{(3,6)}^{(7)}   &(1000,500)   &8.5
&2382&-& - &- %
&8334&-&-  &- %
&24077&- %
&4.1&- %
\\
 \bottomrule
\end{array}$
\end{table*}

The second class is constructed with the same pattern based on a random $(3,6)$-regular $(1000,500)$ LDPC code; the resulting nonbinary codes are denoted $\C^{(3)}_{3,6}$ (for $p=3$), $\C^{(5)}_{3,6}$ (for $p=5$), and $\C^{(7)}_{3,6}$ (for $p=7$), with results shown in Fig.~\ref{fig:RandomLDPC}. This code has a much smaller check-node degree, hence PLP/CLP decoding is possible but extremely slow (see Table~\ref{table:resultsNumbers}). Interestingly, for this code the RPC algorithm \eirikfinal{only improves decoding performance for fairly low error rates for $p=3$. }
\revone{As for the codes above, we also plot the FER performance of the SP decoder  with a maximum of $100$ iterations. Somewhat surprisingly, for $\C_{(3,6)}^{(3)}$, ALP decoding \eirikfinal{significantly} outperforms SP decoding for high SNRs due a \eirikfinal{quite high} error floor. This also seems to be the case for $\C_{(3,6)}^{(5)}$, \eirikfinal{but for a much lower error rate}, while for low SNRs there is some performance loss compared to SP decoding. For  $\C_{(3,6)}^{(7)}$ there is no visible error floor with SP decoding.}

Table~\ref{table:resultsNumbers} shows that the proposed algorithms are clearly favorable in terms of decoding complexity compared with both CLP and PLP; in particular, they scale well with both block length and check node degree. The complexity of the static LP decoders in contrast explodes with increasing check node degree and quickly becomes intractable.

Note that, for ALP and $p$ fixed, the number of inequalities depends on the check node degrees only, such that it scales especially well for sparse codes.
Examining the different LP formulations and the results from Table~\ref{table:resultsNumbers}, one can name various reasons for this performance gain. First, the number $np$ (using constant-weight embedding) of variables in ALP is much smaller even compared with PLP.
Secondly, the number of constraints in the final linear program is virtually negligible as opposed to both static LP formulations. These two observations imply that the cost of a single simplex iteration is much lower in ALP than in PLP or CLP; the effect of this saving increases due to the much smaller number of simplex iterations observed for ALP. Finally, these advantages entirely outweigh the overhead introduced by solving several LP problems instead of one, as this number stays very small even for large codes.
\end{notms}

\section{Conclusion and Future Work} \label{sec:conclu}

In this work, we presented an explicit construction of valid inequalities (using no auxiliary variables) for the codeword polytope (or the convex hull) of the so-called constant-weight embedding of an SPC code over any prime field. The inequalities are assembled from classes of building blocks and can be proved to be facet-defining under some conditions. We observed numerically, for all primes $p \leq 19$, that a valid class gives facet-defining inequalities if and only if it is symmetric, and we conjectured this to be true in general. 
For ternary codes we proved that the inequalities from the construction together with the simplex constraints give a complete and irredundant description of the embedded codeword polytope. For quinary codes, based on extensive numerical evidence, we conjectured this to be the case as well. For $p>5$, there exist other types of facet-defining inequalities besides the ones that can be constructed from basic classes. A complete characterization of such inequalities is left as future work; the similarities of Lemmas~\ref{lem:conditions_theta5} and \ref{lem:conditions_theta6} with Lemma~\ref{lem:conditions} suggest that it might be possible to subsume all three types by a more general unifying form. Our initial numerical results however show that these are not required for achieving close-to-exact LP decoding performance, at least for small $p$.

Building on the explicit form of the inequalities, we presented an efficient (relaxed) ALP decoder for general linear codes over any prime field, in which efficient separation of the underlying inequalities describing the decoding polytope was done through DP. An explicit efficient implementation was also provided for ternary linear codes.
Next, an ACG-ALP decoder was presented, generalizing the corresponding decoding algorithm for binary codes, and we briefly showed how the results can be generalized to fields of size ${p^m}$ where $m > 1$ by introducing a relaxation. Numerical results for both LDPC and HDPC codes showed that our proposed ALP decoder outperforms (in terms of decoding complexity) a static decoder
(using both the plain and the cascaded LP formulation). 

We believe that many of the results in this paper generalize to nonprime fields. In particular, the concept of building blocks and building block classes appears to be universal. In particular, facet-defining inequalities for the codeword polytope of an embedded SPC code over $\F_8$ can be constructed using this principle as shown in \cite{ros16_ita}. In fact, the same \emph{structure} (or, equivalently, the same \emph{types} of building block classes) as for codes over $\F_7$ can be observed. %
An important open problem would be to completely understand how to construct such  building blocks and classes in the general case, with  the ultimate goal of  constructing an efficient algorithm to perform Euclidean projections onto the codeword polytope of an embedded SPC code in order to construct an efficient decoder using the ADMM framework. We believe that the characterization of the facet-defining inequalities (using no auxiliary variables) of the codeword polytope of an embedded nonbinary SPC code is a first step towards this goal.

\appendices

\revtwo{
\section{Two Technical Lemmas} \label{app:intermediate_results}

In addition to Lemma~\ref{lem:affinely}, the following lemma will become important in several proofs later on in the appendices.
\begin{lemma}\label{lem:affInd}
  Let $q = p^m \geq 3$, $m \geq 1$, and assume $\boldc$ and $\boldc^0, \boldc^1,\dotsc,\boldc^k \in \F_q^n$. The following are equivalent:
  \begin{enumerate}
  \item $\Fv(\boldc^0), \dotsc, \Fv(\boldc^k)$ are affinely independent.
  \item $\Fv(\boldc + \boldc^0), \dotsc, \Fv(\boldc + \boldc^k)$ are affinely independent.
  \item $\Fv'(\boldc^1-\boldc^0),\dotsc,\Fv'(\boldc^k-\boldc^0)$ are \emph{linearly} independent.
  \end{enumerate}
\end{lemma}

\begin{IEEEproof}
  By definition of $\Fv$ and $\f$, adding a \emph{fixed} vector $\bm c$ to each $\boldc^j$ results in a \emph{fixed} permutation of the entries in each $q$-block of $\Fv(\boldc^j)$. As this permutation has no effect on the affine independence, the equivalence of 1) and 2) follows.
  
  Assume 1) holds, then by 2) with $\boldc = -\boldc^0$ also $\Fv(\boldc^0-\boldc^0=\bm 0),\dotsc, \Fv(\boldc^k-\boldc^0)$ are affinely independent, hence by Lemma~\ref{lem:affinely} the vectors \[\Fv'(\bm 0),\Fv'(\bm c^1-\bm c^0),\dotsc, \Fv'(\boldc^k-\boldc^0)\] are affinely independent, which by definition of affine independence is equivalent to \[\Fv'(\bm c^1-\bm c^0) - \Fv'(\bm 0),\dotsc, \Fv'(\bm c^k-\bm c^0) - \Fv'(\bm 0)\] being linearly independent. But $\Fv'(\bm 0) = \bm 0$, which concludes the proof.
\end{IEEEproof}

The following result will be used in various proofs in the appendices.
\begin{lemma}\label{lem:bb-hilo-plusi}
For a building block class $\mathcal{T}^{\bm m}$ and any $k, i \in \F_p$,
\ifonecolumn
\begin{align}
&t_{k,(t_{k,\uparrow}+i)} - t_{k,t_{k,\uparrow}} = t_{k,\sigma-k+i} - t_{k,\sigma - k}= t_{\sigma,i} \label{eq:bb-hi-plusi}\\
\text{and}\;&t_{k,(t_{k,\downarrow}+i)} - t_{k,t_{k,\downarrow}} = t_{k,-k+i} - t_{k,-k} = t_{0,i}. \label{eq:bb-lo-plusi}
\end{align}
\else
\begin{align}
&t_{k,(t_{k,\uparrow}+i)} - t_{k,t_{k,\uparrow}} = t_{k,\sigma-k+i} - t_{k,\sigma - k}&&= t_{\sigma,i} \label{eq:bb-hi-plusi}\\
\text{and}\;&t_{k,(t_{k,\downarrow}+i)} - t_{k,t_{k,\downarrow}} = t_{k,-k+i} - t_{k,-k} &&= t_{0,i}. \label{eq:bb-lo-plusi}
\end{align}
\fi
In particular, both expressions  are independent of $k$.
\end{lemma}
\begin{IEEEproof}
  The left-hand equations are by \eqref{eq:zeta-hi} and \eqref{eq:zeta-lo}, respectively, while the right-hand equations follow by Property~\ref{part:bb-property4} of Lemma~\ref{lem:bb-properties} with $l=\sigma - k$  and $l=-k$ for \eqref{eq:bb-hi-plusi} and \eqref{eq:bb-lo-plusi}, respectively.
\end{IEEEproof}
}

\section{How to Switch Between Embeddings $\Fv$ and $\Fv'$}\label{app:embeddings}

Let $\C$ be a nonbinary code of length $d$ over the field $\F_q$, for some prime power $q$, and define $\P=\conv(\Fv(\C))$. Assume
\begin{subequations}
  \label{eq:pdesc}
  \begin{align}
    &\bm a^{\nu T}\bm x \leq \alpha^\nu&&\text{for }\nu \in \range N,
    \label{eq:pdesc-ineq}\\
    &\bm b^{\mu T}\bm x = \beta^\mu&&\text{for }\mu \in \range M, \label{eq:pdesc-eq}\\
      &\sum_{j \in \F_q} x_{i,j} = 1&&\text{for }i\in \range d, \label{eq:simplex-eq}\\
      &x_{i,j} \geq 0&&\text{for }i\in \range d, j \in \F_q \label{eq:simplex-ineq}
  \end{align}
\end{subequations}
is a description of $\P$ by means of $N+ dq$ linear inequalities and $M+d$ linear equations (for some natural numbers $N$ and $M$). Note that the existence of \eqref{eq:simplex-eq} and \eqref{eq:simplex-ineq} in this description can be assumed without loss of generality, because that part exactly specifies $S_{q-1}^d$, and $\P \subseteq S_{q-1}^d$ by definition of $\Fv$.

By adding, for $i \in \range d$, $-a^\nu_{i,0}$ times \eqref{eq:simplex-eq} to each inequality $\bm a^{\nu T} \boldx \leq \alpha^\nu$ of \eqref{eq:pdesc-ineq}, we may further assume that $a^\nu_{i,0} = 0$ for $i \in \range d$, and likewise that $b^\mu_{i,0}=0$ for all equations of \eqref{eq:pdesc-eq}. For $j=0$, the corresponding step turns \eqref{eq:simplex-ineq} into
\[ -\sum_{j \neq 0} x_{i,j} \geq -1 \text{ or equivalently } \sum_{j \neq 0} x_{i,j} \leq 1.\]
Now, for each $\bm a^\nu$, $\nu \in \range N$ (and $\bm b^\mu$ \eiriknew{and $\bm x$} analogously), we define $\bm a'^{\nu} \in \R^{d(q-1)}$ by removing all entries $a^\nu_{i,0}$, $i \in \range d$ (if we extend the definition of $\Pv$ to points outside of $S_{q-1}^d$, this can be written as $\bm a'^\nu = \Pv(\bm a^\nu)$), and define the polytope $\tilde \P \subseteq \R^{d(q-1)}$ by
\begin{subequations}
  \label{eq:ppdesc}
  \begin{align}
    &\bm a'^{\nu T} \boldx' \leq \alpha^\nu &&\text{for }\nu \in \range N,\label{eq:ppdesc-ineq} \\
    &\bm b'^{\mu T} \boldx' = \beta^\nu &&\text{for }\mu \in \range M,\label{eq:ppdesc-eq}\\
    &\sum_{j \neq 0} x_{i,j} \leq 1 &&\text{for } i \in \range d, \label{eq:hsimplex-1}\\
    &x_{i,j} \geq 0 &&\text{for }i \in \range d, j \in \F_q\setminus \{0\} \label{eq:hsimplex-0}
  \end{align}
\end{subequations}
which is obtained from \eqref{eq:pdesc} by removing the equations \eqref{eq:simplex-eq} and all coefficients belonging to field element $\zeta=0$, after they have been set to $0$ as described above.
\begin{proposition}
  $\tilde \P = \P'$, where $\P'= \conv(\Fv'(\C))$.
\end{proposition}
\begin{IEEEproof}
  Let $\boldx \in \P$. By construction of $\bolda'^\nu$ and $\boldb'^\mu$, $\bolda'^{\nu T} \Pv(\boldx) = \bolda^{\nu T} \boldx \leq \kappa$ and $\boldb'^{\mu T} \Pv(\boldx) = \boldb^{\mu T} \boldx$, thus $\Pv(\boldx)$ fulfills \eqref{eq:ppdesc-ineq} and \eqref{eq:ppdesc-eq}. Furthermore, \eqref{eq:hsimplex-0} and \eqref{eq:hsimplex-1} are obviously satisfied by $\Pv(\boldx)$, which shows that $\Pv(\boldx) \in \tilde \P$. Since $\Pv(\P) = \P'$ by Lemma~\ref{lem:embeddings}, we have established $\P' \subseteq \tilde \P$.
  
  Conversely, let $\boldx' \in \tilde\P \subseteq \hat S_{q-1}^d$. Again by construction of $\bolda'^{\nu}$ and $\boldb'^{\nu}$, $\bolda^{\nu T}\Lv(\boldx') \leq \alpha^\nu$ and $\boldb^{\mu T}\Lv(\boldx') = \beta^\mu$, and because further $\Lv(\boldx') \in S_{q-1}^d$, we see that $\Lv(\boldx')$ satisfies all constraints of \eqref{eq:pdesc}, hence $\Lv(\boldx') \in \P$. Using again Lemma~\ref{lem:embeddings}, this implies  that $\Pv(\Lv(\boldx')) = \boldx' \in \P'$, i.e., $\tilde\P \subseteq \eiriknew{\P'}$.
\end{IEEEproof}

The following explicit version of Corollary~\ref{cor:facetEquiv} is a by-product of the above proof.
\begin{corollary}
  If $\bolda^{\nu T} \boldx \leq \alpha^\nu$ induces the face $F$ of $\P$ with $\dim(F) = \delta$, then $\bolda'^{\nu T} \boldx' \leq \alpha^\nu$ induces $\Pv(F)$ (which by Corollary~\ref{cor:facetEquiv} also has dimension $\delta$). This holds in particular if $F$ is a facet.
\end{corollary}

\begin{remark}
  \begin{enumerate}
    \item If $\mathcal C$ is an SPC code with $d\geq 3$, \eqref{eq:simplex-eq} already specifies the affine hull of $\P$ by Proposition~\ref{prop:Pjdim}. Hence, no further equations are necessary, i.e., $M=0$.
    \item By Property~\ref{part:bb-property2} of Lemma~\ref{lem:bb-properties}, $a^{\nu}_{i,0} = 0$ already holds for all inequalities constructed in Section~\ref{sec:bb}. There is no need to add multiples of \eqref{eq:simplex-eq} to those inequalities to establish this assumption.
    \item While the above results show that $\P$ and $\P'$ are practically equivalent with respect to most \emph{polyhedral} properties, they are \emph{geometrically} different because the underlying embeddings $\f$ and $\f'$ are. For example, when $p=3$, $\lVert \f(1) - \f(0) \rVert_2 = \lVert \f(2) - \f(1) \rVert_2$, where $\Vert \cdot \Vert_2$ denotes the Euclidean norm of its argument, while the two distances are different when replacing $\f$ by $\f'$. This has consequences when using nonlinear solvers, such as the penalized ADMM decoder (cf.\ \cite[Sec.~VI]{liu14_1}, \cite{liu14}).
    \item At a first glance, using $\P'$ instead of $\P$ appears to be computationally preferable because \eqref{eq:ppdesc} exhibits less variables than \eqref{eq:pdesc}. However, when solved with the simplex method, the inequalities \eqref{eq:hsimplex-1} will be internally expanded to the form \eqref{eq:simplex-eq} by introducing slack variables; hence, internally, the algorithm will perform \emph{exactly the same steps} no matter which embedding is used.
  \end{enumerate}
\end{remark}

\section{Proof of Proposition~\ref{prop:Pjdim}}\label{app:proofPjdim}
Assume $p \geq 3$ and let $\mathcal A = \{\boldx \in \R^{dp}\colon \eqref{eq:spx-eq} \text{ holds for }i \in \range d\}$ be the affine hull of $S_{p-1}^d$. Because $\Pd \subseteq S_{p-1}^d \subseteq \mathcal A$ by definition and further $\dim(\mathcal A)=d(p-1)$ (as follows immediately from the structure of \eqref{eq:spx-eq}), we obtain
\begin{gather}
  \dim (\Pd) \leq \dim (S_{p-1}^d) = \dim(\mathcal A)= d(p-1) \label{eq:dimUb}\\
  \text{and}\quad\aff(\Pd) \subseteq \mathcal A.\label{eq:dimAff}
\end{gather}
\emph{Part~1):}
  By \eqref{eq:dimUb} it suffices to show that $\dim(\Pd) \geq d(p-1)$, i.e., find $d(p-1) + 1$ affinely independent elements of $\Pd$.
  
  Define the set $S = \{\boldc^0,\dotsc,\boldc^{d(p-1)}\} \subseteq \mathcal C$ of $d(p-1)+1$ codewords of $\mathcal C$ as follows. First, $\boldc^0 = (0,\dotsc,0)\eiriknewnew{^T}$. Then, for $ 0 \leq i < d-1$ and $1 \leq l < p$, $\boldc^{i(p-1) + l}$ is a codeword with two nonzeros only, defined by
   \[c_j^{i(p-1)+l} = \begin{cases}
      [l]_p&\text{if }j = i+1,\\
      -[l]_p&\text{if }j = d,\\
      0&\text{otherwise.} \end{cases}\]
  Finally, there are $p-1$ codewords with three nonzeros, namely $\boldc^{(d-1)(p-1)+l} = (0,\dotsc,0,[l]_p,[l]_p,[-2l]_p)\eiriknewnew{^T}$, $1 \leq l < p$ (note that here we need $p \neq 2$, because $[-2]_2 =[0]_2$).
  
  We now show that the $d(p-1)$ vectors
  \[\Fv'(\boldc^1),\dotsc, \Fv'(\boldc^{d(p-1)})\]
  are linearly independent, from which the claim follows by Lemma~\ref{lem:affInd} since $\bm c^0  = \bm 0$. Let $\bm{M}$ be the real square $0/1$ matrix with rows $\Fv'(\boldc^1)^T, \dotsc, \Fv'(\boldc^{d(p-1)})^T$. Then, $M$ can be written as a $d\times d$ block matrix with blocks of size $(p-1)\times (p-1)$ having the form
  \[ \bm M = \begin{pmatrix}
    \bm I_{p-1} &   & &&\bar{\bm I}_{p-1} \\
      & \ddots & &&\vdots \\
      & & \ddots & &\vdots\\
      & &         &  \bm I_{p-1} & \bar{\bm I}_{p-1} \\
      & &         \bm I_{p-1} & \bm I_{p-1} & \bm C
  \end{pmatrix} \in \R^{d(p-1) \times d(p-1)}\]
  where $\bm I_{p-1}\in \R^{(p-1)\times (p-1)}$ is the identity matrix,
  \begin{equation} \notag %
  \bar{\bm I}_{p-1} = \begin{pmatrix} && 1 \\ & \iddots \\ 1 \end{pmatrix}
   \in \R^{(p-1)\times (p-1)}
  \end{equation}
  is the \enquote{reverse} identity, 
  and $\bm C$ is a permutation matrix with a single one at column index $[-2l]_p$ (because $p$ is prime, these values are distinct for all $l \in \range{p-1}$) for row index $l$, $1 \leq l < p$.  Thus, $\bm{M}$ is almost upper triangular except for the lower right $3\times 3$ block-submatrix, which we now show to have full rank. By elementary row operations,
  \[
    \det\begin{pmatrix} \bm I_{p-1}& &\bar{\bm I}_{p-1} \\ & \bm I_{p-1} & \bar{\bm I}_{p-1} \\ \bm I_{p-1} & \bm I_{p-1} & \bm C \end{pmatrix}
    = \det\begin{pmatrix} 2\bar{\bm I}_{p-1} - \bm C \end{pmatrix};\]
  it hence suffices to show that the latter matrix is nonsingular. To that end, note that, for $j \in \range{p-1}$, the $j$-th row of $2\bar{\bm I}_{p-1}$ has an entry $2$ in column $[-j]_p$, while the $j$-th row of $\bm C$ has an entry $1$ in column $[-2j]_p$ (all other entries are $0$). Because $p$ is prime, $[-j]_p \neq [-2j]_p$ for $j \in  \range{p-1}$. Hence, when reversing the rows of $2\bar{\bm I}_{p-1} - \bm C$, the result is a \emph{strictly diagonally dominant} matrix, which by the Levy-Desplanques theorem (see, e.g., \cite[Cor.~5.6.17]{HornJohnson12Matrix}) implies that it is nonsingular. This concludes the proof.\\
\emph{Part~2):}
  We have just shown that $\dim(\aff(\P)) =\dim(\P) =  \dim(\mathcal A)$. Since both are affine spaces and by \eqref{eq:dimAff} one is contained in the other, they must indeed be equal.\\
\emph{Part~3):}
  Assume wlog.\ that $i=d$ (the rest follows from symmetry), and consider $j=0$ first. In the proof of part~\ref{prop:Pjdim-1}, we have shown that the $d(p-1)$ points $\Fv'(\boldc^1), \dotsc, \Fv'(\boldc^{d(p-1)})$ are linearly and hence also affinely independent, such that, by Lemma~\ref{lem:affinely},  $\Fv(\boldc^1),\dotsc,\Fv(\boldc^{d(p-1)})$ are affinely independent, too. By construction $c^i_d \neq 0$ and thus $(\Fv(\boldc^i))_{d,0} = 0$ for all of them, thus \eqref{eq:spx-geq} is satisfied with equality for $d(p-1)$ affinely independent elements of $\Pd$, such that \eqref{eq:spx-geq} defines a facet of $\Pd$ for $j=0$.
  
  For $j \neq 0$, let $\boldc = (0,\dotsc, 0, -j, j)\eiriknewnew{^T} \in \C$. By the above and Lemma~\ref{lem:affInd}, $\Fv(\boldc + \boldc^1),\dotsc, \Fv(\boldc + \boldc^{d(p-1)})$ are affinely independent. By linearity, all $\boldc + \boldc^i \in \C$, and because $c_d = j$ and $c^i_d \neq 0$ for all $i \in \range{d(p-1)}$, it follows that $(\boldc + \boldc^i)_d \neq j$ for $i \in \range{d(p-1)}$. Hence, for $j \neq 0$, \eqref{eq:spx-geq} is satisfied with equality by $\Fv(\boldc + \boldc^1),\dotsc, \Fv(\boldc + \boldc^{d(p-1)})$, \ie defines a facet of $\Pd$.

\section{Proof of Lemma~\ref{lem:conditions}}\label{app:proofConditions}

Let $\boldtheta^T \boldx \leq \kappa$ with $\boldtheta=(\bm t_{k_1} \mid \dotsc \mid \bm t_{k_d})\eiriknew{^T}$ be contained in $\Theta^{\bm m}$, and denote by $\boldc \in \C$ the canonical codeword corresponding to $\boldtheta$. Then, \eqref{eq:Thetacondition} follows from \eqref{eq:hilocons-ci} and \eqref{eq:hilocons-cd}, since $\bm c \in \C$
\[
  \Leftrightarrow\sum_{i=1}^d c_i =\sum_{i=1}^{d-1}(\sigma-k_i) - k_d = 0 
    \Leftrightarrow \sum_{i=1}^d k_i = [d-1]_p\sigma.
\]
Likewise, by construction
\begin{align*}
  \kappa &= \boldtheta^T \Fv(\boldc)
          = \sum_{i=1}^d t_{k_i, c_i} &&\text{by \eqref{eq:ineqVals}}\\
         &= \sum_{i=1}^{d-1} \max(\bm t_{k_i}) + \min(\bm t_{k_d}) &&\text{by Construction~\ref{constr:hilo}}\\
         &= \sum_{i=1}^{d-1} \left( t_{0,\sigma} - t_{0,k_i}\right) -t_{0,k_d}&&\text{using \eqref{eq:val-k-hi} and \eqref{eq:val-k-lo}}\\
         &= (d-1)t_{0,\sigma} - \sum_{i=1}^d t_{0,k_i}
\end{align*}
which is \eqref{eq:kappacondition}. It is easy to see that the proof works in both directions, \ie if $\boldtheta$ and $\kappa$ fulfill \eqref{eq:hilo-conditions}, then they are covered by Construction~\ref{constr:hilo}.

Finally, the equivalence of \eqref{eq:hilo-conditions} and \eqref{eq:hilo-conditions'} follows immediately by using the equations $\sum_{i=1}^d k_i = \sum_{k\in \F_p} k \left[\abs{V_k^\boldtheta}\right]_p$ and $d = \sum_{k \in \F_p} \abs{V_k^\boldtheta}$ in \eqref{eq:Thetacondition} and \eqref{eq:kappacondition}, and by \eqref{eq:property-3}.
\section{Proofs for Section~\ref{sec:validInvalid}}\label{app:validInvalid}

\begin{IEEEproof}[Proof of Lemma~\ref{lem:validIndependent}]
  By Construction~\ref{constr:hilo}, $\kappa = \boldtheta^T \Fv(\boldc)$.
  \ifonecolumn
  Hence, 
  \begin{align*}
   \boldtheta^T \Fv(\bm c + \bm\xi) - \kappa &=\boldtheta^T \left(\Fv(\bm c + \bm\xi) - \Fv(\bm c)\right)\\
    &=\sum_{i=1}^d (t_{k_i, c_i + \xi_i} - t_{k_i, c_i})&\text{by \eqref{eq:ineqVals}}\\
    &=\sum_{i=1}^{d-1} (t_{\sigma - c_i, c_i + \xi_i} - t_{\sigma-c_i, c_i}) 
      + t_{-c_d, c_d + \xi_d} - t_{-c_d, c_d}&\text{by \eqref{eq:hilocons-ki} and \eqref{eq:hilocons-kd}}\\
    &=\sum_{i=1}^{d-1} t_{\sigma,\xi_i} + t_{0,\xi_d}&\text{by Lemma~\ref{lem:bb-hilo-plusi}.}
  \end{align*}
  \else
  Hence, $\boldtheta^T \Fv(\bm c + \bm\xi) - \kappa$
  \begin{align*}
    &=\boldtheta^T \left(\Fv(\bm c + \bm\xi) - \Fv(\bm c)\right)\\
    &=\sum_{i=1}^d (t_{k_i, c_i + \xi_i} - t_{k_i, c_i})&\text{by \eqref{eq:ineqVals}}\\
    &=\sum_{i=1}^{d-1} (t_{\sigma - c_i, c_i + \xi_i} - t_{\sigma-c_i, c_i}) \\
      &\phantom{=}+ t_{-c_d, c_d + \xi_d} - t_{-c_d, c_d}&\text{by \eqref{eq:hilocons-ki} and \eqref{eq:hilocons-kd}}\\
    &=\sum_{i=1}^{d-1} t_{\sigma,\xi_i} + t_{0,\xi_d}&\text{by Lemma~\ref{lem:bb-hilo-plusi}.}
  \end{align*}
  \fi
\end{IEEEproof}

\begin{IEEEproof}[Proof of Corollary~\ref{cor:allValidOrNot}]
  \begin{enumerate}
    \item Let $(\boldtheta^T \boldx \leq \kappa) \in \Theta^{\bm m}$ with canonical codeword $\boldc^{\boldtheta}$. In order to show that the inequality is valid for $\P = \conv(\Fv(\C))$, it suffices to prove it valid for all vertices of $\P$, which by definition of $\P$ are given by $\Fv(\boldc)$ for $\boldc \in \C$. To that end, let $\boldc \in \C$ be chosen arbitrarily and define $\boldc' = \boldc - \boldc^{\boldtheta} \in \C$. Then,
    \ifonecolumn
    \begin{displaymath}
      \boldtheta^T \Fv(\boldc) - \kappa = \boldtheta^T \Fv(\boldc' + \boldc^\boldtheta) - \kappa 
      = \sum_{i=1}^{d-1} t_{\sigma, c'_i} + t_{0,c'_d} \leq 0
    \end{displaymath}
    \else
    \begin{align*}
      \boldtheta^T \Fv(\boldc) - \kappa &= \boldtheta^T \Fv(\boldc' + \boldc^\boldtheta) - \kappa \\
      &= \sum_{i=1}^{d-1} t_{\sigma, c'_i} + t_{0,c'_d} \leq 0
    \end{align*}
    \fi
    where the last two equations hold because of Lemma~\ref{lem:validIndependent} and the assumption applied to $\boldc'$, respectively.
    \item Let $\boldtheta^T \boldx \leq \kappa$ and $\boldc^\boldtheta$ as above, and define $\boldc' = \boldc^\boldtheta + \boldc \in \C$. Then,
    \ifonecolumn
    \begin{displaymath}
      \boldtheta^T \Fv(\boldc') - \kappa = \boldtheta^T \Fv(\boldc + \boldc^\boldtheta) - \kappa 
      = \sum_{i=1}^{d-1} t_{\sigma, c_i} + t_{0,c_d} > 0;
    \end{displaymath}
    \else
    \begin{align*}
      \boldtheta^T \Fv(\boldc') - \kappa &= \boldtheta^T \Fv(\boldc + \boldc^\boldtheta) - \kappa \\
      &= \sum_{i=1}^{d-1} t_{\sigma, c_i} + t_{0,c_d} > 0;
    \end{align*}
    \fi
    the inequality is violated by $\boldc' \in \C$ and hence invalid for $\P$.
  \end{enumerate}
\end{IEEEproof}

\begin{IEEEproof}[Proof of Theorem~\ref{thm:newValidProgram}]
  Let $\mathcal T^{\bm m}$ be a valid class and $d >0$. We show that \eqref{eq:allValidCondition} holds for all $\bm c\in \C$, which implies the first statement by Corollary~\ref{cor:allValidOrNot}. Let $\bm c \in \C$, and denote the left side of \eqref{eq:allValidCondition} by $\gamma(\bm c) =\sum_{i=1}^{d-1} t_{\sigma, c_i} + t_{0,c_d}$. The following three observations will be used:  
  \begin{enumerate}
  \item By Property~\ref{part:bb-property1} of Lemma~\ref{lem:bb-properties}, $[\gamma(\bm c)]_p = \sum_{i=1}^d c_i = 0$.
  \item By Property~\ref{part:bb-property3} of Lemma~\ref{lem:bb-properties}, $t_{\sigma,i} \leq 0$ for all $i \in \F_p$, \ie the left term in \eqref{eq:allValidCondition} contains nonpositive entries only.
  \item By definition of $\sigma$ in \eqref{eq:sigma}, $t_{0,c_d} \leq \max(\bm t_0) = t_{0,\sigma} <2p$.
  \end{enumerate}
  Assume now  $\gamma(\bm c) > 0$, \ie \eqref{eq:allValidCondition} is violated. By 1), this implies $\gamma(\bm c) \geq p$, while 2) and 3) imply that $\gamma(\bm c) \leq t_{0,\sigma} < 2p$, hence $\gamma(\bm c) = p$. By 2), this shows that $t_{0,c_d} \geq p$, hence $m_{c_d} = 1$. Thus, the equation $\gamma(\bm c) = p$ can be written as
  $\sum_{i=1}^{d-1} t_{\sigma,\eirik{c_i}} + p + [c_d]_\Z = p$, \ie
  \begin{equation}
    \sum_{i=1}^{d-1} t_{\sigma,\eirik{c_i}} = -[c_d]_\Z.
    \label{eq:thmValid-proof}
  \end{equation}
  Further, the result $m_{c_d}=1$ rules out the case $\bm m = (0,\dotsc,0)$, hence $m_{\sigma} = 1$ and thus $[\sigma]_\Z \geq [c_d]_\Z$ by definition of $\sigma$.
  
  In \eqref{eq:thmValid-proof}, all terms on the left are $\leq 0$, hence $t_{\sigma,\eirik{c_i}} \geq -[c_d]_\Z \geq -[\sigma]_\Z$ for $i \in \range{d-1}$.  
  But then with $n_j=\abs{\{i \in \range{d-1}\colon c_i = j\}}$ for $j \in J = \{j \in \F_p\colon 0 > t_{\sigma,j} \geq -[\sigma]_\Z\}$ (observe that $t_{\sigma,0} = 0$ by Property~\ref{part:bb-property2}) of Lemma~\ref{lem:bb-properties}),
  \[\sum_{i=1}^{d-1} t_{\sigma,\eirik{c_i}} +  [c_d]_\Z = \sum_{j\in J} n_jt_{\sigma,j} + [c_d]_\Z = 0\]
  with $m_{c_d} = 1$ and $c_d = -\sum_{j\in J} [n_j]_p\cdot j$, \ie $\mathcal T^{\bm m}$ is not valid, a contradiction.
  
  For the second statement of the theorem, let $\{n_i\}_{i \in I}$ and  $r=-\sum_{i\in I} [n_i]_p\cdot i$ be a solution to \eqref{eq:valid-class-condition} with $m_r=1$. Choose $d \geq d_0 =\sum_{i \in I} n_i + 1$ and let $\bm c \in C$ be any codeword that has, for $i \in I$, $n_i$ $i$-entries among the first $d-1$ entries, $c_d= r$, and all other entries are zero (any such vector is a codeword of $\C$ by the condition on $r$).
  
  Analogously to above, we see that
  \[\sum_{j=1}^{d-1} t_{\sigma,c_j} + t_{0,c_d} = \sum_{i \in I} n_i t_{\sigma,i} + [c_d]_\Z + p = p > 0\]
  such that, by the second part of Corollary~\ref{cor:allValidOrNot}, no inequality in $\Theta^{\bm m}$ is valid for $\P$. Finally, $\eqref{eq:valid-class-condition}$ gives $\sum_{i\in I} n_i \leq [r]_\Z$ because all $t_{\sigma,i} \leq -1$ for $i \in I$ by definition of $I$, such that $d_0 = \sum_{i \in I} n_i  + 1\leq [r]_\Z + 1 \leq [\sigma]_\Z + 1$, which concludes the proof.
\end{IEEEproof}
\begin{IEEEproof}[Proof of Lemma~\ref{lem:valid-symmetric-condition}]
  We show that the system in Definition~\ref{def:valid-class} is solvable if and only if the one from Lemma~\ref{lem:valid-symmetric-condition} is. For the \enquote{only if} part, let $\{n_i\}_{i \in I}$ and $r$ be a solution to \eqref{eq:valid-class-condition} with $r=-\sum_{i\in I} [n_i]_p \cdot i$ and $m_r = 1$.
  
  By Item~\ref{lem:bb-symmetric-property1} of Lemma~\ref{lem:bb-symmetric_properties}, $ I = \{i \in \F_p\colon 0 > t_{\sigma,i} \geq -[\sigma]_\Z\} = \{i \colon 0 > -t_{0,-i} \geq -[\sigma]_\Z\} = \{i\colon 0 < t_{0,-i} \leq [\sigma]_\Z\}$. Because $[\sigma]_\Z<p$, $t_{0,-i} \leq [\sigma]_\Z$ implies $m_{-i} = 0$ for $i \in I$. Furthermore, $m_r=1$ implies $m_\sigma = 1$, such that $t_{0,\sigma} = p + [\sigma]_\Z$ and by Property~\ref{part:bb-property1} of Lemma~\ref{lem:bb-properties} $t_{0,k} \neq [\sigma]_\Z$ for any $k \in \F_p$. Concludingly,
  \[ I = \{i\colon m_{-i} = 0\text{ and }0 < t_{0,-i} < [\sigma]_\Z\}\]
  and thus $J = \{-i\colon i \in I\}$. For $j \in J$, define $\nu_j = n_{-j}$ and let $\rho = r$. Then,
  \ifonecolumn
  \begin{displaymath} \notag
    0 = \sum_{i \in I} n_i \cdot t_{\sigma,i} + [r]_\Z 
    = \sum_{j \in J} \nu_j t_{\sigma,-j} + [\rho]_\Z = -\sum_{j \in J} \nu_j t_{0,j} + [\rho]_\Z = \sum_{j \in J} \nu_j \cdot j + [\rho]_\Z
  \end{displaymath}
  \else
  \begin{multline} \notag
    0 = \sum_{i \in I} n_i \cdot t_{\sigma,i} + [r]_\Z 
    = \sum_{j \in J} \nu_j t_{\sigma,-j} + [\rho]_\Z \\= -\sum_{j \in J} \nu_j t_{0,j} + [\rho]_\Z = \sum_{j \in J} \nu_j \cdot \eirik{[j]_\Z} + [\rho]_\Z \end{multline}
  \fi
  which completes the proof of the \enquote{only if} direction. For the \enquote{if} part, one can see analogously that $I = \{-j\colon j \in J\}$ and use $n_i=-\nu_j$ and $r = \rho$ to construct a solution for \eqref{eq:valid-class-condition}, which completes the proof.
\end{IEEEproof}

\begin{IEEEproof}[Proof of Lemma~\ref{lem:counting_formulas}]
By Lemma~\ref{lem:validIndependent}, a vector $\bm\zeta = \bm c + \bm\xi$ satisfies $\boldtheta^T \Fv(\boldzeta) > \kappa$ if and only if $I^>_{\xi_d,J^{\bm\xi}} = 1$, where the multiset $J^{\bm\xi}$ contains the nonzero entries of $\xi_1,\dotsc,\xi_{d-1}$ (with multiplicity). The first claim follows because, for $\xi_d$ fixed, there are $\binom{d-1}{|J^{\bm\xi}|} \frac{\eirikNew{|J^{\bm\xi}|!}}{n^J_1! \cdots n^J_{k^J}!}$ different vectors $\bm\xi$ that result in the same multiset $J^{\bm\xi}$, where the multinomial coefficient is due to the number possible permutations of the multiset $J$.

The second part is analogous; note that $\bm\zeta$ is a codeword if and only if $\sum_{i=1}^d \xi_i = \xi_d + \|J^{\bm\xi}\|_1=0$, which accounts for the additional condition in the definition \eirik{for} $I^=_{c,J}$.
\end{IEEEproof}

%
%
%
%
%
%
%
%
%
%
%
%
%
%
%
%
%
%
%
%
%
%
%
%
%
%
%
%
%
%
%
%
%
%
%
%
%
%
%
%
%
%
%
%

\section{Proof of Lemma~\ref{lem:facets}}\label{app:prooffacets}
Let $\boldtheta^T \bm x \leq \kappa$ be an inequality from $\Theta^{\bm m}$.
In order to prove the claim, we construct a set $S \subset \C$ of codewords such that $\boldtheta^T \Fv(\bm c) = \kappa$  for any $\bm c \in S$, and at least $1+d(p-1)-1-(p-3)/2$ elements of $S$ have affinely independent embeddings. Let $\bm c^{\boldtheta}$ be the  canonical codeword from Construction~\ref{constr:hilo} that is used to generate the inequality and define, for $s \in \range{(p-1)(d-1)}$, the vector $\bm\xi^s = \bm\xi^{i(p-1)+l} \in \F_p^d$ by
\begin{equation}
  \xi^{i(p-1)+l}_j = \begin{cases}
  \phantom{-}\left[l\right]_p &\text{if \eirik{$j=i+1$}},\\
  -\left[l\right]_p  & \text{if $j=d$},\\
  \phantom{-}0 &\text{otherwise} \end{cases}
  \label{eq:faceproof-xi-type1}
\end{equation}
where $0 \leq i < d-1$ and $1 \leq l < p$, and let $\bm c^s = \bm c^{\boldtheta} + \bm\xi^s$ for $s \in \range{(p-1)(d-1)}$. By Lemma~\ref{lem:validIndependent}, $\boldtheta^T \Fv(\bm c^{i(p-1)+l}) - \kappa = t_{\sigma,[l]_p} + t_{0,[-l]_p} = 0$,
where the second equation is due to Part~\ref{lem:bb-symmetric-property1}) of Lemma~\ref{lem:bb-symmetric_properties}, such that the inequality is tight for the embeddings of these $(d-1)(p-1)$ codewords.

Now, we construct $p-2$ additional codewords that differ from $\bm c^{\boldtheta}$ in the last \emph{three} entries by adding, for each $i \in \F_p \setminus \{0,\sigma\}$, the vector $\bm\zeta^i \in \F_p^d$ defined by
\[ \zeta^i_j = \begin{cases}
  -i&\text{if }j=d-2,\\
  i-\sigma &\text{if }j=d-1,\\
  \sigma&\text{if }j=d,\\
  0&\text{otherwise}
  \end{cases}
\]
to the canonical codeword $\bm c^{\boldtheta}$. Using again Lemma~\ref{lem:validIndependent} we find that, for $i \in \F_p\setminus\{0,\sigma\}$,
\ifonecolumn
\begin{align*}
  \boldtheta^T\Fv(\bm c^{\boldtheta} + \bm\zeta^i) - \kappa
  &=t_{\sigma,-i} + t_{\sigma,i-\sigma} + t_{0,\sigma} \\
  &=-t_{0,i} - t_{0,\sigma-i} + t_{0,\sigma} \\
  &= - [i]_\Z - p m_i - [\sigma - i]_\Z - p m_{\sigma - i} + [\sigma]_\Z + p m_\sigma.
\end{align*}
\else
\begin{align*}
  &\boldtheta^T\Fv(\bm c^{\boldtheta} + \bm\zeta^i) - \kappa\\
  &=t_{\sigma,-i} + t_{\sigma,i-\sigma} + t_{0,\sigma} \\
  &=-t_{0,i} - t_{0,\sigma-i} + t_{0,\sigma} \\
  &= - [i]_\Z - p m_i - [\sigma - i]_\Z - p m_{\sigma - i} + [\sigma]_\Z + p m_\sigma.
\end{align*}
\fi
For $[i]_\Z \leq [\sigma]_\Z$, the above is zero because then $-[i]_\Z - [\sigma-i]_\Z + [\sigma]_\Z = 0$ and $m_i + m_{\sigma - i} = m_\sigma$ by Part~\ref{lem:bb-symmetric-property3} of Lemma~\ref{lem:bb-symmetric_properties}. If $[i]_\Z > [\sigma]_\Z$ then $m_i = 0$ by definition, thus also $m_{\sigma - i} = 0$. Also, this case implies $\sigma \neq p-1$, hence $m_\sigma = 1$, and furthermore $[\sigma - i]_\Z = [\sigma]_\Z - [i]_\Z + p$, such that $\boldtheta^T \Fv(\bm c^{\boldtheta} + \bm\zeta^i) = \kappa$ also for this case. Concludingly, we have constructed a set $S$ of $(p-1)(d-1) + p - 2 = (p-1)d - 1$ codewords of $\C$, the embeddings of which satisfy the inequality with equality.

We now switch to Flanagan's embedding from Definition~\ref{rem:Flanagan} and show that the span of $\Fv'(\boldc^1-\boldc^{\boldtheta}),\dotsc,\Fv'(\boldc^{(p-1)d-1}-\boldc^{\boldtheta})$ has rank $d(p-1)-1-(p-3)/2$, from which the claim follows by Lemma~\ref{lem:affInd}. To that end, consider the real $0/1$ matrix $\bm{M}_F$ whose rows
are $\Fv'(\boldc^1-\boldc^{\boldtheta})^T,\dotsc,\Fv'(\boldc^{(p-1)d-1}-\boldc^{\boldtheta})^T$. This matrix is of the form
  \begin{equation} \label{eq:MF}
    \bm{M}_F = \begin{pmatrix}
    \bm{I}_{p-1} & \bm{0} & & &  \bm{0} & \bar{\bm I}_{p-1}\\
    \bm{0} & \bm{I}_{p-1} & & &  \bm{0} & \bar{\bm I}_{p-1}\\
     & &\ddots&     &&\\
     & &      &    \bm{I}_{p-1}&\bm{0}&\bar{\bm I}_{p-1}\\
     & &      &    \bm{0}&\bm{I}_{p-1}&\bar{\bm I}_{p-1}\\
     & &      &    \bar{\bm{D}}_{p-2}&\bm D_{p-2}&\bm{E}_{p-2}
     \end{pmatrix}
     \end{equation}
where $\bm{I}_{p-1}$ and $\bar{\bm I}_{p-1}$ are defined in Appendix~\ref{app:proofPjdim},
\ifonecolumn
\begin{displaymath}
  \bar{\bm D}_{p-2} =
    \begin{pmatrix}
      &\eiriknew{\bm{0}} 0& \bar{\bm I}_{[\sigma]_\Z-1}\\
      \bar{\bm I}_{[-\sigma]_\Z-1} & \eiriknew{\bm{0}}
    \end{pmatrix}, \;
    \bm D_{p-2} =
      \begin{pmatrix}
        &0&\bm I_{[\sigma]_\Z-1}\\
        \bm I_{[-\sigma]_\Z-1}&0
      \end{pmatrix}
\end{displaymath}
\else
\begin{align*}
  \bar{\bm D}_{p-2} &=
    \begin{pmatrix}
      &\eiriknew{\bm{0}}& \bar{\bm I}_{[\sigma]_\Z-1}\\
      \bar{\bm I}_{[-\sigma]_\Z-1} & \eiriknew{\bm{0}}
    \end{pmatrix},\\
    \bm D_{p-2} &=
      \begin{pmatrix}
        &\eiriknew{\bm{0}}&\bm I_{[\sigma]_\Z-1}\\
        \bm I_{[-\sigma]_\Z-1}&\eiriknew{\bm{0}}
      \end{pmatrix}
\end{align*}
\fi
(note that the $[-\sigma]_\Z$-th column (with $[-\sigma]_\Z = p - [\sigma]_\Z$) of both $\bm D_{p-2}$ and $\bar{\bm D}_{p-2}$ is all-zero), and $\bm E_{p-2}$ has $1$-entries in the $[\sigma]_\Z$-th column and zeros \eirik{elsewhere}.
Now, note that the matrix $\bm{M}_F$ is (similar to the one in the proof of Proposition~\ref{prop:Pjdim}) almost upper triangular except for the lower right $3(p-1)-1 \times 3(p-1)$ block. After turning $\bar{\bm D}_{p-2}$ and $\bm D_{p-2}$ into zero matrices by Gaussian elimination, the lower right $\bm E_{p-2}$ turns into the matrix
\[ 
 \tilde{\bm E}_{p-2} = \begin{pmatrix} -\bm X_{[\eiriknew{\sigma}]_\Z-1} & \eiriknew{\bm{1}} \\
 &\eiriknew{\bm{1}} & \eiriknew{-}\bm X_{[\eiriknew{-\sigma}]_\Z -1}
\end{pmatrix}\]
where $\bm X_l$ is the $l\times l$ \enquote{X}-shaped 0/1 matrix \eiriknew{for $l \geq 2$ and \eirikfinal{$\bm X_1 = (2)$}} (details omitted). Now, $\tilde{\bm E}_{p-2}$ is easily verified to have rank $\frac12(p-1)$, such that the total rank of $\bm M_F$ is $(d-1)(p-1) + \frac12(p-1) = d(p-1) - 1 - \frac12(p-3)$, which concludes the proof.
\begin{remark}\label{rem:numerical-check-facet}
  The above proof implies a simple numerical procedure by which a specific valid symmetric basic building block class $\mathcal T^{\bm m}$ can be verified to be facet-defining: find $\frac12(p-3)$ \emph{additional} nonzero vectors $\bm \xi \in \F_p^3$ with $\sum_{i=1}^3 \xi_i = 0$ that satisfy $t_{\sigma,\xi_1} + t_{\sigma,\xi_2} + t_{0,\xi_3} = 0$ such that their (Flanagan) embeddings, together with the lower right $3(p-1)-1 \times 3(p-1)$ part of $\bm M_F$ above, are linearly independent and thus complete $\bm M_F$ to a matrix of rank $d(p-1)-1$.
  
  While passing this test is only a sufficient condition for a valid class being facet-defining (in theory, there could be classes that are facet-defining only for some $d \geq d_0 > 3$), we conjecture it to be necessary as well, as we did not find any counter-example in numerical experiments.
\end{remark}

\section{Numerical Procedure to Verify the \enquote{Only If}-Part of Conjecture~\ref{conj:facetsymmetric}} \label{app:only-if-part}

\eiriknewnew{In this appendix, for ease of notation, codewords are represented as row vectors as opposed to column vectors in the rest of the paper.} The procedure is based on the following key lemma.

\begin{lemma} \label{lem:d0}
Let $\mathcal{T}^{\bm m}$ be a valid basic building block class and $\boldtheta^T\boldx \leq \kappa$ any inequality from $\Theta^{\bm m}$, where \eirik{$\bm c^{\boldtheta}_d$} is the canonical codeword according to Construction~\ref{constr:hilo}, for a given length $d$ of the SPC code. Denote by $\bm M_F^d$ the matrix whose rows consist of \eirik{$\Fv'(\eiriknewnew{(\bm c - \bm c_d^{\boldtheta})^T})^T$}, where \eirik{$\bm c \neq \eirik{\bm c_d^{\boldtheta}}$} runs over all codewords for which the inequality is tight (as outlined in the proof of Lemma~\ref{lem:facets} in Appendix~\ref{app:prooffacets}; see \eqref{eq:MF}). Note that $\bm M_F^d$ is independent of the chosen inequality. Let $d_0$ be the smallest $d \geq 3$ such that the support of all codewords of $\bm M_F^{d_0}$ (before being embedded) is at most $d_0-2$. Then,
\begin{displaymath}
\rank\left(\bm M_F^{d+1}\right) \leq \rank\left(\bm M_F^{d}\right) + p-1
\end{displaymath}
for all $d \geq d_0$.
\end{lemma}

\begin{IEEEproof}
First note that for any valid basic building block class $\mathcal{T}^{\bm m}$, $d_0$ is finite, since $t_{0,j} \geq 0$ (see Definition~\ref{def:bb}) for all $j \in \F_p$ and $t_{\sigma,j} < 0$ for $j \in \F_p \setminus \{0 \}$ (follows from Property~\ref{part:bb-property3} of Lemma~\ref{lem:bb-properties}), and then the result follows from Lemma~\ref{lem:validIndependent} (see (\ref{eq:tightness})). Furthermore, for Construction~\ref{constr:hilo}, we choose the first coordinate as the constrained coordinate (it can be arbitrarily chosen).

Now, let $\bm M^{d_0}$ be the corresponding matrix before the embedding. 
It follows that
\begin{displaymath}
\bm M^{d_0+1} = \begin{pmatrix} \bm M^{d_0} & \bm {0} \\
\bm A & \bm b
\end{pmatrix}
\end{displaymath}
where $\bm 0$ is an all-zero column vector over $\F_p$, $\bm b$ is a column vector in which  all entries are different from $[0]_p$, and $\bm A$ is a matrix of partial codewords (one partial codeword per row) such that the concantenation $(\bm A \mid  \bm b)$ contains all codewords \eirik{$\bm c - \bm c^{\boldtheta}_{d_0+1}$, $\bm c \neq \bm c^{\boldtheta}_{d_0+1}$, with a nonzero final entry and for which $\bm c$ is tight for the inequality} in $\Theta^{\bm m}$ for $d = d_0+1$  \eirik{with canonical codeword $\bm c^{\boldtheta}_{d_0+1}$}. \eirik{Note that the canonical codeword $\bm c^{\boldtheta}_{d_0+1}$ is arbitrary and that $\bm M^{d_0+1}$ does not depend on the  \eiriknew{specific} corresponding inequality in $\Theta^{\bm m}$.} Now, suppose that there exist two \eiriknew{distinct} codewords (rows) in $(\bm A \mid  \bm b)$ in which the final entry is the same, i.e., the two codewords are of the form \eirik{$\bm c_1 - \bm c^{\boldtheta}_{d_0+1} = (\bm z_1,u,v,0,0,x)$} and \eirik{$\bm c_2 - \bm c^{\boldtheta}_{d_0+1} = (\bm z_2,0,0,w,y,x)$}, respectively, where $x \in \F_p \setminus \{0\}$, $u,v,w,y \in \F_p$, and $\bm z_1$ and $\bm z_2$ are vectors of length $d_0-4$ over $\F_p$. Here, without loss of generality, we have assumed, \eirik{disregarding the last coordinate with value $x \in \F_p \setminus \{0\}$}, that the support of the first codeword is contained within the first $d_0-2$ coordinates, while the support of the second codeword is contained within the first $d_0-4$ coordinates, as well as within the coordinates $d_0-1$ and $d_0$. This assumes $d_0 \geq 4$. If $d_0=3$, then the two codewords would overlap with a zero entry in at least one coordinate. However, \eirikfinal{a similar argument to the one below} can be repeated to prove the lemma in this special case as well. Now, since by assumption both codewords \eirik{$\bm c_1$ and $\bm c_2$}  are tight for the \eirik{given inequality in} $\Theta^{\bm m}$ for $d=d_0+1$, so are \eirik{$(\bm z_1,u,v,0,x,0) + \bm c^{\boldtheta}_{d_0+1}$} and \eirik{$(\bm z_2,x,0,w,y,0)+\bm c^{\boldtheta}_{d_0+1}$}, %
\eirik{since in general if $\bm c + \bm c^{\boldtheta}_{d_0+1}$, $\bm c \neq \bm c^{\boldtheta}_{d_0+1}$, is tight  (for the given inequality), where $\bm c$ denotes a codeword, then $\pi(\bm c) + \bm c^{\boldtheta}_{d_0+1}$,  $\bm \pi(\bm c) \neq \bm c^{\boldtheta}_{d_0+1}$, is also tight, where $\pi(\cdot)$ denotes an arbitrary coordinate permutation not involving the constrained coordinate of Construction~\ref{constr:hilo}  (see (\ref{eq:tightness}) of Lemma~\ref{lem:validIndependent}).} %
Consider the codewords
 $(\bm z_2,x,y,w,0,0)$ and $(\bm z_2,0,y,w,x,0)$. Again, they are permutations (not involving the constrained coordinate of Construction~\ref{constr:hilo}) of $(\bm z_2,x,0,w,y,0)$,  and thus \eirik{$(\bm z_2,x,y,w,0,0)+\bm c^{\boldtheta}_{d_0+1}$ and $(\bm z_2,0,y,w,x,0)+\bm c^{\boldtheta}_{d_0+1}$ are} tight for the \eirik{given inequality} in $\Theta^{\bm m}$ for $d=d_0+1$. %
 Taking the real linear combination
 \ifonecolumn
 \begin{align*}
 &\Fv'\left((\bm z_1,u,v,0,x,0)\eiriknewnew{^T} \right) - \Fv'\left((\bm z_2,x,0,w,y,0)\eiriknewnew{^T} \right)  
 + \Fv'\left((\bm z_2,x,y,w,0,0)\eiriknewnew{^T} \right) - \Fv'\left((\bm z_2,0,y,w,x,0)\eiriknewnew{^T} \right) \\
 &= \Fv'\left((\bm z_1,u,v,0,0,x)\eiriknewnew{^T} \right) - \Fv'\left((\bm z_2,0,0,w,y,x)\eiriknewnew{^T} \right)
\end{align*}
\else
 \begin{align*}
 &\Fv'\left((\bm z_1,u,v,0,x,0)\eiriknewnew{^T} \right) - \Fv'\left((\bm z_2,x,0,w,y,0)\eiriknewnew{^T}\right)  \\
 &+ \Fv'\left((\bm z_2,x,y,w,0,0)\eiriknewnew{^T}\right) - \Fv'\left((\bm z_2,0,y,w,x,0)\eiriknewnew{^T}\right) \\
 &= \Fv'\left((\bm z_1,u,v,0,0,x)\eiriknewnew{^T}\right) - \Fv'\left((\bm z_2,0,0,w,y,x)\eiriknewnew{^T}\right)
\end{align*}
\fi
shows that $\Fv'((\bm z_1,u,v,0,0,x)\eiriknewnew{^T})\eiriknew{^T}$ can be written as a real linear combination of four rows from \eirikNew{$(\bm M_F^{d_0} \mid \bm 0)$} and the row $\Fv'((\bm z_2,0,0,w,y,x)\eiriknewnew{^T})\eiriknew{^T}$. \eiriknewnew{Moreover, if the second codeword $\bm c_2 - \bm c^{\boldtheta}_{d_0+1}$ overlaps with a single zero entry with $\bm c_1 - \bm c^{\boldtheta}_{d_0+1}$, i.e.,  $\bm c_2 - \bm c^{\boldtheta}_{d_0+1}= (\bm z_2,w,0,0,y,x)$, then it can be shown in a similar manner that $\Fv'((\bm z_1,u,v,0,0,x)^T)^T$ can be written as a real linear combination of two rows from $(\bm M_F^{d_0} \mid \bm 0)$ and $\Fv'((\bm z_2,w,0,0,y,x)^T)^T$.} Thus, the set of rows of $(\bm A \mid \bm b)$ can at most increase the rank by $p-1$, and the result of the lemma follows for $d=d_0$. The result for $d > d_0$ follows by induction.
\end{IEEEproof}

The procedure works in the following way and is repeated for each valid basic building block class $\mathcal{T}^{\bm m}$ (indexed by $\bm m$) for a given prime $p$. Choose $d=3$ and build the matrix $\bm M_F^d$. If its rank is less than \eirik{$d(p-1)-1$}, then we know that the class cannot be facet-defining for $d=3$. If this is not the case, stop. Otherwise, compute the reduced row echelon form of $\bm M_F^d$ and remove the all-zero rows. The resulting matrix is denoted by $\bm M_{F, \rm red}^d$. Now, construct the matrix 
\begin{equation} \label{eq:MF_d}
\begin{pmatrix} \bm M_{F, \rm red}^{d} & \bm 0  \\
\Fv'(\eirik{\bm a\eiriknewnew{^T_1}})^T & \f'(1) \\
\vdots & \vdots \\
\Fv'(\eirik{\bm a\eiriknewnew{^T_{p-1}}})^T & \f'(p-1) \\
\Fm'(\eirik{\bm A}\eiriknewnew{^T})^T & \Fm'(\eirik{\bm b}\eiriknewnew{^T})^T 
\end{pmatrix}
\end{equation}
where the matrix-embedding $\Fm'$ is analog to the vector-embedding $\Fv'$, applying the vector-embedding on each \eirik{column of} the matrix individually. In (\ref{eq:MF_d}), $(\bm a_i \mid i)$, \eirik{$i \in \F_p \setminus \{0\}$}, is any (if any exist) codeword for length $d+1$ with at least two zero entries in $\bm a_i$ \eirik{such that  $(\bm a_i \mid i)+\bm c^{\boldtheta}_{d+1}$ is tight} \eiriknew{(for the inequality in $\Theta^{\bm m}$ for length $d+1$ with canonical codeword $\bm c^{\boldtheta}_{d+1}$)},  and each row of the matrix $(\bm A \mid \bm b)$ is a codeword  for length $d+1$ with a nonzero final entry and at most one zero entry among the first $d$ coordinates \eirik{such that its sum with $\bm c^{\boldtheta}_{d+1}$ is tight.} Then, it follows from the proof of Lemma~\ref{lem:d0} that the rank of $\bm M_F^{d+1}$ is equal to the rank of the matrix in (\ref{eq:MF_d}), since any additional codeword for length $d+1$ with at least two zero entries in the coordinates $1$ through $d$ \eirik{and such that its sum with $\bm c^{\boldtheta}_{d+1}$ is tight} can be written as a real linear combination of the rows of $(\bm M_{F, \rm red}^{d} \mid \bm 0)$ \eiriknew{and the rows $(\Fv'(\bm a_1^T)^T \mid  \f'(1)),\dotsc,  (\Fv'(\bm a^T_{p-1})^T \mid \f'(p-1))$}.  
The reduced row echelon form of $\bm M_F^{d+1}$ (after removing the all-zero rows) is \eirik{equivalent} to the reduced row echelon form (again after removing the all-zero rows) of the matrix in (\ref{eq:MF_d}), and the procedure can be repeated for the next value of $d$ if the rank of the matrix in (\ref{eq:MF_d}) is strictly smaller than \eirik{$d(p-1)-1$}. Otherwise, the class would be facet-defining for this particular value of $d$. The procedure is repeated until $d=d_0$, unless the rank at some point reaches \eirik{$d(p-1)-1$}. Now, if $\rank \left( \bm M_F^{d} \right) < \eirik{d(p-1)-1}$ for all $3 \leq d \leq d_0$ (and the procedure does not stop until $d=d_0$), it follows from Lemma~\ref{lem:d0} that the class cannot be facet-defining for any $d \geq 3$.

The complexity of the approach depends on the value of $d_0$, which again depends on the building block class, and in particular on the number of rows of the matrix $(\Fm'(\eirik{\bm A^T})^T \mid \Fm'(\eirik{\bm b^T})^T)$ for each value of $d$. The size of the matrix $(\Fm'(\eirik{\bm A^T})^T \mid \Fm'(\eirik{\bm b^T})^T)$ (for a given $d$) can be determined from the principle behind the counting formula in Lemma~\ref{lem:counting_formulas}. Note that for some classes,  $d_0$ is as large as $p+2$. To address the complexity of the approach, as an example, in the worst case for $p=19$ (when $d_0=p+2=21$), the number of rows of $(\Fm'(\eirik{\bm A^T})^T \mid \Fm'(\eirik{\bm b^T})^T)$ for $d=3,\dotsc,20$ is
\ifonecolumn
\begin{displaymath}
\begin{split}
(&908, 4464, 17400, 53886, 131826, 254652, 390302, 
 477511,468383, 368507, 231336, 114444, 43656, 12393, 
 2466, \\
 &307, 18, 0).
\end{split}
\end{displaymath}
\else
\begin{displaymath}
\begin{split}
(&908, 4464, 17400, 53886, 131826, 254652, 390302, \\
 &477511,468383, 368507, 231336, 114444, 43656, 12393, \\
 &2466, 307, 18, 0).
\end{split}
\end{displaymath}
\fi

\section{Proof of Proposition~\ref{prop:facets_modified}}\label{app:prooffacets_modified}
We use Remark~\ref{rem:numerical-check-facet} to prove the proposition. The following lemma yields the required $\bm\xi$-vectors.

\begin{lemma} \label{lem:bb-doubly_symmetric_properties}
Let $\mathcal T^{\bm m}$ be almost doubly-symmetric and denote by $\sigma_1$ the index of the second-largest entry in $\bm t_0$, \ie $t_{0,\sigma_1} = \smax(\bm t_0)$. Then, there are $(p-3)/2$ different elements $c_1,\dotsc,c_{(p-3)/2} \in \F_p$ such that, for $k \in \range{(p-3)/2}$,  
\begin{equation} \label{eq:double_sym_condition}
t_{\sigma, -c_k} + t_{\sigma, c_k-\sigma_1} + t_{0,\sigma_1} = 0.
\end{equation}
Furthermore, $c_k \neq \sigma_1-c_l$ for all $k,l \in \range{(p-3)/2}$ and all $c_k, \sigma_1-c_l \notin \{0, \sigma_1, \sigma\}$.
\end{lemma}

\begin{IEEEproof}
Choose $i \in \tilde T_0^\mathrm{proj}$ as in \eqref{eq:condition_dsymmetrix_3}. By there is a $c \in \F_p$ such that $i = t_{0,c} = t_{0,\sigma} - t_{0,\sigma - c}$ (cf. \eqref{eq:symmetry-alt} for the second equality).
By \eqref{eq:condition_dsymmetrix_3}, $t_{0,\sigma_1} - t_{0,c} \in T_0^\mathrm{proj}$, so that again by \eqref{eq:symmetry-alt} \eiriknew{and Definition~\ref{def:bb}} we conclude that $t_{0,\sigma_1} - t_{0,c} = t_{0,\sigma} - t_{0,\sigma-\sigma_1+c}$.
Inserting the above expression for $t_{0,c}$ and using Definition~\ref{def:bb} results in
\[ t_{0,\sigma_1} +  t_{\sigma,- c} = -t_{\sigma,c-\sigma_1}\]
which shows \eqref{eq:double_sym_condition} for this particular $c$.
Now by \eqref{eq:condition_dsymmetrix_2} all $t_{0,c_k} \in \tilde T_0^{\rm proj}$, while all $t_{0,\sigma_1-c} \in T_0^{\rm proj} \setminus \tilde T_0^{\rm proj}$, which shows the remaining claims.
\end{IEEEproof}
Let now $c_1,\dotsc, c_{(p-3)/2}$ denote the $(p-3)/2$ $c$-values obtained from the above lemma, and define, for $k \in \range{(p-3)/2}$,
\[\bm\xi_k = (0,\dotsc,0,-c_k, c_k - \sigma_1, \sigma_1)\eiriknewnew{^T}\]
which fulfills the conditions of Remark~\ref{rem:numerical-check-facet} by Lemma~\ref{lem:bb-doubly_symmetric_properties}. The (Flanagan) embeddings of the $\bm \xi_k$ lead to a $(p-3)/2 \times 3(p-1)$ matrix of the form
\[\begin{pmatrix} \bm G_{\eiriknew{(p-3)/2}} &\bar{\bm G}_{\eiriknew{(p-3)/2}} &\bm E'_{(p-3)/2} \end{pmatrix}\]
where $\bm G_{\eiriknew{(p-3)/2}}$ \eiriknew{(of dimensions $(p-3)/2 \times (p-1)$)} contains a permutation matrix in the columns $[\eiriknew{-}c_1]_\Z,\dotsc, [\eiriknew{-}c_{(p-3)/2}]_\Z$ and zeros \eirik{elsewhere}, $\bar{\bm G}_{\eiriknew{(p-3)/2}}$ \eiriknew{(of dimensions $(p-3)/2 \times (p-1)$)} is a permutation matrix in columns $[\eiriknew{c_1-\sigma_1}]_\Z, \dotsc, [\eiriknew{c_{(p-3)/2}-\sigma_1}]_\Z$ with zeros \eiriknew{elsewhere}, and $\bm E'_{(p-3)/2}$  \eiriknew{(of dimensions $(p-3)/2 \times (p-1)$)} has $1$-entries in the $[\sigma_1]_\Z$-th column and zeros \eirik{elsewhere}. Now, as in Appendix~E both $\bm G_{\eiriknew{(p-3)/2}}$ and $\bar{\bm G}_{\eiriknew{(p-3)/2}}$ can trivially be eliminated, and one can readily show that this operation turns $\bm E'_{(p-3)/2}$ into a matrix $\tilde{\bm E}'_{(p-3)/2}$ that has zeros in the $[\eiriknew{\sigma}]_\Z$-th column, ones in the $[\eiriknew{\sigma_1}]_\Z$-th column, and two negated permutation matrices in the remaining columns such that the conditions of Remark~\ref{rem:numerical-check-facet} are fulfilled, which completes the proof.

\section{Proof of Proposition~\ref{prop:equivalent}}\label{app:equivalent}
  
  For $p=2$, there are only two possible $\bm m$-vectors $(0,0)$ and $(0,1)$, so that the first part of the claim is trivial. Because $\mathcal T^{(0,0)} = \{ (0,1), (0,-1) \}$ and $\mathcal T^{(0,1)} = \{(0,3), (0, -3)\}$, it is obvious that $\Theta^{(0,1)}$ is obtained by multiplying each inequality in $\Theta^{(0,0)}$ by $3$, which concludes the proof for $p=2$. Thus, assume $p>2$ in the following.
  
  By Construction~\ref{constr:hilo}, $\boldtheta_1 = \rot_\varphi(\boldt^{\bm m}_{k_1})$ and $\boldtheta'_1 = \rot_{\varphi'}(\boldt^{\bm m'}_{k'_1})$, \ie the first $p$-block of each inequality is a rotated building block of $\mathcal T^{\bm m}$  and $\mathcal T^{\bm m'}$, respectively.  Hence the assumption implies
  $\rot_\varphi(\boldt^{\bm m}_{k_1}) = a \rot_{\varphi'}(\boldt^{\bm m'}_{k'_1})$, and thus
  \[ \set(\boldt^{\bm m}_{k_1}) = a\cdot\set(\boldt^{\bm m'}_{k'_1}) \Rightarrow \set(\boldt^{\bm m}_0) = a\cdot \set(\boldt^{\bm m'}_0) - at^{\bm m'}_{0,k'_1} + t^{\bm m}_{0,k_1}\]
  by \eqref{eq:property-3}. But $\min(\bm t_0^{\bm m}) = 0 = \min(\bm t_0^{\bm m'})$, so that $t_{0,k_1}^{\bm m} - a t_{0,k'_1}^{\bm m'}=0$ must hold, \ie 
  \begin{equation}
    \set(\bm t_0^{\bm m}) = a\cdot\set(\bm t_0^{\bm m'}).
    \label{eq:W0aW0}
  \end{equation}
  As both sets in \eqref{eq:W0aW0} contain integer entries only, $a$ must be rational, i.e., $a=r/s$ where $r,s>0$ ($a=0$ would imply $\set(\bm t_0^{\bm m}) = \{0\}$, contradicting Definition~\ref{def:bb}) with $\gcd(r,s) = 1$. Hence, $s$ divides all $p$ distinct nonnegative entries of $\bm t_0^{\bm m'}$, thus
    $\max(\bm t^{\bm m'}_0) \geq s(p-1)$,
  while by Definition~\ref{def:bb}, $\max(\bm t^{\bm m'}_0) \leq 2p-1$.
  Together, this implies $s \leq 2 + \frac1{p-1}$. As $s$ is integer and we assume $p>2$, this means $s \in \{1,2\}$.\\
  \emph{Case $s=1$:} Because $0 < a = r/s \leq 1$ by assumption, $s=1$ implies $a=1$, which by \eqref{eq:W0aW0} implies $\set(\bm t^{\bm m}_0) = \set(\bm t^{\bm m'}_0)$, contradicting the assumption that $\bm m \neq \bm m'$.\\
  \emph{Case $s=2$:} Because $s$ divides all elements of $\set(\bm t^{\bm m'}_0)$, in this case $t^{\bm m'}_{0,\zeta}$ is even for all $\zeta \in \F_p$. But $t^{\bm m'}_{0,\zeta} = [\zeta]_\Z + m'_\zeta p$ by definition, which implies (because $p$ is odd) that $m'_\zeta = 1$ if and only if $[\zeta]_\Z$ is odd, \ie $\bm m'=(0,1,0,1,\dotsc)$ as claimed.
  
  To see that $\bm m = (0,\dotsc,0)$, note that $s=2$ and $0 < r/s \leq 1$ implies $a = r/s = 1/2$. From \eqref{eq:W0aW0} follows that
  \[\max(\bm t_0^{\bm m}) = \frac12 \max(\bm t^{\bm m'}_0) = \frac12(2p-2) = p-1\]
  by the structure of $\bm m'$, which implies that $\bm m = (0,\dotsc,0)$.
  
  For the proof of the second claim, we use the following lemma.
  \begin{lemma}\label{lem:2tkm}
  For $k \in \{0,\dotsc,p-1\}$ and $\bm m$, $\bm m'$ as above, $2\boldt^{\bm m}_{k} = \rot_2(\boldt^{\bm m'}_{\varphi_2(k)})$.    
  \end{lemma}
  \begin{IEEEproof}
    By definition of $\bm m$, we have
    \begin{equation}
    (2\boldt^{\bm m}_k)_j = 2 t^{\bm m}_{k,j} = 2(t^{\bm m}_{0,k+j} - t^{\bm m}_{0,k}) = 2 ( [k+j]_\Z - [k]_\Z).
    \label{eq:2tkm}
    \end{equation}
    On the other hand,
    \[(\rot_2(\boldt^{\bm m'}_{\varphi_2(k)}))_j = t^{\bm m'}_{2k, 2j} = t^{\bm m'}_{0,2(k+j)} - t^{\bm m'}_{0,2k}.\]
    Now, the structure of $\bm m'$ implies that, for $i\in \F_p$, $t^{\bm m'}_{0,2i} = 2[i]_\Z$. It follows that
    \[t^{\bm m'}_{0,2(k+j)} - t^{\bm m'}_{0,2k} = 2 [k+j]_\Z - 2[k]_\Z\]
    which equals \eqref{eq:2tkm}.
  \end{IEEEproof}

  Now, let $\rot_\varphi(\boldtheta)^T \boldx \leq \kappa$ be in $\varphi(\Theta^{\bm m})$, then by Corollary~\ref{cor:rotation-autfq} also $\boldtheta^T \boldx \leq \kappa$ is in $\Theta^{\bm m}$. Let $\boldtheta = (\boldt^{\bm m}_{k_1} \mid \dotsc \mid \boldt^{\bm m}_{k_d})\eiriknew{^T}$. We now show that $2\boldtheta^T \boldx \leq 2\kappa$ is in $\varphi_2(\Theta^{\bm m'})$.
  
  By Lemma~\ref{lem:2tkm},
  \[2\boldtheta = (\rot_2(\boldt^{\bm m'}_{\varphi_2(k_1)}) \mid \dotsc \mid \rot_2(\boldt^{\bm m'}_{\varphi_2(k_d)}))\eiriknew{^T}.\]
  Using Corollary~\ref{cor:rotation-autfq} again, the claim reduces to showing that $\boldtheta'^T\boldx \leq \kappa$ with $\boldtheta' = (\boldt^{\bm m'}_{\varphi_2(k_1)} \mid \dotsc \mid \boldt^{\bm m'}_{\varphi_2(k_d)})\eiriknew{^T}$ and $\kappa' = 2\kappa$ is contained in $\Theta^{\bm m'}$. We show the latter by verifying \eqref{eq:hilo-conditions} from Lemma~\ref{lem:conditions}; namely, \eqref{eq:Thetacondition} holds because
  \begin{align*}
  \sum_{i=1}^d \varphi_2(k_i) &= [2]_p \sum_{i=1}^d k_i &&\text{(by def.\ of }\varphi_2\text{)}\\
  &= [2]_p [d-1]_p\sigma^{\bm m} &&\text{(\eqref{eq:Thetacondition} for $\boldtheta$)}\\
  &= [d-1]_p\sigma^{\bm m'}&&
  \end{align*}
  where the last step follows because $\sigma^{\bm m} = [p-1]_p = [-1]_p$ and $\sigma^{\bm m'} = [p-2]_p = [-2]_p$, whereas for \eqref{eq:kappacondition} we compute
  \begin{align*}
    \kappa' = 2\kappa &= 2(d-1)t^{\bm m}_{0,\sigma^{\bm m}} - 2\sum_{i=1}^d \eirik{t^{\bm m}_{0,k_i}} &&\text{(\eqref{eq:kappacondition} for $\kappa$)}\\
    &= 2(d-1)(p-1) - \sum_{i=1}^d 2[k_i]_\Z&&\text{(as $\bm m = \bm 0$)} \\
    &= (d-1)t^{\bm m'}_{0,\sigma^{\bm m'}} - \sum_{i=1}^d t^{\bm m'}_{0,\varphi_2(k_i)}&&\text{(by Lemma~\ref{lem:2tkm})}
  \end{align*}
  which proves the claim. Hence, for every inequality in $\Phi(\Theta^{\bm m})$ there is an equivalent inequality in $\Phi(\Theta^{\bm m'})$. Furthermore, by Remark~\ref{rem:phiThetaunique} and because both sets are of the same size, the converse must also hold, which concludes the proof.
  
\section{Proof of Theorem~\ref{thm:facetComplete}}\label{app:proofFacetComplete3}
We will need a technical lemma for this proof.
\begin{lemma}\label{lem:bracketBijection}
  Let a finite field $\F_q$, $d\geq1$, and $\boldxi =( \xi_1,\dotsc,\xi_d)\eiriknewnew{^T} \in \F_q^d$ be given. Then, there is a permutation matrix $\bm P_\boldxi \in \{0,1\}^{dq \times dq}$ with the property that
  \begin{equation} \bm P_\boldxi \Fv(\bm \eta) = \Fv(\bm \eta - \bm\xi)\label{eq:Pxi-condition}\end{equation}
  for any $\bm\eta \in \F_q^d$. In particular, the linear map defined by $\bm P_\boldxi$ is bijective and maps $S^d_{q-1}$ onto itself.
\end{lemma}
\begin{IEEEproof}
  Define $\bm P_\boldxi$ to be the block-diagonal matrix
  \[\bm P_\boldxi = \begin{pmatrix}
    \bm P_{\xi_1} \\
    & \ddots &\\
    && \bm P_{\xi_d}\end{pmatrix}\]
where, for $\xi_i \in \F_q$, $\bm P_{\xi_i}$ is the $q\times q$ permutation matrix defined by
\[\bm P_{\xi_i} \eirik{\f(\eta)^T} = \eirik{\f(\eta - \xi_i)^T}\]
for $\eta \in \F_q$. Note that the above equation defines the image of $\bm P_{\xi_i}$ for all unit vectors in $\R^q$, and the right-hand side runs over all unit vectors as well, such that $\bm P_{\xi_i}$ and hence also $\bm P_\boldxi$ is a permutation matrix.
\end{IEEEproof}

\begin{example}
Let $d=1$ and $q = 3$, and $\xi_1 = 2$. Then, 
\[
\bm P_{\xi_1} =
\begin{pmatrix}
  0&0&1\\
  1&0&0\\
  0&1&0
\end{pmatrix}
\]
and it is easily verified that $\bm P_{\xi_1}\f(\eta)\eiriknew{^T} = \f(\eta - 2)\eiriknew{^T}$ holds.
\end{example}

Let now $\mathcal Q$ denote the polytope defined by the inequalities in $\Theta$. From Proposition~\ref{prop:Theta1Facets} follows that $\Pd \subseteq \mathcal Q$, so it remains to show that $\mathcal Q \subseteq \Pd$. Suppose not. Then there exists $\boldf \in \mathcal Q \setminus \P$. Since the facets in $\Delta_3^d$ by definition describe $S_2^d$  we have $\mathcal Q \subseteq S_2^d$ and hence $\boldf \in S_2^d$. Since $\boldf \notin \P$, there must be a facet $F$ of $\Pd$ that cuts $\boldf$ and hence also some vertex $\bm v$ of $S_2^d$. We will show that $F$ is contained in either $\Theta^{\bm m}$ or $\varphi_2(\Theta^{\bm m})$, which contradicts the assumption that $\boldf \in \mathcal Q$ and hence shows $\Pd = \mathcal Q$.

Note that on the vertices of $S_2^d$ the inverse of $\Fv$, $\Fv^{-1}\colon \R^{2d} \rightarrow \F_3^d$ is defined. Let $\boldxi = \Fv^{-1}(\bm v) \in \F_3^d$. Then, $\boldxi \notin \C$. Assume $\sum_{i=1}^d \xi_i = [2]_3$; the case $\sum_{i=1}^d \xi_i = [1]_3$ is completely symmetric. We will prove that $F \in \Theta^{\bm m} \cup \varphi_2(\Theta^{\bm m})$ by going through several cases, distinguished by the Hamming distance between $\boldxi$ and the codewords for which $F$ is tight. For each case, we will derive a set of vertices of $\Pd$ for which $F$ could potentially be tight; then, we will show that there is an inequality in $\Theta^{\bm m} \cup \varphi_2(\Theta^{\bm m})$ that is tight for \emph{all} of those vertices, hence that inequality must define $F$.

As $F$ is a face of $\P$, there are $\bolda' \in \R^{2d}$ and $b \in \R$ such that
\begin{equation}
\bolda'^T \bm x \geq b
\label{eq:p3proof-facetInequality}
\end{equation}
is valid for $\P$ and induces $F$, i.e., $F = \{\bm x \in \Pd\colon \bolda'^T \bm x = b \}$.
We use the permutation matrix of Lemma~\ref{lem:bracketBijection} to state \eqref{eq:p3proof-facetInequality} in a more convenient form, where we use the short-hand notation $\langle\bm x \rangle_{\boldxi} = \bm P_\boldxi \bm x$ for any $\bm x \in\R^{dq}$.

Define $\bm a = \bm P_{\bm\xi}^{-T} \bm a'$. Then,
\ifonecolumn
\begin{equation}
\bm a^T \langle\bm x\rangle_\boldxi \geq b 
\Leftrightarrow 
\left(\bm P_{\bm\xi}^{-T} \bm a'\right)^T \bm P_\boldxi \bm x = \bm a'^T \bm P_\boldxi^{-1} \bm P_\boldxi \bm x = \bm a'^T \bm x \geq b
\label{eq:p3proof-ineq}
\end{equation}
\else
\begin{multline}
\bm a^T \langle\bm x\rangle_\boldxi \geq b 
\Leftrightarrow \\
\left(\bm P_{\bm\xi}^{-T} \bm a'\right)^T \bm P_\boldxi \bm x = \bm a'^T \bm P_\boldxi^{-1} \bm P_\boldxi \bm x = \bm a'^T \bm x \geq b
\label{eq:p3proof-ineq}
\end{multline}
\fi
\ie \eqref{eq:p3proof-facetInequality} can be restated in terms of $\langle \bm x\rangle_\boldxi$ instead of $\bm x$. The particular form of $\bm P_\boldxi$ further allows us to assume that $a_{i,0} = 0$ for $i \in \range d$, by adding appropriate multiples of \eqref{eq:spx-eq} to the inequality (cf.\ Appendix~\ref{app:embeddings}).
\begin{remark}
  The construction of $\langle\cdot\rangle_{\boldxi}$ and the way it is used above generalizes the definition and usage of the map $[\cdot]$ in the proof of \cite[Thm.~5.15]{fel03}.
\end{remark}

As $\bm v$ is cut by $F$, we have $b > \bolda^T\langle\bm v\rangle_\boldxi = \bolda^T\langle\Fv(\boldxi)\rangle_\boldxi = \bolda^T\Fv(\bm 0) = \sum_{i=1}^d a_{i,0} = 0$ by \eqref{eq:Pxi-condition} and the above assumption. 
Let $\bm e^i$ denote the $i$-th unit \eiriknew{(column)} vector in $\F_3^d$, \ie $\bm e^i_i = [1]_3$ and $\bm e^i_j = [0]_3$ for $j \neq i$. Since $\sum_{i=1}^d \xi_i = [2]_3$, $\boldxi + \bm e^i \in \C$ for $i\in \range d$. Because \eqref{eq:p3proof-ineq} is valid for $\Pd$, $b \leq \bolda^T\langle\Fv(\boldxi+\bm e^i)\rangle_\boldxi = \bolda^T \Fv(\bm e^i) = a_{i,1}$. Likewise $\boldxi - \bm e^i - \bm e^j \in \C$ for $i \neq j$, hence
 \[ b \leq \bolda^T\langle\Fv(\boldxi-\bm e^i  -\bm e^j)\rangle_\boldxi = \bolda^T\Fv(2\bm e^i + 2 \bm e^j) =  a_{i,2} + a_{j,2}\]
and finally from $\boldxi + \bm e^i + \bm e^j - \bm e^k \in \C$ for pairwise different $i,j,k \in \range d$ we conclude analogously that $a_{i,1} + a_{j,1} + a_{k,2} \geq b$. To sum up, for arbitrary but different $i,j,k \in \range d$ holds:
\begin{align}
  b &> 0, \notag \\
  a_{i,1} &\geq b, \label{eq:a1grb}\\
  a_{i,2} + a_{j,2} &\geq b, \label{eq:a2grb}\\
  a_{i,1} + a_{j,1} + a_{k,2} &\geq b. \label{eq:triplegrb}
\end{align}
Furthermore, \eqref{eq:a2grb} implies that $a_{i,2} < b/2$ can hold for at most one $i \in \range d$, which allows the stronger statement that, for $I \subseteq \range d$ with $|I| \geq 2$,
\begin{equation}
  \sum_{i \in I} a_{i,2} \geq |I|\cdot b/2.\label{eq:aIgrb}
\end{equation}
\begin{lemma}\label{lem:p3proof-dh3}
  $F$ contains $\Fv(\boldc)$ for at least one codeword $\boldc$ with $d_{\rm H}(\boldc, \boldxi) \geq 3$.
\end{lemma}
\begin{IEEEproof}
  Assume the contrary. As $F$ is a facet of the (by Proposition~\ref{prop:Pjdim}) $2d$-dimensional polytope $\P$, there must be a set $\mathcal F$ of $2d$ codewords of $\C$ such that $\{\Fv(\bm c)\colon \bm c \in \mathcal F\} \subseteq F$, this set is affinely independent, and by assumption $d_{\rm H}(\boldxi, \boldc) \leq 2$ for $\boldc \in \mathcal F$. We now show that $\Theta^{\bm m}$ contains an inequality $\bm \theta^T \bm x \leq \kappa$ that induces $F$.
  
  Let $\boldc^{\boldxi} = \boldxi + \bm e^d \in \C$. Choose $\boldtheta = (\boldt^{\bm m}_{k_1} \mid \dotsc \mid  \boldt^{\bm m}_{k_d})^T$ according to Construction~\ref{constr:hilo} using $\boldc^{\boldxi}$ as canonical codeword. Using $\sigma = 2$ and  Lemma~\ref{lem:hilo-formulas}, this implies $k_i = t_{\uparrow, c^{\bm\xi}_i} = 2 - c^{\bm\xi}_i =2- \xi_i$ for $i \in \range{d-1}$, while also $k_d = t_{\downarrow,c^{\bm\xi}_d} = -(\xi_d + 1) = 2 - \xi_d$. By Proposition~\ref{prop:Theta1Facets}, $\boldtheta^T \bm x \leq \kappa$ with $\kappa = \boldtheta^T \Fv(\boldc^\boldxi)$ defines a facet $G$ of $\Pd$.  We now show that $\boldtheta^T \Fv(\bm c) = \kappa$ for all $\bm c \in \mathcal F$.

  If $\dH(\boldc,\boldxi)=1$, then $\boldc = \boldxi + \bm e^i$ for some $i \in \range d$. By Lemma~\ref{lem:validIndependent},
  \ifonecolumn
  \begin{displaymath}
      \boldtheta^T\Fv(\bm c) - \kappa 
    = \boldtheta^T\Fv(\bm c^{\bm\xi} - \bm e^d + \bm e^i) - \kappa
    = \begin{cases}
      0&\text{if }i=d,\\
      t_{\sigma,1}+t_{0,2}=-2+2=0&\text{otherwise}\end{cases}
  \end{displaymath}
  \else
    \begin{align*}
      \boldtheta^T\Fv(\bm c) - \kappa 
    &= \boldtheta^T\Fv(\bm c^{\bm\xi} - \bm e^d + \bm e^i) - \kappa\\
    &= \begin{cases}
      0&\text{if }i=d,\\
      t_{\sigma,1}+t_{0,2}=-2+2=0&\text{otherwise}\end{cases}
    \end{align*}
  \fi
  where the explicit values of $\bm t_k$ can be looked up, e.g., in Table~\ref{table:values_tk}. If on the other hand $\dH(\bm c, \bm\xi) = 2$, then $\boldc = \boldxi + 2 \bm e^i + 2 \bm e^j$ for $i\neq j \in \range d$. Using Lemma~\ref{lem:validIndependent} again,
  \ifonecolumn
  \begin{displaymath}
        \boldtheta^T\Fv(\bm c) - \kappa 
      = \boldtheta^T\Fv(\bm c^{\bm\xi} - \bm e^d + 2\bm e^i + 2\bm e^j) - \kappa
      =\begin{cases}
        t_{\sigma,2}+t_{0,1} = 0&\text{if }d \in \{i,j\},\\
        t_{\sigma,2}+t_{\sigma,2} + t_{0,2} =0&\text{otherwise}.\end{cases}
    \end{displaymath}
    \else
    \begin{align*}
        \boldtheta^T\Fv(\bm c) - \kappa 
      &= \boldtheta^T\Fv(\bm c^{\bm\xi} - \bm e^d + 2\bm e^i + 2\bm e^j) - \kappa \\
      &= \begin{cases}
        t_{\sigma,2}+t_{0,1} = 0&\text{if }d \in \{i,j\},\\
        t_{\sigma,2}+t_{\sigma,2} + t_{0,2} =0&\text{otherwise}.\end{cases}
    \end{align*}
    \fi
  In conclusion, $\boldtheta^T\boldx \leq \kappa$ is tight for the embeddings of \emph{all} codewords $\boldc$ with $\dH(\boldc,\boldxi) \leq 2$, and in particular for all of $\boldc \in \mathcal F$. Because the latter by assumption lead to $2d$ affinely independent elements of $F$, which has dimension $2d-1$, they already uniquely specify the facet $F$, which hence must equal $G$, \ie $F \in \Theta^{\bm m}$.
\end{IEEEproof}
\begin{lemma}
  Let $\bolddelta \in \F_3^d$. If $d_{\rm H}(\bolddelta, \boldxi) \geq 4$, $F$ does not contain $\Fv(\bolddelta)$, i.e., $\bolda^T\langle\Fv(\bolddelta)\rangle_{\boldxi} > b$. In particular, $F$ does not contain any codeword $\bm c \in \C$ with $\dH(\bm c, \boldxi) > 3$.
\end{lemma}
\begin{IEEEproof}
  Let $\bolddelta \in \F_3^d$ with $d_{\rm H}(\bolddelta, \boldxi) = w \geq 4$. Then there are disjoint index sets $I_1, I_2 \subseteq \range d$, $|I_1| + |I_2| = w$ such that $I_1 = \{i\colon \delta_i - \xi_i = [1]_3\}$ and $I_2 = \{i\colon \delta_i - \xi_i = [2]_3\}$. 
  Assume \eqref{eq:p3proof-ineq} is not strictly satisfied by $\bolddelta$, \ie
    \[b \geq \bolda^T\langle\Fv(\bolddelta)\rangle_{\boldxi} = \bolda^T\Fv(\bolddelta-\boldxi) = \sum_{i \in I_1} a_{i,1} + \sum_{i \in I_2} a_{i,2}.\]
  Now, \eqref{eq:a1grb} implies $|I_2| > 0$, while \eqref{eq:aIgrb} demands $|I_2| < 2$, but \eqref{eq:triplegrb} forbids $|I_2| = 1$. So the assumption must be false.
\end{IEEEproof}

From the above two lemmas we conclude that there exists a $\boldc^3 \in \C$ with  $\dH(\boldc^3,\boldxi) = 3$ and $\Fv(\boldc^3) \in F$. As $\sum_{i=1}^d \xi_i = 2$, this implies that $\boldc^3 = \boldxi + \bm e^i + \bm e^j + 2 \bm e^k$ for pairwise different $i,j,k$. Hence, $b=\bolda^T\langle\Fv(\boldc^3)\rangle_{\boldxi} = a_{i,1} + a_{j,1} + a_{k,2}$, thus $a_{k,2} \leq -b$ by \eqref{eq:a1grb}, so by \eqref{eq:a2grb} $a_{l,2} \geq 2b$ for any $l\in\range{d}$ with $l \neq k$. This means that $a_{k,2}$ is the unique negative entry of $\bolda$, so that all $\boldc' \in \C$ with $d_{\rm H}(\boldc', \boldxi) \in  \{2, 3\}$ and $\bolda^T\langle\Fv(\boldc')\rangle_{\boldxi} = b$ must have $c'_k = c_k = \xi_k + 2$.

As in the proof of Lemma~\ref{lem:p3proof-dh3}, there is a set $\mathcal F \subset \C$, $\abs{\mathcal F} = 2d$, such that $\bm a^T \Fv(\bm c) = b$ for all $\bm c\in \mathcal F$ and the embeddings are affinely independent. From the above discussion, each  $\boldc \in \mathcal F$ is either of the form $\boldc = \boldxi + \bm e^i$, $\boldc = \boldxi + 2 \bm e^i + 2 \bm e^k$, or $\boldc = \boldxi + \bm e^i + \bm e^j + 2 \bm e^k$ for $i,j \neq k$ and $i \neq j$.

We now assume wlog.\ (in view of Remark~\ref{rem:hilo-arbitrary}) that $k=d$ and use Construction~\ref{constr:hilo} with the canonical codeword $\boldc^{\boldxi} = \boldxi + \bm e^d$ to obtain an inequality $\boldtheta^T \boldx \leq \kappa$ from $\varphi_2(\Theta^{\bm m})$, where $\boldtheta = (\varphi_2(\bm t_{k_1}) \mid \dotsc \mid \varphi_2(\bm t_{k_d}))\eiriknew{^T}$ and $k_i = 2 - \xi_i$, \eirik{$i \neq d$},  as in the proof of Lemma~\ref{lem:p3proof-dh3}. Analogously to above, one can check that this inequality is tight for
\begin{enumerate}
  \item $\boldc^\xi$ itself,
  \item $\boldc^\xi - \bm e^d + \bm e^i = \boldxi+\bm e^i$ for any $i \neq d$,
  \item $\boldc^\xi + \bm e^d - \bm e^i = \boldxi + 2 \bm e^i + 2 \bm e^d$ for any $i \neq d$, and
  \item $\boldc^\xi + \bm e^d + \bm e^i + \bm e^j = \boldxi + \bm e^i + \bm e^j + 2 \bm e^d$ for $i, j  \neq d$ and $i \neq j$,
\end{enumerate}
\ie is tight for \emph{all} potential codewords for which \eqref{eq:p3proof-ineq} is tight, which again shows that $F = \{\boldx \colon \boldtheta^T \boldx \leq \kappa\}$, contradicting the assumption that $F \notin \Theta^{\bm m} \cup \varphi_2(\Theta^{\bm m})$.

As we have gone through all cases, this concludes the proof that $\P = \Q$. Finally, the irredundancy statement follows from Proposition~\ref{prop:equivalent}.

\section{Proof of Proposition~\ref{prop:Theta5_p7} (Outline)} \label{app:proofTheta5}
  The analog of Lemma~\ref{lem:validIndependent} leads to the result that
  \[ \boldtheta^T \Fv(\bm c + \bm\xi) - \kappa = \sum_{i\neq i^{\rm nb}} t^{\rm b}_{\sigma^{\rm b}, \xi_i} + t^{\rm nb}_{0,\xi_{i^{\rm nb}}}\]
  for any $\bm\xi \in \F_7^d$; the adaption of Corollary~\ref{cor:allValidOrNot} shows that the inequalities in $\Theta_5$ are valid for $\P$ if and only if the following holds for all $\bm c \in \C$ (in analogy to \eqref{eq:allValidCondition}):
  \[\sum_{i \neq i^{\rm nb}} t^{\rm b}_{\sigma^{\rm b}, c_i} + t^{\rm nb}_{0,c_{i^{\rm nb}}} \leq 0.\]
  Then, the proper generalization of Definition~\ref{def:valid-class} and Theorem~\ref{thm:newValidProgram} leads to the equivalent condition that $\Theta_5$ is valid if and only if
  \[ \sum_{i \in I} n_i t^{\rm b}_{\sigma^{\rm b},i} + [r]_\Z = 0\]
  with $I= \{i \in \F_7\colon 0 > t^{\rm b}_{\sigma^{\rm b}, i} \geq -[\sigma^{\rm b}]_\Z\}$, \eirik{nonnegative integer variables $n_i$}, and $r = -\sum_{i \in I} [n_i]_7 \cdot i$ has no solution for which $m^{\rm nb}_r = 1$. From $\sigma^{\rm b} = 2$ follows that $I = \emptyset$, such that the system has no solution and hence all inequalities in $\Theta_5$ are valid for $\P$.
  
  Let now $\boldtheta^T\bm x \leq \kappa \in \Theta_5$. In order to show that the inequality defines a facet, we can partially recycle the proof of Lemma~\ref{lem:facets} in Appendix~\ref{app:prooffacets}. For the sake of consistency, assume that $i^{\rm nb} = d$, which is without loss of generality because the role of $d$ is arbitrary (cf.\ Remark~\ref{rem:hilo-arbitrary}).
  
  For $s \in \range{(p-1)(d-1)}$, define $\bm \xi^s$ (and hence $\bm c^s$) as in \eqref{eq:faceproof-xi-type1}. Note that $t_{\sigma^{\rm b}, [l]_7}^{\rm b} + t_{0, [-l]_7}^{\rm nb} = 0$ holds (even though Part~\ref{lem:bb-symmetric-property1} of Lemma~\ref{lem:bb-symmetric_properties}, which is used to show the same result in the original proof, is not applicable here), such that $\boldtheta^T \Fv(\bm c^s) = \kappa$ for $s \in \range{(p-1)(d-1)}$.
  
  The construction of the next $p-2$ codewords cannot be copied from Appendix~\ref{app:prooffacets} because the condition $t^{\rm b}_{\sigma^{\rm b},-i} + t^{\rm b}_{\sigma^{\rm b},i-\sigma^{\rm b}} + t^{\rm nb}_{0,\sigma^{\rm b}} = 0$ does not hold. However, one can check that the following $p-2=5$ additional $\bm\xi$-vectors
  \ifonecolumn
  \begin{displaymath}
  (0,\dotsc,0,3,2,2), (0,\dotsc,0,1,4,2), (0,\dotsc,0,4,2,1), 
  (0,\dotsc,0,4,4,6), (0,\dotsc,0,3,3,1)
  \end{displaymath}
  \else
  \begin{gather*}
  (0,\dotsc,0,3,2,2)\eiriknewnew{^T}, (0,\dotsc,0,1,4,2)\eiriknewnew{^T}, (0,\dotsc,0,4,2,1)\eiriknewnew{^T},\\
  (0,\dotsc,0,4,4,6)\eiriknewnew{^T}, (0,\dotsc,0,3,3,1)\eiriknewnew{^T}
  \end{gather*}
  \fi
  satisfy $\sum_{i=1}^{d-1} t_{\sigma^{\rm b},\xi_i}^{\rm b} + t^{\rm nb}_{0,\xi_d} = 0$, such that the corresponding codewords are tight for the inequality. Furthermore, the corresponding embeddings are linearly independent, such that the counterpart of $\bm M_F$ as defined in \eqref{eq:MF} has full rank $d(p-1) -1$, hence the inequality indeed defines a facet.

\section{Proof of Proposition~\ref{prop:Theta6_p7}} \label{app:proofTheta6}
For the first statement, we again follow the arguments of Section~\ref{sec:validInvalid}, but need to be careful not to rely on features of basic building block classes.
  
  First, observe that Lemma~\ref{lem:bb-hilo-plusi} generalizes to the current case; in particular, \eqref{eq:bb-hi-plusi} holds for $\mathcal T_6^{\rm b}$ and $\mathcal T_6^\hi$, and \eqref{eq:bb-lo-plusi} holds for $\mathcal T_6^\lo$. This allows to generalize Lemma~\ref{lem:validIndependent} (the proof of which relies on Lemma~\ref{lem:bb-hilo-plusi}), resulting in
  \begin{equation}
    \boldtheta^T \Fv(\bm c +\bm\xi)-\kappa = \sum_{i=1}^d t^{l_i}_{0,\xi_i}
    \label{eq:cPlusXiTheta6}
  \end{equation}
  for $\bm\xi \in \F_7^d$. This immediately implies (cf.\ Corollary~\ref{cor:allValidOrNot}) that the inequalities in $\Theta_6$ are valid if and only if all $\bm c \in \C$ satisfy the condition $\sum_{i=1}^d t^{l_i}_{0,c_i}\leq 0$ for all configurations of $i^\lo \neq i^\hi$. Since $\bm t^{\rm b}_0$ and $\bm t^\hi_0$ contain nonpositive entries only, this condition can be violated only if $t^\lo_{0,c_{i^\lo}} = 1$, \ie $c_{i^\lo} \in \{1,2,4\}$ and simultaneously $t^{\rm b}_{0,c_i} = 0$ (\ie $c_i = 0$) for $l_i = {\rm b}$, and also $t^\hi_{0, c_{i^\hi}} =0$, \ie $c_{i^\hi} \in \{0,1,2,4\}$. But then $\sum c_i = c_{i^\lo} + c_{i^\hi} \neq 0$, which contradicts $\bm c \in \C$ and hence concludes the proof of the first claim.
  
  It remains to show that each inequality from $\Theta_6$ defines a facet of $\P$. To that end, let $\bm\theta^T \bm x \leq \kappa$ be such an inequality, where we assume, for the sake of notation and without loss of generality, that $i^\hi = d-1$ and $i^\lo = d$, and assume that $\bm c \in \C$ is the canonical codeword. Analogously to the proof of Lemma~\ref{lem:facets}, we construct $d(p-1)-1 = 6d-1$ codewords  $\bm\xi^s$, $s \in \range{6d-1}$, with the property that $\boldtheta^T\Fv(\bm c + \bm \xi^s) - \kappa =0$ for $s \in \range{6d-1}$ and such that the (Flanagan) embeddings $\Fv'(\bm\xi^s)$ are linearly independent.
  
  For $s \in \range{6(d-2)}$, define vectors $\bm\xi^s = \bm\xi^{6i+l}$ (where $1 \leq l \leq 6$ and $0 \leq i \leq d-3$) by
  \[
    \xi^{6i+l}_j = \begin{cases}
    [l]_7 &\text{if \eirik{$j=i+1$}},\\
    [-l \bmod 3]_7  & \text{if $j=d-1$},\\
    [4]_7&\text{if $j=d$ and $l \in \{1,2,3\}$},\\
    [1]_7&\text{if $j=d$ and $l \in \{4,5,6\}$},\\
    \phantom{-}0 &\text{otherwise} \end{cases}
  \]
  each of which is a codeword and satisfies, by \eqref{eq:cPlusXiTheta6}, that $\boldtheta^T \Fv(\bm c + \bm\xi^s) -\kappa = \sum_{j=1}^d t^{l_j}_{0,\xi^s_j} = 0$ because $\eirik{t^{\rm b}_{0,\xi^s_{i+1}}} = -1$, $\eirik{t^\lo_{0,\xi^s_d}} = 1$, and $\eirik{t^\hi_{0,\xi^s_{d-1}}} = 0$ by construction. The matrix whose rows are the embeddings $\Fv'(\bm \xi^s)$, $s \in \range{6(d-2)}$, then has the form
    \begin{equation} \label{eq:MF-Theta6}
      \begin{pmatrix}
      \bm{I}_6 & & &  \bm A&\bm B\\
       &\ddots&     &\vdots&\vdots\\
        &      &    \bm{I}_6&\bm A&\bm B\\
       \end{pmatrix}
       \end{equation}
   with
   \[\bm A = \left(\begin{smallmatrix}
     0&1&0&0&0&0\\
     1&0&0&0&0&0\\
     0&0&0&0&0&0\\
    0&1&0&0&0&0\\
    1&0&0&0&0&0\\
    0&0&0&0&0&0\\    
   \end{smallmatrix}\right)\text{ and } \bm B = \left(\begin{smallmatrix}
   0&0&0&1&0&0\\
   0&0&0&1&0&0\\
   0&0&0&1&0&0\\   
   1&0&0&0&0&0\\
   1&0&0&0&0&0\\
   1&0&0&0&0&0
   \end{smallmatrix}\right)
 \]
 and hence obviously full row rank $6(d-2)$. The remaining $11$ codewords $\bm\xi^s$, $6(d-2)+1 \leq s \leq 6d-1$, are zero except for the last three entries, which are given by
 \ifonecolumn
 \begin{displaymath} (0,1,6), (0,2,5), (0,3,4), (0,4,3), (0,5,2), (0,6,1), 
 (1,4,2), (2,4,1), (3,2,2), (5,0,2), (6,4,4).
 \end{displaymath}
 \else
 \begin{gather*} (0,1,6), (0,2,5), (0,3,4), (0,4,3), (0,5,2), (0,6,1),\\
 (1,4,2), (2,4,1), (3,2,2), (5,0,2), (6,4,4).
 \end{gather*}
 \fi
 It can be checked by hand that the condition $\bm\theta^T \Fv(\bm c + \bm\xi^s) = \kappa$ holds for these codewords as well, and one can verify numerically (cf. Remark~\ref{rem:numerical-check-facet}) that their Flanagan embeddings, together with the last block-row of \eqref{eq:MF-Theta6}, are linearly independent, such that they  complete \eqref{eq:MF-Theta6} to a matrix of rank $6d-1$, which concludes the proof.

\section{Proof of Proposition~\ref{prop:pm}} \label{app:proofproppm}
The proof is along the same lines as the proof of Lemmas~8 and 12 in \cite{liu14}, first showing that $\sum_{k \in \mathcal{K}} \gamma_k \cdot \mathsf{p}(h_j c_j)_k = \beta$ if and only if 
\[ \left[ g_j^{(\mathcal{K},\boldsymbol{\gamma}, \bm f)} \right]_p =\beta \]
for $\beta \in \F_p$.  %

We first need a technical result.

\begin{lemma} \label{lem:intermediate_result_prop_pm}
For any vector $\bm c \in \F_q^d$ and its embedding $\bm f = \Fv(\bm c)$,
$\left[ g_j^{(\mathcal{K},\boldsymbol{\gamma}, \bm f)} \right]_p= \beta$ if and only if $\sum_{k \in \mathcal{K}} \gamma_k \cdot \mathsf{p}(h_j c_j)_k = \beta$ for all $\emptyset \neq \mathcal{K} \subset \range{m}$, $\boldsymbol{\gamma} \in (\F_p \setminus \{0\})^{|\mathcal{K}|}$, and $j \in \range d$, where $\beta \in \F_p$.
\end{lemma}

\begin{IEEEproof}
Assume that $\beta \in \F_p$ and consider a fixed $j \in \range d$. If $\sum_{k \in \mathcal{K}} \gamma_k \cdot \mathsf{p}(h_j c_j)_k = \beta$, then $c_j \in \mathcal{B}^{(\beta)}(\mathcal{K},\boldsymbol{\gamma},h_j)$ by definition. Since $f_{j,c_j}=1$ and $f_{j,i} = 0$ for all $c_j \neq i \in \F_q$,
\ifonecolumn
\begin{displaymath}
 g_j^{(\mathcal{K},\boldsymbol{\gamma}, \bm f)} = \sum_{\beta \in \F_p \setminus \{0\}} \sum_{i \in \mathcal{B}^{(\beta)}(\mathcal{K},\boldsymbol{\gamma},h_j)} [\beta]_\Z \cdot f_{j,i} 
= [\beta]_\Z \cdot f_{j,c_j} = [\beta]_\Z
\end{displaymath}
\else
\begin{displaymath}
\begin{split}
 g_j^{(\mathcal{K},\boldsymbol{\gamma}, \bm f)} &= \sum_{\beta \in \F_p \setminus \{0\}} \sum_{i \in \mathcal{B}^{(\beta)}(\mathcal{K},\boldsymbol{\gamma},h_j)} [\beta]_\Z \cdot f_{j,i} \\
&= [\beta]_\Z \cdot f_{j,c_j} = [\beta]_\Z
\end{split}
\end{displaymath}
\fi
since the sets $\mathcal{B}^{(\beta)}(\mathcal{K},\boldsymbol{\gamma},h_j)$ are disjoint. Thus, $\left[ g_j^{(\mathcal{K},\boldsymbol{\gamma}, \bm f)} \right]_p = [[\beta]_\Z]_p = \beta$.

For the converse, assume that
\[\left[ g_j^{(\mathcal{K},\boldsymbol{\gamma},\bm f)} \right]_p = \sum_{\eta \in \F_p \setminus \{0\}} \sum_{i \in \mathcal{B}^{(\eta)}(\mathcal{K},\boldsymbol{\gamma},h_j)} \eta  \cdot \left[ f_{j,i} \right]_p=  \beta\]
for $\beta \in \F_p$. This implies that $c_j \in \mathcal{B}^{(\beta)}(\mathcal{K},\boldsymbol{\gamma},h_j)$ and, by definition, $\sum_{k \in \mathcal{K}} \gamma_k \cdot \mathsf{p}(h_j c_j)_k = \beta$. %
\end{IEEEproof}

Now, assume that $\bm f \in \Fv(\C)$. We will show that this implies $\bm f \in \mathcal{E}$. By assumption there exists a codeword $\bm c \in \C$ such that $\bm f = \Fv(\bm c)$. Obviously, the two first conditions of the proposition are satisfied due to the properties of the constant-weight embedding from Definition~\ref{def:Constant} (all symbols are embedded to weight-$1$ vectors of length $q$). Now, since $\bm c$ is codeword, the syndrome $s = \sum_{j=1}^d h_j c_j = [0]_q$. Furthermore, $\mathsf{p}(s)=\eirik{([0]_p,\dotsc,[0]_p)}$ (a vector of length $m$) since we are working in the field $\F_q$ where $q = p^m$. Hence, $\mathsf{p}(s)_k = \sum_{j=1}^d\mathsf{p}(h_j c_j)_k = [0]_p$ and using Lemma~\ref{lem:intermediate_result_prop_pm}, we get 
\ifonecolumn
\begin{displaymath}
 \sum_{j=1}^d \left[g_j^{(\mathcal{K},\boldsymbol{\gamma}, \bm f)} \right]_p = \sum_{j=1}^d \sum_{k \in \mathcal{K}} \gamma_k \cdot \mathsf{p}(h_j c_j)_k 
= \sum_{k \in \mathcal{K}}  \gamma_k \sum_{j=1}^d \mathsf{p}(h_j c_j)_k = [0]_p
\end{displaymath}
\else
\begin{displaymath}
\begin{split}
 \sum_{j=1}^d \left[g_j^{(\mathcal{K},\boldsymbol{\gamma}, \bm f)} \right]_p &= \sum_{j=1}^d \sum_{k \in \mathcal{K}} \gamma_k \cdot \mathsf{p}(h_j c_j)_k \\
&= \sum_{k \in \mathcal{K}}  \gamma_k \sum_{j=1}^d \mathsf{p}(h_j c_j)_k = [0]_p
\end{split}
\end{displaymath}
\fi
which implies that the third condition of the proposition is indeed true.

Conversely, assume that $\bm f \in \mathcal{E}$ fulfills all three conditions of the proposition. From the first two conditions it follows that there exists a unique vector $\bm c \in \F_q^{d}$ such that $\bm f = \Fv(\bm c)$. From the last condition of the proposition and Lemma~\ref{lem:intermediate_result_prop_pm} we know that 
\ifonecolumn
\begin{displaymath}
\sum_{j=1}^d \left[g_j^{(\mathcal{K},\boldsymbol{\gamma},\bm f)} \right]_p = \sum_{j=1}^d \sum_{k \in \mathcal{K}} \gamma_k \cdot \mathsf{p}(h_j c_j)_k 
=  \sum_{k \in \mathcal{K}} \gamma_k \sum_{j=1}^d \mathsf{p}(h_j c_j)_k = [0]_p. 
\end{displaymath}
\else
\begin{displaymath}
\begin{split}
\sum_{j=1}^d \left[g_j^{(\mathcal{K},\boldsymbol{\gamma},\bm f)} \right]_p &= \sum_{j=1}^d \sum_{k \in \mathcal{K}} \gamma_k \cdot \mathsf{p}(h_j c_j)_k \\
&=  \sum_{k \in \mathcal{K}} \gamma_k \sum_{j=1}^d \mathsf{p}(h_j c_j)_k = [0]_p. 
\end{split}
\end{displaymath}
\fi
Fixing $\mathcal{K}=\{k\}$ and $\boldsymbol{\gamma}=(1)$ for any fixed $k \in \range d$, we get
\begin{displaymath}
 \sum_{j=1}^d \mathsf{p}(h_j c_j)_k = \mathsf{p}\left( \sum_{j=1}^d h_j c_j \right)_k = [0]_p
\end{displaymath}
for all $k \in \range d$, which implies that $\sum_{j=1}^d h_j c_j = [0]_q$ and $\bm c$ is indeed a valid codeword. Thus, $\bm f \in \Fv(\C)$.

\balance

\end{document}